# Correlation between Macroscopic and Microscopic Relaxation Dynamics of Water: Evidence for Two Liquid Forms

Nguyen Q. Vinh,[1,2,*] Luan C. Doan,[1,2] Ngoc L. H. Hoang,[1] Jiarong R. Cui,[1] Ben Sindle[1]

Department of Physics and Center for Soft Matter and Biological Physics, Virginia Tech, Blacksburg, VA 24061, USA

Department of Mechanical Engineering, Virginia Tech, Blacksburg, Virginia 24061, USA

*corresponding author: vinh@vt.edu

**Abstract:**

Water is vital for life, and without it biomolecules and cells cannot maintain their structures and functions. The remarkable properties of water originate from its ability to form hydrogen-bonding networks and dynamics, which the connectivity constantly alters because of the orientation rotation of individual water molecules. Experimental investigation of the dynamics of water, however, has proven challenging due to the strong absorption of water at terahertz frequencies. In response, by employing a high-precision terahertz spectrometer, we have measured and characterized the terahertz dielectric response of water from supercooled liquid to near the boiling point to explore the motions. The response reveals dynamic relaxation processes corresponding to the collective orientation, single-molecule rotation, and structural rearrangements resulting from breaking and reforming hydrogen bonds in water. We have observed the direct relationship between the macroscopic and microscopic relaxation dynamics of water, and the results have provided evidence of two liquid forms in water with different transition temperatures and thermal activation energies. The results reported here thus provide an unprecedented opportunity to directly test microscopic computational models of water dynamics.

**INTRODUCTION**

Water is a lubricant for life and affects virtually every aspect of our lives at different levels of complexity from molecules and cells to organisms.[1-6] Specifically, all nucleic acids and proteins are active in water, and interactions between water and biomolecules control their structure, dynamics, flexibility, structural stability, and biological functions.[3] Thus, the dynamics and structure of water establish a central subject in natural sciences.[3-6] In chemical/biological systems, water establishes the environment for chemical/biological activities by mediating and supporting bio-chemical reactions. The interactions between water and biomolecules at the molecular level are also a subject of major interest for understanding biological and chemical processes in aqueous solutions and with the goal of revealing cellular functions. Thus, water is the most studied chemical system with enormous theoretical and experimental studies. However, despite the wide interest and tremendous research efforts on water, we still do not fully understand many roles of its participation at the molecular level.[7, 8]

Water shows many anomalous physical properties, and it has been often speculated that life depends on anomalous properties, including unusually high melting and boiling points, large heat capacity, remarkably high surface tension, self-diffusivity, thermal conductivity, and maximum density at 277 K (only a few examples).[1-3, 9] Although the anomalies occur in the supercooled region, they also appear at ambient conditions where most of important chemical, biological, and physical processes happen. The anomalous properties must be originated in molecular interactions between water molecules and local arrangements. A fundamental question to investigate the properties of water is if the hydrogen-bonding network of water consists of two liquid forms with different densities, including low- and high-density liquid water. The local low-density liquid (LDL) form of water favors through maximizing hydrogen-bond formation in the near tetrahedral configuration, limiting the number of neighbors. The high-density liquid (HDL) configuration occurs in disordered structures, squeezing water molecules tighter by distorting or breaking hydrogen bonds.[5] A liquid-liquid phase transition (LLPT) between the two liquid forms,[4, 5, 10-12] terminating at a liquid-liquid critical point (LLCP) located at high pressures and supercooled regions,[13] has been considered for these anomalies.[4] To test the proposal, we need to perform experiments in a wide range of temperatures. However, the rapid crystallization occurring below the crystal homogeneous nucleation temperature ($T_H \sim 235$ K) has challenged investigations of the liquid or amorphous states beyond this point.[1]

To understand physical and biochemical processes that take place in water, knowledge about the orientation relaxation of water is crucial. The orientation motions of water molecules in liquid forms have been broadly studied, including nuclear magnetic resonance,[14] dielectric relaxation,[15-19] terahertz spectroscopy,[15, 19-21] mid-infrared pump-probe experiments,[22] neutron scattering,[23] Raman-induced, and optical Kerr-effect spectroscopy.[24] These experimental techniques have probed specific orientation motions in a certain range of frequencies and temperatures of the liquid.[15-21] The dielectric spectra from megahertz to terahertz frequencies of water reveal several orientation relaxation processes,[15, 16, 19-21] including the typical signatures of relaxation processes of bulk water, single-water molecular rotation, and the breaking and forming of hydrogen bonds. At high temperatures, the dielectric response of water shifts to terahertz frequencies, resulting in extremely strong absorption of water in this region. Thus, experimental investigation of these dynamics has proven challenging. In response, we have employed a high-precision terahertz frequency-domain spectrometer to explore these motions in a wide range of temperatures from supercooled liquid to near the boiling point of water and a wide frequency range from megahertz to terahertz region. Careful analysis of water dynamics allows us to understand the relationship between the macroscopic and microscopic relaxation processes and obtain evidence of the co-existence of two liquid forms of water. We then employ the results to directly test microscopic computational models of water dynamics at the megahertz to terahertz frequencies.

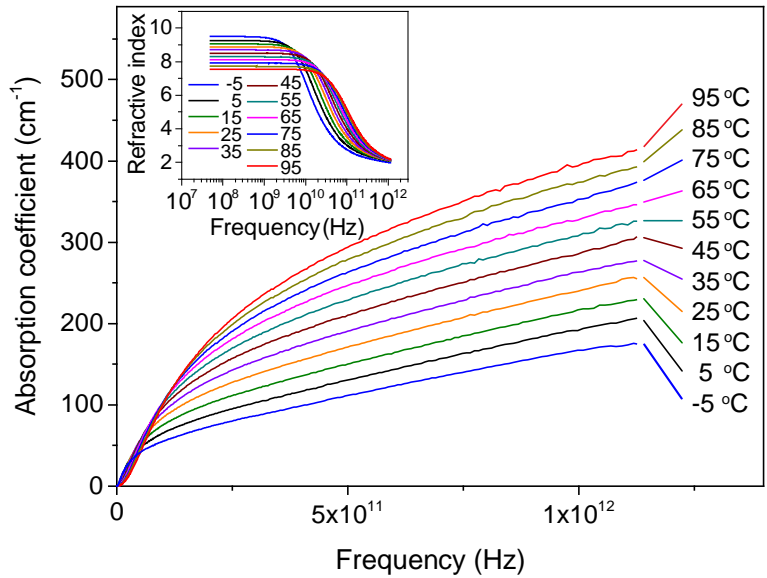

**FIG. 1.** Interaction between water and electromagnetic wave in the megahertz to terahertz frequencies reveals the dynamics of water molecules. Absorption (a) and refractive index (b) spectra of water collected at different temperatures from supercooled liquid (-5 °C) to near the boiling point of water (95 °C) increase and decrease with rising frequency, respectively.

**EXPERIMENTAL METHODS AND RESULTS**

To clarify the relaxation dynamics of water, it is necessary to consider a wide range of frequencies from microwave to terahertz regions. The broadband dielectric spectra of pure, de-ionized water (resistivity of 18 MΩ.cm) have been obtained in a wide range of frequencies from 50 MHz to 1.2 THz (0.002 – 40.03 cm$^{-1}$). By employing a vector network analyzer (Agilent PNA N5225A) together with frequency extenders, our systems allow us to simultaneously measure the absorption and refractive index (i.e., the dielectric dispersion and loss) of pure water. For the megahertz to gigahertz frequency range, an open-end reflection probe (Agilent 85070E) and a transmission test cell combined with the vector network analyzer have been used.[19, 25] The open-end reflection probe was calibrated with three standard measurements, including pure water, air, and mercury for short circuits. The transmission test cell consists of a coaxial line in a circular cylindrical waveguide containing water.[25] The dielectric response, including the real and imaginary components, was obtained directly from these systems. Water temperature can be precisely controlled with an accuracy of ± 0.02 °C from -5 °C to 95 °C using a Lakeshore 336 temperature controller.

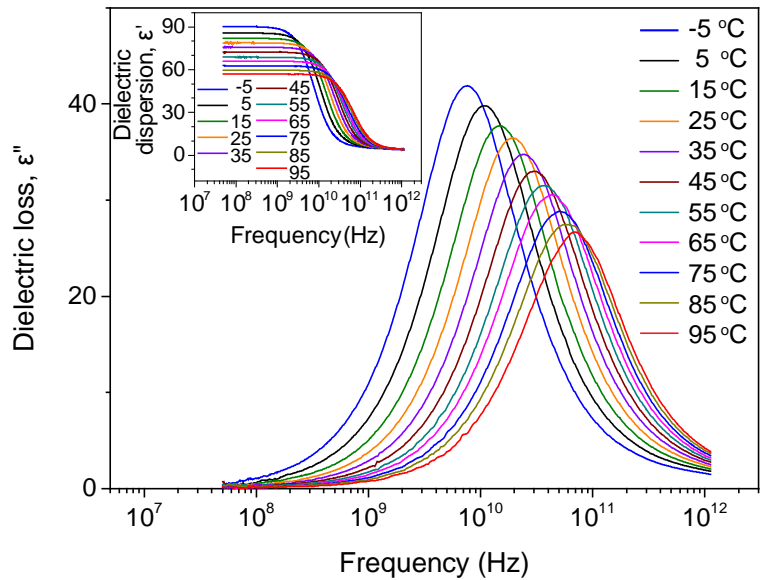

**FIG. 2.** Complex dielectric response, including dielectric loss and dielectric dispersion spectra of water at different temperatures, has been obtained from the absorption and refractive index spectra.

At high frequencies, the dielectric response of water has been studied using a gigahertz to terahertz spectrometer based on frequency extenders from Virginia Diodes together with the above vector network analyzer. In brief, our experimental set up[20, 25, 26] consists of a number of rectangular waveguides, a variable-thickness sample cell, and the above temperature controller. We can accurately measure changes in the transmitted power and phase of a liquid as a function of the path-length, $d$, of a sample. Fitting this data to Beer's law, $I(d,\nu) = I_0(\nu)\exp(-\alpha(\nu).d)$, with $I_0(\nu)$ corresponding to the incident intensity at the frequency, $\nu$, provides precise measurements. At the same time, the refractive index of the sample, $n$, is determined with the fitting of the observed phase shift to a linear function of the sample path-length, $\phi(d,\nu) = \phi_0(\nu) + n(\nu).d.2\pi\nu/c$, with $c$ being the speed of light. Without the need for precise measurements of absolute path-lengths of the sample, standard errors of the mean of replicate measurements are smaller than 0.2% of the absorption coefficient and reflective index of the sample at a specific frequency. Optical properties of water, including absorbance and refractive index, vary strongly with rising frequency over the spectral range, monotonically increasing and decreasing, respectively. The results have been employed to determine the complex index of refraction, $n^*(\nu) = n(\nu) + i\alpha(\nu) \cdot c/(4\pi\nu)$. Fig. 1 shows the absorption coefficient and refractive index of water at selected temperatures from -5 °C to 95 °C. For all investigated temperatures, the results are presented in Fig. S1, supplementary material. We note that the results have been collected by a single research group from our broad-band dielectric spectrometer with improved signal-to-noise ratio for a wide range of temperatures for water, not including any data from the literature. As a resource for researchers investigating the water dynamics at different temperatures, we provide the results from -5 °C to 95 °C with the step of temperature of 5 °C in the supplementary material.

To determine the complete description of interactions of water with electromagnetic wave as a function of temperature, it is proper to represent the complex refractive index of water in the form of the complex dielectric response, $n^*(\nu) = \sqrt{\epsilon^*(\nu)}$, including the real, $\epsilon'_{\text{sol}}(\nu)$ and imaginary components, $\epsilon''_{\text{sol}}(\nu)$, or the dielectric dispersion and loss, respectively:[26]

$$\begin{aligned} \epsilon'_{\text{sol}}(\nu) &= n^2(\nu) - \big(c\alpha(\nu)/(4\pi\nu)\big)^2 \\ \epsilon''_{\text{sol}}(\nu) &= 2n(\nu)c\alpha(\nu)/(4\pi\nu) \end{aligned} \quad (1)$$

Employing this approach, we determine the dielectric dispersion and loss of the complex permittivity of water over a wide range of frequencies from megahertz to terahertz frequencies with unprecedented precision and resolution (Fig. 2). At a certain temperature, the dielectric dispersion reduces with increasing frequency. The curves shift to the terahertz frequency with rising temperature. The dielectric loss has a maximum centered at ~20 GHz at room temperature. The main peak moves to the terahertz frequency when the temperature increases.

**DISCUSSION**

**Terahertz spectroscopy**

The dielectric response of water at megahertz to terahertz frequencies provides information on the dynamics of water. The main peak of the dielectric relaxation spectra obeys the Debye law, corresponding to the collective orientation dynamics or cooperative relaxation of water.[15, 16, 19, 21] At higher frequencies, damped harmonic oscillators at ~60, 180, 400, and 700 cm$^{-1}$ or ~1.8, 5.4, 12, and 21 THz,[17, 27-29] respectively, are outside the scope of this paper. As reported in the literature, one Debye term and the four above oscillators are not enough to describe the dielectric spectra of water. To explain the dielectric response in the sub-terahertz frequency range at room temperature, an additional Debye term with a relaxation time of ~ 1.1 ps has been included.[15-17, 19, 21, 29] However, results from dielectric response spectra collected with the terahertz time-domain spectroscopy (0.1 < ν (THz) < 2) have yielded another fast Debye process of ~ 0.18 ps for water at room temperature.[20, 21] This technique provides insufficient frequency to cover the gigahertz frequency region, neglecting the ~1.1 ps Debye component. To cover the frequency from megahertz to terahertz, the complex dielectric response, $\epsilon^*_{sol}(\nu) = \epsilon'_{sol}(\nu) + i\epsilon''_{sol}(\nu)$, of water can be characterized as a sum of three Debye relaxation components:[15, 19, 30, 31]

$$\epsilon^*_{sol}(\nu) = \epsilon_\infty + \frac{\epsilon_s - \epsilon_1}{1 + i2\pi\nu\tau_D} + \frac{\epsilon_1 - \epsilon_2}{1 + i2\pi\nu\tau_2} + \frac{\epsilon_2 - \epsilon_\infty}{1 + i2\pi\nu\tau_3} \tag{2}$$

where $\Delta\epsilon_1 = \epsilon_s - \epsilon_1$, $\Delta\epsilon_2 = \epsilon_1 - \epsilon_2$, and $\Delta\epsilon_3 = \epsilon_2 - \epsilon_\infty$ are dielectric contributions of individual relaxation processes with corresponding relaxation times, $\tau_D$, $\tau_2$, and $\tau_3$ respectively, to the total dielectric response of water. We employ here $\tau_D$ to denote $\tau_1$ (the first Debye process) for consistency with the expression in the literature. The relaxation process represents the collective orientation motion of water molecules in the liquid state. Two faster dynamic processes have constant times of $\tau_2$ and $\tau_3$. $\epsilon_\infty$ captures contributions to the dielectric response of water from molecular oscillation processes at frequencies much higher than our spectral range, including previously reported modes at ~1.8, 5.4, 12, and 21 THz.[17, 27, 28, 31] $\epsilon_s = \epsilon_\infty + \sum_{i=1}^{3}\Delta\epsilon_i$ is the static dielectric constant, for example, at 25 °C, $\epsilon_s = $ 78.38 for pure water.[15, 16, 18]

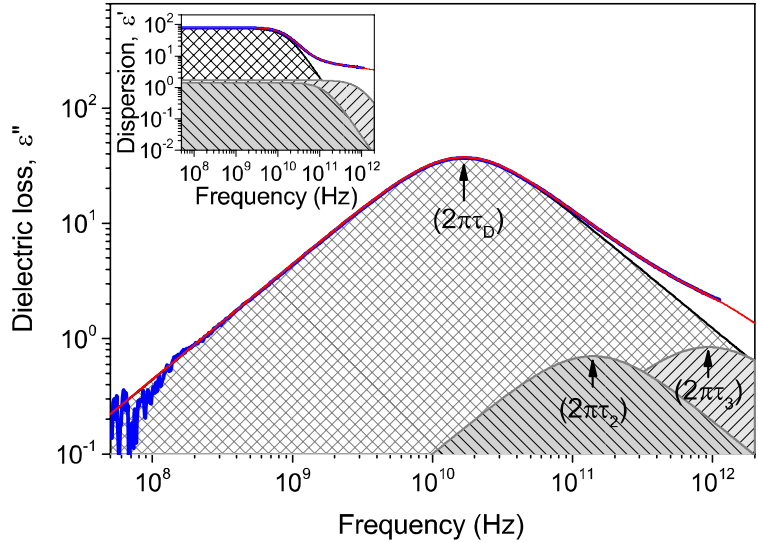

**FIG. 3.** Dielectric response pure water at 20 °C provides insights into the water dynamics over the femtosecond to picosecond timescale. Dielectric dispersion, $\epsilon'_{sol}(\nu)$ in the inset, and the dielectric loss, $\epsilon''_{sol}(\nu)$, spectra are deconvoluted into three dynamic processes of water. The red curves are fits of the real and imaginary components of the complex dielectric response.

Employing the method, the spectra of dielectric response at different temperatures, including the real, $\epsilon'_{sol}(\nu)$, and imaginary parts, $\epsilon''_{sol}(\nu)$, are fitted at the same time to Eq 2. To achieve the best fit, all six parameters are altered concurrently, except the static dielectric constant, $\epsilon_s$, which is changed after a few iterations. As an example, Fig. 3 shows such fit for water at 20 °C, which confirms that Eq 2 can sufficiently describe the dielectric response of water. If we use a similar model to Eq 2 but only one or two Debye terms are used, residuals are obviously distinguishable between the measured response and the fit. As reported in the literature, the two-Debye model is insufficient to describe the relaxation dynamics of water, and a model with three Debye components has been suggested earlier by Vij et al.,[31] Ellison,[15] Beneduci,[30] and our previous work[19]. Therefore, to obtain the strength and relaxation time of each mode, we employ Eq 2 to fit the dielectric response results. At temperatures above 80 °C, the dielectric spectra shift to higher frequencies at the terahertz region. The maximum dielectric response of the fastest water dynamics ($\tau_3$) will be out from our frequency range. Thus, two Debye terms are employed to evaluate the dielectric response of water.

The measured complex dielectric response is fitted to the three-Debye model with the least-square method. Individual data points collected with the frequency in logarithmic scale are weighted equally in the whole spectrum from megahertz to terahertz frequencies. Three constant times, $\tau_D$, $\tau_2$, and $\tau_3$, have been observed, Fig. 3. The slow relaxation time for water, $\tau_D$, at 20 °C is found to be $9.52 \pm 0.05$ ps ($16.72 \pm 0.08$ GHz),[15] representing the collective orientation motion of water molecules in the liquid state. The two faster constant times are $\tau_2 = 1.14 \pm 0.07$ ps ($139.61 \pm 8.61$ GHz) and $\tau_3 = 170 \pm 18$ fs ($936.2 \pm 99.5$ GHz), which are separated for the orientation of single-water molecules and the structural rearrangements as a result of breaking and reforming hydrogen bonds in pure water, respectively.[19, 23, 32-34] The corresponding times of water dynamics charactered as $\tau_2$ and $\tau_3$ are faster when compared with those of the collective orientation motion of water molecules with factors of ~8 and ~40, respectively. Fitting dielectric spectra to Eq 2 also provides the dielectric strength of each relaxation mode, which quantifies the contribution from each group to the overall dielectric response of water. For pure water at 20 °C, we have obtained the contribution of each dynamic process, $\Delta\epsilon_1 = 74.06 \pm 0.25$, $\Delta\epsilon_2 = 1.41 \pm 0.15$, and $\Delta\epsilon_3 = 1.69 \pm 0.15$. The fitted values of the time constant and dielectric strength of water at different temperatures are included in Table 1 and plotted in Figs 4 and 5. Note that the dielectric response of water reported in the literature using the megahertz to gigahertz radiation can only provide the dielectric response up to 65 °C. The temperature dependence of the collective orientation process, $\tau_D$, obtained from our terahertz dielectric response below 65 °C are very similar to previous observations.[16, 35, 36]

Terahertz spectroscopy allows us to characterize the relaxation dynamics of water in a wide range of temperatures. As demonstrated, when the temperature increases, the main peak of the dielectric loss moves to the terahertz frequencies (Fig. 2). Without terahertz radiation, we cannot characterize the dynamics of water at high temperatures. The absorption of water increases significantly at high temperatures, preventing an accurate determination of the relationship between the macroscopic and microscopic dynamics. Thus, relaxation times of single-water molecules have intrinsically large error bars by using a terahertz time-domain system.[20, 21, 37] The high dynamical range of our terahertz frequency-domain setup[25] allows us to explore the dynamics of the fast components at the terahertz frequency. These relaxation processes accelerate and become faster when the temperature increases (Fig. 4), confirming the presence of these fast processes. The dielectric strength of the $\tau_2$ relaxation process reduces at high temperatures, following a similar trend as the collective orientation process. However, the dielectric strength of the $\tau_3$ dynamic process weakly depends on temperature.

**Table 1:** Dielectric parameters of water at different temperatures from supercooled liquid to near the boiling point of water.

| T (°C) | $\tau_D$ (ps) | $\tau_2$ (ps) | $\tau_3$ (ps) | $\epsilon_s$ | $\Delta\epsilon_1$ | $\Delta\epsilon_2$ | $\Delta\epsilon_3$ | $\epsilon_\infty$ |
|---|---|---|---|---|---|---|---|---|
| -5 | 21.05 | 2.35 | 0.22 | 90.16 | 83.81 | 1.61 | 1.70 | 3.10 |
| 5 | 14.62 | 1.80 | 0.19 | 85.76 | 79.09 | 1.53 | 1.83 | 3.31 |
| 10 | 12.59 | 1.53 | 0.18 | 83.68 | 77.10 | 1.55 | 1.83 | 3.22 |
| 15 | 10.79 | 1.26 | 0.18 | 81.89 | 75.51 | 1.40 | 1.68 | 3.37 |
| 20 | 9.52 | 1.14 | 0.17 | 80.29 | 74.06 | 1.41 | 1.69 | 3.27 |
| 25 | 8.35 | 1.07 | 0.14 | 78.81 | 72.86 | 1.29 | 1.78 | 3.10 |
| 30 | 7.41 | 0.93 | 0.16 | 77.12 | 71.44 | 1.08 | 1.65 | 3.28 |
| 35 | 6.64 | 0.81 | 0.16 | 75.65 | 70.19 | 1.03 | 1.58 | 3.29 |
| 40 | 5.95 | 0.74 | 0.14 | 73.97 | 68.73 | 1.03 | 1.52 | 3.24 |
| 45 | 5.35 | 0.69 | 0.16 | 72.25 | 67.21 | 0.93 | 1.43 | 3.36 |
| 50 | 4.85 | 0.62 | 0.15 | 70.58 | 65.82 | 0.87 | 1.39 | 3.31 |
| 55 | 4.40 | 0.58 | 0.13 | 68.70 | 64.22 | 0.82 | 1.43 | 3.19 |
| 60 | 4.03 | 0.54 | 0.13 | 67.18 | 62.95 | 0.72 | 1.43 | 3.18 |
| 65 | 3.80 | 0.53 | 0.13 | 65.95 | 62.06 | 0.63 | 1.51 | 3.00 |
| 70 | 3.51 | 0.47 | 0.13 | 64.22 | 60.53 | 0.68 | 1.46 | 2.95 |
| 75 | 3.20 | 0.44 | 0.15 | 62.56 | 59.16 | 0.40 | 1.42 | 3.11 |
| 80 | 2.98 | 0.38 | | 61.40 | 57.66 | | | 3.57 |
| 85 | 2.78 | 0.36 | | 59.58 | 56.25 | | | 3.55 |
| 90 | 2.56 | 0.34 | | 58.13 | 55.16 | | | 3.60 |
| 95 | 2.37 | 0.29 | | 56.98 | 54.45 | | | 3.65 |

**Relationship between macroscopic and microscopic orientation relaxation dynamics of water**

The relationship between the collective (macroscopic) and single-molecule (microscopic) orientation relaxation times of water has been a subject of much discussion.[38-48] In a strongly polar liquid like water, intermolecular interactions play an important role in identifying the dynamics of collective relaxation processes rather than intramolecular dynamics. However, single-molecular dynamics are essential in determining the microscopic structure of the dense liquid. Many reports on the microscopic theory have been proposed to investigate the relationship between the macro- and microscopic relaxation processes, intermolecular interactions as well as the microscopic structure of the polar liquid. The relationship between modes in a strong polar liquid was first proposed by Debye, using his continuum theory of dielectric relaxation, $\tau_D/\tau_2 = (\epsilon_S + 2)/(\epsilon_\infty + 2)$.[38] The ratio was estimated about 15 for water at room temperature using macroscopic dielectric constants from the literature.[16, 19, 36] However, the Debye method is inappropriate for strongly polar solvent like water. Glarum[48] and Powles[39] provided an alternative relation based on Onsager's model of static dielectric constant and obtained the ratio of 1.5 for water at room temperature. Later, by employing an analysis of Kirkwood's factor for the radial distribution function of dipolar liquids, Madden and Kivelson[42, 44] estimated the ratio of 0.006. The theoretical results do not agree with each other, and high-precision experimental results are essential to test the computational models.

The megahertz to terahertz spectroscopy provides complex dielectric response of water to an oscillating electric field, where intermolecular processes play an essential role. As a result of a finite relaxation process, the electrical field within the dielectric material retards behind the external electric field with underlying exponential dynamics. The strong interactions between water molecules lead to a slow motion of collective relaxation process in a cluster of water molecules, characterized by the megahertz to gigahertz spectroscopy with the time constant of $\tau_D$. The fast relaxation time, $\tau_2$, corresponding to the single-water molecule reorientation can be detected under terahertz radiation. Accordingly, the terahertz time-domain spectroscopy of water yields two components,[20, 21] including a slow component close to, $\tau_D \sim 8$ ps, and a faster time constant ~170 fs at room temperature. The fast relaxation time with large errors in these measurements differs from the single-molecule relaxation time of water closed to ~1 ps reported by the gigahertz as well as terahertz frequency-domain spectroscopy.[15, 16, 19] By employing the terahertz frequency-domain spectroscopy with high precision, we are able to explore the relaxation time processes in water.

We have characterized the relaxation times of the macroscopic and microscopic processes as a function of temperature with high precision (Table 1). Logarithmic ratios (blue open circles) between the collective, $\tau_D$, and single-molecule, $\tau_2$, orientation times, $\tau_D/\tau_2$, are plotted in a wide range of temperatures from supercooled water to 95 °C (Fig. 4, inset). The ratio shows an increase with decreasing temperature (or as a function of $1000/T$). The experimental results for the macroscopic and microscopic relaxation processes have revealed correlation activities, indicating a strong temperature effect. The fast process, $\tau_2$, is coupled to the collective motion through large water clusters, and the temperature dependence is similar to that of $\tau_D$. As temperature increases, the sizes of water clusters reduce, resulting in the decrease of the relaxation times. Thus, the relaxation time ratio, $\tau_D/\tau_2$, between these modes strongly depends on temperature. The red curve is a fitting of the experimental ratios in the logarithmic form to a linear function with a slope of 0.095 and intercept of 0.592.

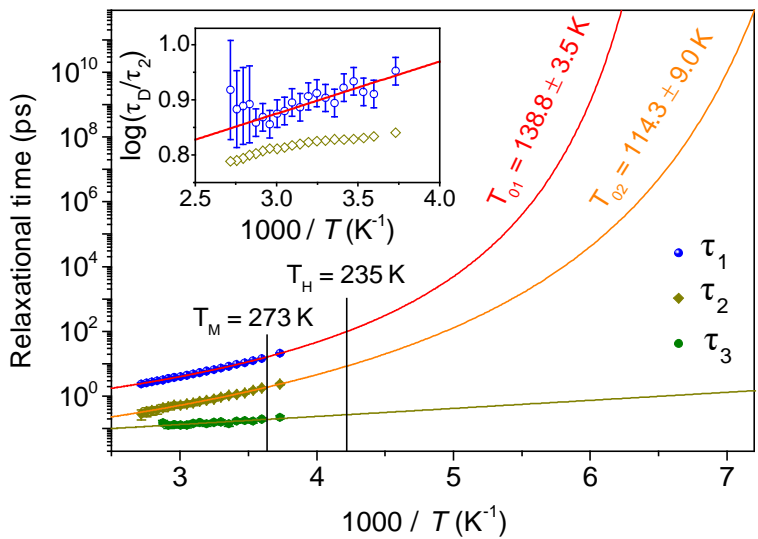

**FIG. 4.** The temperature dependence behaviors of orientation relaxation dynamics of water reveal the relationship between the macroscopic and microscopic dynamics of water and provide evidence of two liquid forms existing in water at ambient conditions. The temperature dependence of collective motions of water follows the VFT function and starts to diverge at ~138 K. The relaxation time of single-water molecules also obeys the VFT temperature with the transition temperature of ~114 K. Vertical lines indicate the melting temperature, $T_M$, and the crystal homogeneous nucleation temperature, $T_H$. (inset) The logarithmic ratios between the collective and single-molecule orientation relaxation times, $\log(\tau_D/\tau_2)$, are plotted as a function of temperature. Dark open yellow diamonds are the $\log(\tau_D/\tau_2)$ ratios estimated from $\epsilon_S$, $\epsilon_\infty$ and μ using Eq. 3. The ratios provide direct evidence of the relationship between the macroscopic and microscopic relaxation processes in water.

Two relaxation processes are sensitive to different microscopic modes, but they connect each other through large water clusters. As mentioned above, a number of theoretical approaches have been proposed to explore the relationship between the macro- and microscopic relaxation processes, but they do not agree with each other. Recently, Chandra et al.,[45] and Arkhipov et al.,[46] suggested a behavior between these relaxation modes by employing microscopic theory. Using the Mori-Zwanzig projection operator formalism, Arkhipov et al.,[46] derived a relationship between the macro- and microscopic processes and obtained an estimation of the number of water molecules involved in the orientation processes.

$$\frac{\tau_D}{\tau_2} = \frac{3k_B T m_0}{4\pi\mu^2 \rho_c} \frac{(\epsilon_S - \epsilon_\infty)(2\epsilon_S + \epsilon_\infty)}{\epsilon_S} \tag{3}$$

where $k_B$ is the Boltzmann constant, $T$ is the absolute temperature in K, $m_0$ is the molecular weight, $\rho_c = Nm_0/V$ is the density of a cluster with volume of $V$ and cluster size of $N$, and $\mu$ is the molecular dipole moment. From the theoretical approach, they estimated a ratio between the collective, $\tau_D$, and the single-molecule, $\tau_2$, orientation times of $\tau_D/\tau_2 \approx 8$ at room temperature for a number of water molecules of ~ 9 in a cluster. The number is derived from the "*tetrahedral displacement mechanism*"[6, 49] and an assumption of a coordination number of 4 around one water molecule, in which the translation involves 4 old and 4 new neighboring water molecules. The collective orientation relaxation time involves the breaking of the hydrogen-bonding network with old water molecules, rotating and reforming hydrogen bonds with new water molecules. At 20 ºC, we have observed the dynamics of the collective orientation time of water of 9.52 ps and the orientation relaxation time of single-water molecules of 1.14 ps. The relaxation time ratio, $\tau_D/\tau_2$, between these modes is 8.35, indicating good agreement between the experimental observation and theoretical estimation.

We use our fitting parameters to test the model and recalculate the $\tau_D/\tau_2$ ratio. Following the above approach, we estimate the $\tau_D/\tau_2$ ratios from $\epsilon_S$, $\epsilon_\infty$ and $\mu = 2.6$ D[50] in cgs unit (Fig. 4, dark yellow open diamonds). The ratios depend only on the macroscopic dielectric constants and do not contain microscopic parameters. The behavior of the relaxation ratios in the logarithmic form has the same trend as the direct ratios estimated from relaxation times. The results indicate that the fast relaxation mode of single-water molecules, $\tau_2$, participates in the slow collective orientation motion, $\tau_D$, through large water clusters, in which the slow motion happens to be localized. The single-water molecules have weak hydrogen bonds to neighboring water molecules, rotating with a small activation energy. In contrast, water molecules with strong hydrogen bonds in the near tetrahedral configuration will have to wait for the hydrogen-bond strength to be weaken before being able to rotate, giving a slow relaxation time. Both types of water molecules correlate with each other but possess their own microscopic structure. In the hydrogen-bonding network and at a certain time, one water molecule can have up to 4 hydrogen bonds. Statically, we can consider two types of water molecules happening in the liquid state. The experimental results show a continuous trend from the liquid state to the supercooled water, and no abruption has been observed. We can expect a continuous rearrangement between two types of water dynamics.

**Dynamic properties of supercooled water**

It has been postulated that bulk water is composed of a mixture of two distinct liquids (LDL and HDL) with different local hydrogen-bonding networks, densities, and microscopic structures.[4, 11, 22] Computational simulations have demonstrated that the liquid form of the LDL occurs in a high degree of the local tetrahedral configuration, which is not very different from crystalline ice, though obviously lacking long-range order.[6] However, the form of the HDL has been experimentally identified as a disordered local configuration with a higher coordination number, in which, on average, each water also has four hydrogen bonds, but a fifth water molecule has entered the first coordination shell.[51] Based on results of molecular dynamics simulations,[4, 5] these liquid forms are metastable with respect to crystallization in free-energy basins, and the existence of a liquid-liquid phase transition between two metastable liquids terminates at the liquid-liquid critical point in deeply supercooled regions. Thus, there is the coexistence line (Widom line) along which the liquid phases of LDL and HDL coexist,[7] providing the fundamental explanation for the anomalous nature of water. The forms of water (LDL and HDL) are related to their counterparts in glassy states of amorphous forms, including the low-density amorphous (LDA) and high-density amorphous (HDA) ices, respectively, of the phase diagram.[11] A large number of methods were employed to produce amorphous ices with densities varying from 0.3 to 1.3 g/cm$^3$,[52] and transformations between amorphous ices have been experimentally reported. The glass-liquid transition, $T_{g1}$, in LDA has been experimentally characterized around ~ 136 K.[10, 53-57] In addition, an identification for the glass–liquid transition in HDA at ambient pressure and elevated pressures has been provided at $T_{g2}$ ~ 116 K.[53, 58]

Monitoring the relaxation times of water, including the collective orientation motions, rotations of single-water molecules, and the structural rearrangements as a result of breaking and reforming hydrogen bonds in the liquid state, reveals the dynamic properties of water.[59-61] The frequency of the dielectric loss provides an estimation of the average relaxation time, enabling a measure for the mobility of water molecules as a function of temperature. We analyze the relaxation times by fitting the data to the empirical Vogel-Fulcher-Tammann (VFT)[62] equation,

$$\nu(T) = \frac{1}{2\pi\tau(T)} = \frac{1}{2\pi\tau_\infty} \exp\left(\frac{-DT_0}{T-T_0}\right), \tag{4}$$

where $\tau_\infty$ is the relaxation time at high-temperature limit, $T_0$ denotes as the fitted temperature parameter or the VFT temperature at which the orientation relaxation time, $\tau$, of water starts to diverge, $D$ describes the deviation from an Arrhenius behavior of the temperature dependence and corresponds to the fragile temperature characteristics,[62-64]

$$\nu(T) = \frac{1}{2\pi\tau(T)} = \frac{1}{2\pi\tau_\infty} \exp\left(\frac{-E_A}{k_B T}\right), \tag{5}$$

where $E_A$ describes the thermal activation energy of the orientation process.

The temperature dependence of the collective orientation dynamics in water obeys the VFT-type or non-Arrhenius behavior. We obtain the best fit with $\tau_\infty = 0.155 \times 10^{-12}$ s, $D = 4.57$, and $T_0 = 138.8 \pm 3.5$ K (Fig. 4, red curve) for the collective orientation relaxation time as a function of temperature. Similarly, the temperature dependence of the orientation relaxation time of single-water molecules is well characterized by the non-Arrhenius relation. The best-fit values of $\tau_\infty = 0.015 \times 10^{-12}$ s, $D = 0.049$, and $T_0 = 114.3 \pm 9.0$ K have been obtained for the temperature dependence of the relaxation time (Fig. 4, orange curve). The observation suggests that the dynamics of the water molecules having a weak interaction with neighboring molecules follows a similar behavior of water molecules in the collective relaxation motion. The $T_0$ temperature for single-water molecules is lower than that of water molecules in the collective arrangement in water. When the temperature of supercooled water decreases, water molecules move more and more slowly, and the viscosity increases. At a certain temperature, the water molecules will move so slowly that they do not have a chance to rearrange their positions, at which the viscosity or the relaxation time diverges. Lastly, the temperature dependence of the fastest dynamics of water follows the Arrhenius behavior with $\tau_\infty = 0.025 \times 10^{-12}$ s and $E_A = 0.049$ meV (Fig. 4, dark yellow curve).

There is not necessarily a one-to-one correspondence between the two relaxation times of two kinds of water and two liquid forms in bulk water because the possibility cannot be excluded that each water in the HDL also has four hydrogen-bonds, but a fifth water molecule has entered the first coordination shell.[65] It is clear that water molecules with strong hydrogen-bonds can only relax through the collective orientation process, including breaking hydrogen bonds, rotating molecules, and reforming hydrogen bonds. The relaxation process happens in both liquid forms. However, in the HDL, the dynamics of the fifth water molecule accommodated in the first coordination shell with some broken or weakened hydrogen bonds[65] will contribute considerably to the single-water molecular process. Thus, regarding the orientation dynamics, two distinct forms of water molecules exist in bulk water, and the dielectric response of the two liquid forms of water differs in the extra contribution of the single-molecule relaxation process. The relaxation times as a function of temperature show a divergence at the supercooled regime for the collective reorientation and single-water molecular rotation processes happened at ~ 138.8 K, and ~ 114.3 K, respectively. The $T_0$ values for these water molecules do not correspond to glass transitions.[63, 64] Specifically, the $T_0$ values should be lower than their $T_g$ values as reported ~ 50 K in the polymer community.[66] However, we observe here a little difference between them, this result may be resulted from dynamics of water molecules in local regions which contain a mixing of two kinds of waters.[7, 8] The $D$ value of the collective relaxation process of water indicates the fragile property of liquid water, additionally, the value for the single-water molecule process suggests a strong fragile property of the water molecules in HDL clusters.[63, 64]

The fastest of the three dynamic processes we observe, which indicates ~170 fs time constant at room temperature, is also reported via terahertz time-domain spectroscopy.[20, 21] Calculations using quantum mechanics or molecular dynamics suggest that this phase arises due to the breaking and reforming of individual hydrogen bonds.[33, 34, 67] This dynamic process may correlate with water molecules that slightly liberate from their most stable geometry in the hydrogen-bonding network, breaking a single hydrogen bond and returning to the same position. This observation has been proposed in the jumping model.[33] Water molecules do not have enough energy to change their position, or the rotation process of water molecules does not have enough amplitude for the jumps. The process follows the Arrhenius behavior, reflected from the temperature dependence of the fastest dynamics of water.

**Temperature effects on the collective relaxation dynamics**

The temperature dependence of the collective orientation relaxation time, $\tau_D$, is expected to connect to the viscosity of water, $\eta$, via the Einstein-Stock-Debye relation. In this model, a water molecule with an electrical dipole is considered as a sphere, in which the rotation of the water molecule in response to an oscillating electric field is opposed by the hydrodynamic friction of surrounding water molecules. The relaxation time of a spherical molecule is given,[49]

$$\tau_D = \frac{4\pi R^3 \eta(T)}{k_B T}, \tag{6}$$

where $R$ is the hydrodynamic radius of a rotating molecule. We have characterized the collective orientation dynamics in a wide range of temperatures from supercooled liquid to near the boiling point of water with high precision. The results provide us a great estimation of the viscosity of water.

To investigate the relationship between the viscosity and collective orientation relaxation time, $\tau_D$, of water, we have plotted our results versus the viscosity divided by temperature, in which the viscosity data were taken from the literature (Fig. S3).[36] Indeed, our data for the collective relaxation time scale linearly with the viscosity of water divided by temperature in a wide range of temperatures from supercooled liquid to near the boiling point. We have obtained the hydrodynamic radius of a rotating molecule, $R$, of 1.42 Å, corresponding to a hydrodynamic volume for a rotating molecule of $1.19 \times 10^{-23}$ cm$^3$. The hydrodynamic volume is smaller than the volume of each water molecule of ~ $2.99 \times 10^{-23}$ cm$^3$ at room temperature. The rotation of water molecules is caused by a tetrahedral displacement,[49] including a rotation and a translation from one site to a neighboring site. For the process, water molecules have less than four hydrogen bonds; thus, effectively, the hydrodynamic volume of water molecules is smaller. Additionally, the viscosity of water is an average of two kinds of water molecules (water molecules with four hydrogen bonds and single-water molecules). We expect here a smaller value of the hydrodynamic volume of water molecules. The view agrees with the postulation that the structure of water has a random network of hydrogen bonds with frequently strained and broken bonds, continuously subject to spontaneous restructuring.[68]

To quantify the thermal activation of the orientation relaxation processes, we employ the Eyring theory,[69] assuming that the relaxation pathway occurs through a thermally activated transition state,

$$\tau_D = \frac{h}{k_B T} \exp\left(\frac{\Delta H^{\neq} - T \Delta S^{\neq}}{RT}\right), \tag{7}$$

where $h$, $R$, $\Delta S^{\neq}$, $\Delta H^{\neq}$ are the Planck, gas constants, entropy, and enthalpy of the activation energy, respectively. Our plot of experimental results, $\ln(\tau_D k_B T/h)$ versus $1/T$, deviates from linear behavior (Fig. S4), indicating the temperature dependence of entropy and enthalpy by the isobaric heat capacity, $\Delta c_p^{\neq}$,[16, 36]

$$\Delta H^{\neq} - T\Delta S^{\neq} = \Delta H_{298}^{\neq} + \Delta c_p^{\neq}(T - T^*) - T\left(\Delta S_{298}^{\neq} + \Delta c_p^{\neq} \ln \frac{T}{T^*}\right) \tag{8}$$

where $\Delta S_{298}^{\neq}$ and $\Delta H_{298}^{\neq}$ are activation parameters at $T^* = 298.15$ K. We have obtained fitted parameters for the enthalpy, entropy, and isobaric heat capacity of thermal activation for the collective orientation and single-water molecule relaxation processes. The extracted values are provided in Table 2. The results for the collective relaxation process are in line with previous literature results for water.[16, 36]

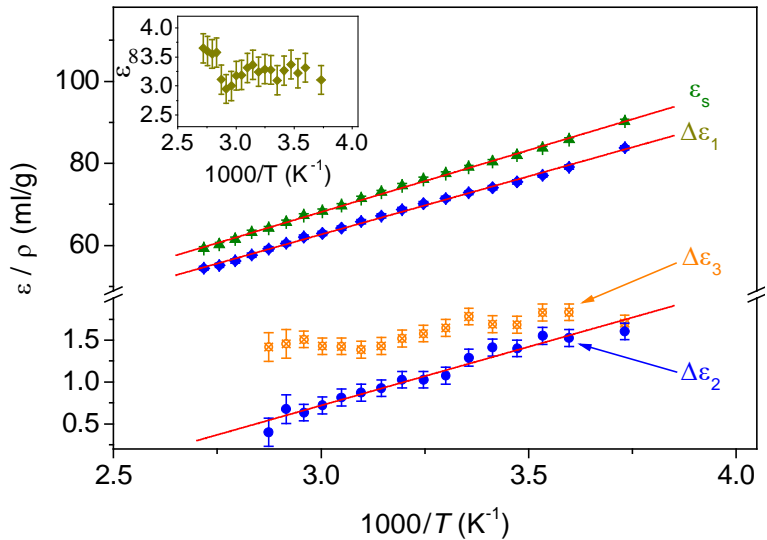

**FIG. 5.** The large dielectric constant of water, $\epsilon_s(T)$, originates from the large dipole moment in each molecule and the strong correlation between dipole moments of its molecules. The values are strongly depending on temperature. When the temperature increases, water increases activities; thus, the dielectric strength as well as the extent of the hydrogen-bonding decrease.

**Temperature dependence of the static dielectric constant**

The large value of the static dielectric constant, $\epsilon_s(T)$, of pure water originates not only from the polarity of individual water molecules and the number of dipoles per unit volume but also from the correlated mutual orientations of these molecules. The static dielectric constants, $\epsilon_s(T)$, of liquid and supercooled water have been reported by several research groups, and the values are consistent with each other.[15, 18] To seek a realistic physical function that describes the interaction of electromagnetic field with water molecules, several functions have been provided. Polynomial functions have been proposed for the temperature dependence of the static dielectric constant over the temperature range 0 – 100 ºC.[15, 18, 70] By employing the most precise measurements, Hamelin et al., yielded a function, $\epsilon_s(t) = 87.9144 - 0.404399t + 9.58726 \times 10^{-4}t^2 - 1.32892 \times 10^{-6}t^3$, with 0 ºC $< t <$ 145 ºC.[70] The estimated standard deviation of the static dielectric constant is small but may not work for temperatures outside the range. In particular, the large dielectric constant of water arises from both the large dipole moment in each molecule and the strong angular correlation between dipole moments of its molecules, strongly depending on temperature. For such a highly polar substance, Kirkwood's equation sheds light on the inverse relationship between the static dielectric constant and temperature that originated from the opposition of the thermal excitation to the alignment of electrical dipole moments in the direction of the applied field. Thus, the random network model can describe the static dielectric property of water, in which hydrogen bonding in water is continuous, although distorted, throughout configurations occupied in the liquid state. The static dielectric constant from 100 ºC down to the supercooled range is a function of the inverse absolute temperature ($1/T$), Kirkwood correlation factor ($g_K$), density of water ($\rho$), and molecular dipole moment ($\mu$).[71] The temperature dependence of the static dielectric constant is described by an empirical equation,[15, 71]

$$\epsilon_s(T) = A + B \frac{\rho g_K \mu^2}{T} \tag{9}$$

Estimating the static dielectric constant of water depends on several parameters, which are also temperature dependent. The parameters are well predicted by theoretical calculations or experimental phenomena fitting functions. The density of water at the standard atmospheric pressure is well known from precise measurements in the range from 239 to 423 K.[72, 73] The density of water can be obtained in the form of rational functions $\rho = \sum_{n=0}^{5} a_n t^n /(1 + bt)$, with $a_0$ = 0.99983952, $a_1$ = 16.945176, $a_2$ = -7.9870401×10$^{-3}$, $a_3$ = -46.170461×10$^{-6}$, $a_4$ = 105.56302×10$^{-9}$, $a_5$ = -280.54253×10$^{-12}$, and $b$ = 16.879850×10$^{-3}$ for -30 ºC $< t <$ 150 ºC.[72] The theoretical Kirkwood correlation factor, $g_K$, of water is basically independent of temperature.[74] Déjardin et al., have found that $g_K$ = 2.73 and 2.72 at $T$ = 0 and 83 ºC, respectively. The dipole moment of water molecules has been estimated using molecular dynamics simulations.[50, 75] We have plotted the static dielectric constant divided by the density, $\epsilon_s(T)/\rho$, as a function of inverse absolute temperature, $1000/T$, and we find a linear dependence (Fig. 5). The red line is fitting of experimental data to Eq 9, and we have obtained $A/\rho$ = -22.34 and $Bg_K\mu^2$ = 30.163. The results in a large range of temperatures are in support a continuity of state of hydrogen bonding between liquid and supercooled water as described in the random network model.[71]

**Table 2:** Fitted values of the enthalpy, $\Delta H^{\neq}_{298}$, entropy, $\Delta S^{\neq}_{298}$, and heat capacity, $\Delta c_p^{\neq}$, of thermal activation for the collective orientation and single-water molecule relaxation processes of water.

| Modes | $T_0$ (K) | $\Delta H^{\neq}_{298}$ (kJ.mol$^{-1}$) | $\Delta S^{\neq}_{298}$ (J.mol$^{-1}$.K) | $\Delta c_p^{\neq}$ (J.mol$^{-1}$.K$^{-1}$) | Ref. |
|---|---|---|---|---|---|
| Collective relaxation | 138.8 ± 3.5 | 16.353 ± 0.085 | 22.1 ± 0.3 | -108 ± 7 | This work |
| | | 15.9 ± 0.2 | 20.4 ± 0.7 | -160 ± 22 | 16 |
| | | 16.1 ± 0.2 | 21.1 ± 0.7 | -94 ± 12 | 36 |
| Single-molecule | 114.3 ± 9.0 | 14.690 ± 0.229 | 33.8 ± 0.8 | -74 ± 17 | This work |

When the temperature increases, water increases the activity; thus, the dielectric strength, as well as the extent of the hydrogen-bonding decrease. Dielectric dispersion and loss spectra of water between -5 and 95 °C from megahertz to terahertz frequencies exhibit the effect of increasing temperature (Fig. 2). The dielectric dispersion of water becomes lower at higher temperature, lessening the difficulty of the movement of the water dipole moment, and so, allowing water molecules to oscillate at higher frequencies. As the result, the force between hydrogen bonds reduces, lowering the friction and hence the dielectric loss. Note that the dielectric constant at high frequencies, $\epsilon_\infty$, changes slightly with temperature (Fig. 5, inset). We have employed the empirical equation (Eq 9) to fit the dielectric strength of each relaxation process and obtained fitting parameters, $A_1/\rho$ = -22.28 and $B_1g_K\mu^2$ = 28.323 for the collective orientation process, together with $A_2/\rho$ = -3.48, $B_2g_K\mu^2$ = 1.401 for the single-molecule rotation process. The dielectric strength for the structural rearrangement due to breaking and reforming hydrogen bonds in the pure water changes slightly with temperature and does not follow the $1/T$ behavior.

**CONCLUSIONS**

In conclusion, we have performed terahertz spectroscopy for water in a wide range of temperatures from supercooled state to near the boiling point. We have observed three dynamic processes, including the collective orientation dynamics, single-water molecule mode, and structural rearrangement resulting from breaking and reforming hydrogen bonds. The observation shows a relationship between the macroscopic and microscopic relaxation dynamics of water, providing evidence of two liquid forms at ambient conditions. The temperature dependence of the dynamics of the two orientation processes follows the VFT behavior, related to the two liquid forms in water of LDL and HDL. The temperature dependence of the fastest process obeys the Arrhenius-type. From the temperature dependence of the relaxation times, we have estimated experimental values for the enthalpy, entropy, and heat capacity of thermal activation for the collective orientation and single-water molecule relaxation processes of water, as well as provided an alternative approach to evaluate the temperature dependence of the static dielectric constant. The results provide direct evidence of water dynamics related to the two liquid forms in different local environments.

## SUPPLEMENTARY MATERIAL

See the supplementary material for details of terahertz spectroscopy of water at different temperatures.

## ACKNOWLEGMENTS

Authors gratefully acknowledge financial support by the Air Force Office of Scientific Research, United States under award number FA9550-18-1-0263 and the National Science Foundation, United States (CHE-1665157).

## AUTHOR DECLARATIONS

**Conflict of Interest.** The authors have no conflicts to disclose.

**Data availability.** The data that support the findings of this study are available within the article and its supplementary material.

# Correlation between Macroscopic and Microscopic Relaxation Dynamics of Water: Evidence for Two Liquid Forms

## Supplementary Material

Nguyen Q. Vinh,[1,2,*] Luan C. Doan,[1,2] N. L. H. Hoang,[1] Jiarong R. Cui,[1] B. Sindle[1]

[1]Department of Physics and Center for Soft Matter and Biological Physics, Virginia Tech, Blacksburg, VA 24061, USA

[2]Department of Mechanical Engineering, Virginia Tech, Blacksburg, Virginia 24061, USA

*corresponding author: vinh@vt.edu

**Absorption and refractive index measurements**

We have employed a highly sensitive spectrometer capable of measuring both the absorption and refractive index of strongly absorbing liquids, such as water, over the range 50 MHz to 1.2 THz (0.002 to 40.030 cm$^{-1}$). Our spectrometer employs a commercial Agilent Vector Network Analyzer, the PNA N5225A, which covers the frequency range from 10 MHz to 50 GHz. The microwave output of the PNA N5225A is translated to the higher frequencies via photomixing in a set of frequency multipliers and then is detected using a matched set of harmonic detectors after photomixing back to lower frequencies. The system comprised of the frequency multipliers and the matched harmonic detectors was developed by Virginia Diodes, Inc. (Charlottesville, VA).[1, 2]

We have used a variable path-length cell setup consisting of two parallel windows, one immobile and the other mounted on an ultra-precise linear translation stage (relative accuracy 80 nm). The cell walls were fixed in aluminum holders to minimize the leakage of stray radiation. These holders were mounted on Peltier temperature control plates, allowing precise control of the temperature of the sample. The absorbance and refractive index of water are temperature dependent and thus all experiments carried out at 25.00 ± 0.02 °C using this device. The liquid sample was held between two parallel windows in the sample cell. The distance between the two windows, as thus the sample pathlength, was adjusted using the linear stage. At each frequency we examined an average of 200 different path-lengths, with increments ranging from 0.5 to 5 µm depending on the absorption strength of the sample. To mitigate problems associated with multiple reflections of the incident light (standing waves, etalon effect), the thickness of our shortest path length was selected to be long enough to insure strong attenuation of the incident radiation. We determined the absorption coefficients, $\alpha$, and refractive index, $n$, of our samples from linear fits of the change in absorbance, $\Delta A$, and the unwrapped phase shift, $\Delta\theta_{transmission}$, respectively, with changing path length, $\Delta l$, as a function of frequency, $v$:[3-14]

$$\begin{cases} \Delta A = \Delta(-\ln[I_{transmission}]) = \alpha \cdot \Delta l \\ \Delta(\theta_{transmission}) = \dfrac{n \cdot 2\pi \cdot v}{c} \cdot \Delta l \end{cases} \quad (1)$$

where $I_{transmission}$, $\theta_{transmission}$ are the transmitted intensity and phase, $c$ is the velocity of light. This method supports the precise determination of absorption coefficients and refractive indexes without the need for precise (and difficult to obtain) measurements of the absolute pathlength and the intrinsic optical properties of the sample cell. All experiments were repeated approximately ten times to estimate confident limits.

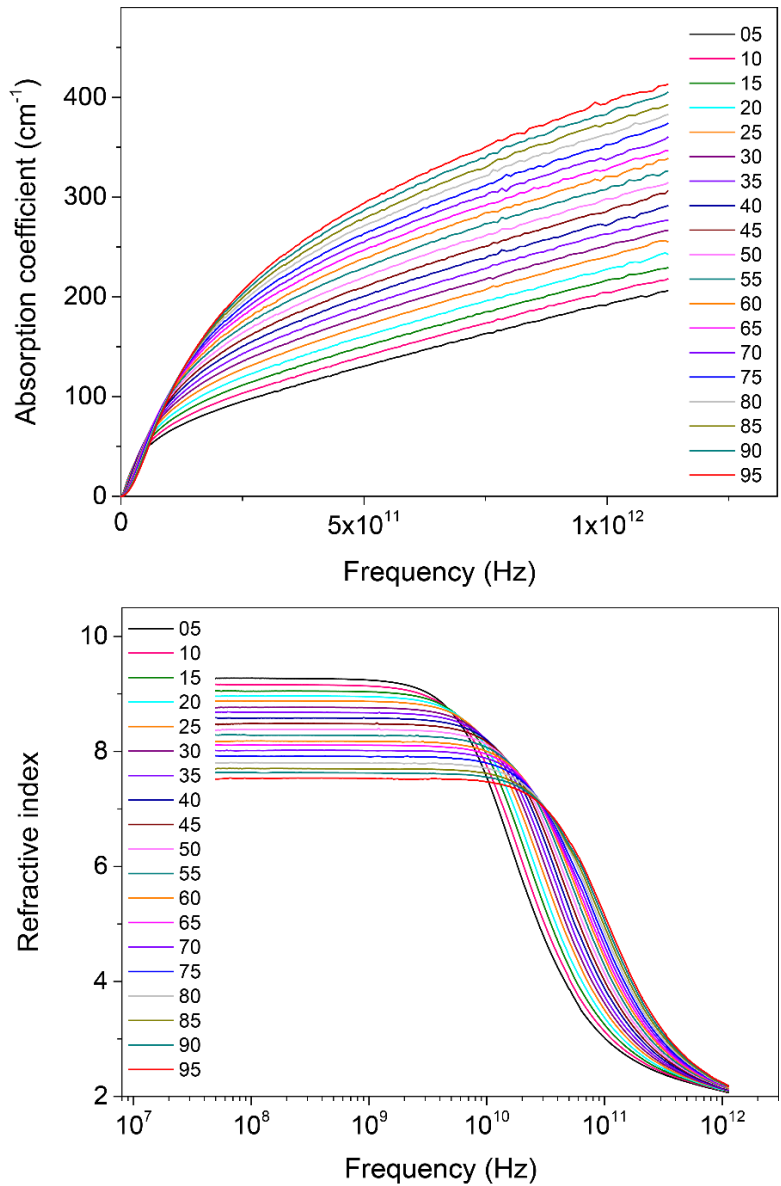

**FIG. S1.** The absorption and refractive index of water at different temperatures from -5 to 95 °C.

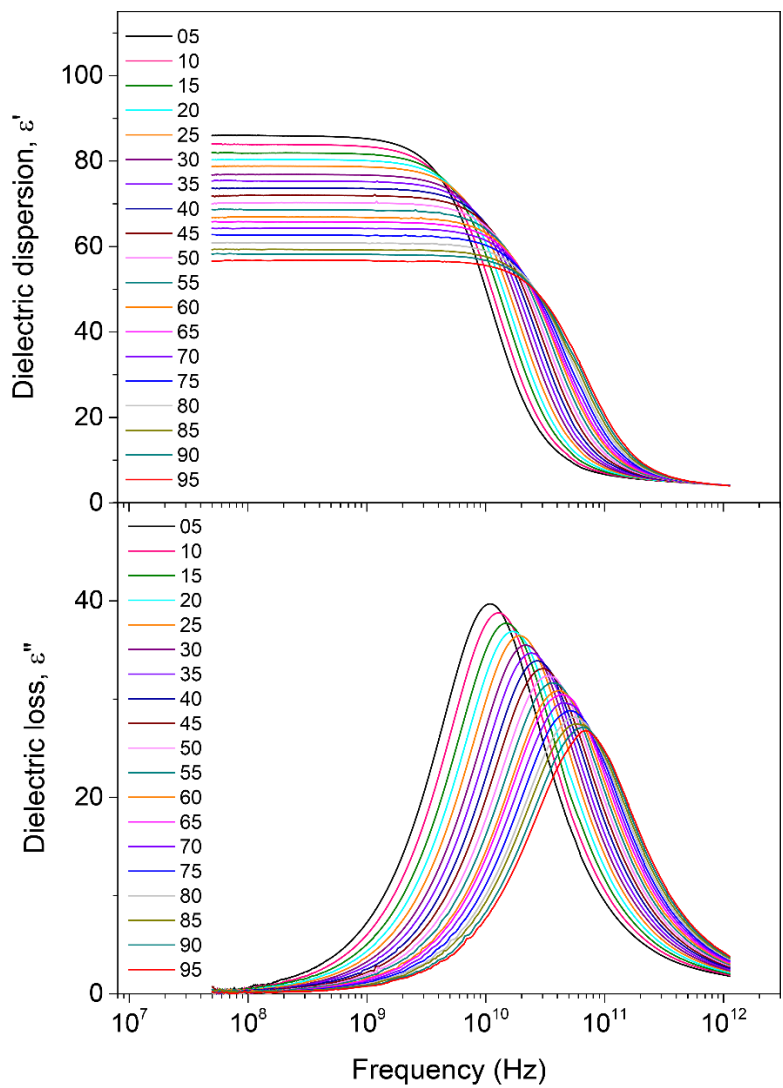

**FIG. S2.** The dielectric response including the dielectric dispersion (real part) and dielectric loss (imaginary part) of liquid water at different temperatures from -5 to 95 ºC.

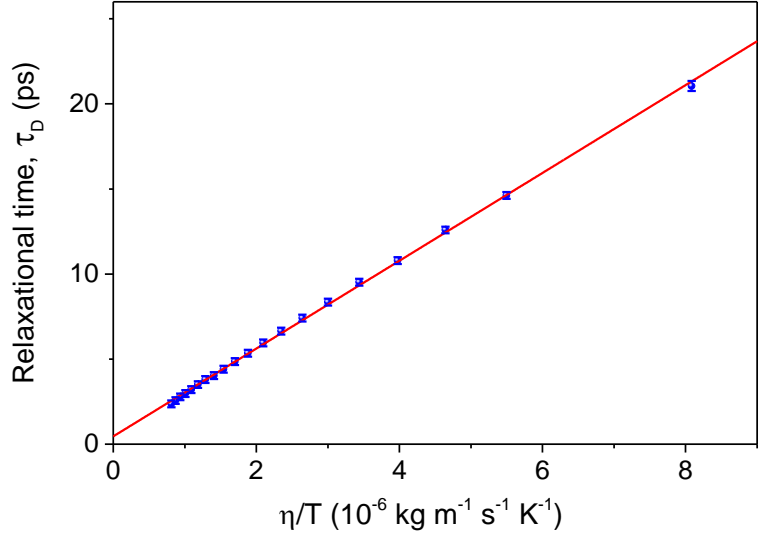

**FIG. S3.** The temperature dependence of the collective orientation relaxation dynamics, $\tau_D$, connects to the viscosity of water, $\eta$, via the Einstein-Stock-Debye relation. The collective orientation relaxation time scale linearly with the viscosity of water divided by temperature, $\eta/T$, from supercooled liquid to near the boiling point. Symbols represent experimental data, while the solid red line is the results of the fits, $\tau_D = a + b\eta/T$, with $a = 4.1 \times 10^{-13}$ (s); $b = 2.58 \times 10^{-6}$ ($s^2 \cdot m \cdot K/kg$).

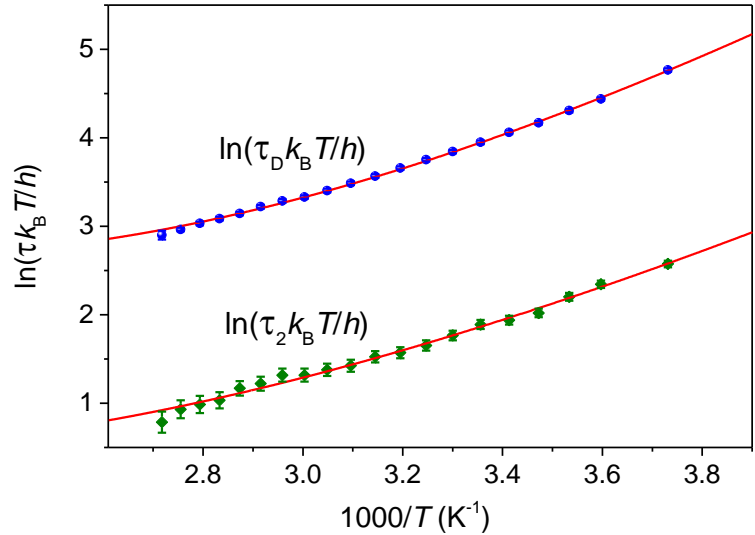

**FIG. S4.** Activation-energy plots for relaxation times of the collective orientations, $\tau_D$, and single-molecule motions, $\tau_2$, deviate from linear behavior. Symbols represent the experimental results, and solid red lines are the results of the fits using Eyring equation.

| Freq (GHz) | Water at -5 ºC a (cm⁻¹) | n | Water at 5 ºC a (cm⁻¹) | n | Water at 10 ºC a (cm⁻¹) | n | Water at 15 ºC a (cm⁻¹) | n | Water at 20 ºC a (cm⁻¹) | n | Water at 25 ºC a (cm⁻¹) | n |
|---|---|---|---|---|---|---|---|---|---|---|---|---|
| 0.05000 | 0.00029 | 9.496 | 0.00028 | 9.259 | 0.00035 | 9.158 | 0.00022 | 9.057 | 0.00016 | 8.960 | 0.00017 | 8.876 |
| 0.05058 | 0.00033 | 9.498 | 0.00027 | 9.260 | 0.00021 | 9.158 | 0.00023 | 9.057 | 0.00017 | 8.960 | 0.00022 | 8.874 |
| 0.05117 | 0.00041 | 9.495 | 0.00027 | 9.259 | 0.00026 | 9.155 | 0.00023 | 9.055 | 0.00016 | 8.962 | 0.00036 | 8.875 |
| 0.05176 | 0.00039 | 9.495 | 0.00034 | 9.259 | 0.00038 | 9.155 | 0.00014 | 9.057 | 0.00024 | 8.960 | 0.00036 | 8.877 |
| 0.05236 | 0.00048 | 9.497 | 0.00025 | 9.259 | 0.00023 | 9.155 | 0.00007 | 9.054 | 0.00034 | 8.960 | 0.00035 | 8.876 |
| 0.05297 | 0.00041 | 9.501 | 0.00023 | 9.260 | 0.00030 | 9.156 | 0.00025 | 9.053 | 0.00037 | 8.960 | 0.00028 | 8.876 |
| 0.05358 | 0.00046 | 9.504 | 0.00035 | 9.260 | 0.00031 | 9.155 | 0.00039 | 9.054 | 0.00029 | 8.962 | 0.00024 | 8.879 |
| 0.05420 | 0.00056 | 9.504 | 0.00030 | 9.259 | 0.00040 | 9.154 | 0.00033 | 9.051 | 0.00025 | 8.959 | 0.00032 | 8.877 |
| 0.05483 | 0.00049 | 9.502 | 0.00021 | 9.259 | 0.00057 | 9.159 | 0.00050 | 9.049 | 0.00034 | 8.959 | 0.00029 | 8.874 |
| 0.05546 | 0.00062 | 9.501 | 0.00034 | 9.259 | 0.00024 | 9.157 | 0.00032 | 9.049 | 0.00041 | 8.962 | 0.00024 | 8.874 |
| 0.05610 | 0.00063 | 9.498 | 0.00038 | 9.260 | 0.00024 | 9.156 | 0.00036 | 9.048 | 0.00030 | 8.962 | 0.00036 | 8.876 |
| 0.05675 | 0.00078 | 9.500 | 0.00042 | 9.261 | 0.00031 | 9.155 | 0.00008 | 9.051 | 0.00016 | 8.962 | 0.00031 | 8.879 |
| 0.05741 | 0.00076 | 9.503 | 0.00051 | 9.260 | 0.00046 | 9.155 | 0.00020 | 9.049 | 0.00012 | 8.961 | 0.00041 | 8.877 |
| 0.05808 | 0.00073 | 9.501 | 0.00043 | 9.259 | 0.00013 | 9.156 | 0.00035 | 9.056 | 0.00013 | 8.961 | 0.00038 | 8.877 |
| 0.05875 | 0.00071 | 9.502 | 0.00036 | 9.260 | 0.00020 | 9.155 | 0.00051 | 9.057 | 0.00027 | 8.963 | 0.00048 | 8.879 |
| 0.05943 | 0.00064 | 9.499 | 0.00032 | 9.262 | 0.00038 | 9.156 | 0.00027 | 9.053 | 0.00025 | 8.964 | 0.00046 | 8.876 |
| 0.06012 | 0.00074 | 9.496 | 0.00039 | 9.261 | 0.00037 | 9.156 | 0.00030 | 9.051 | 0.00031 | 8.962 | 0.00043 | 8.878 |
| 0.06081 | 0.00072 | 9.502 | 0.00032 | 9.260 | 0.00033 | 9.152 | 0.00060 | 9.052 | 0.00044 | 8.964 | 0.00050 | 8.879 |
| 0.06152 | 0.00079 | 9.497 | 0.00036 | 9.262 | 0.00052 | 9.149 | 0.00045 | 9.052 | 0.00049 | 8.963 | 0.00054 | 8.878 |
| 0.06223 | 0.00085 | 9.501 | 0.00044 | 9.261 | 0.00055 | 9.152 | 0.00051 | 9.053 | 0.00052 | 8.963 | 0.00062 | 8.875 |
| 0.06295 | 0.00083 | 9.502 | 0.00032 | 9.260 | 0.00047 | 9.152 | 0.00056 | 9.054 | 0.00045 | 8.962 | 0.00057 | 8.882 |
| 0.06368 | 0.00081 | 9.502 | 0.00021 | 9.261 | 0.00062 | 9.154 | 0.00070 | 9.054 | 0.00036 | 8.960 | 0.00053 | 8.877 |
| 0.06442 | 0.00084 | 9.501 | 0.00037 | 9.261 | 0.00050 | 9.155 | 0.00048 | 9.053 | 0.00051 | 8.962 | 0.00046 | 8.875 |
| 0.06516 | 0.00083 | 9.502 | 0.00042 | 9.260 | 0.00054 | 9.153 | 0.00053 | 9.051 | 0.00050 | 8.964 | 0.00058 | 8.879 |
| 0.06592 | 0.00091 | 9.500 | 0.00040 | 9.260 | 0.00069 | 9.152 | 0.00072 | 9.054 | 0.00036 | 8.962 | 0.00066 | 8.877 |
| 0.06668 | 0.00100 | 9.502 | 0.00052 | 9.261 | 0.00064 | 9.153 | 0.00064 | 9.052 | 0.00041 | 8.964 | 0.00062 | 8.876 |
| 0.06745 | 0.00080 | 9.498 | 0.00054 | 9.261 | 0.00073 | 9.150 | 0.00072 | 9.055 | 0.00022 | 8.965 | 0.00066 | 8.879 |
| 0.06823 | 0.00100 | 9.498 | 0.00048 | 9.260 | 0.00062 | 9.151 | 0.00069 | 9.053 | 0.00003 | 8.964 | 0.00072 | 8.880 |
| 0.06902 | 0.00108 | 9.499 | 0.00049 | 9.260 | 0.00055 | 9.153 | 0.00057 | 9.053 | 0.00019 | 8.963 | 0.00067 | 8.878 |
| 0.06982 | 0.00119 | 9.498 | 0.00043 | 9.260 | 0.00073 | 9.150 | 0.00057 | 9.052 | 0.00026 | 8.963 | 0.00075 | 8.880 |
| 0.07063 | 0.00126 | 9.499 | 0.00050 | 9.261 | 0.00072 | 9.151 | 0.00047 | 9.051 | 0.00025 | 8.963 | 0.00073 | 8.878 |
| 0.07145 | 0.00136 | 9.502 | 0.00061 | 9.261 | 0.00062 | 9.150 | 0.00072 | 9.055 | 0.00031 | 8.964 | 0.00070 | 8.879 |
| 0.07228 | 0.00125 | 9.497 | 0.00049 | 9.261 | 0.00064 | 9.151 | 0.00078 | 9.052 | 0.00020 | 8.962 | 0.00063 | 8.878 |
| 0.07311 | 0.00130 | 9.502 | 0.00061 | 9.260 | 0.00059 | 9.153 | 0.00070 | 9.055 | 0.00019 | 8.961 | 0.00070 | 8.880 |
| 0.07396 | 0.00136 | 9.499 | 0.00078 | 9.260 | 0.00055 | 9.150 | 0.00071 | 9.054 | 0.00039 | 8.963 | 0.00055 | 8.878 |
| 0.07482 | 0.00137 | 9.497 | 0.00083 | 9.260 | 0.00056 | 9.152 | 0.00066 | 9.052 | 0.00043 | 8.962 | 0.00069 | 8.880 |
| 0.07568 | 0.00132 | 9.499 | 0.00096 | 9.260 | 0.00054 | 9.152 | 0.00052 | 9.049 | 0.00031 | 8.961 | 0.00066 | 8.879 |
| 0.07656 | 0.00136 | 9.501 | 0.00100 | 9.261 | 0.00060 | 9.155 | 0.00066 | 9.051 | 0.00022 | 8.961 | 0.00061 | 8.878 |
| 0.07745 | 0.00133 | 9.498 | 0.00084 | 9.260 | 0.00056 | 9.153 | 0.00072 | 9.050 | 0.00034 | 8.960 | 0.00049 | 8.880 |
| 0.07834 | 0.00139 | 9.499 | 0.00079 | 9.260 | 0.00062 | 9.152 | 0.00055 | 9.051 | 0.00049 | 8.962 | 0.00067 | 8.878 |
| 0.07925 | 0.00148 | 9.500 | 0.00081 | 9.260 | 0.00072 | 9.152 | 0.00062 | 9.049 | 0.00054 | 8.963 | 0.00064 | 8.879 |
| 0.08017 | 0.00157 | 9.500 | 0.00083 | 9.260 | 0.00077 | 9.151 | 0.00059 | 9.049 | 0.00048 | 8.963 | 0.00078 | 8.879 |
| 0.08110 | 0.00161 | 9.500 | 0.00079 | 9.260 | 0.00073 | 9.153 | 0.00073 | 9.048 | 0.00055 | 8.963 | 0.00066 | 8.880 |
| 0.08204 | 0.00164 | 9.500 | 0.00079 | 9.259 | 0.00080 | 9.153 | 0.00064 | 9.051 | 0.00057 | 8.963 | 0.00079 | 8.880 |
| 0.08299 | 0.00185 | 9.499 | 0.00077 | 9.260 | 0.00072 | 9.153 | 0.00064 | 9.049 | 0.00057 | 8.963 | 0.00077 | 8.880 |

| | | | | | | | | | | |
|---|---|---|---|---|---|---|---|---|---|---|
|0.08395|0.00197|9.500|0.00070|9.260|0.00080|9.152|0.00079|9.048|0.00056|8.964|0.00083|8.879|
|0.08492|0.00178|9.500|0.00078|9.260|0.00085|9.151|0.00097|9.049|0.00057|8.964|0.00078|8.879|
|0.08590|0.00195|9.501|0.00082|9.260|0.00078|9.150|0.00073|9.049|0.00072|8.963|0.00072|8.877|
|0.08690|0.00197|9.498|0.00092|9.260|0.00085|9.151|0.00079|9.048|0.00059|8.963|0.00068|8.878|
|0.08790|0.00212|9.498|0.00107|9.260|0.00085|9.152|0.00089|9.047|0.00043|8.963|0.00071|8.879|
|0.08892|0.00195|9.497|0.00111|9.260|0.00087|9.152|0.00079|9.047|0.00043|8.963|0.00064|8.878|
|0.08995|0.00208|9.497|0.00105|9.260|0.00079|9.150|0.00087|9.047|0.00054|8.963|0.00072|8.878|
|0.09099|0.00199|9.497|0.00107|9.261|0.00102|9.151|0.00088|9.047|0.00067|8.964|0.00089|8.879|
|0.09204|0.00198|9.497|0.00104|9.261|0.00105|9.151|0.00097|9.048|0.00075|8.965|0.00085|8.879|
|0.09311|0.00192|9.497|0.00103|9.261|0.00128|9.151|0.00101|9.048|0.00078|8.963|0.00099|8.880|
|0.09419|0.00206|9.496|0.00111|9.261|0.00114|9.152|0.00093|9.048|0.00068|8.961|0.00098|8.880|
|0.09528|0.00204|9.498|0.00125|9.260|0.00115|9.151|0.00108|9.048|0.00064|8.962|0.00099|8.879|
|0.09638|0.00219|9.497|0.00136|9.260|0.00106|9.153|0.00105|9.048|0.00062|8.962|0.00093|8.879|
|0.09750|0.00222|9.496|0.00135|9.261|0.00123|9.152|0.00116|9.048|0.00062|8.963|0.00101|8.878|
|0.09863|0.00233|9.498|0.00128|9.261|0.00138|9.151|0.00130|9.049|0.00071|8.963|0.00088|8.879|
|0.09977|0.00235|9.497|0.00130|9.261|0.00143|9.150|0.00125|9.047|0.00081|8.962|0.00090|8.880|
|0.10093|0.00234|9.497|0.00147|9.262|0.00109|9.152|0.00119|9.048|0.00090|8.962|0.00087|8.880|
|0.10209|0.00250|9.497|0.00161|9.261|0.00134|9.150|0.00126|9.048|0.00089|8.962|0.00085|8.879|
|0.10328|0.00250|9.496|0.00168|9.260|0.00142|9.148|0.00115|9.049|0.00082|8.963|0.00094|8.879|
|0.10447|0.00242|9.496|0.00165|9.260|0.00131|9.150|0.00121|9.051|0.00083|8.964|0.00093|8.878|
|0.10568|0.00251|9.498|0.00165|9.259|0.00112|9.149|0.00132|9.048|0.00104|8.964|0.00104|8.879|
|0.10691|0.00258|9.497|0.00183|9.260|0.00126|9.148|0.00129|9.048|0.00104|8.962|0.00108|8.878|
|0.10814|0.00277|9.497|0.00187|9.260|0.00120|9.150|0.00136|9.048|0.00088|8.962|0.00103|8.877|
|0.10940|0.00291|9.497|0.00195|9.259|0.00128|9.149|0.00147|9.048|0.00098|8.962|0.00107|8.878|
|0.11066|0.00296|9.497|0.00221|9.260|0.00149|9.149|0.00147|9.048|0.00123|8.963|0.00092|8.879|
|0.11194|0.00296|9.495|0.00229|9.259|0.00140|9.150|0.00158|9.047|0.00126|8.963|0.00101|8.880|
|0.11324|0.00306|9.496|0.00224|9.259|0.00144|9.150|0.00161|9.048|0.00110|8.962|0.00106|8.879|
|0.11455|0.00317|9.497|0.00231|9.260|0.00145|9.150|0.00166|9.048|0.00104|8.962|0.00111|8.879|
|0.11588|0.00333|9.497|0.00235|9.259|0.00191|9.150|0.00159|9.048|0.00109|8.962|0.00123|8.879|
|0.11722|0.00339|9.498|0.00228|9.258|0.00187|9.150|0.00157|9.048|0.00119|8.962|0.00132|8.879|
|0.11858|0.00351|9.496|0.00236|9.259|0.00202|9.149|0.00175|9.047|0.00123|8.962|0.00134|8.878|
|0.11995|0.00353|9.497|0.00242|9.260|0.00205|9.149|0.00187|9.048|0.00127|8.963|0.00130|8.879|
|0.12134|0.00354|9.497|0.00236|9.260|0.00201|9.151|0.00195|9.049|0.00144|8.963|0.00127|8.878|
|0.12274|0.00349|9.496|0.00239|9.260|0.00220|9.150|0.00199|9.048|0.00150|8.962|0.00130|8.878|
|0.12417|0.00348|9.496|0.00246|9.260|0.00240|9.150|0.00212|9.049|0.00150|8.962|0.00136|8.879|
|0.12560|0.00364|9.495|0.00255|9.260|0.00235|9.148|0.00194|9.048|0.00168|8.962|0.00143|8.879|
|0.12706|0.00371|9.495|0.00263|9.259|0.00237|9.149|0.00207|9.049|0.00176|8.962|0.00136|8.879|
|0.12853|0.00385|9.496|0.00271|9.259|0.00241|9.150|0.00220|9.048|0.00177|8.962|0.00155|8.879|
|0.13002|0.00407|9.495|0.00289|9.259|0.00257|9.149|0.00219|9.049|0.00171|8.962|0.00156|8.880|
|0.13152|0.00426|9.495|0.00289|9.259|0.00232|9.149|0.00230|9.048|0.00161|8.961|0.00156|8.879|
|0.13305|0.00427|9.495|0.00285|9.259|0.00233|9.149|0.00233|9.050|0.00164|8.961|0.00171|8.880|
|0.13459|0.00440|9.494|0.00286|9.259|0.00247|9.150|0.00210|9.049|0.00167|8.962|0.00172|8.879|
|0.13614|0.00455|9.495|0.00284|9.259|0.00263|9.150|0.00211|9.049|0.00179|8.962|0.00175|8.879|
|0.13772|0.00466|9.496|0.00283|9.259|0.00256|9.150|0.00236|9.049|0.00197|8.962|0.00171|8.879|
|0.13932|0.00477|9.494|0.00283|9.259|0.00272|9.151|0.00238|9.049|0.00212|8.963|0.00171|8.880|
|0.14093|0.00483|9.495|0.00287|9.258|0.00257|9.151|0.00245|9.049|0.00216|8.963|0.00182|8.879|
|0.14256|0.00489|9.496|0.00306|9.257|0.00271|9.150|0.00257|9.050|0.00231|8.963|0.00177|8.880|

| | | | | | | | | | | | |
|---|---|---|---|---|---|---|---|---|---|---|---|
| 0.14421 | 0.00492 | 9.495 | 0.00324 | 9.257 | 0.00268 | 9.150 | 0.00290 | 9.049 | 0.00233 | 8.964 | 0.00188 | 8.880 |
| 0.14588 | 0.00516 | 9.495 | 0.00341 | 9.257 | 0.00270 | 9.149 | 0.00279 | 9.049 | 0.00232 | 8.964 | 0.00196 | 8.879 |
| 0.14757 | 0.00530 | 9.495 | 0.00352 | 9.257 | 0.00283 | 9.150 | 0.00269 | 9.050 | 0.00244 | 8.964 | 0.00199 | 8.880 |
| 0.14928 | 0.00553 | 9.495 | 0.00342 | 9.258 | 0.00309 | 9.151 | 0.00282 | 9.050 | 0.00242 | 8.964 | 0.00205 | 8.880 |
| 0.15101 | 0.00565 | 9.494 | 0.00341 | 9.257 | 0.00303 | 9.151 | 0.00294 | 9.050 | 0.00248 | 8.963 | 0.00217 | 8.880 |
| 0.15276 | 0.00567 | 9.494 | 0.00348 | 9.257 | 0.00305 | 9.151 | 0.00295 | 9.050 | 0.00260 | 8.963 | 0.00222 | 8.880 |
| 0.15453 | 0.00573 | 9.494 | 0.00352 | 9.257 | 0.00320 | 9.151 | 0.00281 | 9.050 | 0.00260 | 8.963 | 0.00231 | 8.880 |
| 0.15631 | 0.00600 | 9.495 | 0.00363 | 9.257 | 0.00343 | 9.151 | 0.00297 | 9.050 | 0.00257 | 8.963 | 0.00230 | 8.880 |
| 0.15812 | 0.00618 | 9.495 | 0.00379 | 9.257 | 0.00340 | 9.151 | 0.00305 | 9.051 | 0.00268 | 8.964 | 0.00232 | 8.879 |
| 0.15996 | 0.00623 | 9.495 | 0.00387 | 9.257 | 0.00330 | 9.152 | 0.00303 | 9.050 | 0.00284 | 8.964 | 0.00225 | 8.879 |
| 0.16181 | 0.00632 | 9.494 | 0.00405 | 9.257 | 0.00332 | 9.152 | 0.00291 | 9.051 | 0.00288 | 8.964 | 0.00232 | 8.879 |
| 0.16368 | 0.00660 | 9.494 | 0.00439 | 9.257 | 0.00361 | 9.151 | 0.00306 | 9.051 | 0.00299 | 8.964 | 0.00241 | 8.879 |
| 0.16558 | 0.00677 | 9.495 | 0.00447 | 9.257 | 0.00361 | 9.152 | 0.00325 | 9.050 | 0.00311 | 8.965 | 0.00244 | 8.879 |
| 0.16749 | 0.00693 | 9.495 | 0.00449 | 9.257 | 0.00358 | 9.151 | 0.00330 | 9.050 | 0.00314 | 8.964 | 0.00252 | 8.879 |
| 0.16943 | 0.00703 | 9.495 | 0.00460 | 9.257 | 0.00388 | 9.152 | 0.00330 | 9.050 | 0.00309 | 8.965 | 0.00256 | 8.879 |
| 0.17140 | 0.00719 | 9.494 | 0.00482 | 9.257 | 0.00381 | 9.152 | 0.00334 | 9.049 | 0.00303 | 8.965 | 0.00272 | 8.879 |
| 0.17338 | 0.00740 | 9.494 | 0.00504 | 9.256 | 0.00405 | 9.152 | 0.00357 | 9.050 | 0.00326 | 8.965 | 0.00276 | 8.879 |
| 0.17539 | 0.00757 | 9.493 | 0.00519 | 9.256 | 0.00409 | 9.151 | 0.00368 | 9.050 | 0.00336 | 8.965 | 0.00275 | 8.879 |
| 0.17742 | 0.00774 | 9.494 | 0.00518 | 9.256 | 0.00416 | 9.150 | 0.00389 | 9.050 | 0.00334 | 8.965 | 0.00289 | 8.879 |
| 0.17947 | 0.00780 | 9.495 | 0.00541 | 9.256 | 0.00414 | 9.150 | 0.00387 | 9.050 | 0.00342 | 8.965 | 0.00283 | 8.879 |
| 0.18155 | 0.00794 | 9.495 | 0.00561 | 9.256 | 0.00440 | 9.150 | 0.00396 | 9.050 | 0.00351 | 8.965 | 0.00299 | 8.880 |
| 0.18365 | 0.00823 | 9.495 | 0.00582 | 9.256 | 0.00450 | 9.151 | 0.00395 | 9.051 | 0.00359 | 8.965 | 0.00303 | 8.879 |
| 0.18578 | 0.00841 | 9.494 | 0.00601 | 9.256 | 0.00462 | 9.150 | 0.00413 | 9.051 | 0.00361 | 8.966 | 0.00307 | 8.879 |
| 0.18793 | 0.00857 | 9.494 | 0.00618 | 9.256 | 0.00472 | 9.151 | 0.00424 | 9.051 | 0.00364 | 8.965 | 0.00335 | 8.879 |
| 0.19011 | 0.00880 | 9.495 | 0.00634 | 9.256 | 0.00475 | 9.151 | 0.00460 | 9.050 | 0.00379 | 8.964 | 0.00341 | 8.879 |
| 0.19231 | 0.00904 | 9.494 | 0.00640 | 9.256 | 0.00485 | 9.151 | 0.00455 | 9.050 | 0.00387 | 8.964 | 0.00341 | 8.879 |
| 0.19454 | 0.00913 | 9.494 | 0.00646 | 9.255 | 0.00529 | 9.151 | 0.00465 | 9.050 | 0.00399 | 8.964 | 0.00336 | 8.879 |
| 0.19679 | 0.00946 | 9.494 | 0.00656 | 9.255 | 0.00557 | 9.151 | 0.00472 | 9.051 | 0.00403 | 8.964 | 0.00351 | 8.880 |
| 0.19907 | 0.00962 | 9.494 | 0.00663 | 9.255 | 0.00537 | 9.151 | 0.00487 | 9.051 | 0.00406 | 8.964 | 0.00350 | 8.879 |
| 0.20137 | 0.00986 | 9.494 | 0.00688 | 9.255 | 0.00573 | 9.151 | 0.00513 | 9.051 | 0.00424 | 8.964 | 0.00352 | 8.879 |
| 0.20370 | 0.01008 | 9.494 | 0.00714 | 9.256 | 0.00585 | 9.151 | 0.00530 | 9.052 | 0.00442 | 8.964 | 0.00373 | 8.879 |
| 0.20606 | 0.01046 | 9.494 | 0.00719 | 9.255 | 0.00586 | 9.151 | 0.00536 | 9.052 | 0.00444 | 8.964 | 0.00386 | 8.879 |
| 0.20845 | 0.01068 | 9.494 | 0.00719 | 9.255 | 0.00608 | 9.150 | 0.00551 | 9.052 | 0.00440 | 8.964 | 0.00388 | 8.879 |
| 0.21086 | 0.01083 | 9.494 | 0.00749 | 9.256 | 0.00612 | 9.151 | 0.00551 | 9.052 | 0.00445 | 8.964 | 0.00391 | 8.879 |
| 0.21330 | 0.01099 | 9.494 | 0.00766 | 9.255 | 0.00625 | 9.150 | 0.00561 | 9.051 | 0.00460 | 8.964 | 0.00399 | 8.879 |
| 0.21577 | 0.01137 | 9.494 | 0.00784 | 9.255 | 0.00632 | 9.150 | 0.00586 | 9.051 | 0.00479 | 8.964 | 0.00411 | 8.879 |
| 0.21827 | 0.01165 | 9.494 | 0.00808 | 9.255 | 0.00656 | 9.150 | 0.00585 | 9.051 | 0.00493 | 8.964 | 0.00422 | 8.879 |
| 0.22080 | 0.01179 | 9.493 | 0.00828 | 9.255 | 0.00692 | 9.149 | 0.00602 | 9.051 | 0.00506 | 8.964 | 0.00435 | 8.879 |
| 0.22336 | 0.01203 | 9.494 | 0.00838 | 9.255 | 0.00701 | 9.150 | 0.00611 | 9.050 | 0.00505 | 8.964 | 0.00427 | 8.879 |
| 0.22594 | 0.01241 | 9.494 | 0.00854 | 9.255 | 0.00709 | 9.150 | 0.00634 | 9.051 | 0.00503 | 8.964 | 0.00440 | 8.879 |
| 0.22856 | 0.01284 | 9.493 | 0.00884 | 9.255 | 0.00728 | 9.150 | 0.00644 | 9.052 | 0.00509 | 8.964 | 0.00452 | 8.879 |
| 0.23121 | 0.01302 | 9.493 | 0.00907 | 9.254 | 0.00732 | 9.150 | 0.00648 | 9.052 | 0.00533 | 8.964 | 0.00461 | 8.879 |
| 0.23388 | 0.01323 | 9.493 | 0.00923 | 9.255 | 0.00740 | 9.150 | 0.00659 | 9.051 | 0.00567 | 8.964 | 0.00481 | 8.879 |
| 0.23659 | 0.01344 | 9.493 | 0.00942 | 9.255 | 0.00765 | 9.150 | 0.00697 | 9.051 | 0.00575 | 8.964 | 0.00494 | 8.879 |
| 0.23933 | 0.01377 | 9.493 | 0.00963 | 9.255 | 0.00784 | 9.150 | 0.00693 | 9.051 | 0.00586 | 8.964 | 0.00504 | 8.879 |
| 0.24210 | 0.01425 | 9.493 | 0.00984 | 9.255 | 0.00776 | 9.150 | 0.00719 | 9.051 | 0.00598 | 8.965 | 0.00531 | 8.879 |
| 0.24491 | 0.01455 | 9.493 | 0.01002 | 9.255 | 0.00834 | 9.149 | 0.00741 | 9.051 | 0.00596 | 8.964 | 0.00537 | 8.879 |

| | | | | | | | | | | | |
|---|---|---|---|---|---|---|---|---|---|---|---|
| 0.24774 | 0.01496 | 9.492 | 0.01019 | 9.254 | 0.00841 | 9.149 | 0.00752 | 9.051 | 0.00616 | 8.964 | 0.00556 | 8.878 |
| 0.25061 | 0.01541 | 9.493 | 0.01043 | 9.254 | 0.00862 | 9.148 | 0.00784 | 9.050 | 0.00649 | 8.965 | 0.00572 | 8.879 |
| 0.25351 | 0.01570 | 9.492 | 0.01074 | 9.254 | 0.00871 | 9.149 | 0.00789 | 9.051 | 0.00665 | 8.964 | 0.00590 | 8.879 |
| 0.25645 | 0.01592 | 9.493 | 0.01101 | 9.254 | 0.00913 | 9.149 | 0.00785 | 9.051 | 0.00665 | 8.964 | 0.00604 | 8.878 |
| 0.25942 | 0.01633 | 9.492 | 0.01116 | 9.254 | 0.00927 | 9.149 | 0.00811 | 9.051 | 0.00672 | 8.964 | 0.00611 | 8.879 |
| 0.26242 | 0.01665 | 9.493 | 0.01141 | 9.254 | 0.00954 | 9.149 | 0.00832 | 9.051 | 0.00686 | 8.964 | 0.00625 | 8.879 |
| 0.26546 | 0.01713 | 9.492 | 0.01187 | 9.254 | 0.00968 | 9.149 | 0.00853 | 9.051 | 0.00715 | 8.964 | 0.00642 | 8.878 |
| 0.26853 | 0.01740 | 9.492 | 0.01221 | 9.254 | 0.01007 | 9.148 | 0.00900 | 9.051 | 0.00734 | 8.964 | 0.00647 | 8.878 |
| 0.27164 | 0.01807 | 9.492 | 0.01251 | 9.254 | 0.01009 | 9.149 | 0.00907 | 9.050 | 0.00754 | 8.964 | 0.00668 | 8.879 |
| 0.27479 | 0.01843 | 9.492 | 0.01279 | 9.254 | 0.01043 | 9.149 | 0.00911 | 9.050 | 0.00791 | 8.964 | 0.00687 | 8.878 |
| 0.27797 | 0.01871 | 9.492 | 0.01300 | 9.254 | 0.01056 | 9.150 | 0.00941 | 9.050 | 0.00810 | 8.964 | 0.00692 | 8.878 |
| 0.28119 | 0.01920 | 9.491 | 0.01322 | 9.254 | 0.01104 | 9.149 | 0.00932 | 9.050 | 0.00805 | 8.964 | 0.00710 | 8.878 |
| 0.28445 | 0.01967 | 9.491 | 0.01353 | 9.254 | 0.01123 | 9.148 | 0.00953 | 9.050 | 0.00830 | 8.964 | 0.00717 | 8.878 |
| 0.28774 | 0.02011 | 9.491 | 0.01389 | 9.254 | 0.01126 | 9.149 | 0.00981 | 9.050 | 0.00852 | 8.964 | 0.00732 | 8.878 |
| 0.29107 | 0.02070 | 9.491 | 0.01420 | 9.254 | 0.01173 | 9.149 | 0.00999 | 9.050 | 0.00862 | 8.964 | 0.00752 | 8.878 |
| 0.29444 | 0.02105 | 9.491 | 0.01466 | 9.253 | 0.01176 | 9.149 | 0.01020 | 9.050 | 0.00896 | 8.964 | 0.00776 | 8.878 |
| 0.29785 | 0.02166 | 9.490 | 0.01502 | 9.253 | 0.01198 | 9.149 | 0.01054 | 9.050 | 0.00928 | 8.964 | 0.00801 | 8.879 |
| 0.30130 | 0.02215 | 9.491 | 0.01527 | 9.253 | 0.01253 | 9.149 | 0.01074 | 9.050 | 0.00935 | 8.964 | 0.00816 | 8.879 |
| 0.30479 | 0.02259 | 9.491 | 0.01548 | 9.253 | 0.01248 | 9.149 | 0.01101 | 9.050 | 0.00939 | 8.964 | 0.00836 | 8.878 |
| 0.30832 | 0.02325 | 9.490 | 0.01581 | 9.253 | 0.01300 | 9.149 | 0.01127 | 9.050 | 0.00947 | 8.963 | 0.00871 | 8.879 |
| 0.31189 | 0.02379 | 9.490 | 0.01614 | 9.253 | 0.01320 | 9.148 | 0.01147 | 9.050 | 0.00991 | 8.963 | 0.00883 | 8.879 |
| 0.31550 | 0.02424 | 9.490 | 0.01649 | 9.253 | 0.01338 | 9.148 | 0.01172 | 9.050 | 0.01028 | 8.963 | 0.00909 | 8.878 |
| 0.31915 | 0.02485 | 9.490 | 0.01700 | 9.253 | 0.01389 | 9.148 | 0.01213 | 9.050 | 0.01030 | 8.963 | 0.00938 | 8.878 |
| 0.32285 | 0.02544 | 9.490 | 0.01739 | 9.252 | 0.01424 | 9.149 | 0.01246 | 9.050 | 0.01057 | 8.963 | 0.00941 | 8.878 |
| 0.32659 | 0.02610 | 9.490 | 0.01775 | 9.252 | 0.01452 | 9.148 | 0.01265 | 9.050 | 0.01096 | 8.964 | 0.00961 | 8.878 |
| 0.33037 | 0.02664 | 9.489 | 0.01815 | 9.252 | 0.01506 | 9.149 | 0.01319 | 9.050 | 0.01130 | 8.963 | 0.00982 | 8.878 |
| 0.33420 | 0.02733 | 9.489 | 0.01859 | 9.252 | 0.01543 | 9.148 | 0.01365 | 9.050 | 0.01156 | 8.963 | 0.01009 | 8.879 |
| 0.33806 | 0.02789 | 9.489 | 0.01922 | 9.252 | 0.01571 | 9.148 | 0.01383 | 9.050 | 0.01186 | 8.963 | 0.01018 | 8.879 |
| 0.34198 | 0.02856 | 9.489 | 0.01980 | 9.252 | 0.01629 | 9.148 | 0.01417 | 9.050 | 0.01202 | 8.963 | 0.01054 | 8.878 |
| 0.34594 | 0.02918 | 9.489 | 0.02020 | 9.252 | 0.01647 | 9.148 | 0.01460 | 9.050 | 0.01213 | 8.963 | 0.01071 | 8.878 |
| 0.34995 | 0.03000 | 9.489 | 0.02052 | 9.252 | 0.01666 | 9.148 | 0.01462 | 9.050 | 0.01243 | 8.963 | 0.01099 | 8.878 |
| 0.35400 | 0.03085 | 9.488 | 0.02099 | 9.252 | 0.01716 | 9.148 | 0.01475 | 9.050 | 0.01272 | 8.963 | 0.01121 | 8.878 |
| 0.35810 | 0.03145 | 9.488 | 0.02156 | 9.251 | 0.01764 | 9.148 | 0.01538 | 9.049 | 0.01315 | 8.963 | 0.01147 | 8.878 |
| 0.36224 | 0.03235 | 9.488 | 0.02201 | 9.251 | 0.01810 | 9.148 | 0.01567 | 9.050 | 0.01356 | 8.963 | 0.01182 | 8.878 |
| 0.36644 | 0.03298 | 9.488 | 0.02260 | 9.252 | 0.01841 | 9.147 | 0.01580 | 9.050 | 0.01385 | 8.963 | 0.01208 | 8.878 |
| 0.37068 | 0.03373 | 9.488 | 0.02290 | 9.251 | 0.01879 | 9.148 | 0.01612 | 9.049 | 0.01414 | 8.963 | 0.01224 | 8.878 |
| 0.37497 | 0.03445 | 9.487 | 0.02344 | 9.251 | 0.01930 | 9.148 | 0.01662 | 9.050 | 0.01443 | 8.963 | 0.01264 | 8.878 |
| 0.37931 | 0.03517 | 9.487 | 0.02414 | 9.251 | 0.01984 | 9.147 | 0.01677 | 9.049 | 0.01479 | 8.963 | 0.01293 | 8.878 |
| 0.38371 | 0.03616 | 9.487 | 0.02472 | 9.250 | 0.02030 | 9.147 | 0.01723 | 9.049 | 0.01524 | 8.963 | 0.01332 | 8.878 |
| 0.38815 | 0.03682 | 9.486 | 0.02517 | 9.250 | 0.02063 | 9.147 | 0.01794 | 9.049 | 0.01553 | 8.963 | 0.01352 | 8.878 |
| 0.39264 | 0.03785 | 9.487 | 0.02585 | 9.251 | 0.02118 | 9.147 | 0.01837 | 9.049 | 0.01594 | 8.963 | 0.01388 | 8.878 |
| 0.39719 | 0.03849 | 9.486 | 0.02648 | 9.251 | 0.02184 | 9.147 | 0.01879 | 9.049 | 0.01628 | 8.963 | 0.01427 | 8.878 |
| 0.40179 | 0.03952 | 9.486 | 0.02700 | 9.251 | 0.02215 | 9.147 | 0.01896 | 9.049 | 0.01665 | 8.963 | 0.01456 | 8.878 |
| 0.40644 | 0.04045 | 9.486 | 0.02779 | 9.250 | 0.02260 | 9.146 | 0.01964 | 9.049 | 0.01712 | 8.962 | 0.01485 | 8.878 |
| 0.41115 | 0.04150 | 9.486 | 0.02849 | 9.250 | 0.02321 | 9.146 | 0.01994 | 9.049 | 0.01741 | 8.962 | 0.01524 | 8.878 |
| 0.41591 | 0.04238 | 9.485 | 0.02913 | 9.251 | 0.02383 | 9.147 | 0.02033 | 9.049 | 0.01783 | 8.963 | 0.01550 | 8.877 |
| 0.42073 | 0.04326 | 9.485 | 0.02966 | 9.250 | 0.02432 | 9.146 | 0.02098 | 9.049 | 0.01817 | 8.962 | 0.01586 | 8.878 |

| | | | | | | | | | | | |
|---|---|---|---|---|---|---|---|---|---|---|---|
| 0.42560 | 0.04457 | 9.485 | 0.03030 | 9.250 | 0.02502 | 9.146 | 0.02149 | 9.049 | 0.01858 | 8.962 | 0.01627 | 8.878 |
| 0.43053 | 0.04538 | 9.485 | 0.03112 | 9.250 | 0.02581 | 9.146 | 0.02187 | 9.049 | 0.01903 | 8.962 | 0.01664 | 8.878 |
| 0.43551 | 0.04658 | 9.484 | 0.03204 | 9.250 | 0.02629 | 9.146 | 0.02234 | 9.049 | 0.01941 | 8.962 | 0.01709 | 8.878 |
| 0.44055 | 0.04765 | 9.484 | 0.03280 | 9.250 | 0.02669 | 9.146 | 0.02273 | 9.049 | 0.02015 | 8.962 | 0.01744 | 8.878 |
| 0.44566 | 0.04865 | 9.484 | 0.03349 | 9.249 | 0.02744 | 9.146 | 0.02326 | 9.049 | 0.02073 | 8.962 | 0.01785 | 8.878 |
| 0.45082 | 0.04978 | 9.484 | 0.03435 | 9.249 | 0.02815 | 9.146 | 0.02374 | 9.048 | 0.02108 | 8.962 | 0.01824 | 8.878 |
| 0.45604 | 0.05099 | 9.484 | 0.03522 | 9.249 | 0.02868 | 9.146 | 0.02425 | 9.048 | 0.02150 | 8.962 | 0.01856 | 8.878 |
| 0.46132 | 0.05196 | 9.483 | 0.03582 | 9.249 | 0.02945 | 9.146 | 0.02491 | 9.048 | 0.02197 | 8.962 | 0.01891 | 8.877 |
| 0.46666 | 0.05319 | 9.483 | 0.03663 | 9.249 | 0.03011 | 9.146 | 0.02550 | 9.048 | 0.02244 | 8.962 | 0.01947 | 8.877 |
| 0.47206 | 0.05448 | 9.483 | 0.03752 | 9.249 | 0.03101 | 9.145 | 0.02622 | 9.048 | 0.02297 | 8.962 | 0.01995 | 8.877 |
| 0.47753 | 0.05589 | 9.482 | 0.03842 | 9.249 | 0.03161 | 9.145 | 0.02677 | 9.048 | 0.02345 | 8.962 | 0.02033 | 8.877 |
| 0.48306 | 0.05724 | 9.482 | 0.03927 | 9.249 | 0.03224 | 9.145 | 0.02747 | 9.048 | 0.02378 | 8.962 | 0.02079 | 8.877 |
| 0.48865 | 0.05842 | 9.482 | 0.04009 | 9.248 | 0.03312 | 9.145 | 0.02802 | 9.048 | 0.02438 | 8.962 | 0.02132 | 8.877 |
| 0.49431 | 0.05977 | 9.481 | 0.04090 | 9.248 | 0.03388 | 9.145 | 0.02855 | 9.048 | 0.02496 | 8.962 | 0.02190 | 8.877 |
| 0.50003 | 0.06119 | 9.481 | 0.04185 | 9.248 | 0.03446 | 9.144 | 0.02928 | 9.048 | 0.02535 | 8.961 | 0.02228 | 8.877 |
| 0.50582 | 0.06280 | 9.481 | 0.04281 | 9.248 | 0.03539 | 9.145 | 0.02999 | 9.048 | 0.02601 | 8.961 | 0.02288 | 8.877 |
| 0.51168 | 0.06391 | 9.480 | 0.04385 | 9.248 | 0.03610 | 9.144 | 0.03091 | 9.048 | 0.02661 | 8.961 | 0.02344 | 8.877 |
| 0.51761 | 0.06542 | 9.480 | 0.04482 | 9.247 | 0.03717 | 9.144 | 0.03140 | 9.048 | 0.02733 | 8.962 | 0.02401 | 8.877 |
| 0.52360 | 0.06696 | 9.480 | 0.04588 | 9.247 | 0.03785 | 9.144 | 0.03208 | 9.048 | 0.02799 | 8.961 | 0.02423 | 8.877 |
| 0.52966 | 0.06845 | 9.479 | 0.04704 | 9.247 | 0.03882 | 9.144 | 0.03283 | 9.048 | 0.02877 | 8.961 | 0.02496 | 8.877 |
| 0.53580 | 0.06995 | 9.479 | 0.04800 | 9.246 | 0.03973 | 9.144 | 0.03356 | 9.047 | 0.02944 | 8.961 | 0.02575 | 8.877 |
| 0.54200 | 0.07159 | 9.478 | 0.04906 | 9.246 | 0.04060 | 9.143 | 0.03438 | 9.047 | 0.03009 | 8.961 | 0.02641 | 8.877 |
| 0.54828 | 0.07343 | 9.478 | 0.05023 | 9.246 | 0.04161 | 9.143 | 0.03504 | 9.048 | 0.03082 | 8.961 | 0.02673 | 8.877 |
| 0.55463 | 0.07516 | 9.477 | 0.05149 | 9.246 | 0.04269 | 9.143 | 0.03609 | 9.047 | 0.03142 | 8.961 | 0.02761 | 8.877 |
| 0.56105 | 0.07716 | 9.477 | 0.05268 | 9.246 | 0.04372 | 9.143 | 0.03706 | 9.047 | 0.03218 | 8.961 | 0.02809 | 8.877 |
| 0.56754 | 0.07865 | 9.477 | 0.05389 | 9.246 | 0.04457 | 9.143 | 0.03782 | 9.047 | 0.03314 | 8.961 | 0.02875 | 8.876 |
| 0.57412 | 0.08091 | 9.476 | 0.05507 | 9.246 | 0.04566 | 9.143 | 0.03886 | 9.047 | 0.03367 | 8.961 | 0.02941 | 8.876 |
| 0.58076 | 0.08269 | 9.476 | 0.05644 | 9.246 | 0.04673 | 9.143 | 0.03959 | 9.047 | 0.03431 | 8.961 | 0.03012 | 8.876 |
| 0.58749 | 0.08442 | 9.475 | 0.05777 | 9.245 | 0.04792 | 9.143 | 0.04054 | 9.047 | 0.03523 | 8.960 | 0.03087 | 8.876 |
| 0.59429 | 0.08652 | 9.475 | 0.05908 | 9.245 | 0.04914 | 9.142 | 0.04144 | 9.046 | 0.03609 | 8.960 | 0.03163 | 8.876 |
| 0.60117 | 0.08841 | 9.475 | 0.06038 | 9.245 | 0.04995 | 9.142 | 0.04233 | 9.046 | 0.03700 | 8.960 | 0.03238 | 8.876 |
| 0.60814 | 0.09052 | 9.474 | 0.06184 | 9.245 | 0.05131 | 9.142 | 0.04324 | 9.046 | 0.03783 | 8.960 | 0.03311 | 8.876 |
| 0.61518 | 0.09253 | 9.474 | 0.06332 | 9.245 | 0.05239 | 9.142 | 0.04433 | 9.046 | 0.03874 | 8.960 | 0.03384 | 8.876 |
| 0.62230 | 0.09471 | 9.473 | 0.06462 | 9.244 | 0.05383 | 9.142 | 0.04532 | 9.046 | 0.03957 | 8.960 | 0.03471 | 8.876 |
| 0.62951 | 0.09697 | 9.472 | 0.06611 | 9.244 | 0.05501 | 9.141 | 0.04646 | 9.046 | 0.04043 | 8.960 | 0.03543 | 8.876 |
| 0.63680 | 0.09940 | 9.472 | 0.06780 | 9.243 | 0.05645 | 9.141 | 0.04759 | 9.046 | 0.04135 | 8.960 | 0.03618 | 8.876 |
| 0.64417 | 0.10154 | 9.471 | 0.06928 | 9.243 | 0.05747 | 9.141 | 0.04885 | 9.046 | 0.04237 | 8.960 | 0.03697 | 8.876 |
| 0.65163 | 0.10357 | 9.470 | 0.07090 | 9.243 | 0.05893 | 9.141 | 0.05003 | 9.045 | 0.04338 | 8.959 | 0.03793 | 8.876 |
| 0.65917 | 0.10613 | 9.470 | 0.07269 | 9.243 | 0.06029 | 9.140 | 0.05114 | 9.045 | 0.04429 | 8.959 | 0.03891 | 8.876 |
| 0.66681 | 0.10865 | 9.469 | 0.07440 | 9.243 | 0.06162 | 9.140 | 0.05234 | 9.045 | 0.04532 | 8.959 | 0.03954 | 8.875 |
| 0.67453 | 0.11096 | 9.469 | 0.07615 | 9.242 | 0.06304 | 9.140 | 0.05355 | 9.045 | 0.04629 | 8.959 | 0.04061 | 8.875 |
| 0.68234 | 0.11383 | 9.468 | 0.07779 | 9.242 | 0.06470 | 9.140 | 0.05467 | 9.045 | 0.04738 | 8.959 | 0.04140 | 8.875 |
| 0.69024 | 0.11643 | 9.467 | 0.07972 | 9.242 | 0.06598 | 9.140 | 0.05600 | 9.045 | 0.04876 | 8.959 | 0.04232 | 8.875 |
| 0.69823 | 0.11904 | 9.467 | 0.08158 | 9.241 | 0.06743 | 9.139 | 0.05737 | 9.045 | 0.05014 | 8.959 | 0.04339 | 8.875 |
| 0.70632 | 0.12170 | 9.466 | 0.08342 | 9.241 | 0.06903 | 9.139 | 0.05880 | 9.044 | 0.05115 | 8.959 | 0.04438 | 8.875 |
| 0.71450 | 0.12468 | 9.465 | 0.08549 | 9.241 | 0.07064 | 9.139 | 0.06001 | 9.044 | 0.05221 | 8.958 | 0.04542 | 8.875 |
| 0.72277 | 0.12785 | 9.465 | 0.08723 | 9.240 | 0.07254 | 9.139 | 0.06134 | 9.044 | 0.05344 | 8.958 | 0.04659 | 8.875 |

| | | | | | | | | | | |
|---|---|---|---|---|---|---|---|---|---|---|
| 0.73114 | 0.13026 | 9.464 | 0.08923 | 9.240 | 0.07405 | 9.138 | 0.06276 | 9.044 | 0.05466 | 8.958 | 0.04748 | 8.874 |
| 0.73961 | 0.13341 | 9.463 | 0.09147 | 9.240 | 0.07574 | 9.138 | 0.06417 | 9.044 | 0.05626 | 8.958 | 0.04868 | 8.874 |
| 0.74817 | 0.13670 | 9.463 | 0.09352 | 9.239 | 0.07761 | 9.138 | 0.06579 | 9.044 | 0.05756 | 8.958 | 0.04978 | 8.874 |
| 0.75683 | 0.13962 | 9.462 | 0.09576 | 9.239 | 0.07922 | 9.138 | 0.06725 | 9.043 | 0.05871 | 8.958 | 0.05098 | 8.874 |
| 0.76560 | 0.14290 | 9.461 | 0.09803 | 9.239 | 0.08104 | 9.137 | 0.06872 | 9.043 | 0.05992 | 8.958 | 0.05213 | 8.874 |
| 0.77446 | 0.14614 | 9.460 | 0.10036 | 9.238 | 0.08305 | 9.137 | 0.07028 | 9.043 | 0.06110 | 8.958 | 0.05331 | 8.874 |
| 0.78343 | 0.14952 | 9.459 | 0.10265 | 9.238 | 0.08487 | 9.137 | 0.07204 | 9.043 | 0.06246 | 8.957 | 0.05473 | 8.874 |
| 0.79250 | 0.15323 | 9.459 | 0.10502 | 9.238 | 0.08685 | 9.136 | 0.07376 | 9.043 | 0.06404 | 8.957 | 0.05562 | 8.874 |
| 0.80168 | 0.15660 | 9.458 | 0.10756 | 9.237 | 0.08909 | 9.136 | 0.07549 | 9.042 | 0.06564 | 8.957 | 0.05695 | 8.873 |
| 0.81096 | 0.16026 | 9.457 | 0.11000 | 9.237 | 0.09111 | 9.136 | 0.07723 | 9.042 | 0.06717 | 8.957 | 0.05829 | 8.873 |
| 0.82035 | 0.16387 | 9.456 | 0.11234 | 9.236 | 0.09294 | 9.135 | 0.07907 | 9.042 | 0.06852 | 8.957 | 0.05960 | 8.873 |
| 0.82985 | 0.16755 | 9.455 | 0.11501 | 9.236 | 0.09553 | 9.135 | 0.08094 | 9.042 | 0.07023 | 8.957 | 0.06108 | 8.874 |
| 0.83946 | 0.17152 | 9.454 | 0.11774 | 9.235 | 0.09747 | 9.135 | 0.08285 | 9.041 | 0.07201 | 8.956 | 0.06230 | 8.874 |
| 0.84918 | 0.17536 | 9.453 | 0.12034 | 9.235 | 0.09988 | 9.134 | 0.08472 | 9.041 | 0.07362 | 8.956 | 0.06361 | 8.875 |
| 0.85901 | 0.17962 | 9.452 | 0.12308 | 9.234 | 0.10229 | 9.134 | 0.08688 | 9.041 | 0.07531 | 8.956 | 0.06548 | 8.875 |
| 0.86896 | 0.18341 | 9.451 | 0.12593 | 9.234 | 0.10471 | 9.134 | 0.08876 | 9.041 | 0.07702 | 8.956 | 0.06708 | 8.875 |
| 0.87902 | 0.18768 | 9.450 | 0.12883 | 9.234 | 0.10699 | 9.133 | 0.09074 | 9.040 | 0.07876 | 8.956 | 0.06897 | 8.876 |
| 0.88920 | 0.19227 | 9.449 | 0.13171 | 9.233 | 0.10967 | 9.133 | 0.09292 | 9.040 | 0.08065 | 8.955 | 0.07142 | 8.876 |
| 0.89950 | 0.19664 | 9.448 | 0.13467 | 9.233 | 0.11199 | 9.133 | 0.09513 | 9.040 | 0.08253 | 8.955 | 0.07314 | 8.874 |
| 0.90991 | 0.20110 | 9.447 | 0.13786 | 9.232 | 0.11443 | 9.132 | 0.09715 | 9.040 | 0.08435 | 8.955 | 0.07476 | 8.873 |
| 0.92045 | 0.20589 | 9.446 | 0.14112 | 9.232 | 0.11739 | 9.132 | 0.09934 | 9.039 | 0.08665 | 8.955 | 0.07645 | 8.872 |
| 0.93111 | 0.21040 | 9.445 | 0.14425 | 9.231 | 0.12020 | 9.131 | 0.10169 | 9.039 | 0.08873 | 8.954 | 0.07795 | 8.871 |
| 0.94189 | 0.21524 | 9.443 | 0.14754 | 9.230 | 0.12282 | 9.131 | 0.10389 | 9.039 | 0.09054 | 8.954 | 0.07941 | 8.871 |
| 0.95280 | 0.22063 | 9.442 | 0.15114 | 9.230 | 0.12559 | 9.130 | 0.10652 | 9.038 | 0.09256 | 8.954 | 0.08124 | 8.871 |
| 0.96383 | 0.22547 | 9.441 | 0.15474 | 9.229 | 0.12843 | 9.130 | 0.10902 | 9.038 | 0.09472 | 8.954 | 0.08308 | 8.871 |
| 0.97499 | 0.23080 | 9.440 | 0.15835 | 9.229 | 0.13143 | 9.129 | 0.11125 | 9.038 | 0.09688 | 8.954 | 0.08482 | 8.871 |
| 0.98628 | 0.23614 | 9.439 | 0.16207 | 9.228 | 0.13482 | 9.129 | 0.11376 | 9.037 | 0.09928 | 8.953 | 0.08684 | 8.871 |
| 0.99770 | 0.24149 | 9.437 | 0.16560 | 9.227 | 0.13782 | 9.129 | 0.11648 | 9.037 | 0.10180 | 8.953 | 0.08903 | 8.870 |
| 1.00925 | 0.24701 | 9.436 | 0.16938 | 9.227 | 0.14081 | 9.128 | 0.11940 | 9.037 | 0.10417 | 8.953 | 0.09089 | 8.870 |
| 1.02094 | 0.25280 | 9.435 | 0.17351 | 9.226 | 0.14446 | 9.128 | 0.12246 | 9.036 | 0.10640 | 8.953 | 0.09317 | 8.870 |
| 1.03276 | 0.25869 | 9.433 | 0.17750 | 9.226 | 0.14780 | 9.127 | 0.12505 | 9.036 | 0.10881 | 8.952 | 0.09521 | 8.870 |
| 1.04472 | 0.26465 | 9.432 | 0.18159 | 9.225 | 0.15107 | 9.126 | 0.12779 | 9.036 | 0.11129 | 8.952 | 0.09729 | 8.870 |
| 1.05682 | 0.27079 | 9.430 | 0.18586 | 9.224 | 0.15427 | 9.126 | 0.13085 | 9.035 | 0.11381 | 8.951 | 0.09981 | 8.869 |
| 1.06905 | 0.27692 | 9.429 | 0.19015 | 9.224 | 0.15794 | 9.125 | 0.13393 | 9.035 | 0.11642 | 8.951 | 0.10207 | 8.869 |
| 1.08143 | 0.28328 | 9.427 | 0.19439 | 9.223 | 0.16193 | 9.125 | 0.13663 | 9.035 | 0.11910 | 8.951 | 0.10441 | 8.869 |
| 1.09396 | 0.28998 | 9.426 | 0.19877 | 9.222 | 0.16575 | 9.124 | 0.13975 | 9.034 | 0.12183 | 8.951 | 0.10680 | 8.869 |
| 1.10662 | 0.29620 | 9.424 | 0.20345 | 9.221 | 0.16944 | 9.123 | 0.14325 | 9.034 | 0.12455 | 8.950 | 0.10914 | 8.869 |
| 1.11944 | 0.30290 | 9.422 | 0.20797 | 9.221 | 0.17338 | 9.123 | 0.14566 | 9.034 | 0.12740 | 8.950 | 0.11172 | 8.868 |
| 1.13240 | 0.31002 | 9.421 | 0.21289 | 9.220 | 0.17750 | 9.122 | 0.14963 | 9.035 | 0.13045 | 8.950 | 0.11430 | 8.868 |
| 1.14551 | 0.31725 | 9.419 | 0.21798 | 9.219 | 0.18172 | 9.121 | 0.15368 | 9.035 | 0.13323 | 8.949 | 0.11689 | 8.868 |
| 1.15878 | 0.32445 | 9.417 | 0.22308 | 9.218 | 0.18570 | 9.121 | 0.15813 | 9.036 | 0.13600 | 8.950 | 0.11982 | 8.868 |
| 1.17220 | 0.33190 | 9.416 | 0.22835 | 9.218 | 0.19001 | 9.121 | 0.16164 | 9.033 | 0.13922 | 8.950 | 0.12249 | 8.867 |
| 1.18577 | 0.33966 | 9.414 | 0.23343 | 9.217 | 0.19435 | 9.120 | 0.16591 | 9.032 | 0.14297 | 8.951 | 0.12533 | 8.867 |
| 1.19950 | 0.34739 | 9.412 | 0.23874 | 9.216 | 0.19882 | 9.120 | 0.16933 | 9.031 | 0.14684 | 8.951 | 0.12822 | 8.867 |
| 1.21339 | 0.35534 | 9.410 | 0.24420 | 9.215 | 0.20366 | 9.118 | 0.17291 | 9.030 | 0.15084 | 8.950 | 0.13132 | 8.867 |
| 1.22744 | 0.36307 | 9.408 | 0.24979 | 9.214 | 0.20847 | 9.119 | 0.17692 | 9.029 | 0.15452 | 8.948 | 0.13413 | 8.866 |
| 1.24165 | 0.37154 | 9.406 | 0.25570 | 9.213 | 0.21344 | 9.117 | 0.18064 | 9.029 | 0.15821 | 8.947 | 0.13704 | 8.866 |

| | | | | | | | | | | | |
|---|---|---|---|---|---|---|---|---|---|---|---|
| 1.25603 | 0.38027 | 9.404 | 0.26158 | 9.212 | 0.21837 | 9.117 | 0.18531 | 9.028 | 0.16201 | 8.946 | 0.14032 | 8.866 |
| 1.27057 | 0.38904 | 9.402 | 0.26754 | 9.211 | 0.22327 | 9.115 | 0.18943 | 9.028 | 0.16532 | 8.946 | 0.14368 | 8.865 |
| 1.28529 | 0.39800 | 9.400 | 0.27398 | 9.210 | 0.22944 | 9.114 | 0.19365 | 9.028 | 0.16892 | 8.945 | 0.14693 | 8.865 |
| 1.30017 | 0.40699 | 9.398 | 0.28037 | 9.209 | 0.23428 | 9.113 | 0.19814 | 9.027 | 0.17298 | 8.945 | 0.15009 | 8.865 |
| 1.31522 | 0.41612 | 9.395 | 0.28677 | 9.208 | 0.23960 | 9.112 | 0.20244 | 9.026 | 0.17695 | 8.944 | 0.15377 | 8.865 |
| 1.33045 | 0.42556 | 9.393 | 0.29333 | 9.207 | 0.24485 | 9.112 | 0.20742 | 9.026 | 0.18087 | 8.944 | 0.15751 | 8.865 |
| 1.34586 | 0.43538 | 9.391 | 0.29985 | 9.205 | 0.25050 | 9.111 | 0.21224 | 9.025 | 0.18490 | 8.943 | 0.16149 | 8.865 |
| 1.36144 | 0.44557 | 9.388 | 0.30650 | 9.204 | 0.25614 | 9.110 | 0.21713 | 9.025 | 0.18907 | 8.943 | 0.16543 | 8.864 |
| 1.37721 | 0.45567 | 9.386 | 0.31368 | 9.203 | 0.26186 | 9.109 | 0.22178 | 9.024 | 0.19349 | 8.943 | 0.16929 | 8.863 |
| 1.39316 | 0.46592 | 9.383 | 0.32088 | 9.202 | 0.26787 | 9.108 | 0.22693 | 9.024 | 0.19779 | 8.942 | 0.17334 | 8.863 |
| 1.40929 | 0.47669 | 9.381 | 0.32827 | 9.200 | 0.27440 | 9.107 | 0.23214 | 9.023 | 0.20237 | 8.942 | 0.17714 | 8.862 |
| 1.42561 | 0.48801 | 9.378 | 0.33589 | 9.199 | 0.28062 | 9.106 | 0.23753 | 9.022 | 0.20725 | 8.941 | 0.18112 | 8.862 |
| 1.44212 | 0.49885 | 9.375 | 0.34350 | 9.198 | 0.28692 | 9.105 | 0.24291 | 9.022 | 0.21238 | 8.941 | 0.18527 | 8.861 |
| 1.45881 | 0.51010 | 9.373 | 0.35133 | 9.197 | 0.29405 | 9.104 | 0.24856 | 9.021 | 0.21712 | 8.940 | 0.18955 | 8.861 |
| 1.47571 | 0.52191 | 9.370 | 0.35936 | 9.196 | 0.30043 | 9.103 | 0.25457 | 9.020 | 0.22193 | 8.940 | 0.19380 | 8.860 |
| 1.49279 | 0.53345 | 9.367 | 0.36769 | 9.194 | 0.30756 | 9.102 | 0.26021 | 9.019 | 0.22692 | 8.939 | 0.19847 | 8.860 |
| 1.51008 | 0.54592 | 9.364 | 0.37620 | 9.193 | 0.31460 | 9.101 | 0.26627 | 9.019 | 0.23216 | 8.938 | 0.20290 | 8.859 |
| 1.52757 | 0.55799 | 9.361 | 0.38479 | 9.192 | 0.32174 | 9.100 | 0.27234 | 9.018 | 0.23755 | 8.938 | 0.20743 | 8.859 |
| 1.54525 | 0.57092 | 9.358 | 0.39359 | 9.190 | 0.32934 | 9.099 | 0.27861 | 9.017 | 0.24289 | 8.937 | 0.21218 | 8.859 |
| 1.56315 | 0.58377 | 9.355 | 0.40273 | 9.189 | 0.33693 | 9.097 | 0.28496 | 9.016 | 0.24853 | 8.937 | 0.21716 | 8.858 |
| 1.58125 | 0.59683 | 9.352 | 0.41213 | 9.187 | 0.34480 | 9.096 | 0.29185 | 9.015 | 0.25425 | 8.936 | 0.22201 | 8.858 |
| 1.59956 | 0.61052 | 9.349 | 0.42145 | 9.186 | 0.35282 | 9.095 | 0.29857 | 9.015 | 0.26012 | 8.935 | 0.22720 | 8.857 |
| 1.61808 | 0.62417 | 9.346 | 0.43091 | 9.184 | 0.36081 | 9.094 | 0.30520 | 9.014 | 0.26625 | 8.935 | 0.23216 | 8.857 |
| 1.63682 | 0.63827 | 9.342 | 0.44089 | 9.183 | 0.36910 | 9.092 | 0.31248 | 9.013 | 0.27251 | 8.934 | 0.23764 | 8.857 |
| 1.65577 | 0.65278 | 9.339 | 0.45107 | 9.181 | 0.37772 | 9.091 | 0.31969 | 9.012 | 0.27887 | 8.933 | 0.24300 | 8.857 |
| 1.67494 | 0.66757 | 9.335 | 0.46122 | 9.179 | 0.38609 | 9.090 | 0.32719 | 9.011 | 0.28525 | 8.933 | 0.24861 | 8.856 |
| 1.69434 | 0.68252 | 9.332 | 0.47167 | 9.178 | 0.39494 | 9.089 | 0.33492 | 9.010 | 0.29185 | 8.932 | 0.25448 | 8.856 |
| 1.71396 | 0.69798 | 9.328 | 0.48238 | 9.176 | 0.40411 | 9.087 | 0.34249 | 9.009 | 0.29863 | 8.931 | 0.26083 | 8.856 |
| 1.73380 | 0.71366 | 9.324 | 0.49346 | 9.174 | 0.41329 | 9.086 | 0.35011 | 9.008 | 0.30558 | 8.930 | 0.26750 | 8.856 |
| 1.75388 | 0.72985 | 9.320 | 0.50486 | 9.172 | 0.42271 | 9.084 | 0.35818 | 9.007 | 0.31284 | 8.930 | 0.27444 | 8.855 |
| 1.77419 | 0.74637 | 9.316 | 0.51652 | 9.170 | 0.43265 | 9.083 | 0.36667 | 9.006 | 0.32005 | 8.929 | 0.28115 | 8.854 |
| 1.79473 | 0.76318 | 9.312 | 0.52809 | 9.169 | 0.44272 | 9.081 | 0.37496 | 9.005 | 0.32732 | 8.928 | 0.28757 | 8.852 |
| 1.81552 | 0.78027 | 9.308 | 0.54032 | 9.167 | 0.45299 | 9.080 | 0.38367 | 9.004 | 0.33487 | 8.927 | 0.29405 | 8.851 |
| 1.83654 | 0.79762 | 9.304 | 0.55283 | 9.165 | 0.46304 | 9.078 | 0.39272 | 9.003 | 0.34272 | 8.926 | 0.30043 | 8.850 |
| 1.85780 | 0.81573 | 9.300 | 0.56550 | 9.163 | 0.47363 | 9.076 | 0.40180 | 9.002 | 0.35048 | 8.925 | 0.30733 | 8.849 |
| 1.87932 | 0.83404 | 9.295 | 0.57855 | 9.161 | 0.48485 | 9.075 | 0.41086 | 9.001 | 0.35847 | 8.924 | 0.31428 | 8.849 |
| 1.90108 | 0.85233 | 9.291 | 0.59167 | 9.159 | 0.49590 | 9.073 | 0.42038 | 9.000 | 0.36709 | 8.923 | 0.32144 | 8.848 |
| 1.92309 | 0.87188 | 9.286 | 0.60500 | 9.157 | 0.50714 | 9.071 | 0.43029 | 8.998 | 0.37555 | 8.922 | 0.32883 | 8.847 |
| 1.94536 | 0.89125 | 9.282 | 0.61878 | 9.154 | 0.51896 | 9.069 | 0.43990 | 8.997 | 0.38393 | 8.922 | 0.33623 | 8.846 |
| 1.96789 | 0.91109 | 9.277 | 0.63295 | 9.152 | 0.53100 | 9.067 | 0.45005 | 8.996 | 0.39270 | 8.921 | 0.34408 | 8.846 |
| 1.99067 | 0.93173 | 9.272 | 0.64741 | 9.150 | 0.54309 | 9.066 | 0.46056 | 8.995 | 0.40182 | 8.920 | 0.35195 | 8.845 |
| 2.01372 | 0.95245 | 9.267 | 0.66207 | 9.147 | 0.55561 | 9.064 | 0.47108 | 8.994 | 0.41104 | 8.919 | 0.36001 | 8.844 |
| 2.03704 | 0.97339 | 9.262 | 0.67731 | 9.145 | 0.56836 | 9.062 | 0.48170 | 8.992 | 0.42071 | 8.918 | 0.36832 | 8.843 |
| 2.06063 | 0.99508 | 9.257 | 0.69271 | 9.143 | 0.58116 | 9.060 | 0.49308 | 8.991 | 0.43038 | 8.917 | 0.37677 | 8.843 |
| 2.08449 | 1.01721 | 9.252 | 0.70833 | 9.140 | 0.59477 | 9.058 | 0.50461 | 8.990 | 0.44031 | 8.916 | 0.38547 | 8.842 |
| 2.10863 | 1.03991 | 9.246 | 0.72442 | 9.138 | 0.60821 | 9.056 | 0.51627 | 8.988 | 0.45054 | 8.915 | 0.39420 | 8.841 |
| 2.13304 | 1.06311 | 9.241 | 0.74085 | 9.135 | 0.62196 | 9.053 | 0.52828 | 8.987 | 0.46071 | 8.913 | 0.40337 | 8.840 |

| | | | | | | | | | | | |
|---|---|---|---|---|---|---|---|---|---|---|---|
| 2.15774 | 1.08671 | 9.235 | 0.75757 | 9.132 | 0.63625 | 9.051 | 0.54042 | 8.985 | 0.47141 | 8.912 | 0.41232 | 8.839 |
| 2.18273 | 1.11081 | 9.229 | 0.77475 | 9.130 | 0.65062 | 9.049 | 0.55301 | 8.984 | 0.48238 | 8.911 | 0.42190 | 8.839 |
| 2.20800 | 1.13545 | 9.223 | 0.79230 | 9.127 | 0.66571 | 9.047 | 0.56571 | 8.982 | 0.49324 | 8.910 | 0.43181 | 8.838 |
| 2.23357 | 1.16039 | 9.218 | 0.81039 | 9.124 | 0.68130 | 9.045 | 0.57865 | 8.981 | 0.50471 | 8.909 | 0.44185 | 8.837 |
| 2.25944 | 1.18604 | 9.211 | 0.82899 | 9.121 | 0.69684 | 9.042 | 0.59209 | 8.979 | 0.51652 | 8.907 | 0.45195 | 8.836 |
| 2.28560 | 1.21214 | 9.205 | 0.84782 | 9.118 | 0.71253 | 9.040 | 0.60553 | 8.977 | 0.52844 | 8.906 | 0.46262 | 8.835 |
| 2.31206 | 1.23756 | 9.199 | 0.86690 | 9.115 | 0.72869 | 9.037 | 0.61950 | 8.976 | 0.54053 | 8.905 | 0.47324 | 8.834 |
| 2.33884 | 1.26449 | 9.192 | 0.88633 | 9.112 | 0.74559 | 9.035 | 0.63370 | 8.974 | 0.55285 | 8.903 | 0.48405 | 8.833 |
| 2.36592 | 1.29245 | 9.186 | 0.90628 | 9.109 | 0.76270 | 9.032 | 0.64802 | 8.972 | 0.56567 | 8.902 | 0.49548 | 8.832 |
| 2.39332 | 1.32070 | 9.179 | 0.92661 | 9.105 | 0.77996 | 9.029 | 0.66313 | 8.970 | 0.57893 | 8.901 | 0.50666 | 8.830 |
| 2.42103 | 1.34959 | 9.172 | 0.94733 | 9.102 | 0.79771 | 9.027 | 0.67857 | 8.968 | 0.59235 | 8.899 | 0.51802 | 8.829 |
| 2.44906 | 1.37774 | 9.165 | 0.96869 | 9.099 | 0.81617 | 9.024 | 0.69380 | 8.966 | 0.60613 | 8.898 | 0.53002 | 8.828 |
| 2.47742 | 1.40812 | 9.158 | 0.99071 | 9.095 | 0.83443 | 9.021 | 0.71001 | 8.964 | 0.61998 | 8.896 | 0.54218 | 8.827 |
| 2.50611 | 1.43733 | 9.150 | 1.01302 | 9.092 | 0.85346 | 9.018 | 0.72621 | 8.962 | 0.63416 | 8.894 | 0.55469 | 8.826 |
| 2.53513 | 1.46840 | 9.143 | 1.03578 | 9.088 | 0.87278 | 9.015 | 0.74274 | 8.960 | 0.64873 | 8.893 | 0.56735 | 8.825 |
| 2.56448 | 1.50037 | 9.135 | 1.05895 | 9.084 | 0.89291 | 9.012 | 0.75958 | 8.958 | 0.66351 | 8.891 | 0.58077 | 8.824 |
| 2.59418 | 1.53304 | 9.127 | 1.08252 | 9.080 | 0.91323 | 9.009 | 0.77669 | 8.956 | 0.67875 | 8.889 | 0.59447 | 8.823 |
| 2.62422 | 1.56622 | 9.120 | 1.10680 | 9.076 | 0.93385 | 9.006 | 0.79458 | 8.954 | 0.69445 | 8.887 | 0.60833 | 8.822 |
| 2.65461 | 1.60051 | 9.111 | 1.13170 | 9.072 | 0.95519 | 9.002 | 0.81272 | 8.951 | 0.71042 | 8.886 | 0.62250 | 8.820 |
| 2.68534 | 1.63504 | 9.103 | 1.15690 | 9.068 | 0.97646 | 8.999 | 0.83127 | 8.949 | 0.72657 | 8.884 | 0.63723 | 8.819 |
| 2.71644 | 1.67010 | 9.095 | 1.18263 | 9.064 | 0.99876 | 8.996 | 0.85026 | 8.947 | 0.74316 | 8.882 | 0.65202 | 8.817 |
| 2.74789 | 1.70596 | 9.086 | 1.20910 | 9.060 | 1.02125 | 8.992 | 0.86940 | 8.944 | 0.76004 | 8.880 | 0.66684 | 8.815 |
| 2.77971 | 1.74261 | 9.078 | 1.23613 | 9.056 | 1.04459 | 8.989 | 0.88893 | 8.942 | 0.77719 | 8.878 | 0.68222 | 8.814 |
| 2.81190 | 1.78029 | 9.069 | 1.26361 | 9.051 | 1.06794 | 8.985 | 0.90928 | 8.939 | 0.79515 | 8.876 | 0.69748 | 8.812 |
| 2.84446 | 1.81861 | 9.060 | 1.29171 | 9.047 | 1.09280 | 8.981 | 0.92985 | 8.937 | 0.81331 | 8.874 | 0.71349 | 8.811 |
| 2.87740 | 1.85762 | 9.050 | 1.32024 | 9.043 | 1.11678 | 8.978 | 0.95136 | 8.934 | 0.83198 | 8.872 | 0.73011 | 8.809 |
| 2.91072 | 1.89720 | 9.041 | 1.34970 | 9.038 | 1.14241 | 8.974 | 0.97272 | 8.931 | 0.85116 | 8.870 | 0.74647 | 8.807 |
| 2.94442 | 1.93797 | 9.032 | 1.37970 | 9.033 | 1.16780 | 8.970 | 0.99487 | 8.929 | 0.87049 | 8.868 | 0.76357 | 8.806 |
| 2.97852 | 1.97882 | 9.022 | 1.41050 | 9.029 | 1.19397 | 8.966 | 1.01798 | 8.926 | 0.89013 | 8.866 | 0.78121 | 8.804 |
| 3.01301 | 2.01944 | 9.012 | 1.44186 | 9.024 | 1.22107 | 8.962 | 1.04116 | 8.923 | 0.91050 | 8.864 | 0.79933 | 8.802 |
| 3.04789 | 2.06232 | 9.001 | 1.47382 | 9.019 | 1.24844 | 8.958 | 1.06443 | 8.920 | 0.93143 | 8.861 | 0.81733 | 8.801 |
| 3.08319 | 2.10626 | 8.991 | 1.50659 | 9.014 | 1.27661 | 8.953 | 1.08837 | 8.917 | 0.95256 | 8.859 | 0.83624 | 8.799 |
| 3.11889 | 2.15047 | 8.981 | 1.53987 | 9.008 | 1.30538 | 8.949 | 1.11314 | 8.914 | 0.97445 | 8.857 | 0.85521 | 8.797 |
| 3.15500 | 2.19609 | 8.970 | 1.57400 | 9.003 | 1.33477 | 8.945 | 1.13871 | 8.911 | 0.99651 | 8.854 | 0.87462 | 8.795 |
| 3.19154 | 2.24230 | 8.959 | 1.60891 | 8.997 | 1.36470 | 8.940 | 1.16445 | 8.908 | 1.01883 | 8.852 | 0.89470 | 8.793 |
| 3.22849 | 2.28966 | 8.948 | 1.64449 | 8.992 | 1.39492 | 8.935 | 1.19052 | 8.905 | 1.04220 | 8.849 | 0.91521 | 8.791 |
| 3.26588 | 2.33717 | 8.937 | 1.68055 | 8.986 | 1.42647 | 8.931 | 1.21710 | 8.901 | 1.06594 | 8.846 | 0.93605 | 8.789 |
| 3.30370 | 2.38611 | 8.926 | 1.71724 | 8.980 | 1.45832 | 8.926 | 1.24507 | 8.898 | 1.09024 | 8.844 | 0.95743 | 8.787 |
| 3.34195 | 2.43633 | 8.914 | 1.75482 | 8.974 | 1.49095 | 8.921 | 1.27262 | 8.894 | 1.11485 | 8.841 | 0.97923 | 8.785 |
| 3.38065 | 2.48711 | 8.902 | 1.79368 | 8.968 | 1.52387 | 8.916 | 1.30101 | 8.891 | 1.13984 | 8.838 | 1.00183 | 8.783 |
| 3.41979 | 2.53827 | 8.890 | 1.83319 | 8.961 | 1.55780 | 8.912 | 1.33093 | 8.887 | 1.16611 | 8.835 | 1.02466 | 8.781 |
| 3.45939 | 2.59046 | 8.878 | 1.87312 | 8.955 | 1.59183 | 8.910 | 1.36099 | 8.883 | 1.19285 | 8.833 | 1.04804 | 8.779 |
| 3.49945 | 2.64323 | 8.865 | 1.91391 | 8.948 | 1.62687 | 8.906 | 1.39159 | 8.880 | 1.21964 | 8.830 | 1.07225 | 8.776 |
| 3.53997 | 2.69763 | 8.852 | 1.95542 | 8.942 | 1.66357 | 8.900 | 1.42291 | 8.876 | 1.24731 | 8.827 | 1.09660 | 8.774 |
| 3.58096 | 2.75069 | 8.839 | 1.99805 | 8.935 | 1.70026 | 8.896 | 1.45502 | 8.872 | 1.27543 | 8.824 | 1.12163 | 8.772 |
| 3.62243 | 2.80687 | 8.826 | 2.04175 | 8.928 | 1.73791 | 8.890 | 1.48732 | 8.868 | 1.30428 | 8.821 | 1.14723 | 8.769 |
| 3.66438 | 2.86415 | 8.813 | 2.08609 | 8.921 | 1.77678 | 8.883 | 1.52050 | 8.864 | 1.33386 | 8.817 | 1.17332 | 8.767 |

| | | | | | | | | | | | |
|---|---|---|---|---|---|---|---|---|---|---|---|
| 3.70681 | 2.92252 | 8.799 | 2.13136 | 8.914 | 1.81577 | 8.878 | 1.55515 | 8.860 | 1.36405 | 8.814 | 1.19998 | 8.764 |
| 3.74973 | 2.98202 | 8.785 | 2.17732 | 8.906 | 1.85604 | 8.872 | 1.58997 | 8.855 | 1.39515 | 8.811 | 1.22687 | 8.762 |
| 3.79315 | 3.04236 | 8.771 | 2.22430 | 8.899 | 1.89687 | 8.868 | 1.62554 | 8.851 | 1.42667 | 8.807 | 1.25477 | 8.759 |
| 3.83707 | 3.10419 | 8.757 | 2.27211 | 8.891 | 1.93846 | 8.863 | 1.66176 | 8.847 | 1.45863 | 8.804 | 1.28380 | 8.756 |
| 3.88150 | 3.16661 | 8.742 | 2.32077 | 8.883 | 1.98088 | 8.856 | 1.69829 | 8.842 | 1.49144 | 8.800 | 1.31245 | 8.753 |
| 3.92645 | 3.23022 | 8.728 | 2.37051 | 8.876 | 2.02437 | 8.851 | 1.73669 | 8.837 | 1.52544 | 8.796 | 1.34228 | 8.750 |
| 3.97192 | 3.29505 | 8.713 | 2.42136 | 8.867 | 2.06885 | 8.845 | 1.77496 | 8.833 | 1.55968 | 8.792 | 1.37277 | 8.748 |
| 4.01791 | 3.35803 | 8.697 | 2.47337 | 8.859 | 2.11469 | 8.837 | 1.81456 | 8.828 | 1.59485 | 8.789 | 1.40346 | 8.745 |
| 4.06443 | 3.42500 | 8.682 | 2.52674 | 8.851 | 2.16145 | 8.830 | 1.85575 | 8.823 | 1.63084 | 8.785 | 1.43582 | 8.742 |
| 4.11150 | 3.49285 | 8.666 | 2.58051 | 8.842 | 2.20774 | 8.826 | 1.89624 | 8.818 | 1.66757 | 8.781 | 1.46786 | 8.739 |
| 4.15911 | 3.56252 | 8.650 | 2.63519 | 8.833 | 2.25613 | 8.819 | 1.93861 | 8.813 | 1.70504 | 8.776 | 1.50114 | 8.735 |
| 4.20727 | 3.63132 | 8.634 | 2.69100 | 8.824 | 2.30533 | 8.811 | 1.98218 | 8.807 | 1.74295 | 8.772 | 1.53513 | 8.732 |
| 4.25598 | 3.70245 | 8.617 | 2.74802 | 8.815 | 2.35560 | 8.804 | 2.02553 | 8.802 | 1.78162 | 8.768 | 1.56986 | 8.729 |
| 4.30527 | 3.77564 | 8.600 | 2.80620 | 8.806 | 2.40742 | 8.796 | 2.07024 | 8.796 | 1.82169 | 8.764 | 1.60538 | 8.725 |
| 4.35512 | 3.84944 | 8.584 | 2.86570 | 8.796 | 2.46003 | 8.788 | 2.11642 | 8.790 | 1.86321 | 8.759 | 1.64151 | 8.722 |
| 4.40555 | 3.92224 | 8.566 | 2.92626 | 8.787 | 2.51330 | 8.780 | 2.16207 | 8.785 | 1.90477 | 8.755 | 1.67854 | 8.718 |
| 4.45656 | 3.99826 | 8.548 | 2.98785 | 8.777 | 2.56766 | 8.772 | 2.21050 | 8.779 | 1.94674 | 8.750 | 1.71629 | 8.715 |
| 4.50817 | 4.07623 | 8.531 | 3.05087 | 8.767 | 2.62229 | 8.767 | 2.25928 | 8.773 | 1.99020 | 8.745 | 1.75519 | 8.711 |
| 4.56037 | 4.15441 | 8.513 | 3.11445 | 8.757 | 2.67864 | 8.759 | 2.30871 | 8.767 | 2.03439 | 8.740 | 1.79466 | 8.707 |
| 4.61318 | 4.23441 | 8.495 | 3.17949 | 8.746 | 2.73676 | 8.750 | 2.35911 | 8.761 | 2.07943 | 8.736 | 1.83495 | 8.703 |
| 4.66659 | 4.31530 | 8.476 | 3.24586 | 8.736 | 2.79606 | 8.741 | 2.41128 | 8.755 | 2.12564 | 8.731 | 1.87618 | 8.700 |
| 4.72063 | 4.39559 | 8.457 | 3.31323 | 8.725 | 2.85598 | 8.733 | 2.46423 | 8.748 | 2.17269 | 8.725 | 1.91861 | 8.695 |
| 4.77529 | 4.47857 | 8.438 | 3.38163 | 8.714 | 2.91746 | 8.724 | 2.51799 | 8.742 | 2.22041 | 8.720 | 1.96104 | 8.691 |
| 4.83059 | 4.56360 | 8.419 | 3.45144 | 8.703 | 2.97923 | 8.717 | 2.57329 | 8.735 | 2.26985 | 8.715 | 2.00490 | 8.686 |
| 4.88652 | 4.65004 | 8.400 | 3.52278 | 8.692 | 3.04337 | 8.705 | 2.62925 | 8.728 | 2.32068 | 8.709 | 2.05023 | 8.682 |
| 4.94311 | 4.73628 | 8.380 | 3.59547 | 8.680 | 3.10841 | 8.698 | 2.68664 | 8.721 | 2.37258 | 8.704 | 2.09611 | 8.677 |
| 5.00035 | 4.82437 | 8.360 | 3.66937 | 8.669 | 3.17510 | 8.686 | 2.74480 | 8.714 | 2.42465 | 8.698 | 2.14304 | 8.672 |
| 5.05825 | 4.91488 | 8.340 | 3.74436 | 8.657 | 3.24279 | 8.676 | 2.80454 | 8.706 | 2.47798 | 8.692 | 2.19054 | 8.668 |
| 5.11682 | 5.00600 | 8.319 | 3.82099 | 8.644 | 3.31014 | 8.669 | 2.86581 | 8.699 | 2.53257 | 8.686 | 2.23950 | 8.663 |
| 5.17607 | 5.09810 | 8.298 | 3.89883 | 8.632 | 3.38008 | 8.660 | 2.92801 | 8.691 | 2.58793 | 8.680 | 2.28954 | 8.657 |
| 5.23600 | 5.19144 | 8.277 | 3.97809 | 8.620 | 3.45188 | 8.650 | 2.99106 | 8.684 | 2.64516 | 8.674 | 2.34054 | 8.653 |
| 5.29663 | 5.28627 | 8.256 | 4.05877 | 8.607 | 3.52453 | 8.639 | 3.05540 | 8.676 | 2.70333 | 8.667 | 2.39269 | 8.648 |
| 5.35797 | 5.38284 | 8.235 | 4.14059 | 8.594 | 3.59979 | 8.628 | 3.12184 | 8.668 | 2.76239 | 8.661 | 2.44607 | 8.642 |
| 5.42001 | 5.47999 | 8.213 | 4.22420 | 8.581 | 3.67499 | 8.617 | 3.18826 | 8.659 | 2.82255 | 8.654 | 2.50070 | 8.637 |
| 5.48277 | 5.57867 | 8.191 | 4.30906 | 8.567 | 3.75313 | 8.605 | 3.25680 | 8.651 | 2.88397 | 8.648 | 2.55583 | 8.631 |
| 5.54626 | 5.67905 | 8.169 | 4.39445 | 8.554 | 3.83006 | 8.597 | 3.32636 | 8.643 | 2.94698 | 8.641 | 2.61212 | 8.626 |
| 5.61048 | 5.77974 | 8.146 | 4.48215 | 8.540 | 3.91093 | 8.584 | 3.39813 | 8.634 | 3.01147 | 8.634 | 2.66999 | 8.620 |
| 5.67545 | 5.88185 | 8.123 | 4.57168 | 8.526 | 3.99129 | 8.575 | 3.47071 | 8.625 | 3.07722 | 8.627 | 2.72887 | 8.615 |
| 5.74116 | 5.98747 | 8.100 | 4.66243 | 8.512 | 4.07273 | 8.564 | 3.54518 | 8.616 | 3.14403 | 8.620 | 2.78900 | 8.609 |
| 5.80764 | 6.09113 | 8.077 | 4.75430 | 8.497 | 4.15713 | 8.551 | 3.62007 | 8.607 | 3.21218 | 8.612 | 2.85090 | 8.603 |
| 5.87489 | 6.19743 | 8.053 | 4.84834 | 8.483 | 4.24419 | 8.537 | 3.69674 | 8.598 | 3.28100 | 8.605 | 2.91319 | 8.597 |
| 5.94292 | 6.30451 | 8.029 | 4.94431 | 8.468 | 4.33133 | 8.526 | 3.77450 | 8.588 | 3.35171 | 8.597 | 2.97948 | 8.590 |
| 6.01174 | 6.41368 | 8.005 | 5.04113 | 8.453 | 4.41903 | 8.515 | 3.85458 | 8.578 | 3.42449 | 8.589 | 3.04469 | 8.584 |
| 6.08135 | 6.52593 | 7.981 | 5.13949 | 8.437 | 4.51051 | 8.502 | 3.93629 | 8.569 | 3.49848 | 8.581 | 3.11231 | 8.577 |
| 6.15177 | 6.63802 | 7.956 | 5.23921 | 8.421 | 4.60285 | 8.488 | 4.01975 | 8.559 | 3.57402 | 8.573 | 3.18030 | 8.571 |
| 6.22300 | 6.75034 | 7.932 | 5.34087 | 8.405 | 4.69689 | 8.474 | 4.10370 | 8.549 | 3.65109 | 8.565 | 3.25001 | 8.564 |
| 6.29506 | 6.86529 | 7.906 | 5.44447 | 8.389 | 4.79268 | 8.461 | 4.19002 | 8.538 | 3.72939 | 8.556 | 3.32066 | 8.557 |

| | | | | | | | | | | | |
|---|---|---|---|---|---|---|---|---|---|---|---|
| 6.36796 | 6.98129 | 7.881 | 5.54968 | 8.373 | 4.89006 | 8.447 | 4.27692 | 8.528 | 3.80868 | 8.548 | 3.39387 | 8.549 |
| 6.44169 | 7.09769 | 7.856 | 5.65630 | 8.356 | 4.98891 | 8.433 | 4.36661 | 8.517 | 3.88980 | 8.539 | 3.46787 | 8.542 |
| 6.51628 | 7.21692 | 7.830 | 5.76442 | 8.339 | 5.08926 | 8.420 | 4.45750 | 8.506 | 3.97295 | 8.530 | 3.54296 | 8.535 |
| 6.59174 | 7.33693 | 7.804 | 5.87452 | 8.322 | 5.19204 | 8.407 | 4.55027 | 8.495 | 4.05777 | 8.521 | 3.61938 | 8.528 |
| 6.66807 | 7.45846 | 7.778 | 5.98635 | 8.304 | 5.29700 | 8.390 | 4.64381 | 8.483 | 4.14368 | 8.512 | 3.69851 | 8.520 |
| 6.74528 | 7.58077 | 7.751 | 6.09959 | 8.287 | 5.40260 | 8.375 | 4.74088 | 8.472 | 4.23103 | 8.502 | 3.77904 | 8.512 |
| 6.82339 | 7.70458 | 7.725 | 6.21448 | 8.269 | 5.50952 | 8.361 | 4.83928 | 8.460 | 4.32071 | 8.493 | 3.85980 | 8.505 |
| 6.90240 | 7.82985 | 7.698 | 6.33097 | 8.251 | 5.61834 | 8.348 | 4.93860 | 8.448 | 4.41235 | 8.483 | 3.94367 | 8.498 |
| 6.98232 | 7.95639 | 7.671 | 6.44916 | 8.232 | 5.72950 | 8.332 | 5.03998 | 8.436 | 4.50514 | 8.473 | 4.02812 | 8.490 |
| 7.06318 | 8.08405 | 7.644 | 6.56886 | 8.213 | 5.84410 | 8.314 | 5.14394 | 8.423 | 4.60005 | 8.463 | 4.11487 | 8.481 |
| 7.14496 | 8.21327 | 7.617 | 6.69056 | 8.195 | 5.95888 | 8.299 | 5.24779 | 8.411 | 4.69644 | 8.452 | 4.20504 | 8.474 |
| 7.22770 | 8.34302 | 7.589 | 6.81435 | 8.175 | 6.07503 | 8.283 | 5.35500 | 8.398 | 4.79497 | 8.442 | 4.29466 | 8.465 |
| 7.31139 | 8.47536 | 7.561 | 6.93983 | 8.156 | 6.19500 | 8.266 | 5.46384 | 8.385 | 4.89528 | 8.431 | 4.38671 | 8.456 |
| 7.39605 | 8.60787 | 7.533 | 7.06729 | 8.136 | 6.31402 | 8.251 | 5.57479 | 8.372 | 4.99723 | 8.420 | 4.48002 | 8.447 |
| 7.48170 | 8.74201 | 7.505 | 7.19625 | 8.116 | 6.43731 | 8.234 | 5.68774 | 8.359 | 5.10144 | 8.409 | 4.57557 | 8.439 |
| 7.56833 | 8.87825 | 7.477 | 7.32676 | 8.096 | 6.56216 | 8.216 | 5.80191 | 8.345 | 5.20771 | 8.398 | 4.67309 | 8.429 |
| 7.65597 | 9.01469 | 7.448 | 7.45922 | 8.075 | 6.68778 | 8.199 | 5.91926 | 8.331 | 5.31581 | 8.386 | 4.77254 | 8.420 |
| 7.74462 | 9.15162 | 7.420 | 7.59364 | 8.055 | 6.81610 | 8.182 | 6.03724 | 8.317 | 5.42494 | 8.375 | 4.87315 | 8.411 |
| 7.83430 | 9.29154 | 7.391 | 7.73013 | 8.034 | 6.94488 | 8.166 | 6.15823 | 8.302 | 5.53668 | 8.363 | 4.97413 | 8.401 |
| 7.92501 | 9.43232 | 7.361 | 7.86795 | 8.013 | 7.07777 | 8.148 | 6.28096 | 8.288 | 5.65068 | 8.351 | 5.07915 | 8.391 |
| 8.01678 | 9.57274 | 7.333 | 8.00670 | 7.991 | 7.20570 | 8.136 | 6.40584 | 8.273 | 5.76642 | 8.338 | 5.18677 | 8.381 |
| 8.10961 | 9.71521 | 7.303 | 8.14815 | 7.969 | 7.34720 | 8.113 | 6.53188 | 8.258 | 5.88377 | 8.326 | 5.29725 | 8.371 |
| 8.20352 | 9.85950 | 7.273 | 8.29135 | 7.948 | 7.48468 | 8.095 | 6.66083 | 8.243 | 6.00372 | 8.313 | 5.40673 | 8.361 |
| 8.29851 | 10.0049 | 7.244 | 8.43592 | 7.925 | 7.62455 | 8.077 | 6.79059 | 8.227 | 6.12644 | 8.300 | 5.51933 | 8.350 |
| 8.39460 | 10.1501 | 7.214 | 8.58327 | 7.903 | 7.76610 | 8.058 | 6.92536 | 8.212 | 6.25121 | 8.287 | 5.63589 | 8.340 |
| 8.49180 | 10.2978 | 7.184 | 8.73171 | 7.880 | 7.91320 | 8.035 | 7.06064 | 8.196 | 6.37681 | 8.274 | 5.75065 | 8.328 |
| 8.59014 | 10.4466 | 7.154 | 8.88202 | 7.857 | 8.05932 | 8.017 | 7.19760 | 8.180 | 6.50398 | 8.260 | 5.86854 | 8.317 |
| 8.68960 | 10.5957 | 7.124 | 9.03391 | 7.834 | 8.21568 | 7.989 | 7.33635 | 8.163 | 6.63456 | 8.246 | 5.99010 | 8.306 |
| 8.79023 | 10.7464 | 7.094 | 9.18765 | 7.811 | 8.36252 | 7.974 | 7.47742 | 8.146 | 6.76863 | 8.232 | 6.11562 | 8.294 |
| 8.89201 | 10.8987 | 7.063 | 9.34364 | 7.787 | 8.51458 | 7.953 | 7.62202 | 8.129 | 6.90489 | 8.218 | 6.24076 | 8.282 |
| 8.99498 | 11.0517 | 7.033 | 9.50068 | 7.763 | 8.67133 | 7.931 | 7.76970 | 8.112 | 7.04183 | 8.204 | 6.36975 | 8.271 |
| 9.09913 | 11.2057 | 7.002 | 9.66019 | 7.739 | 8.82462 | 7.914 | 7.91791 | 8.095 | 7.18014 | 8.189 | 6.50008 | 8.258 |
| 9.20450 | 11.3601 | 6.972 | 9.82184 | 7.715 | 8.97889 | 7.897 | 8.06799 | 8.078 | 7.32238 | 8.174 | 6.63298 | 8.246 |
| 9.31108 | 11.5165 | 6.940 | 9.98503 | 7.690 | 9.13534 | 7.880 | 8.21981 | 8.059 | 7.46737 | 8.158 | 6.76884 | 8.234 |
| 9.41890 | 11.6742 | 6.910 | 10.1500 | 7.665 | 9.29599 | 7.861 | 8.37422 | 8.041 | 7.61423 | 8.143 | 6.90631 | 8.221 |
| 9.52796 | 11.8320 | 6.879 | 10.3167 | 7.640 | 9.46401 | 7.837 | 8.53241 | 8.023 | 7.76380 | 8.128 | 7.04682 | 8.208 |
| 9.63829 | 11.9915 | 6.848 | 10.4852 | 7.615 | 9.63226 | 7.813 | 8.69393 | 8.005 | 7.91433 | 8.112 | 7.18879 | 8.195 |
| 9.74990 | 12.1511 | 6.816 | 10.6554 | 7.589 | 9.80269 | 7.791 | 8.85585 | 7.986 | 8.06801 | 8.096 | 7.33470 | 8.181 |
| 9.86279 | 12.3121 | 6.785 | 10.8268 | 7.563 | 9.97543 | 7.767 | 9.01988 | 7.966 | 8.22628 | 8.079 | 7.48185 | 8.168 |
| 9.97700 | 12.4745 | 6.754 | 11.0008 | 7.537 | 10.1498 | 7.745 | 9.18587 | 7.947 | 8.38546 | 8.063 | 7.63230 | 8.154 |
| 10.0925 | 12.6372 | 6.723 | 11.1767 | 7.511 | 10.3276 | 7.721 | 9.35704 | 7.928 | 8.54592 | 8.046 | 7.78589 | 8.140 |
| 10.2094 | 12.8007 | 6.691 | 11.3544 | 7.485 | 10.5003 | 7.702 | 9.52861 | 7.908 | 8.70976 | 8.029 | 7.94121 | 8.126 |
| 10.3276 | 12.9697 | 6.660 | 11.5335 | 7.459 | 10.6813 | 7.677 | 9.70145 | 7.888 | 8.87654 | 8.011 | 8.09909 | 8.111 |
| 10.4472 | 13.1352 | 6.629 | 11.7136 | 7.432 | 10.8643 | 7.653 | 9.87813 | 7.867 | 9.04611 | 7.994 | 8.25868 | 8.097 |
| 10.5682 | 13.3091 | 6.598 | 11.8966 | 7.405 | 11.0496 | 7.629 | 10.0573 | 7.847 | 9.21780 | 7.976 | 8.42378 | 8.081 |
| 10.6905 | 13.4780 | 6.567 | 12.0812 | 7.378 | 11.2390 | 7.604 | 10.2396 | 7.826 | 9.39127 | 7.958 | 8.59065 | 8.066 |
| 10.8143 | 13.6456 | 6.535 | 12.2668 | 7.350 | 11.4287 | 7.579 | 10.4221 | 7.806 | 9.56696 | 7.939 | 8.75863 | 8.051 |

| 10.9396 | 13.8158 | 6.504 | 12.4543 | 7.323 | 11.6194 | 7.554 | 10.6069 | 7.784 | 9.74754 | 7.921 | 8.93003 | 8.035 |
|---|---|---|---|---|---|---|---|---|---|---|---|---|
| 11.0662 | 13.9859 | 6.472 | 12.6442 | 7.295 | 11.8150 | 7.529 | 10.7946 | 7.762 | 9.93037 | 7.903 | 9.10412 | 8.019 |
| 11.1944 | 14.1567 | 6.440 | 12.8357 | 7.267 | 12.0098 | 7.504 | 10.9895 | 7.741 | 10.1135 | 7.884 | 9.28101 | 8.003 |
| 11.3240 | 14.3284 | 6.409 | 13.0291 | 7.239 | 12.2083 | 7.478 | 11.1826 | 7.719 | 10.2995 | 7.864 | 9.46152 | 7.987 |
| 11.4551 | 14.5010 | 6.377 | 13.2234 | 7.211 | 12.3947 | 7.460 | 11.3766 | 7.697 | 10.4907 | 7.845 | 9.64472 | 7.970 |
| 11.5878 | 14.6824 | 6.346 | 13.4192 | 7.182 | 12.5970 | 7.434 | 11.5732 | 7.674 | 10.6851 | 7.825 | 9.83026 | 7.954 |
| 11.7220 | 14.8539 | 6.315 | 13.6173 | 7.154 | 12.8036 | 7.407 | 11.7767 | 7.651 | 10.8799 | 7.806 | ###### | 7.936 |
| 11.8577 | 15.0280 | 6.283 | 13.8159 | 7.125 | 13.0084 | 7.382 | 11.9811 | 7.629 | 11.0770 | 7.785 | 10.2096 | 7.919 |
| 11.9950 | 15.2037 | 6.251 | 14.0169 | 7.096 | 13.2178 | 7.355 | 12.1829 | 7.606 | 11.2781 | 7.765 | 10.4031 | 7.901 |
| 12.1339 | 15.3833 | 6.220 | 14.2194 | 7.067 | 13.4284 | 7.328 | 12.3901 | 7.582 | 11.4825 | 7.744 | 10.6008 | 7.884 |
| 12.2744 | 15.5605 | 6.189 | 14.4237 | 7.038 | 13.6412 | 7.301 | 12.6042 | 7.559 | 11.6873 | 7.724 | 10.8006 | 7.866 |
| 12.4165 | 15.7385 | 6.158 | 14.6290 | 7.009 | 13.8544 | 7.275 | 12.8179 | 7.535 | 11.8950 | 7.702 | 11.0071 | 7.847 |
| 12.5603 | 15.9211 | 6.126 | 14.8358 | 6.980 | 14.0722 | 7.247 | 13.0302 | 7.511 | 12.1074 | 7.681 | 11.2105 | 7.829 |
| 12.7057 | 16.0997 | 6.095 | 15.0439 | 6.950 | 14.2911 | 7.219 | 13.2469 | 7.486 | 12.3235 | 7.659 | 11.4212 | 7.810 |
| 12.8529 | 16.2817 | 6.063 | 15.2537 | 6.921 | 14.5129 | 7.192 | 13.4706 | 7.462 | 12.5401 | 7.638 | 11.6324 | 7.791 |
| 13.0017 | 16.4604 | 6.032 | 15.4656 | 6.891 | 14.7350 | 7.164 | 13.6945 | 7.438 | 12.7581 | 7.615 | 11.8473 | 7.772 |
| 13.1522 | 16.6435 | 6.001 | 15.6789 | 6.861 | 14.9587 | 7.136 | 13.9135 | 7.413 | 12.9829 | 7.593 | 12.0653 | 7.753 |
| 13.3045 | 16.8253 | 5.970 | 15.8937 | 6.831 | 15.1854 | 7.108 | 14.1397 | 7.387 | 13.2096 | 7.571 | 12.2879 | 7.733 |
| 13.4586 | 17.0100 | 5.939 | 16.1099 | 6.801 | 15.4147 | 7.080 | 14.3754 | 7.363 | 13.4354 | 7.548 | 12.5102 | 7.713 |
| 13.6144 | 17.1992 | 5.909 | 16.3271 | 6.771 | 15.6286 | 7.060 | 14.6049 | 7.337 | 13.6677 | 7.524 | 12.7382 | 7.693 |
| 13.7721 | 17.3885 | 5.878 | 16.5446 | 6.741 | 15.8616 | 7.030 | 14.8365 | 7.311 | 13.9039 | 7.501 | 12.9704 | 7.673 |
| 13.9316 | 17.5743 | 5.847 | 16.7642 | 6.711 | 16.0945 | 7.002 | 15.0787 | 7.285 | 14.1369 | 7.478 | 13.2030 | 7.652 |
| 14.0929 | 17.7594 | 5.817 | 16.9859 | 6.680 | 16.3301 | 6.974 | 15.3197 | 7.260 | 14.3726 | 7.454 | 13.4402 | 7.631 |
| 14.2561 | 17.9518 | 5.787 | 17.2106 | 6.650 | 16.5679 | 6.945 | 15.5537 | 7.233 | 14.6173 | 7.430 | 13.6797 | 7.610 |
| 14.4212 | 18.1405 | 5.757 | 17.4340 | 6.619 | 16.8086 | 6.916 | 15.8003 | 7.207 | 14.8618 | 7.407 | 13.9230 | 7.589 |
| 14.5881 | 18.3365 | 5.727 | 17.6580 | 6.588 | 17.0314 | 6.894 | 16.0519 | 7.181 | 15.1061 | 7.382 | 14.1682 | 7.567 |
| 14.7571 | 18.5195 | 5.697 | 17.8851 | 6.558 | 17.2916 | 6.857 | 16.2941 | 7.154 | 15.3592 | 7.357 | 14.4180 | 7.545 |
| 14.9279 | 18.7082 | 5.666 | 18.1134 | 6.527 | 17.5172 | 6.837 | 16.5443 | 7.126 | 15.6126 | 7.333 | 14.6722 | 7.523 |
| 15.1008 | 18.9056 | 5.637 | 18.3431 | 6.496 | 17.7626 | 6.807 | 16.8052 | 7.100 | 15.8642 | 7.308 | 14.9283 | 7.500 |
| 15.2757 | 19.0947 | 5.607 | 18.5823 | 6.466 | 18.0117 | 6.778 | 17.0537 | 7.073 | 16.1235 | 7.283 | 15.1866 | 7.478 |
| 15.4525 | 19.2855 | 5.577 | 18.8253 | 6.436 | 18.2601 | 6.748 | 17.3083 | 7.045 | 16.3872 | 7.258 | 15.4478 | 7.455 |
| 15.6315 | 19.4758 | 5.548 | 19.0579 | 6.405 | 18.5126 | 6.719 | 17.5745 | 7.018 | 16.6489 | 7.232 | 15.7138 | 7.432 |
| 15.8125 | 19.6749 | 5.519 | 19.2912 | 6.375 | 18.7659 | 6.689 | 17.8335 | 6.991 | 16.9163 | 7.206 | 15.9826 | 7.409 |
| 15.9956 | 19.8636 | 5.489 | 19.5247 | 6.344 | 19.0183 | 6.660 | 18.0922 | 6.962 | 17.1861 | 7.181 | 16.2537 | 7.386 |
| 16.1808 | 20.0623 | 5.460 | 19.7604 | 6.313 | 19.2738 | 6.630 | 18.3672 | 6.934 | 17.4534 | 7.155 | 16.5281 | 7.362 |
| 16.3682 | 20.2571 | 5.431 | 20.0092 | 6.283 | 19.5337 | 6.600 | 18.6311 | 6.907 | 17.7275 | 7.129 | 16.8053 | 7.338 |
| 16.5577 | 20.4483 | 5.402 | 20.2627 | 6.252 | 19.7926 | 6.570 | 18.8975 | 6.878 | 18.0081 | 7.103 | 17.0841 | 7.315 |
| 16.7494 | 20.6440 | 5.373 | 20.5033 | 6.222 | 20.0544 | 6.541 | 19.1790 | 6.850 | 18.2852 | 7.076 | 17.3667 | 7.291 |
| 16.9434 | 20.8429 | 5.345 | 20.7495 | 6.191 | 20.3158 | 6.511 | 19.4471 | 6.822 | 18.5681 | 7.048 | 17.6557 | 7.267 |
| 17.1396 | 21.0364 | 5.316 | 21.0044 | 6.162 | 20.5787 | 6.481 | 19.7192 | 6.792 | 18.8537 | 7.022 | 17.9459 | 7.242 |
| 17.3380 | 21.2307 | 5.287 | 21.2582 | 6.132 | 20.8451 | 6.451 | 20.0079 | 6.764 | 19.1351 | 6.994 | 18.2372 | 7.217 |
| 17.5388 | 21.4274 | 5.259 | 21.5012 | 6.101 | 21.1132 | 6.421 | 20.2792 | 6.736 | 19.4274 | 6.967 | 18.5369 | 7.192 |
| 17.7419 | 21.6273 | 5.231 | 21.7382 | 6.069 | 21.3795 | 6.391 | 20.5621 | 6.706 | 19.7176 | 6.940 | 18.8311 | 7.169 |
| 17.9473 | 21.8245 | 5.203 | 21.9732 | 6.038 | 21.6503 | 6.361 | 20.8565 | 6.678 | 20.0115 | 6.912 | 19.1347 | 7.143 |
| 18.1552 | 22.0168 | 5.176 | 22.2218 | 6.008 | 21.9225 | 6.330 | 21.1276 | 6.648 | 20.3141 | 6.885 | 19.4413 | 7.117 |
| 18.3654 | 22.2185 | 5.148 | 22.4888 | 5.978 | 22.1927 | 6.300 | 21.4240 | 6.619 | 20.6053 | 6.857 | 19.7449 | 7.092 |
| 18.5780 | 22.4201 | 5.120 | 22.7350 | 5.947 | 22.4658 | 6.270 | 21.7176 | 6.591 | 20.9085 | 6.829 | 20.0581 | 7.066 |

| | | | | | | | | | | | |
|---|---|---|---|---|---|---|---|---|---|---|---|
| 18.7932 | 22.6152 | 5.093 | 22.9700 | 5.916 | 22.7413 | 6.240 | 21.9968 | 6.560 | 21.2149 | 6.802 | 20.3714 | 7.041 |
| 19.0108 | 22.8132 | 5.065 | 23.2471 | 5.887 | 23.0164 | 6.209 | 22.3070 | 6.532 | 21.5174 | 6.773 | 20.6885 | 7.014 |
| 19.2309 | 23.0196 | 5.039 | 23.5081 | 5.858 | 23.2916 | 6.180 | 22.5829 | 6.503 | 21.8336 | 6.745 | 21.0059 | 6.989 |
| 19.4536 | 23.2250 | 5.012 | 23.7466 | 5.829 | 23.5692 | 6.150 | 22.8880 | 6.472 | 22.1385 | 6.716 | 21.3252 | 6.963 |
| 19.6789 | 23.4933 | 4.990 | 24.0101 | 5.801 | 23.8495 | 6.120 | 23.1898 | 6.444 | 22.4549 | 6.688 | 21.6528 | 6.936 |
| 19.9067 | 23.6942 | 4.963 | 24.2574 | 5.770 | 24.1285 | 6.090 | 23.4741 | 6.413 | 22.7687 | 6.660 | 21.9804 | 6.910 |
| 20.1372 | 23.9019 | 4.938 | 24.5184 | 5.740 | 24.4128 | 6.060 | 23.8009 | 6.384 | 23.0827 | 6.630 | 22.3112 | 6.884 |
| 20.3704 | 24.1079 | 4.912 | 24.7821 | 5.710 | 24.6931 | 6.030 | 24.0785 | 6.355 | 23.4065 | 6.602 | 22.6417 | 6.857 |
| 20.6063 | 24.3074 | 4.886 | 25.0280 | 5.682 | 24.9765 | 6.000 | 24.4077 | 6.324 | 23.7233 | 6.573 | 22.9808 | 6.829 |
| 20.8449 | 24.5206 | 4.860 | 25.2938 | 5.654 | 25.2618 | 5.970 | 24.7032 | 6.296 | 24.0543 | 6.544 | 23.3178 | 6.803 |
| 21.0863 | 24.7243 | 4.835 | 25.5668 | 5.628 | 25.5462 | 5.940 | 25.0191 | 6.264 | 24.3762 | 6.515 | 23.6640 | 6.775 |
| 21.3304 | 24.9299 | 4.809 | 25.8395 | 5.601 | 25.8327 | 5.911 | 25.3325 | 6.236 | 24.7073 | 6.486 | 24.0036 | 6.748 |
| 21.5774 | 25.1504 | 4.785 | 26.1150 | 5.571 | 26.1192 | 5.881 | 25.6285 | 6.205 | 25.0403 | 6.457 | 24.3533 | 6.721 |
| 21.8273 | 25.3523 | 4.760 | 26.3920 | 5.541 | 26.4099 | 5.851 | 25.9631 | 6.176 | 25.3773 | 6.428 | 24.7061 | 6.693 |
| 22.0800 | 25.5784 | 4.736 | 26.6788 | 5.512 | 26.6983 | 5.822 | 26.2589 | 6.146 | 25.7112 | 6.398 | 25.0553 | 6.665 |
| 22.3357 | 25.7870 | 4.711 | 26.9475 | 5.484 | 26.9884 | 5.792 | 26.5929 | 6.116 | 26.0487 | 6.369 | 25.4126 | 6.636 |
| 22.5944 | 25.9916 | 4.687 | 27.2252 | 5.457 | 27.2795 | 5.762 | 26.8820 | 6.087 | 26.3880 | 6.339 | 25.7658 | 6.608 |
| 22.8560 | 26.2069 | 4.663 | 27.4925 | 5.433 | 27.5709 | 5.733 | 27.2275 | 6.057 | 26.7271 | 6.310 | 26.1221 | 6.581 |
| 23.1206 | 26.4274 | 4.639 | 27.7487 | 5.406 | 27.8579 | 5.704 | 27.5333 | 6.028 | 27.0687 | 6.281 | 26.4881 | 6.552 |
| 23.3884 | 26.6305 | 4.615 | 28.0212 | 5.377 | 28.1594 | 5.674 | 27.8654 | 5.997 | 27.4120 | 6.251 | 26.8550 | 6.524 |
| 23.6592 | 26.8354 | 4.591 | 28.2978 | 5.347 | 28.4506 | 5.645 | 28.1727 | 5.969 | 27.7587 | 6.222 | 27.2252 | 6.494 |
| 23.9332 | 27.0545 | 4.568 | 28.5759 | 5.317 | 28.7461 | 5.616 | 28.5051 | 5.938 | 28.1076 | 6.192 | 27.5880 | 6.467 |
| 24.2103 | 27.2673 | 4.544 | 28.8336 | 5.290 | 29.0438 | 5.588 | 28.8239 | 5.909 | 28.4552 | 6.162 | 27.9701 | 6.438 |
| 24.4906 | 27.4780 | 4.521 | 29.0930 | 5.264 | 29.3395 | 5.559 | 29.1599 | 5.878 | 28.8045 | 6.133 | 28.3406 | 6.409 |
| 24.7742 | 27.6870 | 4.498 | 29.3535 | 5.237 | 29.6319 | 5.530 | 29.4779 | 5.850 | 29.1617 | 6.103 | 28.7098 | 6.382 |
| 25.0611 | 27.8890 | 4.475 | 29.6162 | 5.209 | 29.9311 | 5.501 | 29.8153 | 5.819 | 29.5123 | 6.074 | 29.0827 | 6.355 |
| 25.3513 | 28.0973 | 4.452 | 29.8781 | 5.183 | 30.2404 | 5.472 | 30.1363 | 5.791 | 29.8646 | 6.044 | 29.4641 | 6.326 |
| 25.6448 | 28.3140 | 4.429 | 30.1492 | 5.155 | 30.5321 | 5.445 | 30.4756 | 5.761 | 30.2280 | 6.014 | 29.8550 | 6.297 |
| 25.9418 | 28.5161 | 4.406 | 30.4249 | 5.127 | 30.8334 | 5.416 | 30.7965 | 5.733 | 30.5760 | 5.985 | 30.2268 | 6.271 |
| 26.2422 | 28.7463 | 4.385 | 30.6728 | 5.104 | 31.1299 | 5.388 | 31.1512 | 5.702 | 30.9462 | 5.956 | 30.6201 | 6.241 |
| 26.5461 | 28.9444 | 4.362 | 30.9358 | 5.078 | 31.4334 | 5.360 | 31.4471 | 5.674 | 31.2926 | 5.926 | 30.9976 | 6.212 |
| 26.8534 | 29.1592 | 4.341 | 31.2058 | 5.050 | 31.7288 | 5.332 | 31.8258 | 5.644 | 31.6641 | 5.897 | 31.3940 | 6.182 |
| 27.1644 | 29.3888 | 4.320 | 31.4759 | 5.023 | 32.0165 | 5.304 | 32.1112 | 5.616 | 32.0298 | 5.867 | 31.7871 | 6.153 |
| 27.4789 | 29.5928 | 4.298 | 31.7494 | 4.996 | 32.3281 | 5.276 | 32.4931 | 5.588 | 32.3897 | 5.838 | 32.1915 | 6.124 |
| 27.7971 | 29.8194 | 4.277 | 32.0364 | 4.967 | 32.6278 | 5.249 | 32.7995 | 5.557 | 32.7720 | 5.808 | 32.5900 | 6.094 |
| 28.1190 | 30.0261 | 4.255 | 32.3111 | 4.940 | 32.9307 | 5.221 | 33.1380 | 5.532 | 33.1213 | 5.779 | 32.9942 | 6.064 |
| 28.4446 | 30.2558 | 4.235 | 32.5698 | 4.916 | 33.2345 | 5.194 | 33.5117 | 5.499 | 33.5092 | 5.750 | 33.3978 | 6.036 |
| 28.7740 | 30.4679 | 4.214 | 32.8047 | 4.895 | 33.5308 | 5.166 | 33.8027 | 5.472 | 33.8775 | 5.719 | 33.7961 | 6.006 |
| 29.1072 | 30.6697 | 4.192 | 33.0639 | 4.871 | 33.8427 | 5.140 | 34.1973 | 5.444 | 34.2401 | 5.691 | 34.2013 | 5.977 |
| 29.4442 | 30.8975 | 4.172 | 33.3438 | 4.844 | 34.1392 | 5.113 | 34.5057 | 5.413 | 34.6370 | 5.662 | 34.6069 | 5.949 |
| 29.7852 | 31.1416 | 4.152 | 33.6195 | 4.818 | 34.4407 | 5.086 | 34.8407 | 5.388 | 34.9938 | 5.633 | 35.0162 | 5.920 |
| 30.1301 | 31.3432 | 4.132 | 33.8826 | 4.794 | 34.7515 | 5.060 | 35.2249 | 5.358 | 35.3735 | 5.605 | 35.4171 | 5.891 |
| 30.4789 | 31.6039 | 4.114 | 34.1494 | 4.770 | 35.1299 | 5.036 | 35.5165 | 5.330 | 35.7648 | 5.575 | 35.8396 | 5.864 |
| 30.8319 | 31.7947 | 4.093 | 34.4216 | 4.744 | 35.4403 | 5.011 | 35.8947 | 5.304 | 36.1215 | 5.546 | 36.2486 | 5.835 |
| 31.1889 | 32.0266 | 4.074 | 34.7054 | 4.718 | 35.7429 | 4.984 | 36.2439 | 5.273 | 36.5164 | 5.518 | 36.6725 | 5.806 |
| 31.5500 | 32.2527 | 4.055 | 34.9762 | 4.693 | 36.0461 | 4.958 | 36.5463 | 5.248 | 36.8923 | 5.488 | 37.0771 | 5.777 |
| 31.9154 | 32.4793 | 4.034 | 35.2778 | 4.670 | 36.3604 | 4.932 | 36.9431 | 5.220 | 37.2441 | 5.461 | 37.4941 | 5.749 |

| | | | | | | | | | | | |
|---|---|---|---|---|---|---|---|---|---|---|---|
| 32.2849 | 32.6999 | 4.015 | 35.6034 | 4.648 | 36.6667 | 4.906 | 37.2773 | 5.190 | 37.6664 | 5.433 | 37.9023 | 5.720 |
| 32.6588 | 32.9119 | 3.996 | 35.8718 | 4.624 | 36.9575 | 4.880 | 37.5750 | 5.166 | 38.0529 | 5.403 | 38.3163 | 5.692 |
| 33.0370 | 33.1470 | 3.978 | 36.1461 | 4.599 | 37.2741 | 4.855 | 37.9864 | 5.139 | 38.3874 | 5.376 | 38.7570 | 5.662 |
| 33.4195 | 33.3560 | 3.958 | 36.4333 | 4.575 | 37.5921 | 4.830 | 38.3095 | 5.108 | 38.7993 | 5.349 | 39.1774 | 5.634 |
| 33.8065 | 33.5654 | 3.939 | 36.7248 | 4.553 | 37.8922 | 4.805 | 38.5813 | 5.085 | 39.2225 | 5.319 | 39.6250 | 5.605 |
| 34.1979 | 33.7971 | 3.921 | 37.0136 | 4.531 | 38.2046 | 4.780 | 38.9992 | 5.060 | 39.5636 | 5.290 | 40.0552 | 5.577 |
| 34.5939 | 34.0190 | 3.903 | 37.3079 | 4.509 | 38.6048 | 4.758 | 39.3946 | 5.029 | 39.9274 | 5.264 | 40.4886 | 5.548 |
| 34.9945 | 34.2699 | 3.885 | 37.6138 | 4.487 | 38.9085 | 4.734 | 39.6588 | 5.003 | 40.3699 | 5.237 | 40.9292 | 5.520 |
| 35.3997 | 34.4809 | 3.867 | 37.8807 | 4.463 | 39.2134 | 4.709 | 39.9978 | 4.980 | 40.7755 | 5.207 | 41.3554 | 5.491 |
| 35.8096 | 34.6979 | 3.849 | 38.1799 | 4.441 | 39.5230 | 4.685 | 40.4316 | 4.952 | 41.1119 | 5.181 | 41.7947 | 5.462 |
| 36.2243 | 34.9094 | 3.832 | 38.4917 | 4.420 | 39.8298 | 4.662 | 40.7723 | 4.923 | 41.5108 | 5.155 | 42.2115 | 5.434 |
| 36.6438 | 35.1258 | 3.814 | 38.7516 | 4.398 | 40.1440 | 4.637 | 41.0559 | 4.899 | 41.9632 | 5.127 | 42.6359 | 5.407 |
| 37.0681 | 35.3665 | 3.798 | 39.0299 | 4.376 | 40.4522 | 4.614 | 41.4256 | 4.877 | 42.3515 | 5.098 | 43.0817 | 5.379 |
| 37.4973 | 35.5889 | 3.781 | 39.3505 | 4.355 | 40.7605 | 4.590 | 41.8507 | 4.849 | 42.6718 | 5.071 | 43.5065 | 5.351 |
| 37.9315 | 35.8045 | 3.762 | 39.6295 | 4.333 | 41.0730 | 4.567 | 42.2108 | 4.819 | 43.0185 | 5.048 | 43.9582 | 5.323 |
| 38.3707 | 36.0475 | 3.747 | 39.8830 | 4.311 | 41.4924 | 4.548 | 42.4714 | 4.794 | 43.4531 | 5.024 | 44.3890 | 5.296 |
| 38.8150 | 36.2707 | 3.730 | 40.1895 | 4.291 | 41.7980 | 4.524 | 42.7544 | 4.773 | 43.9287 | 4.995 | 44.8414 | 5.269 |
| 39.2645 | 36.5219 | 3.714 | 40.5083 | 4.271 | 42.1121 | 4.501 | 43.1430 | 4.750 | 44.3636 | 4.964 | 45.2858 | 5.240 |
| 39.7192 | 36.7609 | 3.697 | 40.7831 | 4.249 | 42.4268 | 4.478 | 43.5931 | 4.725 | 44.7165 | 4.937 | 45.7388 | 5.214 |
| 40.1791 | 36.9761 | 3.679 | 41.0891 | 4.229 | 42.7443 | 4.456 | 44.0007 | 4.697 | 45.0381 | 4.912 | 46.2000 | 5.187 |
| 40.6443 | 37.2048 | 3.663 | 41.4038 | 4.209 | 43.0406 | 4.434 | 44.3105 | 4.670 | 45.4113 | 4.889 | 46.6399 | 5.159 |
| 41.1150 | 37.4114 | 3.647 | 41.6794 | 4.188 | 43.4814 | 4.415 | 44.6136 | 4.646 | 45.8384 | 4.864 | 47.1083 | 5.132 |
| 41.5911 | 37.6359 | 3.631 | 41.9803 | 4.169 | 43.7862 | 4.393 | 44.9257 | 4.624 | 46.2921 | 4.839 | 47.5634 | 5.105 |
| 42.0727 | 37.8805 | 3.616 | 42.2318 | 4.148 | 44.1047 | 4.371 | 45.2915 | 4.603 | 46.7383 | 4.812 | 48.0583 | 5.079 |
| 42.5598 | 38.1140 | 3.600 | 42.4910 | 4.128 | 44.4109 | 4.350 | 45.6834 | 4.580 | 47.1478 | 4.784 | 48.4884 | 5.052 |
| 43.0527 | 38.3685 | 3.586 | 42.8579 | 4.111 | 44.7210 | 4.328 | 46.1061 | 4.554 | 47.5226 | 4.757 | 48.9474 | 5.026 |
| 43.5512 | 38.5812 | 3.570 | 43.1732 | 4.092 | 45.0310 | 4.307 | 46.4884 | 4.528 | 47.8887 | 4.732 | 49.4069 | 4.999 |
| 44.0555 | 38.8147 | 3.554 | 43.4324 | 4.071 | 45.3414 | 4.286 | 46.8380 | 4.503 | 48.2744 | 4.708 | 49.8497 | 4.973 |
| 44.5656 | 39.0312 | 3.538 | 43.6907 | 4.051 | 45.6479 | 4.265 | 47.1741 | 4.478 | 48.6551 | 4.684 | 50.3137 | 4.947 |
| 45.0817 | 39.2685 | 3.523 | 44.0377 | 4.034 | 45.9617 | 4.244 | 47.4925 | 4.456 | 49.0465 | 4.662 | 50.7590 | 4.920 |
| 45.6037 | 39.5341 | 3.509 | 44.3128 | 4.015 | 46.2724 | 4.223 | 47.8182 | 4.433 | 49.4371 | 4.639 | 51.2016 | 4.894 |
| 46.1318 | 39.7599 | 3.494 | 44.5476 | 3.994 | 46.5910 | 4.203 | 48.1318 | 4.412 | 49.8443 | 4.616 | 51.6799 | 4.869 |
| 46.6659 | 39.9693 | 3.479 | 44.8257 | 3.976 | 46.8865 | 4.182 | 48.4726 | 4.390 | 50.2726 | 4.592 | 52.1317 | 4.844 |
| 47.2063 | 40.2557 | 3.464 | 45.1938 | 3.958 | 47.2118 | 4.162 | 48.8071 | 4.368 | 50.6963 | 4.569 | 52.6134 | 4.820 |
| 47.7529 | 40.4878 | 3.451 | 45.4992 | 3.935 | 47.5116 | 4.142 | 49.1830 | 4.346 | 51.1035 | 4.546 | 53.0529 | 4.795 |
| 48.3059 | 40.6796 | 3.435 | 45.7531 | 3.914 | 47.8369 | 4.122 | 49.5132 | 4.324 | 51.5052 | 4.521 | 53.5085 | 4.773 |
| 48.8652 | 40.9194 | 3.421 | 46.0550 | 3.896 | 48.1608 | 4.102 | 49.8878 | 4.301 | 51.9356 | 4.496 | 53.9473 | 4.748 |
| 49.4311 | 41.1823 | 3.407 | 46.3484 | 3.878 | 48.5259 | 4.078 | 50.2610 | 4.279 | 52.3744 | 4.471 | 54.4119 | 4.725 |
| 50.0035 | 41.3918 | 3.393 | 46.6472 | 3.864 | 48.9747 | 4.044 | 50.5167 | 4.256 | 52.8951 | 4.439 | 54.8581 | 4.700 |
| 51.1682 | 41.8192 | 3.364 | 47.3165 | 3.845 | 49.6950 | 4.012 | 51.3967 | 4.216 | 53.9391 | 4.405 | 55.8468 | 4.653 |
| 52.3600 | 42.3024 | 3.337 | 47.8746 | 3.800 | 50.3206 | 3.969 | 52.2109 | 4.170 | 54.8959 | 4.361 | 56.7167 | 4.609 |
| 53.5797 | 42.7651 | 3.311 | 48.6267 | 3.762 | 51.0236 | 3.932 | 53.1407 | 4.129 | 55.8244 | 4.319 | 57.6492 | 4.561 |
| 54.8277 | 43.2994 | 3.285 | 49.3323 | 3.726 | 51.7828 | 3.892 | 54.0037 | 4.080 | 56.6889 | 4.274 | 58.5810 | 4.515 |
| 56.1048 | 43.7313 | 3.261 | 50.0623 | 3.694 | 52.4982 | 3.858 | 54.9227 | 4.040 | 57.6963 | 4.234 | 59.5367 | 4.468 |
| 57.4116 | 44.1673 | 3.235 | 50.7111 | 3.654 | 53.2821 | 3.818 | 55.7535 | 4.003 | 58.5251 | 4.195 | 60.4626 | 4.423 |
| 58.7489 | 44.6046 | 3.209 | 51.1088 | 3.620 | 54.0567 | 3.780 | 56.6192 | 3.962 | 59.3857 | 4.152 | 61.4269 | 4.378 |
| 60.1174 | 44.9367 | 3.184 | 51.7249 | 3.584 | 54.8575 | 3.741 | 57.5601 | 3.923 | 60.3084 | 4.111 | 62.3870 | 4.330 |

| | | | | | | | | | | | |
|---|---|---|---|---|---|---|---|---|---|---|---|
| 61.5177 | 45.4005 | 3.160 | 52.3523 | 3.551 | 55.4760 | 3.709 | 58.6063 | 3.892 | 61.2234 | 4.076 | 63.3933 | 4.284 |
| 62.2300 | 45.7812 | 3.151 | 52.3661 | 3.527 | 55.8088 | 3.692 | 58.9495 | 3.870 | 61.6174 | 4.059 | 63.8696 | 4.262 |
| 62.9506 | 45.9802 | 3.139 | 52.6552 | 3.506 | 56.1625 | 3.679 | 59.2674 | 3.858 | 62.0775 | 4.040 | 64.3606 | 4.240 |
| 63.6796 | 46.1756 | 3.127 | 52.9882 | 3.491 | 56.5356 | 3.663 | 59.7468 | 3.839 | 62.5442 | 4.020 | 64.9008 | 4.218 |
| 64.4169 | 46.3492 | 3.115 | 53.2982 | 3.477 | 56.8798 | 3.645 | 60.1222 | 3.821 | 62.9625 | 3.999 | 65.3597 | 4.197 |
| 65.1628 | 46.5466 | 3.104 | 53.5940 | 3.464 | 57.2086 | 3.632 | 60.4885 | 3.806 | 63.3487 | 3.980 | 65.8613 | 4.174 |
| 65.9174 | 46.7834 | 3.094 | 53.9228 | 3.448 | 57.5750 | 3.614 | 60.9047 | 3.786 | 63.8178 | 3.963 | 66.3818 | 4.151 |
| 66.6807 | 46.8808 | 3.082 | 54.2227 | 3.434 | 57.9573 | 3.598 | 61.3394 | 3.768 | 64.2894 | 3.942 | 66.9146 | 4.128 |
| 67.4528 | 47.1576 | 3.073 | 54.4852 | 3.420 | 58.2328 | 3.582 | 61.6447 | 3.751 | 64.6638 | 3.923 | 67.3796 | 4.107 |
| 68.2339 | 47.3035 | 3.062 | 54.8171 | 3.410 | 58.6000 | 3.570 | 62.0540 | 3.737 | 65.1393 | 3.907 | 67.8213 | 4.090 |
| 69.0240 | 47.5685 | 3.052 | 55.1689 | 3.396 | 58.9902 | 3.554 | 62.4923 | 3.720 | 65.5838 | 3.889 | 68.3671 | 4.068 |
| 69.8232 | 47.8123 | 3.043 | 55.4554 | 3.380 | 59.3019 | 3.536 | 62.8572 | 3.700 | 66.0275 | 3.869 | 68.8798 | 4.046 |
| 70.6318 | 48.0285 | 3.033 | 55.7152 | 3.369 | 59.6177 | 3.523 | 63.2040 | 3.685 | 66.4065 | 3.851 | 69.3258 | 4.027 |
| 71.4496 | 48.2530 | 3.022 | 56.0294 | 3.356 | 59.9718 | 3.509 | 63.6090 | 3.670 | 66.8785 | 3.833 | 69.8216 | 4.008 |
| 72.2770 | 48.4052 | 3.012 | 56.3786 | 3.341 | 60.3714 | 3.493 | 64.0484 | 3.651 | 67.3507 | 3.815 | 70.3358 | 3.987 |
| 73.1139 | 48.6452 | 3.002 | 56.6329 | 3.329 | 60.6627 | 3.479 | 64.3808 | 3.636 | 67.7319 | 3.797 | 70.8269 | 3.970 |
| 73.9605 | 48.8140 | 2.992 | 56.9715 | 3.317 | 61.0359 | 3.465 | 64.8007 | 3.621 | 68.1866 | 3.780 | 71.3005 | 3.951 |
| 74.8170 | 49.0393 | 2.981 | 57.3178 | 3.303 | 61.4188 | 3.450 | 65.2251 | 3.604 | 68.6563 | 3.762 | 71.8142 | 3.931 |
| 75.6833 | 49.2137 | 2.971 | 57.5768 | 3.290 | 61.7240 | 3.436 | 65.5596 | 3.589 | 69.0673 | 3.744 | 72.2973 | 3.912 |
| 76.5597 | 49.4758 | 2.962 | 57.9120 | 3.278 | 62.0946 | 3.423 | 65.9734 | 3.574 | 69.5059 | 3.728 | 72.7989 | 3.896 |
| 77.4462 | 49.7082 | 2.954 | 58.2402 | 3.266 | 62.4749 | 3.408 | 66.4139 | 3.557 | 69.9585 | 3.711 | 73.3284 | 3.874 |
| 78.3430 | 49.8980 | 2.944 | 58.5092 | 3.254 | 62.7739 | 3.395 | 66.7500 | 3.543 | 70.3451 | 3.694 | 73.8031 | 3.859 |
| 79.2501 | 50.1841 | 2.937 | 58.8337 | 3.242 | 63.1322 | 3.381 | 67.1593 | 3.528 | 70.8369 | 3.678 | 74.3186 | 3.841 |
| 80.1678 | 50.4547 | 2.925 | 59.1700 | 3.231 | 63.5285 | 3.368 | 67.5736 | 3.512 | 71.2448 | 3.660 | 74.8510 | 3.823 |
| 81.0961 | 50.7478 | 2.916 | 59.4482 | 3.217 | 63.8284 | 3.354 | 67.9370 | 3.497 | 71.6846 | 3.645 | 75.3194 | 3.807 |
| 82.0352 | 50.9836 | 2.907 | 59.7561 | 3.205 | 64.1781 | 3.342 | 68.3030 | 3.484 | 72.0058 | 3.630 | 75.7532 | 3.790 |
| 82.9851 | 51.2311 | 2.897 | 60.0729 | 3.193 | 64.5556 | 3.328 | 68.7582 | 3.469 | 72.6025 | 3.613 | 76.2361 | 3.774 |
| 83.9460 | 51.5379 | 2.888 | 60.4137 | 3.182 | 64.9384 | 3.315 | 69.1787 | 3.454 | 73.0763 | 3.596 | 76.7495 | 3.757 |
| 84.9180 | 51.8275 | 2.878 | 60.7416 | 3.169 | 65.2955 | 3.301 | 69.5770 | 3.439 | 73.5271 | 3.580 | 77.3030 | 3.736 |
| 85.9014 | 51.9321 | 2.868 | 61.0242 | 3.159 | 65.6356 | 3.290 | 69.9683 | 3.426 | 73.9435 | 3.565 | 77.7809 | 3.722 |
| 86.8960 | 52.3037 | 2.858 | 61.3531 | 3.148 | 65.9728 | 3.277 | 70.3795 | 3.411 | 74.3279 | 3.548 | 78.2986 | 3.703 |
| 87.9023 | 52.4836 | 2.849 | 61.6655 | 3.136 | 66.3185 | 3.264 | 70.7142 | 3.397 | 74.8023 | 3.534 | 78.7817 | 3.685 |
| 88.9201 | 52.7487 | 2.840 | 61.9326 | 3.126 | 66.6574 | 3.253 | 71.1269 | 3.385 | 75.2281 | 3.519 | 79.2255 | 3.673 |
| 89.9498 | 52.9157 | 2.829 | 62.3535 | 3.116 | 67.1499 | 3.242 | 71.5853 | 3.374 | 75.6480 | 3.504 | 79.7869 | 3.652 |
| 90.9913 | 53.1467 | 2.824 | 62.5784 | 3.103 | 67.3629 | 3.227 | 71.9222 | 3.356 | 76.1122 | 3.487 | 80.2913 | 3.636 |
| 92.0450 | 53.2934 | 2.815 | 62.9259 | 3.094 | 67.7758 | 3.216 | 72.3795 | 3.344 | 76.6017 | 3.474 | 80.7815 | 3.622 |
| 93.1108 | 53.6528 | 2.808 | 63.2510 | 3.084 | 68.1489 | 3.205 | 72.7785 | 3.331 | 77.0576 | 3.460 | 81.2919 | 3.606 |
| 94.1890 | 53.9307 | 2.798 | 63.5526 | 3.072 | 68.4627 | 3.192 | 73.1214 | 3.317 | 77.4778 | 3.446 | 81.8041 | 3.589 |
| 95.2796 | 54.2422 | 2.789 | 63.8942 | 3.062 | 68.8701 | 3.181 | 73.6004 | 3.304 | 77.9505 | 3.430 | 82.3396 | 3.573 |
| 96.3829 | 54.2769 | 2.782 | 64.2933 | 3.053 | 69.2937 | 3.171 | 74.0558 | 3.294 | 78.4736 | 3.417 | 82.8589 | 3.557 |
| 97.4990 | 54.5516 | 2.775 | 64.5327 | 3.041 | 69.5973 | 3.157 | 74.4182 | 3.278 | 78.8821 | 3.402 | 83.3280 | 3.542 |
| 98.6279 | 54.8924 | 2.765 | 64.9119 | 3.031 | 69.9962 | 3.146 | 74.8790 | 3.266 | 79.2983 | 3.388 | 83.8812 | 3.526 |
| 99.7700 | 55.1041 | 2.759 | 65.3221 | 3.021 | 70.4429 | 3.134 | 75.3504 | 3.252 | 79.8757 | 3.372 | 84.3949 | 3.508 |
| 100.925 | 55.3286 | 2.750 | 65.5202 | 3.011 | 70.6976 | 3.123 | 75.6209 | 3.240 | 80.2258 | 3.359 | 84.8367 | 3.495 |
| 102.094 | 55.6959 | 2.745 | 65.8972 | 2.999 | 71.1211 | 3.110 | 76.0919 | 3.226 | 80.8033 | 3.344 | 85.3781 | 3.478 |
| 103.276 | 55.9171 | 2.737 | 66.2483 | 2.991 | 71.5046 | 3.100 | 76.5158 | 3.215 | 81.1759 | 3.333 | 85.9180 | 3.462 |
| 104.472 | 56.1504 | 2.726 | 66.5302 | 2.981 | 71.8164 | 3.090 | 76.8848 | 3.203 | 81.5890 | 3.319 | 86.3675 | 3.450 |

| | | | | | | | | | | | |
|---|---|---|---|---|---|---|---|---|---|---|---|
| 105.682 | 56.3808 | 2.718 | 66.8829 | 2.971 | 72.2286 | 3.079 | 77.3280 | 3.191 | 82.0953 | 3.305 | 86.8935 | 3.434 |
| 106.905 | 56.6123 | 2.716 | 67.2297 | 2.962 | 72.5776 | 3.068 | 77.7106 | 3.179 | 82.5181 | 3.292 | 87.4127 | 3.419 |
| 108.143 | 56.9453 | 2.710 | 67.5110 | 2.953 | 72.9143 | 3.059 | 78.1356 | 3.168 | 83.0298 | 3.279 | 87.8701 | 3.406 |
| 109.396 | 57.1176 | 2.700 | 67.8628 | 2.942 | 73.3154 | 3.046 | 78.5631 | 3.154 | 83.4950 | 3.265 | 88.3963 | 3.390 |
| 110.662 | 57.3720 | 2.693 | 68.1788 | 2.935 | 73.6180 | 3.034 | 78.8770 | 3.142 | 83.7327 | 3.253 | 88.7273 | 3.374 |
| 111.944 | 57.7840 | 2.686 | 68.5587 | 2.927 | 74.0253 | 3.025 | 79.3704 | 3.131 | 84.2576 | 3.242 | 89.3125 | 3.359 |
| 113.240 | 57.9622 | 2.680 | 68.8080 | 2.918 | 74.3107 | 3.015 | 79.7032 | 3.120 | 84.6333 | 3.228 | 89.7698 | 3.347 |
| 114.551 | 58.2447 | 2.673 | 69.0957 | 2.909 | 74.6450 | 3.004 | 80.0760 | 3.108 | 85.0617 | 3.216 | 90.2742 | 3.333 |
| 115.878 | 58.4525 | 2.666 | 69.4787 | 2.899 | 75.0918 | 2.993 | 80.5433 | 3.096 | 85.5512 | 3.203 | 90.7080 | 3.320 |
| 117.220 | 58.7763 | 2.659 | 69.7981 | 2.892 | 75.4275 | 2.985 | 80.9169 | 3.087 | 85.9874 | 3.192 | 91.2419 | 3.307 |
| 118.577 | 59.0315 | 2.651 | 70.0717 | 2.883 | 75.7545 | 2.975 | 81.3107 | 3.075 | 86.4150 | 3.181 | 91.7524 | 3.294 |
| 119.950 | 59.2959 | 2.644 | 70.3812 | 2.875 | 76.0924 | 2.965 | 81.6815 | 3.064 | 86.8573 | 3.168 | 92.2368 | 3.282 |
| 121.339 | 59.5363 | 2.638 | 70.7406 | 2.866 | 76.5018 | 2.956 | 82.1383 | 3.054 | 87.3425 | 3.156 | 92.8088 | 3.269 |
| 122.744 | 59.7363 | 2.631 | 71.0998 | 2.858 | 76.9012 | 2.947 | 82.5994 | 3.044 | 87.8409 | 3.145 | 93.2928 | 3.256 |
| 124.165 | 60.0097 | 2.624 | 71.4947 | 2.852 | 77.3452 | 2.938 | 83.0524 | 3.034 | 88.4133 | 3.134 | 93.9120 | 3.242 |
| 125.603 | 60.2169 | 2.608 | 71.7138 | 2.843 | 77.7305 | 2.930 | 83.5083 | 3.026 | 88.8576 | 3.126 | 94.3653 | 3.231 |
| 127.057 | 60.5171 | 2.602 | 72.0787 | 2.834 | 77.9648 | 2.921 | 83.7952 | 3.014 | 89.2155 | 3.114 | 94.9417 | 3.218 |
| 128.529 | 60.7667 | 2.596 | 72.4767 | 2.827 | 78.4048 | 2.912 | 84.2845 | 3.004 | 89.7088 | 3.101 | 95.4137 | 3.205 |
| 130.017 | 60.9979 | 2.591 | 72.7813 | 2.819 | 78.7774 | 2.903 | 84.6691 | 2.994 | 90.1742 | 3.090 | 95.9650 | 3.193 |
| 131.522 | 61.2292 | 2.583 | 73.0436 | 2.810 | 79.0434 | 2.893 | 85.1337 | 2.984 | 90.6689 | 3.079 | 96.4513 | 3.181 |
| 133.045 | 61.6667 | 2.578 | 73.5215 | 2.803 | 79.5835 | 2.885 | 85.5560 | 2.974 | 91.1583 | 3.067 | 96.9996 | 3.168 |
| 134.586 | 61.9274 | 2.570 | 73.8631 | 2.795 | 79.9794 | 2.876 | 86.0255 | 2.964 | 91.6531 | 3.056 | 97.5133 | 3.156 |
| 136.144 | 62.2461 | 2.565 | 74.2388 | 2.788 | 80.3642 | 2.868 | 86.4324 | 2.955 | 92.1354 | 3.047 | 98.0745 | 3.145 |
| 137.721 | 62.5029 | 2.559 | 74.5600 | 2.779 | 80.6919 | 2.859 | 86.8453 | 2.946 | 92.5572 | 3.036 | 98.5954 | 3.132 |
| 139.316 | 62.8917 | 2.553 | 74.9982 | 2.773 | 81.2390 | 2.851 | 87.3547 | 2.936 | 93.0966 | 3.022 | 99.1713 | 3.121 |
| 140.929 | 63.1357 | 2.547 | 75.3333 | 2.765 | 81.5744 | 2.842 | 87.8066 | 2.926 | 93.5747 | 3.013 | 99.7165 | 3.109 |
| 142.561 | 63.3418 | 2.541 | 75.6584 | 2.757 | 81.9462 | 2.833 | 88.1593 | 2.917 | 93.9886 | 3.004 | 100.204 | 3.097 |
| 144.212 | 63.7662 | 2.536 | 75.9926 | 2.750 | 82.3279 | 2.826 | 88.6100 | 2.908 | 94.4789 | 2.994 | 100.736 | 3.086 |
| 145.881 | 64.1039 | 2.531 | 76.3673 | 2.743 | 82.7638 | 2.817 | 89.1056 | 2.898 | 95.0415 | 2.983 | 101.277 | 3.075 |
| 147.571 | 64.4091 | 2.524 | 76.6858 | 2.736 | 83.1426 | 2.810 | 89.5297 | 2.890 | 95.4535 | 2.973 | 101.781 | 3.064 |
| 149.279 | 64.8042 | 2.518 | 77.0624 | 2.729 | 83.5426 | 2.801 | 89.9535 | 2.881 | 95.9463 | 2.963 | 102.335 | 3.053 |
| 151.008 | 64.9163 | 2.513 | 77.4165 | 2.722 | 83.9343 | 2.793 | 90.4169 | 2.872 | 96.4519 | 2.953 | 102.848 | 3.042 |
| 152.757 | 65.3577 | 2.508 | 77.7928 | 2.715 | 84.3221 | 2.785 | 90.7894 | 2.863 | 96.9063 | 2.944 | 103.365 | 3.031 |
| 154.525 | 65.6678 | 2.502 | 78.1753 | 2.708 | 84.7523 | 2.778 | 91.2877 | 2.854 | 97.4062 | 2.934 | 103.928 | 3.021 |
| 156.315 | 65.9410 | 2.497 | 78.5460 | 2.702 | 85.1556 | 2.770 | 91.7372 | 2.846 | 97.8835 | 2.925 | 104.526 | 3.010 |
| 158.125 | 66.3176 | 2.491 | 78.8980 | 2.694 | 85.5439 | 2.762 | 92.1517 | 2.837 | 98.4095 | 2.915 | 105.019 | 3.000 |
| 159.956 | 66.4692 | 2.487 | 79.2236 | 2.688 | 85.9482 | 2.755 | 92.5930 | 2.829 | 98.8570 | 2.906 | 105.585 | 2.989 |
| 161.808 | 66.8083 | 2.480 | 79.7592 | 2.682 | 86.5467 | 2.750 | 93.2029 | 2.821 | 99.6342 | 2.896 | 106.127 | 2.978 |
| 163.682 | 67.3567 | 2.475 | 80.0815 | 2.675 | 86.8815 | 2.743 | 93.4794 | 2.813 | 100.016 | 2.888 | 106.639 | 2.968 |
| 165.577 | 67.5673 | 2.470 | 80.4072 | 2.668 | 87.3065 | 2.734 | 94.0605 | 2.803 | 100.614 | 2.877 | 107.189 | 2.957 |
| 167.494 | 67.7857 | 2.465 | 80.9583 | 2.661 | 87.8074 | 2.727 | 94.5101 | 2.794 | 101.040 | 2.867 | 107.752 | 2.949 |
| 169.434 | 68.1752 | 2.461 | 81.1570 | 2.657 | 88.1568 | 2.723 | 95.0018 | 2.789 | 101.697 | 2.861 | 108.278 | 2.938 |
| 171.396 | 68.3863 | 2.456 | 81.5614 | 2.650 | 88.5500 | 2.714 | 95.3962 | 2.780 | 102.111 | 2.850 | 108.876 | 2.928 |
| 173.380 | 68.8199 | 2.451 | 81.9601 | 2.645 | 88.9287 | 2.709 | 95.9001 | 2.775 | 102.649 | 2.844 | 109.390 | 2.919 |
| 175.388 | 69.0973 | 2.445 | 82.3499 | 2.637 | 89.3834 | 2.700 | 96.2991 | 2.764 | 103.058 | 2.833 | 109.923 | 2.908 |
| 177.419 | 69.4775 | 2.441 | 82.7307 | 2.631 | 89.8258 | 2.693 | 96.8258 | 2.757 | 103.644 | 2.825 | 110.469 | 2.899 |
| 179.473 | 69.8207 | 2.436 | 83.1051 | 2.625 | 90.2462 | 2.686 | 97.2519 | 2.749 | 104.105 | 2.816 | 111.019 | 2.890 |

| | | | | | | | | | | | |
|---|---|---|---|---|---|---|---|---|---|---|---|
| 181.552 | 70.0683 | 2.432 | 83.4354 | 2.619 | 90.5732 | 2.679 | 97.6516 | 2.741 | 104.560 | 2.808 | 111.557 | 2.880 |
| 183.654 | 70.5899 | 2.427 | 83.8935 | 2.613 | 91.0998 | 2.673 | 98.1948 | 2.734 | 105.151 | 2.800 | 112.142 | 2.870 |
| 185.780 | 70.7725 | 2.422 | 84.2468 | 2.607 | 91.5074 | 2.666 | 98.6494 | 2.727 | 105.673 | 2.791 | 112.662 | 2.861 |
| 187.932 | 71.2744 | 2.416 | 84.6646 | 2.601 | 91.9055 | 2.660 | 99.0640 | 2.719 | 106.088 | 2.783 | 113.200 | 2.853 |
| 190.108 | 71.4933 | 2.412 | 84.8379 | 2.597 | 92.2300 | 2.654 | 99.5060 | 2.712 | 106.575 | 2.774 | 113.797 | 2.844 |
| 192.309 | 71.9423 | 2.407 | 85.4493 | 2.591 | 92.7815 | 2.648 | 100.076 | 2.705 | 107.147 | 2.767 | 114.385 | 2.835 |
| 194.536 | 72.2973 | 2.404 | 85.7991 | 2.585 | 93.1108 | 2.642 | 100.361 | 2.699 | 107.556 | 2.760 | 114.894 | 2.826 |
| 196.789 | 72.6161 | 2.400 | 86.1306 | 2.579 | 93.5865 | 2.634 | 100.985 | 2.691 | 108.205 | 2.751 | 115.494 | 2.818 |
| 199.067 | 73.0722 | 2.396 | 86.6996 | 2.574 | 94.1629 | 2.628 | 101.618 | 2.684 | 108.916 | 2.744 | 116.084 | 2.809 |
| 201.372 | 73.3478 | 2.392 | 87.0991 | 2.569 | 94.4750 | 2.623 | 101.785 | 2.678 | 109.035 | 2.737 | 116.644 | 2.800 |
| 203.704 | 73.8049 | 2.388 | 87.4967 | 2.563 | 94.9752 | 2.616 | 102.392 | 2.671 | 109.629 | 2.729 | 117.225 | 2.792 |
| 206.063 | 74.2281 | 2.383 | 87.9505 | 2.557 | 95.4771 | 2.610 | 102.984 | 2.663 | 110.300 | 2.721 | 117.788 | 2.783 |
| 208.449 | 74.4166 | 2.379 | 88.3237 | 2.551 | 95.9297 | 2.603 | 103.519 | 2.656 | 110.906 | 2.713 | 118.396 | 2.775 |
| 210.863 | 74.7045 | 2.376 | 88.7682 | 2.546 | 96.4136 | 2.597 | 103.944 | 2.649 | 111.410 | 2.705 | 118.965 | 2.766 |
| 213.304 | 75.0332 | 2.371 | 89.2023 | 2.541 | 96.8477 | 2.591 | 104.441 | 2.643 | 111.821 | 2.698 | 119.484 | 2.758 |
| 215.774 | 75.5860 | 2.366 | 89.6726 | 2.536 | 97.2788 | 2.585 | 104.822 | 2.635 | 112.392 | 2.690 | 120.068 | 2.751 |
| 218.273 | 75.9019 | 2.364 | 90.0787 | 2.531 | 97.7524 | 2.578 | 105.381 | 2.628 | 112.974 | 2.682 | 120.600 | 2.743 |
| 220.800 | 76.2926 | 2.361 | 90.4704 | 2.526 | 98.1933 | 2.572 | 105.848 | 2.621 | 113.512 | 2.675 | 121.196 | 2.735 |
| 223.357 | 76.6480 | 2.356 | 90.9272 | 2.521 | 98.6658 | 2.567 | 106.405 | 2.615 | 114.080 | 2.668 | 121.783 | 2.727 |
| 225.944 | 76.9449 | 2.354 | 91.3203 | 2.515 | 99.1584 | 2.561 | 106.938 | 2.609 | 114.752 | 2.661 | 122.372 | 2.719 |
| 228.560 | 77.3222 | 2.349 | 91.7958 | 2.510 | 99.6792 | 2.554 | 107.416 | 2.601 | 115.138 | 2.653 | 122.983 | 2.711 |
| 231.206 | 77.7199 | 2.345 | 92.2106 | 2.505 | 100.099 | 2.549 | 107.941 | 2.596 | 115.756 | 2.646 | 123.560 | 2.704 |
| 233.884 | 78.0592 | 2.341 | 92.5805 | 2.501 | 100.469 | 2.544 | 108.326 | 2.589 | 116.208 | 2.640 | 124.117 | 2.697 |
| 236.592 | 78.3420 | 2.339 | 93.0732 | 2.496 | 100.993 | 2.538 | 108.838 | 2.584 | 116.692 | 2.634 | 124.709 | 2.689 |
| 239.332 | 78.8272 | 2.335 | 93.4564 | 2.491 | 101.393 | 2.533 | 109.331 | 2.577 | 117.292 | 2.625 | 125.302 | 2.681 |
| 242.103 | 79.1659 | 2.330 | 93.9766 | 2.486 | 101.999 | 2.527 | 110.003 | 2.571 | 118.033 | 2.619 | 125.858 | 2.674 |
| 244.906 | 79.5548 | 2.327 | 94.3766 | 2.481 | 102.422 | 2.523 | 110.405 | 2.566 | 118.426 | 2.613 | 126.466 | 2.667 |
| 247.742 | 79.8879 | 2.323 | 94.9000 | 2.476 | 103.041 | 2.516 | 111.087 | 2.559 | 119.095 | 2.605 | 127.034 | 2.660 |
| 250.611 | 80.2156 | 2.319 | 95.3974 | 2.472 | 103.524 | 2.512 | 111.555 | 2.554 | 119.634 | 2.600 | 127.690 | 2.653 |
| 253.513 | 80.7648 | 2.315 | 95.7623 | 2.467 | 103.868 | 2.507 | 111.926 | 2.549 | 120.061 | 2.594 | 128.249 | 2.646 |
| 256.448 | 81.0363 | 2.312 | 96.1427 | 2.461 | 104.319 | 2.500 | 112.472 | 2.541 | 120.791 | 2.585 | 128.891 | 2.639 |
| 259.418 | 81.4930 | 2.308 | 96.6107 | 2.458 | 104.832 | 2.495 | 113.005 | 2.535 | 121.288 | 2.579 | 129.443 | 2.632 |
| 262.422 | 81.9256 | 2.305 | 97.1186 | 2.453 | 105.322 | 2.491 | 113.527 | 2.531 | 121.839 | 2.574 | 129.949 | 2.625 |
| 265.461 | 82.2834 | 2.302 | 97.5569 | 2.448 | 105.821 | 2.485 | 114.030 | 2.524 | 122.423 | 2.567 | 130.616 | 2.618 |
| 268.534 | 82.7284 | 2.298 | 97.9649 | 2.443 | 106.218 | 2.481 | 114.467 | 2.520 | 122.839 | 2.561 | 131.149 | 2.613 |
| 271.644 | 83.1997 | 2.294 | 98.2821 | 2.439 | 106.679 | 2.476 | 114.994 | 2.513 | 123.359 | 2.555 | 131.744 | 2.606 |
| 274.789 | 83.6254 | 2.291 | 98.9960 | 2.436 | 107.212 | 2.472 | 115.356 | 2.510 | 123.844 | 2.551 | 132.397 | 2.600 |
| 277.971 | 84.0432 | 2.287 | 99.3663 | 2.431 | 107.722 | 2.466 | 116.082 | 2.503 | 124.597 | 2.543 | 132.875 | 2.593 |
| 281.190 | 84.4727 | 2.284 | 99.8067 | 2.426 | 108.180 | 2.461 | 116.596 | 2.497 | 125.117 | 2.538 | 133.565 | 2.588 |
| 284.446 | 84.8202 | 2.281 | 100.250 | 2.423 | 108.655 | 2.458 | 117.024 | 2.494 | 125.637 | 2.533 | 134.325 | 2.582 |
| 287.740 | 85.3194 | 2.279 | 100.757 | 2.420 | 109.220 | 2.454 | 117.735 | 2.489 | 126.352 | 2.528 | 134.711 | 2.576 |
| 291.072 | 85.7422 | 2.276 | 100.953 | 2.416 | 109.678 | 2.448 | 118.287 | 2.484 | 127.070 | 2.522 | 135.550 | 2.570 |
| 294.442 | 86.0603 | 2.272 | 101.657 | 2.412 | 110.057 | 2.444 | 118.733 | 2.478 | 127.635 | 2.516 | 136.322 | 2.564 |
| 297.852 | 86.0280 | 2.270 | 102.264 | 2.407 | 110.846 | 2.440 | 119.478 | 2.473 | 128.212 | 2.511 | 136.826 | 2.560 |
| 301.301 | 86.5791 | 2.267 | 102.818 | 2.404 | 111.349 | 2.436 | 119.888 | 2.470 | 128.671 | 2.506 | 137.557 | 2.551 |
| 304.789 | 87.0259 | 2.263 | 103.352 | 2.400 | 111.948 | 2.432 | 120.463 | 2.465 | 129.210 | 2.502 | 138.187 | 2.546 |
| 308.319 | 87.4469 | 2.260 | 103.786 | 2.397 | 112.389 | 2.428 | 121.096 | 2.460 | 130.111 | 2.497 | 138.858 | 2.541 |

| | | | | | | | | | | | |
|---|---|---|---|---|---|---|---|---|---|---|---|
| 311.889 | 87.9268 | 2.257 | 104.254 | 2.392 | 112.843 | 2.422 | 121.616 | 2.454 | 130.641 | 2.489 | 139.824 | 2.536 |
| 315.500 | 88.2862 | 2.253 | 105.058 | 2.388 | 113.708 | 2.418 | 122.445 | 2.450 | 131.511 | 2.485 | 140.266 | 2.531 |
| 319.154 | 89.0217 | 2.250 | 105.357 | 2.385 | 114.164 | 2.415 | 122.950 | 2.445 | 132.100 | 2.478 | 140.790 | 2.524 |
| 322.849 | 89.3637 | 2.247 | 105.339 | 2.380 | 114.225 | 2.413 | 123.176 | 2.445 | 131.560 | 2.478 | 141.523 | 2.509 |
| 326.588 | 89.8755 | 2.244 | 105.749 | 2.373 | 114.391 | 2.404 | 123.390 | 2.437 | 132.076 | 2.470 | 142.252 | 2.506 |
| 330.370 | 90.1558 | 2.240 | 106.745 | 2.373 | 115.568 | 2.402 | 124.557 | 2.436 | 133.396 | 2.469 | 142.630 | 2.500 |
| 334.195 | 90.7514 | 2.237 | 107.578 | 2.367 | 116.303 | 2.397 | 125.450 | 2.430 | 134.261 | 2.462 | 143.384 | 2.494 |
| 338.065 | 91.0987 | 2.234 | 108.139 | 2.363 | 116.881 | 2.392 | 126.017 | 2.424 | 135.115 | 2.455 | 144.364 | 2.491 |
| 341.979 | 91.5901 | 2.231 | 108.703 | 2.358 | 117.646 | 2.387 | 126.840 | 2.419 | 135.758 | 2.449 | 145.047 | 2.486 |
| 345.939 | 91.9649 | 2.228 | 109.000 | 2.356 | 118.293 | 2.385 | 127.327 | 2.417 | 136.122 | 2.449 | 145.742 | 2.478 |
| 349.945 | 92.5835 | 2.224 | 109.901 | 2.352 | 118.787 | 2.379 | 127.661 | 2.410 | 136.985 | 2.442 | 146.455 | 2.474 |
| 353.997 | 93.0270 | 2.221 | 109.889 | 2.349 | 119.100 | 2.376 | 128.318 | 2.406 | 137.238 | 2.437 | 147.164 | 2.469 |
| 358.096 | 93.4444 | 2.218 | 110.277 | 2.345 | 119.292 | 2.373 | 128.461 | 2.402 | 137.905 | 2.431 | 147.959 | 2.462 |
| 362.243 | 94.0642 | 2.215 | 111.506 | 2.345 | 120.647 | 2.372 | 129.886 | 2.401 | 139.330 | 2.429 | 148.644 | 2.457 |
| 366.438 | 94.4413 | 2.212 | 112.170 | 2.339 | 121.136 | 2.367 | 130.665 | 2.395 | 139.789 | 2.425 | 149.359 | 2.452 |
| 370.681 | 95.1500 | 2.210 | 112.563 | 2.336 | 121.677 | 2.364 | 130.906 | 2.392 | 140.358 | 2.420 | 149.915 | 2.449 |
| 374.973 | 95.8147 | 2.207 | 113.099 | 2.332 | 122.385 | 2.359 | 131.628 | 2.388 | 140.803 | 2.416 | 150.846 | 2.443 |
| 379.315 | 96.3229 | 2.204 | 114.003 | 2.329 | 123.286 | 2.355 | 132.687 | 2.384 | 141.976 | 2.410 | 151.655 | 2.437 |
| 383.707 | 97.0427 | 2.200 | 114.476 | 2.325 | 123.801 | 2.351 | 133.144 | 2.379 | 142.553 | 2.406 | 152.253 | 2.432 |
| 388.150 | 97.5639 | 2.198 | 114.988 | 2.320 | 124.311 | 2.346 | 133.520 | 2.374 | 142.825 | 2.400 | 153.033 | 2.428 |
| 392.645 | 98.0556 | 2.195 | 115.861 | 2.318 | 125.136 | 2.344 | 134.664 | 2.371 | 144.319 | 2.396 | 153.875 | 2.422 |
| 397.192 | 98.5034 | 2.192 | 116.442 | 2.313 | 125.673 | 2.338 | 135.012 | 2.365 | 144.583 | 2.391 | 154.398 | 2.418 |
| 401.791 | 99.3549 | 2.189 | 116.640 | 2.315 | 126.017 | 2.337 | 135.605 | 2.365 | 145.044 | 2.392 | 155.391 | 2.412 |
| 406.443 | 99.6806 | 2.186 | 117.601 | 2.308 | 127.023 | 2.332 | 136.576 | 2.358 | 146.384 | 2.383 | 155.957 | 2.409 |
| 411.150 | 100.177 | 2.184 | 118.481 | 2.307 | 127.945 | 2.330 | 137.399 | 2.357 | 146.757 | 2.382 | 156.889 | 2.403 |
| 415.911 | 100.824 | 2.181 | 118.718 | 2.302 | 128.099 | 2.325 | 137.790 | 2.351 | 147.243 | 2.375 | 157.656 | 2.400 |
| 420.727 | 101.429 | 2.178 | 119.751 | 2.300 | 129.204 | 2.322 | 138.893 | 2.348 | 148.315 | 2.371 | 158.232 | 2.393 |
| 425.598 | 102.085 | 2.176 | 120.405 | 2.296 | 129.767 | 2.317 | 139.378 | 2.343 | 148.736 | 2.368 | 159.292 | 2.389 |
| 430.527 | 102.688 | 2.173 | 120.182 | 2.291 | 129.672 | 2.311 | 139.481 | 2.335 | 149.797 | 2.359 | 159.993 | 2.385 |
| 435.512 | 103.057 | 2.170 | 121.490 | 2.288 | 131.185 | 2.309 | 140.717 | 2.335 | 150.698 | 2.358 | 160.735 | 2.381 |
| 440.555 | 103.730 | 2.167 | 122.159 | 2.286 | 131.685 | 2.308 | 141.653 | 2.332 | 151.428 | 2.355 | 161.556 | 2.376 |
| 445.656 | 104.535 | 2.165 | 122.764 | 2.283 | 132.575 | 2.305 | 142.429 | 2.329 | 151.984 | 2.353 | 162.492 | 2.371 |
| 450.817 | 105.199 | 2.162 | 123.989 | 2.282 | 133.659 | 2.302 | 143.492 | 2.325 | 153.641 | 2.348 | 163.429 | 2.367 |
| 456.037 | 106.081 | 2.159 | 125.021 | 2.278 | 134.459 | 2.299 | 144.129 | 2.321 | 154.309 | 2.344 | 164.221 | 2.363 |
| 461.318 | 106.694 | 2.157 | 125.305 | 2.275 | 135.079 | 2.294 | 145.023 | 2.317 | 155.091 | 2.340 | 164.974 | 2.358 |
| 466.659 | 107.455 | 2.153 | 126.028 | 2.271 | 135.925 | 2.292 | 145.798 | 2.314 | 155.604 | 2.335 | 165.906 | 2.354 |
| 472.063 | 108.025 | 2.152 | 126.807 | 2.268 | 136.254 | 2.288 | 146.113 | 2.310 | 156.220 | 2.330 | 166.705 | 2.349 |
| 477.529 | 108.699 | 2.148 | 127.424 | 2.265 | 137.133 | 2.284 | 147.181 | 2.306 | 157.334 | 2.326 | 167.648 | 2.345 |
| 483.059 | 109.078 | 2.146 | 128.412 | 2.263 | 137.988 | 2.284 | 148.366 | 2.303 | 158.066 | 2.322 | 168.336 | 2.341 |
| 488.652 | 109.772 | 2.143 | 129.060 | 2.259 | 139.118 | 2.279 | 148.856 | 2.300 | 158.903 | 2.319 | 169.197 | 2.337 |
| 494.311 | 110.787 | 2.140 | 129.615 | 2.257 | 139.512 | 2.276 | 148.990 | 2.297 | 159.304 | 2.314 | 170.365 | 2.335 |
| 500.035 | 111.294 | 2.137 | 130.359 | 2.253 | 140.308 | 2.273 | 149.878 | 2.292 | 160.269 | 2.311 | 171.006 | 2.333 |
| 505.825 | 112.108 | 2.135 | 131.138 | 2.250 | 140.873 | 2.269 | 150.962 | 2.289 | 160.908 | 2.308 | 172.060 | 2.327 |
| 511.682 | 112.570 | 2.124 | 131.762 | 2.247 | 142.074 | 2.266 | 151.969 | 2.285 | 161.640 | 2.303 | 172.787 | 2.323 |
| 517.607 | 113.245 | 2.122 | 132.819 | 2.244 | 142.624 | 2.263 | 152.688 | 2.281 | 163.022 | 2.299 | 173.708 | 2.320 |
| 523.600 | 113.978 | 2.119 | 133.520 | 2.241 | 143.603 | 2.259 | 153.569 | 2.278 | 163.601 | 2.296 | 174.381 | 2.316 |
| 529.663 | 114.865 | 2.117 | 134.619 | 2.238 | 144.241 | 2.256 | 154.696 | 2.274 | 164.636 | 2.292 | 175.607 | 2.312 |

| | | | | | | | | | | | |
|---|---|---|---|---|---|---|---|---|---|---|---|
| 535.797 | 115.482 | 2.115 | 135.323 | 2.236 | 145.198 | 2.253 | 155.296 | 2.272 | 165.496 | 2.289 | 176.555 | 2.307 |
| 542.001 | 116.356 | 2.113 | 135.860 | 2.232 | 146.195 | 2.250 | 156.034 | 2.268 | 166.396 | 2.284 | 177.342 | 2.303 |
| 548.277 | 117.075 | 2.111 | 137.002 | 2.230 | 147.047 | 2.247 | 156.684 | 2.264 | 167.282 | 2.280 | 178.296 | 2.300 |
| 554.626 | 117.814 | 2.109 | 137.942 | 2.227 | 147.859 | 2.243 | 157.931 | 2.261 | 168.108 | 2.276 | 179.372 | 2.296 |
| 561.048 | 118.593 | 2.106 | 138.553 | 2.224 | 148.674 | 2.240 | 158.844 | 2.257 | 168.770 | 2.273 | 180.211 | 2.293 |
| 567.545 | 119.264 | 2.104 | 139.631 | 2.221 | 149.322 | 2.237 | 160.180 | 2.255 | 169.981 | 2.269 | 181.195 | 2.288 |
| 574.116 | 120.139 | 2.101 | 139.985 | 2.218 | 150.063 | 2.234 | 160.316 | 2.252 | 171.192 | 2.266 | 182.333 | 2.284 |
| 580.764 | 120.952 | 2.099 | 141.263 | 2.215 | 151.605 | 2.231 | 161.792 | 2.248 | 172.083 | 2.262 | 183.291 | 2.280 |
| 587.489 | 121.738 | 2.097 | 141.754 | 2.213 | 152.098 | 2.228 | 162.664 | 2.245 | 172.754 | 2.258 | 184.154 | 2.277 |
| 594.292 | 122.545 | 2.094 | 142.772 | 2.209 | 153.480 | 2.226 | 162.763 | 2.242 | 173.488 | 2.255 | 184.902 | 2.273 |
| 601.174 | 123.460 | 2.092 | 143.812 | 2.206 | 154.022 | 2.223 | 164.227 | 2.238 | 174.659 | 2.251 | 186.046 | 2.270 |
| 608.135 | 124.439 | 2.088 | 144.371 | 2.204 | 154.862 | 2.219 | 165.387 | 2.235 | 176.175 | 2.248 | 187.187 | 2.266 |
| 615.177 | 124.964 | 2.087 | 145.791 | 2.201 | 155.876 | 2.217 | 166.208 | 2.232 | 176.270 | 2.245 | 188.225 | 2.262 |
| 622.300 | 126.004 | 2.085 | 146.663 | 2.198 | 156.794 | 2.214 | 167.630 | 2.229 | 177.818 | 2.241 | 189.138 | 2.259 |
| 629.506 | 126.817 | 2.083 | 147.644 | 2.196 | 157.908 | 2.210 | 168.498 | 2.226 | 178.853 | 2.238 | 189.949 | 2.255 |
| 636.796 | 127.624 | 2.081 | 148.194 | 2.192 | 158.703 | 2.209 | 169.154 | 2.223 | 179.924 | 2.235 | 191.281 | 2.252 |
| 644.169 | 128.840 | 2.079 | 149.736 | 2.190 | 159.545 | 2.205 | 170.624 | 2.220 | 180.813 | 2.231 | 192.295 | 2.248 |
| 651.628 | 129.501 | 2.076 | 150.266 | 2.187 | 160.859 | 2.202 | 171.266 | 2.218 | 182.259 | 2.229 | 193.515 | 2.245 |
| 659.174 | 130.094 | 2.073 | 151.026 | 2.184 | 161.680 | 2.199 | 171.675 | 2.213 | 183.164 | 2.224 | 194.372 | 2.241 |
| 666.807 | 131.144 | 2.071 | 152.463 | 2.181 | 162.953 | 2.197 | 173.475 | 2.211 | 183.598 | 2.222 | 195.556 | 2.237 |
| 674.528 | 132.184 | 2.068 | 153.850 | 2.179 | 163.665 | 2.194 | 174.047 | 2.208 | 185.303 | 2.217 | 196.863 | 2.234 |
| 682.339 | 132.927 | 2.066 | 154.315 | 2.176 | 165.042 | 2.190 | 175.425 | 2.204 | 186.150 | 2.214 | 198.046 | 2.232 |
| 690.240 | 133.489 | 2.065 | 154.939 | 2.174 | 165.609 | 2.188 | 176.677 | 2.202 | 187.935 | 2.212 | 198.926 | 2.228 |
| 698.232 | 134.648 | 2.063 | 156.613 | 2.171 | 166.981 | 2.184 | 177.810 | 2.198 | 188.880 | 2.209 | 199.926 | 2.225 |
| 706.318 | 135.408 | 2.061 | 157.253 | 2.168 | 168.406 | 2.181 | 178.355 | 2.195 | 190.087 | 2.206 | 201.528 | 2.220 |
| 714.496 | 136.567 | 2.059 | 158.160 | 2.166 | 168.883 | 2.179 | 180.411 | 2.193 | 190.970 | 2.202 | 202.248 | 2.217 |
| 722.770 | 137.217 | 2.058 | 159.038 | 2.163 | 170.299 | 2.177 | 180.489 | 2.189 | 192.310 | 2.199 | 203.139 | 2.214 |
| 731.139 | 138.346 | 2.055 | 160.554 | 2.159 | 171.168 | 2.174 | 182.754 | 2.187 | 192.968 | 2.195 | 204.784 | 2.211 |
| 739.605 | 139.365 | 2.053 | 161.970 | 2.158 | 172.237 | 2.171 | 183.053 | 2.185 | 194.057 | 2.193 | 205.475 | 2.207 |
| 748.170 | 139.985 | 2.051 | 163.059 | 2.154 | 173.417 | 2.168 | 184.835 | 2.181 | 195.556 | 2.188 | 206.835 | 2.204 |
| 756.833 | 140.919 | 2.048 | 163.778 | 2.151 | 174.183 | 2.165 | 185.162 | 2.178 | 196.772 | 2.188 | 208.704 | 2.200 |
| 765.597 | 141.447 | 2.047 | 163.806 | 2.152 | 175.049 | 2.166 | 185.872 | 2.177 | 197.129 | 2.188 | 210.863 | 2.199 |
| 774.462 | 142.717 | 2.045 | 166.447 | 2.147 | 177.586 | 2.162 | 187.977 | 2.171 | 199.327 | 2.179 | 210.318 | 2.195 |
| 783.430 | 143.622 | 2.043 | 166.605 | 2.148 | 177.159 | 2.161 | 188.724 | 2.167 | 199.265 | 2.179 | 211.926 | 2.193 |
| 792.501 | 144.694 | 2.041 | 168.143 | 2.144 | 178.748 | 2.154 | 189.929 | 2.161 | 201.328 | 2.176 | 213.851 | 2.189 |
| 801.678 | 146.029 | 2.038 | 169.277 | 2.142 | 180.127 | 2.156 | 191.447 | 2.161 | 202.617 | 2.174 | 214.287 | 2.187 |
| 810.961 | 146.835 | 2.036 | 169.973 | 2.138 | 181.147 | 2.151 | 192.661 | 2.160 | 203.052 | 2.170 | 215.914 | 2.184 |
| 820.352 | 148.095 | 2.033 | 171.592 | 2.136 | 182.977 | 2.148 | 193.849 | 2.155 | 204.543 | 2.167 | 217.096 | 2.179 |
| 829.851 | 148.916 | 2.031 | 173.312 | 2.133 | 184.140 | 2.147 | 195.236 | 2.153 | 205.952 | 2.164 | 218.048 | 2.177 |
| 839.460 | 149.833 | 2.028 | 174.180 | 2.130 | 185.882 | 2.142 | 196.053 | 2.148 | 207.171 | 2.157 | 219.453 | 2.174 |
| 849.180 | 150.956 | 2.027 | 174.966 | 2.128 | 186.290 | 2.141 | 197.391 | 2.147 | 207.274 | 2.158 | 220.860 | 2.172 |
| 859.014 | 152.244 | 2.025 | 175.792 | 2.125 | 187.712 | 2.138 | 198.367 | 2.145 | 209.571 | 2.153 | 222.163 | 2.168 |
| 868.960 | 153.128 | 2.023 | 177.215 | 2.123 | 188.446 | 2.135 | 199.247 | 2.142 | 210.861 | 2.149 | 223.211 | 2.166 |
| 879.023 | 154.454 | 2.020 | 179.005 | 2.120 | 191.012 | 2.132 | 201.667 | 2.139 | 212.286 | 2.147 | 224.752 | 2.163 |
| 889.201 | 155.571 | 2.019 | 179.964 | 2.119 | 191.070 | 2.131 | 202.419 | 2.137 | 213.657 | 2.144 | 226.690 | 2.158 |
| 899.498 | 156.509 | 2.017 | 181.540 | 2.115 | 193.015 | 2.127 | 203.762 | 2.133 | 215.073 | 2.140 | 227.754 | 2.155 |
| 909.913 | 157.889 | 2.015 | 182.316 | 2.113 | 193.632 | 2.126 | 205.246 | 2.132 | 216.059 | 2.137 | 229.011 | 2.153 |

| | | | | | | | | | | | |
|---|---|---|---|---|---|---|---|---|---|---|---|
| 920.450 | 159.150 | 2.013 | 183.784 | 2.112 | 195.417 | 2.123 | 206.678 | 2.127 | 217.675 | 2.135 | 230.345 | 2.150 |
| 931.108 | 160.256 | 2.011 | 184.892 | 2.108 | 196.493 | 2.120 | 207.894 | 2.126 | 218.878 | 2.132 | 231.798 | 2.148 |
| 941.890 | 161.380 | 2.009 | 187.004 | 2.106 | 198.007 | 2.118 | 209.302 | 2.123 | 220.940 | 2.130 | 233.112 | 2.144 |
| 952.796 | 162.482 | 2.007 | 188.156 | 2.104 | 199.367 | 2.115 | 210.462 | 2.118 | 222.290 | 2.125 | 234.838 | 2.140 |
| 963.829 | 163.356 | 2.005 | 188.801 | 2.101 | 200.348 | 2.112 | 211.559 | 2.117 | 222.675 | 2.125 | 236.374 | 2.138 |
| 974.990 | 164.636 | 2.003 | 190.397 | 2.099 | 200.779 | 2.110 | 213.173 | 2.117 | 223.601 | 2.123 | 237.327 | 2.135 |
| 986.279 | 165.966 | 2.001 | 191.623 | 2.097 | 203.726 | 2.108 | 214.416 | 2.113 | 225.661 | 2.119 | 238.361 | 2.133 |
| 997.700 | 167.040 | 1.998 | 191.451 | 2.093 | 204.549 | 2.106 | 216.485 | 2.111 | 227.413 | 2.119 | 239.783 | 2.129 |
| 1009.25 | 166.962 | 1.996 | 193.891 | 2.092 | 204.511 | 2.103 | 217.746 | 2.110 | 228.879 | 2.118 | 241.333 | 2.129 |
| 1020.94 | 168.378 | 1.994 | 195.156 | 2.089 | 206.369 | 2.101 | 217.923 | 2.107 | 230.011 | 2.114 | 242.921 | 2.123 |
| 1032.76 | 169.463 | 1.992 | 196.220 | 2.084 | 207.889 | 2.096 | 218.976 | 2.102 | 230.365 | 2.109 | 245.216 | 2.123 |
| 1044.72 | 171.006 | 1.989 | 197.824 | 2.084 | 209.319 | 2.096 | 222.305 | 2.100 | 234.113 | 2.108 | 246.531 | 2.119 |
| 1056.82 | 171.483 | 1.987 | 198.953 | 2.081 | 211.431 | 2.093 | 222.561 | 2.098 | 234.993 | 2.108 | 247.929 | 2.118 |
| 1069.05 | 172.528 | 1.985 | 200.741 | 2.077 | 212.313 | 2.088 | 223.223 | 2.099 | 234.609 | 2.109 | 249.341 | 2.115 |
| 1081.43 | 171.666 | 1.981 | 200.113 | 2.073 | 213.385 | 2.082 | 224.599 | 2.094 | 236.967 | 2.103 | 250.986 | 2.108 |
| 1093.96 | 172.245 | 1.979 | 202.175 | 2.073 | 214.211 | 2.087 | 226.168 | 2.090 | 239.381 | 2.104 | 254.390 | 2.109 |
| 1106.62 | 173.935 | 1.977 | 204.592 | 2.069 | 216.052 | 2.083 | 227.808 | 2.093 | 242.203 | 2.101 | 255.830 | 2.106 |
| 1119.44 | 175.342 | 1.974 | 205.376 | 2.069 | 216.863 | 2.077 | 228.331 | 2.088 | 244.021 | 2.096 | 256.347 | 2.099 |
| 1124.60 | 175.088 | 1.973 | 206.291 | 2.066 | 218.008 | 2.081 | 229.245 | 2.085 | 242.705 | 2.094 | 255.156 | 2.100 |

| Freq (GHz) | Water at 30 ºC a (cm$^{-1}$) | n | Water at 35 ºC a (cm$^{-1}$) | n | Water at 40 ºC a (cm$^{-1}$) | n | Water at 45 ºC a (cm$^{-1}$) | n | Water at 50 ºC a (cm$^{-1}$) | n | Water at 55 ºC a (cm$^{-1}$) | n | Water at 60 ºC a (cm$^{-1}$) | n |
|---|---|---|---|---|---|---|---|---|---|---|---|---|---|---|
| 0.05000 | 0.00010 | 8.778 | 0.00048 | 8.703 | 0.00010 | 8.608 | 0.00090 | 8.497 | 0.00017 | 8.389 | 0.00055 | 8.295 | 0.00013 | 8.193 |
| 0.05058 | 0.00013 | 8.779 | 0.00050 | 8.703 | 0.00014 | 8.607 | 0.00080 | 8.496 | 0.00028 | 8.387 | 0.00047 | 8.294 | 0.00010 | 8.194 |
| 0.05117 | 0.00016 | 8.780 | 0.00035 | 8.703 | 0.00011 | 8.605 | 0.00087 | 8.497 | 0.00026 | 8.388 | 0.00053 | 8.292 | 0.00007 | 8.195 |
| 0.05176 | 0.00012 | 8.780 | 0.00035 | 8.706 | 0.00014 | 8.602 | 0.00070 | 8.496 | 0.00008 | 8.393 | 0.00044 | 8.295 | 0.00004 | 8.196 |
| 0.05236 | 0.00008 | 8.780 | 0.00042 | 8.710 | 0.00018 | 8.599 | 0.00044 | 8.486 | 0.00007 | 8.397 | 0.00030 | 8.293 | 0.00007 | 8.195 |
| 0.05297 | 0.00013 | 8.784 | 0.00054 | 8.706 | 0.00011 | 8.597 | 0.00059 | 8.489 | 0.00005 | 8.396 | 0.00012 | 8.291 | 0.00013 | 8.191 |
| 0.05358 | 0.00014 | 8.783 | 0.00055 | 8.708 | 0.00007 | 8.600 | 0.00045 | 8.495 | 0.00006 | 8.391 | 0.00027 | 8.294 | 0.00011 | 8.192 |
| 0.05420 | 0.00019 | 8.780 | 0.00038 | 8.708 | 0.00001 | 8.602 | 0.00050 | 8.498 | 0.00011 | 8.388 | 0.00044 | 8.295 | 0.00006 | 8.196 |
| 0.05483 | 0.00022 | 8.785 | 0.00025 | 8.706 | 0.00004 | 8.600 | 0.00038 | 8.493 | 0.00005 | 8.395 | 0.00054 | 8.295 | 0.00007 | 8.197 |
| 0.05546 | 0.00011 | 8.786 | 0.00038 | 8.708 | 0.00005 | 8.600 | 0.00060 | 8.496 | 0.00015 | 8.397 | 0.00060 | 8.294 | 0.00006 | 8.195 |
| 0.05610 | 0.00007 | 8.784 | 0.00033 | 8.710 | 0.00003 | 8.599 | 0.00053 | 8.490 | 0.00012 | 8.397 | 0.00053 | 8.297 | 0.00008 | 8.193 |
| 0.05675 | 0.00009 | 8.783 | 0.00029 | 8.704 | 0.00006 | 8.597 | 0.00050 | 8.493 | 0.00028 | 8.402 | 0.00053 | 8.299 | 0.00015 | 8.193 |
| 0.05741 | 0.00008 | 8.783 | 0.00029 | 8.704 | 0.00008 | 8.597 | 0.00054 | 8.502 | 0.00025 | 8.402 | 0.00039 | 8.296 | 0.00016 | 8.195 |
| 0.05808 | 0.00003 | 8.780 | 0.00033 | 8.706 | 0.00003 | 8.601 | 0.00064 | 8.492 | 0.00044 | 8.397 | 0.00049 | 8.297 | 0.00015 | 8.196 |
| 0.05875 | 0.00003 | 8.782 | 0.00038 | 8.703 | 0.00003 | 8.604 | 0.00043 | 8.497 | 0.00009 | 8.398 | 0.00047 | 8.301 | 0.00014 | 8.197 |
| 0.05943 | 0.00011 | 8.783 | 0.00046 | 8.707 | 0.00004 | 8.601 | 0.00041 | 8.498 | 0.00027 | 8.396 | 0.00050 | 8.300 | 0.00015 | 8.196 |
| 0.06012 | 0.00018 | 8.782 | 0.00060 | 8.709 | 0.00005 | 8.603 | 0.00032 | 8.494 | 0.00015 | 8.396 | 0.00039 | 8.299 | 0.00019 | 8.197 |
| 0.06081 | 0.00022 | 8.784 | 0.00055 | 8.707 | 0.00006 | 8.603 | 0.00032 | 8.498 | 0.00010 | 8.400 | 0.00043 | 8.298 | 0.00019 | 8.197 |
| 0.06152 | 0.00018 | 8.784 | 0.00051 | 8.706 | 0.00011 | 8.603 | 0.00030 | 8.498 | 0.00027 | 8.403 | 0.00058 | 8.296 | 0.00017 | 8.196 |
| 0.06223 | 0.00029 | 8.784 | 0.00037 | 8.707 | 0.00019 | 8.605 | 0.00046 | 8.503 | 0.00039 | 8.396 | 0.00061 | 8.302 | 0.00011 | 8.194 |
| 0.06295 | 0.00021 | 8.782 | 0.00034 | 8.712 | 0.00015 | 8.605 | 0.00011 | 8.493 | 0.00011 | 8.400 | 0.00056 | 8.301 | 0.00009 | 8.192 |
| 0.06368 | 0.00022 | 8.784 | 0.00029 | 8.711 | 0.00014 | 8.605 | 0.00021 | 8.495 | 0.00023 | 8.401 | 0.00054 | 8.298 | 0.00016 | 8.192 |
| 0.06442 | 0.00015 | 8.785 | 0.00034 | 8.710 | 0.00019 | 8.603 | 0.00008 | 8.501 | 0.00063 | 8.397 | 0.00056 | 8.297 | 0.00033 | 8.194 |
| 0.06516 | 0.00016 | 8.784 | 0.00028 | 8.710 | 0.00032 | 8.604 | 0.00018 | 8.498 | 0.00035 | 8.401 | 0.00065 | 8.298 | 0.00045 | 8.195 |
| 0.06592 | 0.00022 | 8.783 | 0.00039 | 8.711 | 0.00036 | 8.605 | 0.00037 | 8.497 | 0.00023 | 8.403 | 0.00056 | 8.298 | 0.00044 | 8.196 |
| 0.06668 | 0.00028 | 8.785 | 0.00042 | 8.708 | 0.00031 | 8.603 | 0.00025 | 8.501 | 0.00032 | 8.404 | 0.00051 | 8.297 | 0.00032 | 8.198 |
| 0.06745 | 0.00024 | 8.784 | 0.00045 | 8.708 | 0.00018 | 8.597 | 0.00027 | 8.499 | 0.00054 | 8.401 | 0.00051 | 8.297 | 0.00028 | 8.201 |
| 0.06823 | 0.00027 | 8.784 | 0.00049 | 8.709 | 0.00011 | 8.599 | 0.00051 | 8.501 | 0.00035 | 8.400 | 0.00051 | 8.299 | 0.00025 | 8.200 |
| 0.06902 | 0.00029 | 8.785 | 0.00046 | 8.706 | 0.00012 | 8.600 | 0.00045 | 8.502 | 0.00033 | 8.401 | 0.00053 | 8.296 | 0.00011 | 8.198 |
| 0.06982 | 0.00014 | 8.783 | 0.00051 | 8.707 | 0.00033 | 8.601 | 0.00015 | 8.498 | 0.00035 | 8.401 | 0.00046 | 8.297 | 0.00008 | 8.199 |
| 0.07063 | 0.00009 | 8.784 | 0.00037 | 8.707 | 0.00038 | 8.601 | 0.00025 | 8.499 | 0.00033 | 8.399 | 0.00062 | 8.299 | 0.00017 | 8.200 |
| 0.07145 | 0.00021 | 8.785 | 0.00037 | 8.708 | 0.00027 | 8.600 | 0.00023 | 8.501 | 0.00035 | 8.401 | 0.00051 | 8.299 | 0.00023 | 8.200 |
| 0.07228 | 0.00024 | 8.784 | 0.00044 | 8.706 | 0.00020 | 8.602 | 0.00022 | 8.501 | 0.00044 | 8.396 | 0.00046 | 8.297 | 0.00021 | 8.200 |
| 0.07311 | 0.00021 | 8.785 | 0.00052 | 8.707 | 0.00028 | 8.601 | 0.00027 | 8.501 | 0.00040 | 8.398 | 0.00049 | 8.297 | 0.00019 | 8.200 |
| 0.07396 | 0.00024 | 8.783 | 0.00058 | 8.709 | 0.00045 | 8.602 | 0.00052 | 8.501 | 0.00032 | 8.401 | 0.00069 | 8.297 | 0.00010 | 8.201 |
| 0.07482 | 0.00031 | 8.785 | 0.00065 | 8.705 | 0.00047 | 8.603 | 0.00034 | 8.503 | 0.00020 | 8.398 | 0.00077 | 8.297 | 0.00006 | 8.201 |
| 0.07568 | 0.00023 | 8.784 | 0.00047 | 8.706 | 0.00049 | 8.603 | 0.00044 | 8.501 | 0.00012 | 8.399 | 0.00084 | 8.296 | 0.00011 | 8.203 |
| 0.07656 | 0.00037 | 8.783 | 0.00046 | 8.707 | 0.00057 | 8.601 | 0.00051 | 8.502 | 0.00035 | 8.399 | 0.00077 | 8.294 | 0.00014 | 8.202 |
| 0.07745 | 0.00041 | 8.784 | 0.00066 | 8.706 | 0.00063 | 8.600 | 0.00048 | 8.502 | 0.00018 | 8.400 | 0.00071 | 8.295 | 0.00010 | 8.201 |
| 0.07834 | 0.00037 | 8.783 | 0.00064 | 8.707 | 0.00046 | 8.600 | 0.00033 | 8.504 | 0.00011 | 8.400 | 0.00067 | 8.296 | 0.00017 | 8.201 |
| 0.07925 | 0.00032 | 8.781 | 0.00070 | 8.706 | 0.00032 | 8.600 | 0.00036 | 8.500 | 0.00030 | 8.401 | 0.00072 | 8.296 | 0.00031 | 8.202 |
| 0.08017 | 0.00034 | 8.782 | 0.00072 | 8.705 | 0.00040 | 8.602 | 0.00028 | 8.501 | 0.00033 | 8.401 | 0.00077 | 8.296 | 0.00027 | 8.200 |
| 0.08110 | 0.00034 | 8.782 | 0.00055 | 8.704 | 0.00043 | 8.603 | 0.00031 | 8.501 | 0.00022 | 8.403 | 0.00074 | 8.295 | 0.00013 | 8.200 |
| 0.08204 | 0.00045 | 8.784 | 0.00041 | 8.704 | 0.00035 | 8.602 | 0.00046 | 8.501 | 0.00071 | 8.401 | 0.00059 | 8.296 | 0.00014 | 8.201 |
| 0.08299 | 0.00038 | 8.785 | 0.00055 | 8.704 | 0.00024 | 8.602 | 0.00036 | 8.501 | 0.00055 | 8.400 | 0.00051 | 8.296 | 0.00015 | 8.200 |

| | | | | | | | | | | | | | |
|---|---|---|---|---|---|---|---|---|---|---|---|---|---|
| 0.08395 | 0.00049 | 8.785 | 0.00068 | 8.705 | 0.00027 | 8.601 | 0.00038 | 8.504 | 0.00035 | 8.402 | 0.00058 | 8.297 | 0.00011 | 8.200 |
| 0.08492 | 0.00042 | 8.784 | 0.00051 | 8.705 | 0.00040 | 8.601 | 0.00048 | 8.503 | 0.00032 | 8.403 | 0.00048 | 8.297 | 0.00015 | 8.200 |
| 0.08590 | 0.00048 | 8.784 | 0.00045 | 8.704 | 0.00045 | 8.601 | 0.00029 | 8.505 | 0.00023 | 8.403 | 0.00043 | 8.297 | 0.00022 | 8.201 |
| 0.08690 | 0.00054 | 8.783 | 0.00046 | 8.703 | 0.00040 | 8.601 | 0.00039 | 8.506 | 0.00058 | 8.400 | 0.00045 | 8.297 | 0.00020 | 8.201 |
| 0.08790 | 0.00057 | 8.784 | 0.00063 | 8.704 | 0.00035 | 8.600 | 0.00031 | 8.505 | 0.00045 | 8.403 | 0.00051 | 8.297 | 0.00016 | 8.201 |
| 0.08892 | 0.00056 | 8.784 | 0.00064 | 8.704 | 0.00049 | 8.602 | 0.00026 | 8.504 | 0.00041 | 8.402 | 0.00055 | 8.296 | 0.00020 | 8.202 |
| 0.08995 | 0.00047 | 8.784 | 0.00056 | 8.702 | 0.00052 | 8.602 | 0.00029 | 8.505 | 0.00043 | 8.402 | 0.00078 | 8.296 | 0.00023 | 8.203 |
| 0.09099 | 0.00052 | 8.784 | 0.00062 | 8.702 | 0.00057 | 8.602 | 0.00030 | 8.504 | 0.00035 | 8.402 | 0.00069 | 8.295 | 0.00024 | 8.203 |
| 0.09204 | 0.00048 | 8.783 | 0.00067 | 8.703 | 0.00072 | 8.602 | 0.00030 | 8.503 | 0.00033 | 8.402 | 0.00057 | 8.292 | 0.00026 | 8.202 |
| 0.09311 | 0.00052 | 8.783 | 0.00055 | 8.701 | 0.00079 | 8.601 | 0.00030 | 8.506 | 0.00018 | 8.402 | 0.00054 | 8.293 | 0.00025 | 8.202 |
| 0.09419 | 0.00060 | 8.783 | 0.00069 | 8.702 | 0.00071 | 8.601 | 0.00032 | 8.505 | 0.00051 | 8.400 | 0.00047 | 8.292 | 0.00025 | 8.202 |
| 0.09528 | 0.00075 | 8.783 | 0.00087 | 8.702 | 0.00061 | 8.601 | 0.00030 | 8.505 | 0.00060 | 8.401 | 0.00046 | 8.293 | 0.00032 | 8.202 |
| 0.09638 | 0.00068 | 8.783 | 0.00090 | 8.702 | 0.00071 | 8.601 | 0.00056 | 8.507 | 0.00049 | 8.400 | 0.00052 | 8.292 | 0.00042 | 8.202 |
| 0.09750 | 0.00066 | 8.784 | 0.00095 | 8.701 | 0.00075 | 8.601 | 0.00050 | 8.508 | 0.00048 | 8.400 | 0.00057 | 8.291 | 0.00044 | 8.202 |
| 0.09863 | 0.00077 | 8.783 | 0.00091 | 8.701 | 0.00079 | 8.601 | 0.00051 | 8.508 | 0.00062 | 8.399 | 0.00069 | 8.290 | 0.00043 | 8.203 |
| 0.09977 | 0.00077 | 8.784 | 0.00092 | 8.701 | 0.00072 | 8.601 | 0.00057 | 8.509 | 0.00045 | 8.401 | 0.00074 | 8.291 | 0.00046 | 8.203 |
| 0.10093 | 0.00065 | 8.785 | 0.00081 | 8.700 | 0.00062 | 8.601 | 0.00068 | 8.508 | 0.00051 | 8.400 | 0.00080 | 8.291 | 0.00044 | 8.204 |
| 0.10209 | 0.00060 | 8.785 | 0.00087 | 8.701 | 0.00069 | 8.601 | 0.00046 | 8.507 | 0.00052 | 8.401 | 0.00067 | 8.290 | 0.00041 | 8.203 |
| 0.10328 | 0.00068 | 8.784 | 0.00083 | 8.701 | 0.00079 | 8.602 | 0.00061 | 8.509 | 0.00058 | 8.400 | 0.00059 | 8.290 | 0.00056 | 8.202 |
| 0.10447 | 0.00080 | 8.783 | 0.00088 | 8.702 | 0.00076 | 8.602 | 0.00037 | 8.507 | 0.00055 | 8.399 | 0.00068 | 8.290 | 0.00073 | 8.203 |
| 0.10568 | 0.00086 | 8.784 | 0.00089 | 8.700 | 0.00070 | 8.602 | 0.00068 | 8.507 | 0.00050 | 8.400 | 0.00065 | 8.290 | 0.00078 | 8.203 |
| 0.10691 | 0.00082 | 8.784 | 0.00106 | 8.701 | 0.00073 | 8.601 | 0.00051 | 8.508 | 0.00047 | 8.400 | 0.00055 | 8.291 | 0.00081 | 8.202 |
| 0.10814 | 0.00075 | 8.785 | 0.00109 | 8.702 | 0.00077 | 8.601 | 0.00055 | 8.507 | 0.00058 | 8.400 | 0.00056 | 8.290 | 0.00085 | 8.202 |
| 0.10940 | 0.00080 | 8.784 | 0.00102 | 8.701 | 0.00084 | 8.602 | 0.00040 | 8.507 | 0.00071 | 8.398 | 0.00066 | 8.291 | 0.00083 | 8.203 |
| 0.11066 | 0.00102 | 8.785 | 0.00093 | 8.701 | 0.00096 | 8.601 | 0.00046 | 8.506 | 0.00062 | 8.399 | 0.00069 | 8.291 | 0.00074 | 8.202 |
| 0.11194 | 0.00112 | 8.784 | 0.00104 | 8.703 | 0.00112 | 8.603 | 0.00039 | 8.508 | 0.00069 | 8.399 | 0.00084 | 8.291 | 0.00075 | 8.203 |
| 0.11324 | 0.00111 | 8.784 | 0.00089 | 8.703 | 0.00116 | 8.604 | 0.00061 | 8.507 | 0.00064 | 8.401 | 0.00094 | 8.291 | 0.00084 | 8.203 |
| 0.11455 | 0.00105 | 8.783 | 0.00084 | 8.703 | 0.00111 | 8.604 | 0.00054 | 8.506 | 0.00053 | 8.402 | 0.00092 | 8.291 | 0.00080 | 8.203 |
| 0.11588 | 0.00094 | 8.783 | 0.00094 | 8.703 | 0.00111 | 8.603 | 0.00048 | 8.504 | 0.00065 | 8.400 | 0.00096 | 8.291 | 0.00072 | 8.203 |
| 0.11722 | 0.00106 | 8.784 | 0.00105 | 8.703 | 0.00117 | 8.603 | 0.00037 | 8.506 | 0.00054 | 8.402 | 0.00100 | 8.291 | 0.00079 | 8.203 |
| 0.11858 | 0.00107 | 8.784 | 0.00100 | 8.702 | 0.00118 | 8.603 | 0.00060 | 8.504 | 0.00063 | 8.401 | 0.00108 | 8.292 | 0.00088 | 8.203 |
| 0.11995 | 0.00113 | 8.784 | 0.00105 | 8.702 | 0.00120 | 8.604 | 0.00085 | 8.504 | 0.00081 | 8.401 | 0.00104 | 8.290 | 0.00085 | 8.203 |
| 0.12134 | 0.00119 | 8.783 | 0.00108 | 8.703 | 0.00132 | 8.605 | 0.00068 | 8.506 | 0.00067 | 8.401 | 0.00096 | 8.290 | 0.00081 | 8.204 |
| 0.12274 | 0.00120 | 8.783 | 0.00117 | 8.703 | 0.00135 | 8.605 | 0.00063 | 8.505 | 0.00071 | 8.401 | 0.00103 | 8.291 | 0.00076 | 8.203 |
| 0.12417 | 0.00110 | 8.783 | 0.00122 | 8.702 | 0.00127 | 8.604 | 0.00084 | 8.505 | 0.00057 | 8.401 | 0.00117 | 8.291 | 0.00073 | 8.203 |
| 0.12560 | 0.00113 | 8.783 | 0.00116 | 8.702 | 0.00119 | 8.603 | 0.00078 | 8.505 | 0.00075 | 8.401 | 0.00105 | 8.291 | 0.00075 | 8.202 |
| 0.12706 | 0.00113 | 8.784 | 0.00120 | 8.702 | 0.00114 | 8.604 | 0.00081 | 8.505 | 0.00074 | 8.401 | 0.00096 | 8.292 | 0.00075 | 8.202 |
| 0.12853 | 0.00118 | 8.783 | 0.00124 | 8.701 | 0.00121 | 8.604 | 0.00070 | 8.505 | 0.00094 | 8.401 | 0.00096 | 8.292 | 0.00071 | 8.203 |
| 0.13002 | 0.00132 | 8.784 | 0.00114 | 8.702 | 0.00124 | 8.604 | 0.00075 | 8.504 | 0.00090 | 8.401 | 0.00099 | 8.292 | 0.00065 | 8.202 |
| 0.13152 | 0.00130 | 8.784 | 0.00114 | 8.701 | 0.00117 | 8.604 | 0.00081 | 8.504 | 0.00081 | 8.400 | 0.00100 | 8.292 | 0.00054 | 8.202 |
| 0.13305 | 0.00134 | 8.784 | 0.00126 | 8.701 | 0.00118 | 8.603 | 0.00073 | 8.504 | 0.00060 | 8.400 | 0.00107 | 8.291 | 0.00051 | 8.201 |
| 0.13459 | 0.00138 | 8.784 | 0.00144 | 8.702 | 0.00122 | 8.604 | 0.00092 | 8.503 | 0.00087 | 8.401 | 0.00098 | 8.291 | 0.00058 | 8.201 |
| 0.13614 | 0.00158 | 8.784 | 0.00141 | 8.702 | 0.00125 | 8.604 | 0.00091 | 8.503 | 0.00095 | 8.401 | 0.00111 | 8.291 | 0.00067 | 8.201 |
| 0.13772 | 0.00167 | 8.784 | 0.00146 | 8.702 | 0.00124 | 8.603 | 0.00091 | 8.504 | 0.00096 | 8.401 | 0.00124 | 8.291 | 0.00074 | 8.201 |
| 0.13932 | 0.00164 | 8.784 | 0.00142 | 8.701 | 0.00124 | 8.603 | 0.00096 | 8.504 | 0.00095 | 8.402 | 0.00117 | 8.292 | 0.00080 | 8.202 |
| 0.14093 | 0.00173 | 8.784 | 0.00148 | 8.702 | 0.00117 | 8.604 | 0.00106 | 8.504 | 0.00104 | 8.402 | 0.00123 | 8.292 | 0.00080 | 8.202 |
| 0.14256 | 0.00168 | 8.784 | 0.00143 | 8.702 | 0.00120 | 8.604 | 0.00124 | 8.505 | 0.00099 | 8.402 | 0.00129 | 8.292 | 0.00080 | 8.201 |

| | | | | | | | | | | | | | |
|---|---|---|---|---|---|---|---|---|---|---|---|---|---|
| 0.14421 | 0.00167 | 8.783 | 0.00144 | 8.702 | 0.00140 | 8.604 | 0.00109 | 8.505 | 0.00111 | 8.402 | 0.00110 | 8.292 | 0.00079 | 8.201 |
| 0.14588 | 0.00178 | 8.784 | 0.00146 | 8.701 | 0.00153 | 8.604 | 0.00108 | 8.505 | 0.00128 | 8.402 | 0.00106 | 8.292 | 0.00078 | 8.201 |
| 0.14757 | 0.00180 | 8.784 | 0.00158 | 8.701 | 0.00164 | 8.603 | 0.00113 | 8.504 | 0.00133 | 8.402 | 0.00117 | 8.292 | 0.00086 | 8.201 |
| 0.14928 | 0.00173 | 8.784 | 0.00167 | 8.702 | 0.00175 | 8.603 | 0.00138 | 8.505 | 0.00138 | 8.402 | 0.00112 | 8.291 | 0.00087 | 8.201 |
| 0.15101 | 0.00180 | 8.784 | 0.00162 | 8.701 | 0.00172 | 8.604 | 0.00150 | 8.506 | 0.00142 | 8.402 | 0.00113 | 8.292 | 0.00074 | 8.200 |
| 0.15276 | 0.00189 | 8.785 | 0.00161 | 8.701 | 0.00174 | 8.604 | 0.00108 | 8.506 | 0.00132 | 8.402 | 0.00115 | 8.292 | 0.00074 | 8.200 |
| 0.15453 | 0.00198 | 8.785 | 0.00165 | 8.701 | 0.00186 | 8.604 | 0.00135 | 8.506 | 0.00119 | 8.401 | 0.00120 | 8.291 | 0.00086 | 8.200 |
| 0.15631 | 0.00198 | 8.784 | 0.00174 | 8.701 | 0.00196 | 8.604 | 0.00125 | 8.506 | 0.00131 | 8.402 | 0.00132 | 8.292 | 0.00099 | 8.201 |
| 0.15812 | 0.00207 | 8.784 | 0.00180 | 8.702 | 0.00196 | 8.603 | 0.00124 | 8.507 | 0.00146 | 8.404 | 0.00122 | 8.292 | 0.00106 | 8.201 |
| 0.15996 | 0.00214 | 8.784 | 0.00177 | 8.701 | 0.00195 | 8.603 | 0.00136 | 8.507 | 0.00136 | 8.403 | 0.00116 | 8.291 | 0.00102 | 8.200 |
| 0.16181 | 0.00211 | 8.783 | 0.00172 | 8.701 | 0.00189 | 8.603 | 0.00136 | 8.508 | 0.00143 | 8.403 | 0.00121 | 8.291 | 0.00104 | 8.200 |
| 0.16368 | 0.00219 | 8.784 | 0.00174 | 8.701 | 0.00175 | 8.603 | 0.00163 | 8.508 | 0.00131 | 8.403 | 0.00124 | 8.291 | 0.00114 | 8.200 |
| 0.16558 | 0.00234 | 8.784 | 0.00176 | 8.700 | 0.00171 | 8.603 | 0.00165 | 8.508 | 0.00143 | 8.403 | 0.00121 | 8.291 | 0.00113 | 8.201 |
| 0.16749 | 0.00227 | 8.784 | 0.00189 | 8.700 | 0.00180 | 8.604 | 0.00173 | 8.509 | 0.00136 | 8.404 | 0.00139 | 8.292 | 0.00113 | 8.201 |
| 0.16943 | 0.00225 | 8.784 | 0.00197 | 8.700 | 0.00189 | 8.603 | 0.00177 | 8.507 | 0.00145 | 8.403 | 0.00146 | 8.292 | 0.00114 | 8.200 |
| 0.17140 | 0.00224 | 8.784 | 0.00191 | 8.699 | 0.00196 | 8.603 | 0.00171 | 8.507 | 0.00162 | 8.404 | 0.00154 | 8.292 | 0.00123 | 8.200 |
| 0.17338 | 0.00223 | 8.785 | 0.00192 | 8.699 | 0.00206 | 8.603 | 0.00174 | 8.507 | 0.00166 | 8.405 | 0.00172 | 8.292 | 0.00127 | 8.200 |
| 0.17539 | 0.00225 | 8.785 | 0.00197 | 8.700 | 0.00210 | 8.604 | 0.00205 | 8.507 | 0.00173 | 8.405 | 0.00172 | 8.291 | 0.00121 | 8.200 |
| 0.17742 | 0.00227 | 8.785 | 0.00213 | 8.700 | 0.00210 | 8.603 | 0.00162 | 8.507 | 0.00173 | 8.404 | 0.00169 | 8.292 | 0.00117 | 8.200 |
| 0.17947 | 0.00227 | 8.785 | 0.00212 | 8.700 | 0.00208 | 8.603 | 0.00184 | 8.507 | 0.00189 | 8.405 | 0.00179 | 8.292 | 0.00128 | 8.200 |
| 0.18155 | 0.00247 | 8.784 | 0.00212 | 8.700 | 0.00214 | 8.603 | 0.00186 | 8.507 | 0.00181 | 8.404 | 0.00174 | 8.292 | 0.00137 | 8.200 |
| 0.18365 | 0.00267 | 8.785 | 0.00211 | 8.700 | 0.00222 | 8.604 | 0.00202 | 8.507 | 0.00161 | 8.404 | 0.00183 | 8.291 | 0.00132 | 8.200 |
| 0.18578 | 0.00274 | 8.785 | 0.00216 | 8.700 | 0.00218 | 8.604 | 0.00204 | 8.508 | 0.00176 | 8.404 | 0.00188 | 8.292 | 0.00138 | 8.200 |
| 0.18793 | 0.00268 | 8.784 | 0.00248 | 8.700 | 0.00222 | 8.604 | 0.00206 | 8.507 | 0.00167 | 8.405 | 0.00187 | 8.292 | 0.00148 | 8.200 |
| 0.19011 | 0.00266 | 8.785 | 0.00249 | 8.700 | 0.00240 | 8.604 | 0.00216 | 8.508 | 0.00183 | 8.406 | 0.00177 | 8.291 | 0.00153 | 8.200 |
| 0.19231 | 0.00277 | 8.785 | 0.00252 | 8.700 | 0.00243 | 8.603 | 0.00223 | 8.508 | 0.00177 | 8.405 | 0.00185 | 8.292 | 0.00154 | 8.200 |
| 0.19454 | 0.00292 | 8.785 | 0.00258 | 8.700 | 0.00236 | 8.603 | 0.00232 | 8.508 | 0.00195 | 8.405 | 0.00188 | 8.292 | 0.00152 | 8.200 |
| 0.19679 | 0.00295 | 8.784 | 0.00260 | 8.701 | 0.00234 | 8.603 | 0.00231 | 8.507 | 0.00203 | 8.404 | 0.00189 | 8.291 | 0.00150 | 8.200 |
| 0.19907 | 0.00300 | 8.785 | 0.00251 | 8.700 | 0.00246 | 8.602 | 0.00247 | 8.507 | 0.00200 | 8.404 | 0.00202 | 8.291 | 0.00155 | 8.200 |
| 0.20137 | 0.00315 | 8.785 | 0.00264 | 8.700 | 0.00261 | 8.602 | 0.00244 | 8.507 | 0.00185 | 8.404 | 0.00188 | 8.292 | 0.00165 | 8.200 |
| 0.20370 | 0.00318 | 8.785 | 0.00272 | 8.700 | 0.00269 | 8.602 | 0.00251 | 8.507 | 0.00200 | 8.404 | 0.00183 | 8.292 | 0.00180 | 8.200 |
| 0.20606 | 0.00329 | 8.784 | 0.00280 | 8.700 | 0.00263 | 8.602 | 0.00256 | 8.507 | 0.00220 | 8.403 | 0.00190 | 8.292 | 0.00187 | 8.199 |
| 0.20845 | 0.00323 | 8.785 | 0.00279 | 8.700 | 0.00264 | 8.602 | 0.00273 | 8.507 | 0.00217 | 8.404 | 0.00194 | 8.292 | 0.00181 | 8.199 |
| 0.21086 | 0.00334 | 8.785 | 0.00293 | 8.700 | 0.00273 | 8.603 | 0.00267 | 8.506 | 0.00199 | 8.405 | 0.00189 | 8.292 | 0.00180 | 8.199 |
| 0.21330 | 0.00353 | 8.785 | 0.00300 | 8.700 | 0.00285 | 8.603 | 0.00279 | 8.507 | 0.00242 | 8.405 | 0.00199 | 8.292 | 0.00187 | 8.199 |
| 0.21577 | 0.00365 | 8.785 | 0.00317 | 8.700 | 0.00294 | 8.603 | 0.00302 | 8.507 | 0.00235 | 8.404 | 0.00207 | 8.292 | 0.00189 | 8.199 |
| 0.21827 | 0.00367 | 8.785 | 0.00325 | 8.700 | 0.00297 | 8.603 | 0.00299 | 8.507 | 0.00231 | 8.405 | 0.00206 | 8.291 | 0.00186 | 8.198 |
| 0.22080 | 0.00385 | 8.785 | 0.00339 | 8.700 | 0.00297 | 8.604 | 0.00303 | 8.508 | 0.00249 | 8.404 | 0.00210 | 8.291 | 0.00185 | 8.198 |
| 0.22336 | 0.00383 | 8.785 | 0.00357 | 8.700 | 0.00302 | 8.603 | 0.00298 | 8.507 | 0.00244 | 8.404 | 0.00227 | 8.291 | 0.00183 | 8.199 |
| 0.22594 | 0.00391 | 8.785 | 0.00372 | 8.699 | 0.00321 | 8.603 | 0.00301 | 8.506 | 0.00254 | 8.404 | 0.00240 | 8.291 | 0.00183 | 8.200 |
| 0.22856 | 0.00395 | 8.785 | 0.00377 | 8.700 | 0.00340 | 8.604 | 0.00308 | 8.506 | 0.00265 | 8.404 | 0.00253 | 8.292 | 0.00186 | 8.199 |
| 0.23121 | 0.00414 | 8.785 | 0.00389 | 8.701 | 0.00350 | 8.604 | 0.00318 | 8.506 | 0.00276 | 8.405 | 0.00252 | 8.291 | 0.00191 | 8.199 |
| 0.23388 | 0.00432 | 8.785 | 0.00396 | 8.701 | 0.00366 | 8.605 | 0.00302 | 8.506 | 0.00298 | 8.405 | 0.00255 | 8.292 | 0.00197 | 8.200 |
| 0.23659 | 0.00430 | 8.785 | 0.00402 | 8.700 | 0.00378 | 8.605 | 0.00318 | 8.506 | 0.00284 | 8.405 | 0.00268 | 8.292 | 0.00197 | 8.199 |
| 0.23933 | 0.00438 | 8.785 | 0.00420 | 8.700 | 0.00379 | 8.604 | 0.00324 | 8.507 | 0.00293 | 8.405 | 0.00280 | 8.292 | 0.00188 | 8.199 |
| 0.24210 | 0.00453 | 8.785 | 0.00422 | 8.700 | 0.00388 | 8.604 | 0.00328 | 8.507 | 0.00298 | 8.405 | 0.00277 | 8.292 | 0.00182 | 8.199 |
| 0.24491 | 0.00455 | 8.785 | 0.00443 | 8.700 | 0.00396 | 8.604 | 0.00333 | 8.506 | 0.00317 | 8.405 | 0.00282 | 8.292 | 0.00194 | 8.199 |

| | | | | | | | | | | | | | |
|---|---|---|---|---|---|---|---|---|---|---|---|---|---|
| 0.24774 | 0.00469 | 8.785 | 0.00451 | 8.701 | 0.00387 | 8.604 | 0.00332 | 8.507 | 0.00310 | 8.405 | 0.00290 | 8.291 | 0.00211 | 8.199 |
| 0.25061 | 0.00479 | 8.785 | 0.00451 | 8.701 | 0.00399 | 8.604 | 0.00368 | 8.507 | 0.00332 | 8.405 | 0.00301 | 8.291 | 0.00225 | 8.199 |
| 0.25351 | 0.00491 | 8.785 | 0.00453 | 8.701 | 0.00427 | 8.604 | 0.00373 | 8.507 | 0.00326 | 8.406 | 0.00302 | 8.292 | 0.00232 | 8.199 |
| 0.25645 | 0.00502 | 8.785 | 0.00473 | 8.701 | 0.00435 | 8.604 | 0.00384 | 8.507 | 0.00335 | 8.405 | 0.00306 | 8.292 | 0.00230 | 8.199 |
| 0.25942 | 0.00508 | 8.785 | 0.00479 | 8.701 | 0.00446 | 8.604 | 0.00391 | 8.507 | 0.00363 | 8.405 | 0.00319 | 8.292 | 0.00217 | 8.200 |
| 0.26242 | 0.00523 | 8.785 | 0.00480 | 8.701 | 0.00469 | 8.604 | 0.00390 | 8.506 | 0.00372 | 8.405 | 0.00335 | 8.292 | 0.00220 | 8.199 |
| 0.26546 | 0.00552 | 8.785 | 0.00503 | 8.701 | 0.00469 | 8.603 | 0.00408 | 8.507 | 0.00354 | 8.405 | 0.00341 | 8.292 | 0.00245 | 8.199 |
| 0.26853 | 0.00554 | 8.785 | 0.00510 | 8.701 | 0.00472 | 8.604 | 0.00401 | 8.506 | 0.00349 | 8.405 | 0.00326 | 8.292 | 0.00254 | 8.199 |
| 0.27164 | 0.00567 | 8.785 | 0.00514 | 8.701 | 0.00481 | 8.604 | 0.00401 | 8.507 | 0.00365 | 8.405 | 0.00334 | 8.292 | 0.00262 | 8.199 |
| 0.27479 | 0.00588 | 8.785 | 0.00538 | 8.700 | 0.00494 | 8.604 | 0.00420 | 8.506 | 0.00399 | 8.405 | 0.00349 | 8.292 | 0.00279 | 8.199 |
| 0.27797 | 0.00593 | 8.785 | 0.00564 | 8.700 | 0.00511 | 8.604 | 0.00410 | 8.506 | 0.00366 | 8.405 | 0.00351 | 8.292 | 0.00287 | 8.199 |
| 0.28119 | 0.00616 | 8.785 | 0.00567 | 8.700 | 0.00528 | 8.604 | 0.00426 | 8.506 | 0.00378 | 8.405 | 0.00361 | 8.292 | 0.00298 | 8.200 |
| 0.28445 | 0.00621 | 8.785 | 0.00567 | 8.700 | 0.00542 | 8.604 | 0.00461 | 8.506 | 0.00374 | 8.405 | 0.00379 | 8.291 | 0.00314 | 8.200 |
| 0.28774 | 0.00633 | 8.785 | 0.00585 | 8.700 | 0.00552 | 8.604 | 0.00464 | 8.506 | 0.00394 | 8.405 | 0.00390 | 8.291 | 0.00325 | 8.199 |
| 0.29107 | 0.00649 | 8.785 | 0.00607 | 8.700 | 0.00555 | 8.604 | 0.00459 | 8.506 | 0.00417 | 8.405 | 0.00392 | 8.292 | 0.00330 | 8.200 |
| 0.29444 | 0.00672 | 8.785 | 0.00587 | 8.700 | 0.00553 | 8.604 | 0.00469 | 8.506 | 0.00409 | 8.405 | 0.00403 | 8.291 | 0.00327 | 8.199 |
| 0.29785 | 0.00684 | 8.784 | 0.00602 | 8.700 | 0.00579 | 8.604 | 0.00485 | 8.506 | 0.00420 | 8.405 | 0.00427 | 8.291 | 0.00326 | 8.199 |
| 0.30130 | 0.00705 | 8.785 | 0.00621 | 8.700 | 0.00606 | 8.604 | 0.00504 | 8.506 | 0.00430 | 8.405 | 0.00432 | 8.291 | 0.00340 | 8.199 |
| 0.30479 | 0.00713 | 8.785 | 0.00644 | 8.700 | 0.00608 | 8.604 | 0.00517 | 8.506 | 0.00473 | 8.405 | 0.00427 | 8.292 | 0.00356 | 8.199 |
| 0.30832 | 0.00728 | 8.785 | 0.00658 | 8.700 | 0.00616 | 8.604 | 0.00538 | 8.506 | 0.00481 | 8.405 | 0.00437 | 8.292 | 0.00362 | 8.199 |
| 0.31189 | 0.00747 | 8.785 | 0.00649 | 8.701 | 0.00639 | 8.604 | 0.00550 | 8.506 | 0.00494 | 8.405 | 0.00432 | 8.291 | 0.00368 | 8.199 |
| 0.31550 | 0.00764 | 8.785 | 0.00682 | 8.700 | 0.00659 | 8.603 | 0.00552 | 8.505 | 0.00506 | 8.404 | 0.00434 | 8.291 | 0.00374 | 8.199 |
| 0.31915 | 0.00781 | 8.785 | 0.00701 | 8.700 | 0.00666 | 8.603 | 0.00594 | 8.506 | 0.00501 | 8.404 | 0.00441 | 8.291 | 0.00382 | 8.199 |
| 0.32285 | 0.00802 | 8.785 | 0.00722 | 8.700 | 0.00669 | 8.604 | 0.00565 | 8.506 | 0.00524 | 8.405 | 0.00457 | 8.291 | 0.00388 | 8.199 |
| 0.32659 | 0.00831 | 8.785 | 0.00724 | 8.700 | 0.00688 | 8.604 | 0.00584 | 8.506 | 0.00521 | 8.405 | 0.00486 | 8.291 | 0.00392 | 8.199 |
| 0.33037 | 0.00845 | 8.785 | 0.00755 | 8.700 | 0.00721 | 8.604 | 0.00602 | 8.506 | 0.00528 | 8.404 | 0.00497 | 8.291 | 0.00405 | 8.199 |
| 0.33420 | 0.00870 | 8.785 | 0.00780 | 8.700 | 0.00738 | 8.604 | 0.00627 | 8.506 | 0.00527 | 8.405 | 0.00511 | 8.291 | 0.00419 | 8.199 |
| 0.33806 | 0.00904 | 8.785 | 0.00823 | 8.700 | 0.00746 | 8.604 | 0.00643 | 8.506 | 0.00566 | 8.405 | 0.00524 | 8.291 | 0.00425 | 8.199 |
| 0.34198 | 0.00911 | 8.785 | 0.00837 | 8.700 | 0.00764 | 8.604 | 0.00635 | 8.507 | 0.00566 | 8.405 | 0.00529 | 8.291 | 0.00433 | 8.199 |
| 0.34594 | 0.00922 | 8.784 | 0.00852 | 8.700 | 0.00778 | 8.604 | 0.00645 | 8.506 | 0.00586 | 8.405 | 0.00537 | 8.291 | 0.00441 | 8.199 |
| 0.34995 | 0.00944 | 8.785 | 0.00870 | 8.700 | 0.00788 | 8.604 | 0.00658 | 8.506 | 0.00594 | 8.405 | 0.00555 | 8.291 | 0.00451 | 8.199 |
| 0.35400 | 0.00980 | 8.784 | 0.00904 | 8.700 | 0.00802 | 8.604 | 0.00688 | 8.506 | 0.00607 | 8.405 | 0.00564 | 8.291 | 0.00471 | 8.199 |
| 0.35810 | 0.01009 | 8.785 | 0.00922 | 8.700 | 0.00829 | 8.604 | 0.00713 | 8.506 | 0.00628 | 8.405 | 0.00594 | 8.291 | 0.00488 | 8.199 |
| 0.36224 | 0.01028 | 8.785 | 0.00938 | 8.700 | 0.00845 | 8.604 | 0.00730 | 8.505 | 0.00626 | 8.405 | 0.00609 | 8.291 | 0.00504 | 8.199 |
| 0.36644 | 0.01043 | 8.785 | 0.00963 | 8.700 | 0.00863 | 8.604 | 0.00722 | 8.505 | 0.00646 | 8.405 | 0.00609 | 8.291 | 0.00523 | 8.199 |
| 0.37068 | 0.01078 | 8.785 | 0.00977 | 8.700 | 0.00876 | 8.604 | 0.00764 | 8.506 | 0.00656 | 8.405 | 0.00620 | 8.291 | 0.00528 | 8.199 |
| 0.37497 | 0.01082 | 8.785 | 0.00995 | 8.700 | 0.00898 | 8.604 | 0.00779 | 8.506 | 0.00659 | 8.405 | 0.00632 | 8.291 | 0.00533 | 8.199 |
| 0.37931 | 0.01125 | 8.785 | 0.01014 | 8.700 | 0.00912 | 8.604 | 0.00795 | 8.506 | 0.00696 | 8.405 | 0.00656 | 8.291 | 0.00556 | 8.199 |
| 0.38371 | 0.01138 | 8.784 | 0.01037 | 8.699 | 0.00926 | 8.604 | 0.00827 | 8.506 | 0.00723 | 8.405 | 0.00673 | 8.291 | 0.00572 | 8.199 |
| 0.38815 | 0.01169 | 8.784 | 0.01054 | 8.699 | 0.00962 | 8.604 | 0.00829 | 8.506 | 0.00728 | 8.405 | 0.00696 | 8.291 | 0.00575 | 8.199 |
| 0.39264 | 0.01192 | 8.784 | 0.01073 | 8.699 | 0.00999 | 8.604 | 0.00853 | 8.506 | 0.00775 | 8.405 | 0.00705 | 8.291 | 0.00592 | 8.199 |
| 0.39719 | 0.01221 | 8.784 | 0.01108 | 8.699 | 0.01013 | 8.604 | 0.00859 | 8.506 | 0.00799 | 8.405 | 0.00723 | 8.291 | 0.00627 | 8.199 |
| 0.40179 | 0.01243 | 8.784 | 0.01130 | 8.699 | 0.01023 | 8.604 | 0.00875 | 8.506 | 0.00783 | 8.405 | 0.00743 | 8.291 | 0.00639 | 8.199 |
| 0.40644 | 0.01282 | 8.784 | 0.01170 | 8.699 | 0.01051 | 8.604 | 0.00902 | 8.506 | 0.00835 | 8.405 | 0.00743 | 8.291 | 0.00643 | 8.199 |
| 0.41115 | 0.01302 | 8.784 | 0.01184 | 8.699 | 0.01068 | 8.604 | 0.00931 | 8.505 | 0.00857 | 8.405 | 0.00753 | 8.291 | 0.00663 | 8.199 |
| 0.41591 | 0.01331 | 8.784 | 0.01212 | 8.699 | 0.01088 | 8.604 | 0.00941 | 8.506 | 0.00869 | 8.405 | 0.00778 | 8.291 | 0.00687 | 8.199 |
| 0.42073 | 0.01362 | 8.784 | 0.01239 | 8.699 | 0.01116 | 8.604 | 0.00978 | 8.505 | 0.00850 | 8.405 | 0.00797 | 8.291 | 0.00707 | 8.199 |

| | | | | | | | | | | | | | |
|---|---|---|---|---|---|---|---|---|---|---|---|---|---|
|0.42560|0.01400|8.784|0.01281|8.699|0.01148|8.604|0.00998|8.505|0.00897|8.405|0.00808|8.291|0.00730|8.199|
|0.43053|0.01441|8.784|0.01301|8.699|0.01169|8.604|0.01020|8.505|0.00912|8.404|0.00817|8.291|0.00748|8.199|
|0.43551|0.01469|8.784|0.01334|8.699|0.01194|8.604|0.01049|8.506|0.00946|8.404|0.00840|8.291|0.00761|8.199|
|0.44055|0.01501|8.784|0.01382|8.699|0.01234|8.604|0.01059|8.506|0.00986|8.405|0.00859|8.291|0.00774|8.199|
|0.44566|0.01545|8.784|0.01405|8.699|0.01263|8.604|0.01052|8.505|0.01008|8.405|0.00886|8.291|0.00790|8.199|
|0.45082|0.01576|8.784|0.01431|8.699|0.01289|8.604|0.01103|8.505|0.01012|8.404|0.00907|8.291|0.00817|8.199|
|0.45604|0.01629|8.784|0.01480|8.699|0.01328|8.604|0.01139|8.505|0.01020|8.405|0.00941|8.291|0.00836|8.199|
|0.46132|0.01652|8.784|0.01505|8.699|0.01357|8.604|0.01163|8.505|0.01058|8.405|0.00957|8.291|0.00859|8.199|
|0.46666|0.01688|8.784|0.01524|8.699|0.01391|8.604|0.01201|8.505|0.01073|8.405|0.00977|8.291|0.00881|8.199|
|0.47206|0.01733|8.784|0.01571|8.699|0.01413|8.604|0.01200|8.505|0.01098|8.404|0.00997|8.291|0.00896|8.199|
|0.47753|0.01773|8.784|0.01606|8.699|0.01431|8.604|0.01256|8.505|0.01104|8.404|0.01018|8.291|0.00913|8.199|
|0.48306|0.01801|8.784|0.01655|8.699|0.01453|8.604|0.01285|8.505|0.01130|8.404|0.01040|8.291|0.00933|8.199|
|0.48865|0.01851|8.784|0.01688|8.699|0.01487|8.604|0.01277|8.505|0.01141|8.404|0.01085|8.290|0.00955|8.199|
|0.49431|0.01906|8.784|0.01731|8.699|0.01524|8.604|0.01352|8.504|0.01155|8.404|0.01104|8.291|0.00971|8.199|
|0.50003|0.01971|8.784|0.01755|8.699|0.01561|8.603|0.01345|8.504|0.01195|8.404|0.01116|8.290|0.00998|8.199|
|0.50582|0.02008|8.784|0.01801|8.699|0.01592|8.603|0.01396|8.505|0.01221|8.404|0.01139|8.290|0.01029|8.199|
|0.51168|0.02039|8.784|0.01833|8.699|0.01617|8.603|0.01422|8.505|0.01249|8.404|0.01169|8.291|0.01052|8.199|
|0.51761|0.02080|8.783|0.01881|8.699|0.01660|8.603|0.01469|8.504|0.01301|8.404|0.01187|8.291|0.01074|8.200|
|0.52360|0.02132|8.783|0.01939|8.698|0.01709|8.603|0.01490|8.505|0.01324|8.404|0.01213|8.290|0.01104|8.200|
|0.52966|0.02181|8.783|0.01985|8.698|0.01751|8.603|0.01539|8.505|0.01356|8.404|0.01239|8.291|0.01136|8.199|
|0.53580|0.02238|8.783|0.02013|8.699|0.01797|8.603|0.01559|8.504|0.01372|8.404|0.01273|8.291|0.01162|8.199|
|0.54200|0.02284|8.783|0.02053|8.699|0.01832|8.603|0.01592|8.504|0.01414|8.404|0.01296|8.291|0.01178|8.199|
|0.54828|0.02329|8.783|0.02104|8.698|0.01858|8.603|0.01636|8.505|0.01442|8.404|0.01333|8.291|0.01211|8.199|
|0.55463|0.02382|8.783|0.02155|8.698|0.01903|8.603|0.01669|8.504|0.01483|8.404|0.01350|8.291|0.01255|8.199|
|0.56105|0.02436|8.783|0.02208|8.698|0.01963|8.603|0.01729|8.504|0.01497|8.404|0.01365|8.291|0.01266|8.199|
|0.56754|0.02513|8.783|0.02270|8.698|0.02003|8.603|0.01740|8.504|0.01565|8.404|0.01400|8.291|0.01278|8.199|
|0.57412|0.02568|8.783|0.02313|8.698|0.02035|8.603|0.01792|8.504|0.01591|8.404|0.01428|8.291|0.01310|8.199|
|0.58076|0.02623|8.783|0.02357|8.698|0.02087|8.603|0.01836|8.504|0.01634|8.404|0.01475|8.291|0.01348|8.200|
|0.58749|0.02692|8.783|0.02402|8.698|0.02131|8.603|0.01866|8.504|0.01673|8.403|0.01530|8.291|0.01373|8.200|
|0.59429|0.02753|8.783|0.02469|8.698|0.02188|8.603|0.01903|8.504|0.01709|8.404|0.01563|8.291|0.01408|8.199|
|0.60117|0.02820|8.783|0.02520|8.698|0.02243|8.603|0.01953|8.504|0.01774|8.404|0.01615|8.291|0.01457|8.199|
|0.60814|0.02876|8.782|0.02599|8.698|0.02303|8.603|0.02004|8.504|0.01791|8.404|0.01649|8.291|0.01501|8.199|
|0.61518|0.02953|8.782|0.02632|8.698|0.02347|8.603|0.02033|8.504|0.01842|8.404|0.01680|8.291|0.01541|8.199|
|0.62230|0.03019|8.782|0.02703|8.698|0.02388|8.602|0.02111|8.504|0.01885|8.404|0.01701|8.291|0.01567|8.199|
|0.62951|0.03093|8.782|0.02772|8.698|0.02437|8.602|0.02142|8.504|0.01923|8.404|0.01730|8.291|0.01582|8.199|
|0.63680|0.03164|8.782|0.02818|8.698|0.02501|8.602|0.02212|8.504|0.01974|8.404|0.01778|8.291|0.01615|8.199|
|0.64417|0.03231|8.782|0.02889|8.698|0.02566|8.602|0.02220|8.503|0.02017|8.403|0.01830|8.291|0.01649|8.199|
|0.65163|0.03303|8.782|0.02960|8.698|0.02612|8.602|0.02314|8.503|0.02077|8.404|0.01878|8.290|0.01677|8.199|
|0.65917|0.03371|8.782|0.03032|8.698|0.02653|8.602|0.02378|8.504|0.02094|8.403|0.01925|8.290|0.01714|8.199|
|0.66681|0.03461|8.782|0.03092|8.697|0.02718|8.602|0.02438|8.504|0.02160|8.404|0.01963|8.290|0.01756|8.199|
|0.67453|0.03547|8.782|0.03178|8.697|0.02797|8.602|0.02455|8.503|0.02197|8.403|0.02008|8.290|0.01806|8.199|
|0.68234|0.03621|8.782|0.03258|8.697|0.02869|8.602|0.02520|8.503|0.02265|8.403|0.02058|8.290|0.01851|8.199|
|0.69024|0.03707|8.782|0.03325|8.697|0.02936|8.602|0.02596|8.503|0.02327|8.403|0.02093|8.290|0.01888|8.199|
|0.69823|0.03793|8.781|0.03412|8.697|0.03002|8.602|0.02648|8.503|0.02345|8.403|0.02146|8.290|0.01932|8.199|
|0.70632|0.03901|8.781|0.03489|8.697|0.03065|8.602|0.02715|8.503|0.02435|8.403|0.02204|8.290|0.01980|8.199|
|0.71450|0.03980|8.781|0.03557|8.697|0.03137|8.602|0.02748|8.503|0.02469|8.403|0.02247|8.290|0.02026|8.199|
|0.72277|0.04066|8.781|0.03645|8.697|0.03209|8.602|0.02813|8.503|0.02540|8.403|0.02298|8.290|0.02071|8.199|

| | | | | | | | | | | | | |
|---|---|---|---|---|---|---|---|---|---|---|---|---|
| 0.73114 | 0.04169 | 8.781 | 0.03742 | 8.697 | 0.03282 | 8.602 | 0.02871 | 8.503 | 0.02616 | 8.403 | 0.02356 | 8.290 | 0.02113 | 8.198 |
| 0.73961 | 0.04256 | 8.781 | 0.03822 | 8.697 | 0.03351 | 8.602 | 0.02935 | 8.503 | 0.02686 | 8.403 | 0.02401 | 8.290 | 0.02158 | 8.198 |
| 0.74817 | 0.04358 | 8.781 | 0.03921 | 8.697 | 0.03423 | 8.602 | 0.03038 | 8.503 | 0.02721 | 8.403 | 0.02451 | 8.290 | 0.02210 | 8.198 |
| 0.75683 | 0.04455 | 8.781 | 0.04015 | 8.697 | 0.03503 | 8.601 | 0.03096 | 8.503 | 0.02794 | 8.403 | 0.02522 | 8.290 | 0.02255 | 8.198 |
| 0.76560 | 0.04569 | 8.781 | 0.04097 | 8.697 | 0.03588 | 8.601 | 0.03185 | 8.503 | 0.02835 | 8.403 | 0.02592 | 8.290 | 0.02305 | 8.198 |
| 0.77446 | 0.04675 | 8.781 | 0.04165 | 8.697 | 0.03675 | 8.601 | 0.03267 | 8.503 | 0.02890 | 8.403 | 0.02636 | 8.290 | 0.02356 | 8.198 |
| 0.78343 | 0.04780 | 8.780 | 0.04266 | 8.697 | 0.03755 | 8.601 | 0.03339 | 8.503 | 0.02976 | 8.403 | 0.02689 | 8.290 | 0.02406 | 8.198 |
| 0.79250 | 0.04888 | 8.780 | 0.04379 | 8.696 | 0.03850 | 8.601 | 0.03421 | 8.503 | 0.03018 | 8.402 | 0.02742 | 8.290 | 0.02460 | 8.197 |
| 0.80168 | 0.04993 | 8.780 | 0.04486 | 8.696 | 0.03940 | 8.601 | 0.03483 | 8.502 | 0.03148 | 8.402 | 0.02805 | 8.290 | 0.02499 | 8.197 |
| 0.81096 | 0.05114 | 8.781 | 0.04580 | 8.696 | 0.04008 | 8.601 | 0.03559 | 8.502 | 0.03200 | 8.403 | 0.02894 | 8.290 | 0.02532 | 8.197 |
| 0.82035 | 0.05232 | 8.781 | 0.04668 | 8.696 | 0.04100 | 8.601 | 0.03646 | 8.502 | 0.03272 | 8.402 | 0.02959 | 8.290 | 0.02579 | 8.197 |
| 0.82985 | 0.05336 | 8.781 | 0.04800 | 8.696 | 0.04193 | 8.601 | 0.03711 | 8.502 | 0.03337 | 8.402 | 0.03007 | 8.290 | 0.02622 | 8.196 |
| 0.83946 | 0.05476 | 8.781 | 0.04897 | 8.696 | 0.04273 | 8.601 | 0.03834 | 8.502 | 0.03407 | 8.402 | 0.03079 | 8.290 | 0.02672 | 8.196 |
| 0.84918 | 0.05610 | 8.781 | 0.05007 | 8.696 | 0.04362 | 8.601 | 0.03920 | 8.502 | 0.03489 | 8.402 | 0.03168 | 8.290 | 0.02732 | 8.196 |
| 0.85901 | 0.05740 | 8.781 | 0.05106 | 8.696 | 0.04435 | 8.601 | 0.04022 | 8.502 | 0.03538 | 8.402 | 0.03254 | 8.290 | 0.02810 | 8.196 |
| 0.86896 | 0.05900 | 8.781 | 0.05225 | 8.696 | 0.04501 | 8.602 | 0.04103 | 8.502 | 0.03632 | 8.402 | 0.03322 | 8.289 | 0.02900 | 8.196 |
| 0.87902 | 0.06058 | 8.781 | 0.05367 | 8.696 | 0.04572 | 8.604 | 0.04224 | 8.502 | 0.03746 | 8.402 | 0.03386 | 8.289 | 0.02976 | 8.196 |
| 0.88920 | 0.06203 | 8.780 | 0.05499 | 8.696 | 0.04720 | 8.607 | 0.04302 | 8.502 | 0.03812 | 8.402 | 0.03465 | 8.289 | 0.03044 | 8.196 |
| 0.89950 | 0.06350 | 8.779 | 0.05596 | 8.695 | 0.04953 | 8.609 | 0.04393 | 8.502 | 0.03910 | 8.402 | 0.03536 | 8.289 | 0.03132 | 8.196 |
| 0.90991 | 0.06500 | 8.779 | 0.05742 | 8.695 | 0.05196 | 8.607 | 0.04480 | 8.502 | 0.03999 | 8.402 | 0.03633 | 8.289 | 0.03231 | 8.195 |
| 0.92045 | 0.06640 | 8.779 | 0.05886 | 8.695 | 0.05375 | 8.604 | 0.04590 | 8.502 | 0.04114 | 8.402 | 0.03717 | 8.289 | 0.03328 | 8.195 |
| 0.93111 | 0.06794 | 8.779 | 0.06015 | 8.695 | 0.05485 | 8.602 | 0.04679 | 8.502 | 0.04190 | 8.402 | 0.03792 | 8.289 | 0.03430 | 8.195 |
| 0.94189 | 0.06932 | 8.779 | 0.06155 | 8.695 | 0.05603 | 8.601 | 0.04783 | 8.502 | 0.04300 | 8.402 | 0.03901 | 8.289 | 0.03523 | 8.195 |
| 0.95280 | 0.07096 | 8.779 | 0.06303 | 8.695 | 0.05733 | 8.600 | 0.04901 | 8.502 | 0.04425 | 8.401 | 0.03996 | 8.289 | 0.03605 | 8.195 |
| 0.96383 | 0.07257 | 8.778 | 0.06467 | 8.695 | 0.05847 | 8.599 | 0.05038 | 8.502 | 0.04493 | 8.401 | 0.04070 | 8.289 | 0.03703 | 8.194 |
| 0.97499 | 0.07422 | 8.778 | 0.06599 | 8.694 | 0.05931 | 8.599 | 0.05128 | 8.501 | 0.04583 | 8.401 | 0.04152 | 8.289 | 0.03818 | 8.194 |
| 0.98628 | 0.07590 | 8.778 | 0.06752 | 8.694 | 0.06027 | 8.598 | 0.05281 | 8.501 | 0.04702 | 8.401 | 0.04240 | 8.289 | 0.03938 | 8.193 |
| 0.99770 | 0.07771 | 8.778 | 0.06887 | 8.694 | 0.06141 | 8.599 | 0.05392 | 8.501 | 0.04820 | 8.401 | 0.04351 | 8.289 | 0.04056 | 8.193 |
| 1.00925 | 0.07962 | 8.778 | 0.07068 | 8.694 | 0.06275 | 8.599 | 0.05499 | 8.501 | 0.04900 | 8.401 | 0.04450 | 8.288 | 0.04159 | 8.193 |
| 1.02094 | 0.08144 | 8.778 | 0.07233 | 8.694 | 0.06413 | 8.599 | 0.05635 | 8.501 | 0.05027 | 8.401 | 0.04549 | 8.288 | 0.04252 | 8.193 |
| 1.03276 | 0.08335 | 8.778 | 0.07383 | 8.694 | 0.06567 | 8.599 | 0.05764 | 8.501 | 0.05158 | 8.401 | 0.04629 | 8.288 | 0.04335 | 8.193 |
| 1.04472 | 0.08549 | 8.777 | 0.07568 | 8.694 | 0.06724 | 8.599 | 0.05889 | 8.500 | 0.05264 | 8.401 | 0.04761 | 8.288 | 0.04409 | 8.193 |
| 1.05682 | 0.08737 | 8.777 | 0.07745 | 8.693 | 0.06867 | 8.599 | 0.06028 | 8.500 | 0.05358 | 8.400 | 0.04892 | 8.288 | 0.04490 | 8.193 |
| 1.06905 | 0.08930 | 8.777 | 0.07918 | 8.693 | 0.07032 | 8.599 | 0.06157 | 8.500 | 0.05506 | 8.400 | 0.05009 | 8.288 | 0.04583 | 8.194 |
| 1.08143 | 0.09146 | 8.777 | 0.08093 | 8.693 | 0.07195 | 8.598 | 0.06240 | 8.499 | 0.05621 | 8.400 | 0.05120 | 8.287 | 0.04684 | 8.194 |
| 1.09396 | 0.09362 | 8.777 | 0.08285 | 8.693 | 0.07354 | 8.598 | 0.06365 | 8.499 | 0.05747 | 8.399 | 0.05213 | 8.287 | 0.04774 | 8.194 |
| 1.10662 | 0.09564 | 8.777 | 0.08456 | 8.693 | 0.07519 | 8.598 | 0.06478 | 8.499 | 0.05877 | 8.399 | 0.05323 | 8.287 | 0.04871 | 8.194 |
| 1.11944 | 0.09790 | 8.776 | 0.08655 | 8.692 | 0.07692 | 8.598 | 0.06523 | 8.499 | 0.06000 | 8.399 | 0.05429 | 8.286 | 0.04978 | 8.195 |
| 1.13240 | 0.10025 | 8.776 | 0.08861 | 8.692 | 0.07882 | 8.598 | 0.06565 | 8.501 | 0.06089 | 8.398 | 0.05558 | 8.286 | 0.05078 | 8.195 |
| 1.14551 | 0.10253 | 8.776 | 0.09054 | 8.692 | 0.08064 | 8.598 | 0.06568 | 8.507 | 0.06212 | 8.398 | 0.05667 | 8.285 | 0.05198 | 8.195 |
| 1.15878 | 0.10488 | 8.776 | 0.09248 | 8.692 | 0.08253 | 8.598 | 0.06921 | 8.516 | 0.06259 | 8.398 | 0.05762 | 8.284 | 0.05320 | 8.195 |
| 1.17220 | 0.10745 | 8.775 | 0.09446 | 8.692 | 0.08455 | 8.598 | 0.07637 | 8.517 | 0.06286 | 8.400 | 0.05858 | 8.283 | 0.05443 | 8.195 |
| 1.18577 | 0.10982 | 8.775 | 0.09604 | 8.692 | 0.08634 | 8.598 | 0.08253 | 8.510 | 0.06257 | 8.408 | 0.05917 | 8.282 | 0.05590 | 8.195 |
| 1.19950 | 0.11238 | 8.775 | 0.09785 | 8.694 | 0.08828 | 8.598 | 0.08355 | 8.501 | 0.06204 | 8.420 | 0.05979 | 8.281 | 0.05720 | 8.195 |
| 1.21339 | 0.11505 | 8.775 | 0.10075 | 8.697 | 0.09037 | 8.597 | 0.08393 | 8.497 | 0.06631 | 8.422 | 0.06160 | 8.279 | 0.05837 | 8.195 |
| 1.22744 | 0.11781 | 8.775 | 0.10487 | 8.700 | 0.09245 | 8.597 | 0.08423 | 8.497 | 0.07648 | 8.410 | 0.06366 | 8.278 | 0.05971 | 8.195 |
| 1.24165 | 0.12011 | 8.775 | 0.10919 | 8.697 | 0.09463 | 8.597 | 0.08496 | 8.497 | 0.08303 | 8.400 | 0.06566 | 8.279 | 0.06121 | 8.195 |

| | | | | | | | | | | | | | |
|---|---|---|---|---|---|---|---|---|---|---|---|---|---|
| 1.25603 | 0.12358 | 8.775 | 0.11222 | 8.693 | 0.09696 | 8.597 | 0.08630 | 8.497 | 0.08314 | 8.396 | 0.06823 | 8.281 | 0.06260 | 8.195 |
| 1.27057 | 0.12685 | 8.775 | 0.11425 | 8.691 | 0.09928 | 8.597 | 0.08836 | 8.497 | 0.08225 | 8.396 | 0.07065 | 8.281 | 0.06395 | 8.195 |
| 1.28529 | 0.12951 | 8.773 | 0.11599 | 8.690 | 0.10150 | 8.597 | 0.09009 | 8.497 | 0.08233 | 8.396 | 0.07280 | 8.281 | 0.06547 | 8.195 |
| 1.30017 | 0.13237 | 8.773 | 0.11835 | 8.689 | 0.10385 | 8.597 | 0.09196 | 8.498 | 0.08299 | 8.397 | 0.07496 | 8.281 | 0.06687 | 8.195 |
| 1.31522 | 0.13528 | 8.773 | 0.12060 | 8.689 | 0.10621 | 8.597 | 0.09443 | 8.498 | 0.08441 | 8.397 | 0.07693 | 8.281 | 0.06829 | 8.195 |
| 1.33045 | 0.13834 | 8.773 | 0.12304 | 8.689 | 0.10855 | 8.596 | 0.09630 | 8.498 | 0.08606 | 8.397 | 0.07902 | 8.280 | 0.06996 | 8.195 |
| 1.34586 | 0.14142 | 8.772 | 0.12584 | 8.689 | 0.11104 | 8.596 | 0.09861 | 8.498 | 0.08767 | 8.398 | 0.08064 | 8.277 | 0.07162 | 8.194 |
| 1.36144 | 0.14475 | 8.772 | 0.12860 | 8.689 | 0.11366 | 8.596 | 0.10086 | 8.498 | 0.08967 | 8.398 | 0.08275 | 8.277 | 0.07327 | 8.194 |
| 1.37721 | 0.14818 | 8.772 | 0.13179 | 8.689 | 0.11624 | 8.596 | 0.10285 | 8.498 | 0.09142 | 8.398 | 0.08568 | 8.279 | 0.07510 | 8.194 |
| 1.39316 | 0.15173 | 8.772 | 0.13500 | 8.689 | 0.11906 | 8.596 | 0.10536 | 8.497 | 0.09376 | 8.398 | 0.08814 | 8.281 | 0.07694 | 8.194 |
| 1.40929 | 0.15509 | 8.771 | 0.13815 | 8.689 | 0.12187 | 8.595 | 0.10751 | 8.497 | 0.09568 | 8.398 | 0.08982 | 8.281 | 0.07881 | 8.194 |
| 1.42561 | 0.15868 | 8.771 | 0.14117 | 8.688 | 0.12448 | 8.595 | 0.11003 | 8.497 | 0.09813 | 8.398 | 0.09127 | 8.282 | 0.08073 | 8.194 |
| 1.44212 | 0.16250 | 8.771 | 0.14443 | 8.688 | 0.12721 | 8.595 | 0.11295 | 8.497 | 0.10025 | 8.398 | 0.09269 | 8.283 | 0.08255 | 8.194 |
| 1.45881 | 0.16626 | 8.770 | 0.14777 | 8.688 | 0.13024 | 8.595 | 0.11541 | 8.497 | 0.10271 | 8.397 | 0.09446 | 8.283 | 0.08436 | 8.194 |
| 1.47571 | 0.17006 | 8.770 | 0.15119 | 8.688 | 0.13340 | 8.594 | 0.11821 | 8.497 | 0.10492 | 8.397 | 0.09648 | 8.283 | 0.08630 | 8.193 |
| 1.49279 | 0.17411 | 8.769 | 0.15484 | 8.687 | 0.13652 | 8.594 | 0.12045 | 8.497 | 0.10780 | 8.397 | 0.09874 | 8.284 | 0.08833 | 8.193 |
| 1.51008 | 0.17820 | 8.769 | 0.15811 | 8.687 | 0.13965 | 8.594 | 0.12372 | 8.496 | 0.11000 | 8.397 | 0.10087 | 8.284 | 0.09026 | 8.193 |
| 1.52757 | 0.18230 | 8.769 | 0.16173 | 8.687 | 0.14294 | 8.594 | 0.12658 | 8.496 | 0.11259 | 8.397 | 0.10325 | 8.284 | 0.09211 | 8.193 |
| 1.54525 | 0.18624 | 8.768 | 0.16555 | 8.687 | 0.14620 | 8.593 | 0.12949 | 8.496 | 0.11539 | 8.397 | 0.10553 | 8.284 | 0.09413 | 8.192 |
| 1.56315 | 0.19063 | 8.768 | 0.16954 | 8.686 | 0.14947 | 8.593 | 0.13260 | 8.496 | 0.11792 | 8.397 | 0.10803 | 8.284 | 0.09613 | 8.192 |
| 1.58125 | 0.19525 | 8.768 | 0.17332 | 8.686 | 0.15305 | 8.593 | 0.13563 | 8.496 | 0.12051 | 8.397 | 0.11052 | 8.284 | 0.09808 | 8.192 |
| 1.59956 | 0.19969 | 8.767 | 0.17747 | 8.686 | 0.15658 | 8.592 | 0.13873 | 8.496 | 0.12368 | 8.396 | 0.11275 | 8.283 | 0.10021 | 8.192 |
| 1.61808 | 0.20436 | 8.767 | 0.18155 | 8.685 | 0.16017 | 8.592 | 0.14173 | 8.495 | 0.12637 | 8.396 | 0.11506 | 8.283 | 0.10248 | 8.192 |
| 1.63682 | 0.20911 | 8.767 | 0.18566 | 8.685 | 0.16370 | 8.592 | 0.14510 | 8.495 | 0.12913 | 8.396 | 0.11752 | 8.283 | 0.10464 | 8.191 |
| 1.65577 | 0.21393 | 8.767 | 0.19009 | 8.685 | 0.16710 | 8.592 | 0.14841 | 8.495 | 0.13235 | 8.396 | 0.12035 | 8.283 | 0.10671 | 8.191 |
| 1.67494 | 0.21894 | 8.767 | 0.19459 | 8.684 | 0.17061 | 8.591 | 0.15177 | 8.495 | 0.13550 | 8.396 | 0.12298 | 8.283 | 0.10875 | 8.190 |
| 1.69434 | 0.22409 | 8.766 | 0.19901 | 8.684 | 0.17398 | 8.592 | 0.15509 | 8.495 | 0.13840 | 8.396 | 0.12556 | 8.283 | 0.11037 | 8.190 |
| 1.71396 | 0.22962 | 8.766 | 0.20369 | 8.684 | 0.17733 | 8.592 | 0.15905 | 8.494 | 0.14151 | 8.395 | 0.12848 | 8.283 | 0.11151 | 8.189 |
| 1.73380 | 0.23508 | 8.765 | 0.20854 | 8.683 | 0.18136 | 8.594 | 0.16260 | 8.494 | 0.14503 | 8.395 | 0.13144 | 8.283 | 0.11259 | 8.189 |
| 1.75388 | 0.24074 | 8.764 | 0.21326 | 8.683 | 0.18648 | 8.597 | 0.16638 | 8.494 | 0.14836 | 8.395 | 0.13436 | 8.283 | 0.11446 | 8.189 |
| 1.77419 | 0.24624 | 8.763 | 0.21797 | 8.683 | 0.19329 | 8.598 | 0.17014 | 8.494 | 0.15159 | 8.395 | 0.13743 | 8.283 | 0.11779 | 8.190 |
| 1.79473 | 0.25174 | 8.763 | 0.22319 | 8.682 | 0.20066 | 8.596 | 0.17419 | 8.493 | 0.15502 | 8.395 | 0.14055 | 8.282 | 0.12160 | 8.190 |
| 1.81552 | 0.25758 | 8.762 | 0.22872 | 8.682 | 0.20618 | 8.592 | 0.17795 | 8.493 | 0.15880 | 8.394 | 0.14380 | 8.282 | 0.12573 | 8.190 |
| 1.83654 | 0.26329 | 8.762 | 0.23395 | 8.681 | 0.21050 | 8.590 | 0.18236 | 8.493 | 0.16223 | 8.394 | 0.14724 | 8.282 | 0.13142 | 8.189 |
| 1.85780 | 0.26958 | 8.761 | 0.23898 | 8.681 | 0.21439 | 8.588 | 0.18656 | 8.492 | 0.16641 | 8.394 | 0.15062 | 8.282 | 0.13755 | 8.189 |
| 1.87932 | 0.27583 | 8.761 | 0.24480 | 8.680 | 0.21818 | 8.587 | 0.19085 | 8.492 | 0.16965 | 8.394 | 0.15420 | 8.282 | 0.14205 | 8.187 |
| 1.90108 | 0.28202 | 8.760 | 0.25031 | 8.680 | 0.22252 | 8.587 | 0.19512 | 8.492 | 0.17386 | 8.393 | 0.15752 | 8.281 | 0.14586 | 8.184 |
| 1.92309 | 0.28867 | 8.759 | 0.25607 | 8.680 | 0.22735 | 8.587 | 0.19937 | 8.491 | 0.17785 | 8.393 | 0.16129 | 8.281 | 0.15017 | 8.183 |
| 1.94536 | 0.29528 | 8.759 | 0.26222 | 8.679 | 0.23235 | 8.587 | 0.20447 | 8.491 | 0.18193 | 8.393 | 0.16507 | 8.281 | 0.15437 | 8.183 |
| 1.96789 | 0.30186 | 8.758 | 0.26821 | 8.679 | 0.23763 | 8.586 | 0.20854 | 8.490 | 0.18637 | 8.392 | 0.16877 | 8.281 | 0.15740 | 8.184 |
| 1.99067 | 0.30892 | 8.758 | 0.27448 | 8.678 | 0.24313 | 8.586 | 0.21327 | 8.490 | 0.19042 | 8.392 | 0.17282 | 8.280 | 0.15955 | 8.185 |
| 2.01372 | 0.31611 | 8.757 | 0.28078 | 8.678 | 0.24864 | 8.586 | 0.21774 | 8.490 | 0.19511 | 8.392 | 0.17679 | 8.280 | 0.16215 | 8.186 |
| 2.03704 | 0.32334 | 8.757 | 0.28706 | 8.677 | 0.25426 | 8.585 | 0.22252 | 8.490 | 0.19912 | 8.391 | 0.18068 | 8.280 | 0.16532 | 8.186 |
| 2.06063 | 0.33104 | 8.756 | 0.29353 | 8.677 | 0.25988 | 8.585 | 0.22687 | 8.491 | 0.20341 | 8.391 | 0.18460 | 8.279 | 0.16894 | 8.186 |
| 2.08449 | 0.33852 | 8.755 | 0.30032 | 8.677 | 0.26590 | 8.585 | 0.23273 | 8.492 | 0.20784 | 8.391 | 0.18825 | 8.279 | 0.17262 | 8.187 |
| 2.10863 | 0.34666 | 8.755 | 0.30730 | 8.676 | 0.27206 | 8.584 | 0.23903 | 8.492 | 0.21204 | 8.392 | 0.19190 | 8.278 | 0.17644 | 8.187 |
| 2.13304 | 0.35462 | 8.754 | 0.31439 | 8.675 | 0.27802 | 8.584 | 0.24611 | 8.491 | 0.21663 | 8.393 | 0.19509 | 8.278 | 0.18034 | 8.187 |

| | | | | | | | | | | | | | |
|---|---|---|---|---|---|---|---|---|---|---|---|---|---|
| 2.15774 | 0.36280 | 8.753 | 0.32157 | 8.675 | 0.28447 | 8.583 | 0.25179 | 8.490 | 0.22274 | 8.393 | 0.19855 | 8.280 | 0.18424 | 8.187 |
| 2.18273 | 0.37103 | 8.753 | 0.32914 | 8.674 | 0.29103 | 8.583 | 0.25743 | 8.489 | 0.22910 | 8.392 | 0.20333 | 8.282 | 0.18816 | 8.186 |
| 2.20800 | 0.37977 | 8.752 | 0.33691 | 8.674 | 0.29769 | 8.583 | 0.26326 | 8.489 | 0.23443 | 8.391 | 0.20981 | 8.283 | 0.19208 | 8.186 |
| 2.23357 | 0.38872 | 8.751 | 0.34469 | 8.673 | 0.30487 | 8.583 | 0.26905 | 8.488 | 0.24032 | 8.390 | 0.21604 | 8.282 | 0.19594 | 8.186 |
| 2.25944 | 0.39758 | 8.750 | 0.35274 | 8.672 | 0.31230 | 8.582 | 0.27513 | 8.488 | 0.24545 | 8.390 | 0.22173 | 8.281 | 0.19964 | 8.186 |
| 2.28560 | 0.40668 | 8.750 | 0.36070 | 8.672 | 0.31957 | 8.582 | 0.28155 | 8.488 | 0.25095 | 8.390 | 0.22698 | 8.280 | 0.20344 | 8.186 |
| 2.31206 | 0.41607 | 8.749 | 0.36877 | 8.671 | 0.32693 | 8.581 | 0.28817 | 8.488 | 0.25652 | 8.389 | 0.23218 | 8.279 | 0.20781 | 8.187 |
| 2.33884 | 0.42580 | 8.748 | 0.37743 | 8.671 | 0.33447 | 8.581 | 0.29488 | 8.488 | 0.26232 | 8.389 | 0.23697 | 8.279 | 0.21290 | 8.189 |
| 2.36592 | 0.43561 | 8.747 | 0.38604 | 8.671 | 0.34217 | 8.580 | 0.30210 | 8.488 | 0.26833 | 8.390 | 0.24201 | 8.278 | 0.21882 | 8.189 |
| 2.39332 | 0.44556 | 8.746 | 0.39537 | 8.670 | 0.35014 | 8.580 | 0.31042 | 8.488 | 0.27399 | 8.391 | 0.24707 | 8.277 | 0.22510 | 8.189 |
| 2.42103 | 0.45583 | 8.745 | 0.40526 | 8.670 | 0.35816 | 8.579 | 0.31834 | 8.486 | 0.28039 | 8.392 | 0.25139 | 8.277 | 0.23107 | 8.189 |
| 2.44906 | 0.46645 | 8.745 | 0.41524 | 8.669 | 0.36650 | 8.578 | 0.32615 | 8.485 | 0.28828 | 8.391 | 0.25563 | 8.277 | 0.23673 | 8.188 |
| 2.47742 | 0.47731 | 8.744 | 0.42497 | 8.667 | 0.37499 | 8.578 | 0.33354 | 8.483 | 0.29743 | 8.389 | 0.26131 | 8.279 | 0.24229 | 8.188 |
| 2.50611 | 0.48834 | 8.743 | 0.43469 | 8.666 | 0.38342 | 8.577 | 0.34064 | 8.483 | 0.30543 | 8.387 | 0.26861 | 8.284 | 0.24784 | 8.187 |
| 2.53513 | 0.49970 | 8.742 | 0.44452 | 8.665 | 0.39206 | 8.576 | 0.34770 | 8.482 | 0.31243 | 8.385 | 0.27626 | 8.291 | 0.25364 | 8.187 |
| 2.56448 | 0.51107 | 8.741 | 0.45468 | 8.665 | 0.40112 | 8.576 | 0.35554 | 8.481 | 0.31859 | 8.385 | 0.28267 | 8.295 | 0.25966 | 8.186 |
| 2.59418 | 0.52275 | 8.740 | 0.46495 | 8.664 | 0.41041 | 8.575 | 0.36379 | 8.481 | 0.32517 | 8.384 | 0.28990 | 8.291 | 0.26559 | 8.186 |
| 2.62422 | 0.53512 | 8.739 | 0.47560 | 8.663 | 0.41962 | 8.575 | 0.37196 | 8.480 | 0.33248 | 8.384 | 0.29909 | 8.283 | 0.27154 | 8.185 |
| 2.65461 | 0.54764 | 8.738 | 0.48672 | 8.662 | 0.42908 | 8.574 | 0.37996 | 8.480 | 0.33952 | 8.383 | 0.30833 | 8.277 | 0.27741 | 8.185 |
| 2.68534 | 0.56024 | 8.737 | 0.49807 | 8.661 | 0.43925 | 8.574 | 0.38879 | 8.479 | 0.34735 | 8.383 | 0.31764 | 8.274 | 0.28297 | 8.185 |
| 2.71644 | 0.57335 | 8.735 | 0.50935 | 8.660 | 0.44978 | 8.574 | 0.39772 | 8.479 | 0.35517 | 8.383 | 0.32669 | 8.273 | 0.28823 | 8.185 |
| 2.74789 | 0.58654 | 8.734 | 0.52088 | 8.659 | 0.46106 | 8.574 | 0.40704 | 8.478 | 0.36346 | 8.382 | 0.33375 | 8.273 | 0.29339 | 8.186 |
| 2.77971 | 0.59971 | 8.733 | 0.53282 | 8.658 | 0.47285 | 8.573 | 0.41630 | 8.478 | 0.37186 | 8.382 | 0.34023 | 8.272 | 0.29911 | 8.188 |
| 2.81190 | 0.61344 | 8.732 | 0.54506 | 8.658 | 0.48424 | 8.571 | 0.42581 | 8.477 | 0.38025 | 8.381 | 0.34710 | 8.272 | 0.30747 | 8.192 |
| 2.84446 | 0.62764 | 8.731 | 0.55771 | 8.657 | 0.49556 | 8.570 | 0.43565 | 8.477 | 0.38885 | 8.381 | 0.35458 | 8.272 | 0.32041 | 8.194 |
| 2.87740 | 0.64220 | 8.729 | 0.57077 | 8.656 | 0.50664 | 8.569 | 0.44565 | 8.476 | 0.39752 | 8.380 | 0.36265 | 8.272 | 0.33471 | 8.192 |
| 2.91072 | 0.65687 | 8.728 | 0.58375 | 8.655 | 0.51801 | 8.568 | 0.45597 | 8.476 | 0.40699 | 8.380 | 0.37057 | 8.272 | 0.34512 | 8.187 |
| 2.94442 | 0.67190 | 8.727 | 0.59714 | 8.654 | 0.52981 | 8.567 | 0.46663 | 8.475 | 0.41618 | 8.379 | 0.37903 | 8.271 | 0.35155 | 8.183 |
| 2.97852 | 0.68709 | 8.725 | 0.61085 | 8.653 | 0.54168 | 8.566 | 0.47750 | 8.474 | 0.42574 | 8.379 | 0.38782 | 8.271 | 0.35714 | 8.180 |
| 3.01301 | 0.70303 | 8.724 | 0.62477 | 8.652 | 0.55410 | 8.565 | 0.48868 | 8.474 | 0.43602 | 8.378 | 0.39686 | 8.271 | 0.36335 | 8.179 |
| 3.04789 | 0.71955 | 8.723 | 0.63961 | 8.650 | 0.56691 | 8.564 | 0.49952 | 8.473 | 0.44625 | 8.378 | 0.40618 | 8.271 | 0.37045 | 8.179 |
| 3.08319 | 0.73570 | 8.721 | 0.65444 | 8.649 | 0.57976 | 8.563 | 0.51168 | 8.472 | 0.45648 | 8.377 | 0.41614 | 8.270 | 0.37838 | 8.179 |
| 3.11889 | 0.75277 | 8.720 | 0.66951 | 8.648 | 0.59299 | 8.562 | 0.52343 | 8.471 | 0.46705 | 8.376 | 0.42672 | 8.270 | 0.38668 | 8.178 |
| 3.15500 | 0.76996 | 8.718 | 0.68468 | 8.647 | 0.60644 | 8.561 | 0.53546 | 8.471 | 0.47750 | 8.376 | 0.43679 | 8.268 | 0.39509 | 8.178 |
| 3.19154 | 0.78766 | 8.717 | 0.70008 | 8.646 | 0.62028 | 8.560 | 0.54770 | 8.470 | 0.48877 | 8.375 | 0.44680 | 8.267 | 0.40365 | 8.177 |
| 3.22849 | 0.80593 | 8.715 | 0.71658 | 8.644 | 0.63456 | 8.559 | 0.56025 | 8.469 | 0.49993 | 8.374 | 0.45676 | 8.266 | 0.41252 | 8.177 |
| 3.26588 | 0.82443 | 8.714 | 0.73331 | 8.643 | 0.64938 | 8.558 | 0.57346 | 8.468 | 0.51159 | 8.373 | 0.46653 | 8.265 | 0.42188 | 8.176 |
| 3.30370 | 0.84299 | 8.712 | 0.75015 | 8.642 | 0.66431 | 8.557 | 0.58614 | 8.467 | 0.52346 | 8.373 | 0.47677 | 8.265 | 0.43134 | 8.176 |
| 3.34195 | 0.86256 | 8.711 | 0.76696 | 8.640 | 0.67963 | 8.556 | 0.59998 | 8.466 | 0.53550 | 8.372 | 0.48714 | 8.264 | 0.44069 | 8.175 |
| 3.38065 | 0.88241 | 8.709 | 0.78426 | 8.639 | 0.69531 | 8.555 | 0.61368 | 8.466 | 0.54746 | 8.371 | 0.49742 | 8.264 | 0.44983 | 8.174 |
| 3.41979 | 0.90273 | 8.707 | 0.80242 | 8.638 | 0.71110 | 8.554 | 0.62736 | 8.465 | 0.56032 | 8.371 | 0.50869 | 8.263 | 0.45883 | 8.174 |
| 3.45939 | 0.92339 | 8.705 | 0.82093 | 8.636 | 0.72726 | 8.554 | 0.64179 | 8.464 | 0.57300 | 8.370 | 0.52037 | 8.262 | 0.46805 | 8.174 |
| 3.49945 | 0.94443 | 8.704 | 0.83990 | 8.635 | 0.74415 | 8.553 | 0.65601 | 8.463 | 0.58604 | 8.369 | 0.53164 | 8.262 | 0.47770 | 8.175 |
| 3.53997 | 0.96608 | 8.702 | 0.85901 | 8.634 | 0.76143 | 8.552 | 0.67106 | 8.463 | 0.59952 | 8.369 | 0.54336 | 8.261 | 0.48836 | 8.176 |
| 3.58096 | 0.98821 | 8.700 | 0.87865 | 8.632 | 0.77909 | 8.550 | 0.68741 | 8.462 | 0.61248 | 8.368 | 0.55555 | 8.260 | 0.50028 | 8.177 |
| 3.62243 | 1.01069 | 8.698 | 0.89913 | 8.631 | 0.79736 | 8.549 | 0.70425 | 8.461 | 0.62660 | 8.368 | 0.56773 | 8.260 | 0.51317 | 8.177 |
| 3.66438 | 1.03380 | 8.696 | 0.91994 | 8.629 | 0.81621 | 8.548 | 0.72088 | 8.460 | 0.64157 | 8.367 | 0.57987 | 8.259 | 0.52676 | 8.177 |

| | | | | | | | | | | | | | |
|---|---|---|---|---|---|---|---|---|---|---|---|---|---|
| 3.70681 | 1.05774 | 8.694 | 0.94069 | 8.627 | 0.83505 | 8.546 | 0.73770 | 8.459 | 0.65705 | 8.366 | 0.59272 | 8.259 | 0.54088 | 8.177 |
| 3.74973 | 1.08167 | 8.692 | 0.96237 | 8.626 | 0.85422 | 8.545 | 0.75439 | 8.458 | 0.67230 | 8.366 | 0.60601 | 8.259 | 0.55555 | 8.177 |
| 3.79315 | 1.10665 | 8.690 | 0.98423 | 8.624 | 0.87382 | 8.543 | 0.77214 | 8.457 | 0.68801 | 8.365 | 0.61868 | 8.259 | 0.57040 | 8.176 |
| 3.83707 | 1.13194 | 8.688 | 1.00731 | 8.622 | 0.89416 | 8.542 | 0.79059 | 8.456 | 0.70411 | 8.364 | 0.63216 | 8.259 | 0.58495 | 8.174 |
| 3.88150 | 1.15779 | 8.685 | 1.03078 | 8.621 | 0.91482 | 8.541 | 0.80902 | 8.455 | 0.72114 | 8.364 | 0.64705 | 8.259 | 0.59893 | 8.173 |
| 3.92645 | 1.18423 | 8.683 | 1.05424 | 8.619 | 0.93569 | 8.539 | 0.82815 | 8.453 | 0.73859 | 8.362 | 0.66233 | 8.260 | 0.61256 | 8.171 |
| 3.97192 | 1.21163 | 8.681 | 1.07880 | 8.617 | 0.95738 | 8.538 | 0.84748 | 8.451 | 0.75593 | 8.361 | 0.67864 | 8.260 | 0.62617 | 8.170 |
| 4.01791 | 1.23917 | 8.678 | 1.10411 | 8.615 | 0.97951 | 8.536 | 0.86667 | 8.450 | 0.77502 | 8.360 | 0.69636 | 8.261 | 0.63992 | 8.169 |
| 4.06443 | 1.26748 | 8.676 | 1.12934 | 8.613 | 1.00205 | 8.534 | 0.88642 | 8.448 | 0.79266 | 8.358 | 0.71493 | 8.262 | 0.65428 | 8.168 |
| 4.11150 | 1.29603 | 8.674 | 1.15506 | 8.611 | 1.02520 | 8.532 | 0.90677 | 8.446 | 0.81203 | 8.356 | 0.73500 | 8.261 | 0.66941 | 8.167 |
| 4.15911 | 1.32542 | 8.671 | 1.18095 | 8.609 | 1.04846 | 8.531 | 0.92699 | 8.445 | 0.83043 | 8.355 | 0.75672 | 8.260 | 0.68473 | 8.165 |
| 4.20727 | 1.35530 | 8.668 | 1.20765 | 8.606 | 1.07174 | 8.529 | 0.94788 | 8.444 | 0.84887 | 8.353 | 0.77779 | 8.257 | 0.69933 | 8.164 |
| 4.25598 | 1.38596 | 8.666 | 1.23547 | 8.604 | 1.09598 | 8.527 | 0.96905 | 8.442 | 0.86727 | 8.352 | 0.79802 | 8.254 | 0.71340 | 8.163 |
| 4.30527 | 1.41780 | 8.663 | 1.26357 | 8.602 | 1.12116 | 8.526 | 0.99051 | 8.441 | 0.88607 | 8.351 | 0.81554 | 8.250 | 0.72797 | 8.163 |
| 4.35512 | 1.44967 | 8.660 | 1.29210 | 8.600 | 1.14693 | 8.524 | 1.01297 | 8.440 | 0.90614 | 8.350 | 0.83312 | 8.248 | 0.74379 | 8.163 |
| 4.40555 | 1.48263 | 8.657 | 1.32140 | 8.598 | 1.17307 | 8.522 | 1.03659 | 8.438 | 0.92625 | 8.348 | 0.84994 | 8.246 | 0.76108 | 8.163 |
| 4.45656 | 1.51619 | 8.655 | 1.35125 | 8.595 | 1.20004 | 8.521 | 1.05943 | 8.437 | 0.94710 | 8.347 | 0.86801 | 8.244 | 0.77952 | 8.163 |
| 4.50817 | 1.55043 | 8.652 | 1.38191 | 8.593 | 1.22737 | 8.519 | 1.08334 | 8.436 | 0.96874 | 8.346 | 0.88642 | 8.243 | 0.79814 | 8.162 |
| 4.56037 | 1.58561 | 8.649 | 1.41358 | 8.591 | 1.25516 | 8.517 | 1.10788 | 8.435 | 0.99049 | 8.345 | 0.90528 | 8.241 | 0.81640 | 8.161 |
| 4.61318 | 1.62164 | 8.646 | 1.44536 | 8.588 | 1.28416 | 8.515 | 1.13399 | 8.434 | 1.01223 | 8.344 | 0.92371 | 8.240 | 0.83473 | 8.160 |
| 4.66659 | 1.65838 | 8.642 | 1.47804 | 8.586 | 1.31376 | 8.513 | 1.16100 | 8.432 | 1.03508 | 8.344 | 0.94274 | 8.238 | 0.85360 | 8.159 |
| 4.72063 | 1.69539 | 8.639 | 1.51160 | 8.583 | 1.34391 | 8.511 | 1.18831 | 8.431 | 1.05825 | 8.343 | 0.96193 | 8.237 | 0.87338 | 8.159 |
| 4.77529 | 1.73367 | 8.636 | 1.54592 | 8.581 | 1.37469 | 8.509 | 1.21613 | 8.430 | 1.08273 | 8.342 | 0.98152 | 8.237 | 0.89461 | 8.160 |
| 4.83059 | 1.77331 | 8.633 | 1.58146 | 8.578 | 1.40610 | 8.507 | 1.24554 | 8.427 | 1.10867 | 8.341 | 1.00205 | 8.237 | 0.91747 | 8.159 |
| 4.88652 | 1.81312 | 8.629 | 1.61750 | 8.575 | 1.43801 | 8.504 | 1.27478 | 8.425 | 1.13579 | 8.339 | 1.02407 | 8.237 | 0.94178 | 8.158 |
| 4.94311 | 1.85395 | 8.626 | 1.65442 | 8.573 | 1.47060 | 8.502 | 1.30389 | 8.423 | 1.16285 | 8.337 | 1.04807 | 8.237 | 0.96593 | 8.156 |
| 5.00035 | 1.89622 | 8.622 | 1.69200 | 8.570 | 1.50391 | 8.499 | 1.33320 | 8.421 | 1.19115 | 8.336 | 1.07430 | 8.237 | 0.98837 | 8.153 |
| 5.05825 | 1.93886 | 8.619 | 1.73085 | 8.567 | 1.53777 | 8.497 | 1.36358 | 8.419 | 1.21834 | 8.334 | 1.10063 | 8.236 | 1.00966 | 8.151 |
| 5.11682 | 1.98255 | 8.615 | 1.76972 | 8.564 | 1.57285 | 8.495 | 1.39450 | 8.416 | 1.24677 | 8.332 | 1.12739 | 8.235 | 1.03147 | 8.150 |
| 5.17607 | 2.02715 | 8.611 | 1.80963 | 8.560 | 1.60842 | 8.492 | 1.42642 | 8.414 | 1.27608 | 8.330 | 1.15432 | 8.234 | 1.05417 | 8.148 |
| 5.23600 | 2.07231 | 8.607 | 1.85052 | 8.557 | 1.64462 | 8.490 | 1.45734 | 8.412 | 1.30449 | 8.328 | 1.18215 | 8.233 | 1.07738 | 8.147 |
| 5.29663 | 2.11908 | 8.603 | 1.89246 | 8.554 | 1.68178 | 8.487 | 1.48995 | 8.410 | 1.33441 | 8.326 | 1.21040 | 8.231 | 1.10064 | 8.146 |
| 5.35797 | 2.16644 | 8.599 | 1.93500 | 8.551 | 1.72023 | 8.485 | 1.52359 | 8.408 | 1.36457 | 8.324 | 1.23959 | 8.230 | 1.12443 | 8.145 |
| 5.42001 | 2.21481 | 8.595 | 1.97801 | 8.547 | 1.75975 | 8.482 | 1.55816 | 8.406 | 1.39536 | 8.322 | 1.26927 | 8.228 | 1.14922 | 8.144 |
| 5.48277 | 2.26448 | 8.591 | 2.02287 | 8.544 | 1.79939 | 8.479 | 1.59421 | 8.404 | 1.42625 | 8.320 | 1.29983 | 8.226 | 1.17478 | 8.143 |
| 5.54626 | 2.31497 | 8.586 | 2.06850 | 8.540 | 1.84007 | 8.477 | 1.62981 | 8.402 | 1.45897 | 8.318 | 1.33046 | 8.224 | 1.20155 | 8.143 |
| 5.61048 | 2.36638 | 8.582 | 2.11487 | 8.537 | 1.88199 | 8.474 | 1.66766 | 8.400 | 1.49170 | 8.317 | 1.36057 | 8.221 | 1.22992 | 8.142 |
| 5.67545 | 2.41951 | 8.578 | 2.16232 | 8.533 | 1.92477 | 8.471 | 1.70569 | 8.397 | 1.52526 | 8.315 | 1.38993 | 8.219 | 1.25965 | 8.141 |
| 5.74116 | 2.47315 | 8.573 | 2.21053 | 8.530 | 1.96791 | 8.468 | 1.74448 | 8.395 | 1.55983 | 8.313 | 1.42076 | 8.218 | 1.29065 | 8.140 |
| 5.80764 | 2.52830 | 8.568 | 2.26075 | 8.526 | 2.01263 | 8.465 | 1.78482 | 8.393 | 1.59585 | 8.311 | 1.45338 | 8.216 | 1.32189 | 8.138 |
| 5.87489 | 2.58484 | 8.563 | 2.31112 | 8.522 | 2.05884 | 8.461 | 1.82536 | 8.390 | 1.63261 | 8.309 | 1.48571 | 8.214 | 1.35297 | 8.137 |
| 5.94292 | 2.64215 | 8.559 | 2.36258 | 8.518 | 2.10479 | 8.458 | 1.86723 | 8.388 | 1.66995 | 8.307 | 1.51848 | 8.212 | 1.38463 | 8.135 |
| 6.01174 | 2.70082 | 8.554 | 2.41608 | 8.514 | 2.15229 | 8.455 | 1.90955 | 8.385 | 1.70850 | 8.305 | 1.55209 | 8.210 | 1.41657 | 8.133 |
| 6.08135 | 2.76080 | 8.549 | 2.46993 | 8.510 | 2.20146 | 8.451 | 1.95284 | 8.382 | 1.74792 | 8.303 | 1.58737 | 8.209 | 1.44844 | 8.131 |
| 6.15177 | 2.82202 | 8.543 | 2.52526 | 8.506 | 2.25106 | 8.448 | 1.99732 | 8.379 | 1.78836 | 8.300 | 1.62283 | 8.207 | 1.48072 | 8.129 |
| 6.22300 | 2.88439 | 8.538 | 2.58128 | 8.501 | 2.30119 | 8.445 | 2.04370 | 8.376 | 1.82884 | 8.298 | 1.65980 | 8.206 | 1.51337 | 8.127 |
| 6.29506 | 2.94821 | 8.533 | 2.63948 | 8.497 | 2.35278 | 8.441 | 2.08833 | 8.373 | 1.87130 | 8.295 | 1.69959 | 8.204 | 1.54673 | 8.126 |

| | | | | | | | | | | | | | |
|---|---|---|---|---|---|---|---|---|---|---|---|---|---|
| 6.36796 | 3.01302 | 8.527 | 2.69834 | 8.492 | 2.40679 | 8.437 | 2.13581 | 8.370 | 1.91322 | 8.293 | 1.73846 | 8.202 | 1.58153 | 8.124 |
| 6.44169 | 3.07964 | 8.522 | 2.75877 | 8.488 | 2.46103 | 8.434 | 2.18432 | 8.367 | 1.95643 | 8.290 | 1.77770 | 8.200 | 1.61789 | 8.123 |
| 6.51628 | 3.14743 | 8.516 | 2.82078 | 8.483 | 2.51628 | 8.430 | 2.23414 | 8.364 | 2.00239 | 8.288 | 1.81923 | 8.198 | 1.65527 | 8.121 |
| 6.59174 | 3.21646 | 8.510 | 2.88305 | 8.478 | 2.57304 | 8.426 | 2.28438 | 8.361 | 2.04730 | 8.285 | 1.86109 | 8.196 | 1.69331 | 8.120 |
| 6.66807 | 3.28730 | 8.504 | 2.94695 | 8.473 | 2.63081 | 8.422 | 2.33496 | 8.358 | 2.09354 | 8.282 | 1.90438 | 8.194 | 1.73231 | 8.118 |
| 6.74528 | 3.35909 | 8.498 | 3.01244 | 8.468 | 2.68912 | 8.418 | 2.38841 | 8.355 | 2.14061 | 8.280 | 1.94911 | 8.191 | 1.77233 | 8.116 |
| 6.82339 | 3.43343 | 8.492 | 3.07939 | 8.463 | 2.74887 | 8.414 | 2.44169 | 8.351 | 2.18871 | 8.277 | 1.99287 | 8.189 | 1.81367 | 8.114 |
| 6.90240 | 3.50874 | 8.485 | 3.14821 | 8.458 | 2.81099 | 8.410 | 2.49687 | 8.348 | 2.23815 | 8.274 | 2.03760 | 8.186 | 1.85586 | 8.112 |
| 6.98232 | 3.58514 | 8.479 | 3.21773 | 8.453 | 2.87399 | 8.405 | 2.55349 | 8.344 | 2.28930 | 8.271 | 2.08313 | 8.184 | 1.89825 | 8.110 |
| 7.06318 | 3.66355 | 8.472 | 3.28848 | 8.447 | 2.93822 | 8.401 | 2.61144 | 8.341 | 2.34190 | 8.268 | 2.12925 | 8.181 | 1.94164 | 8.108 |
| 7.14496 | 3.74331 | 8.466 | 3.36100 | 8.442 | 3.00400 | 8.396 | 2.67114 | 8.337 | 2.39412 | 8.265 | 2.17760 | 8.179 | 1.98576 | 8.106 |
| 7.22770 | 3.82521 | 8.459 | 3.43561 | 8.436 | 3.07097 | 8.392 | 2.73005 | 8.333 | 2.44894 | 8.262 | 2.22739 | 8.176 | 2.03035 | 8.103 |
| 7.31139 | 3.90808 | 8.452 | 3.51068 | 8.430 | 3.13895 | 8.387 | 2.79187 | 8.330 | 2.50422 | 8.259 | 2.27814 | 8.174 | 2.07616 | 8.101 |
| 7.39605 | 3.99267 | 8.445 | 3.58830 | 8.424 | 3.20841 | 8.382 | 2.85403 | 8.326 | 2.56018 | 8.256 | 2.32903 | 8.171 | 2.12314 | 8.099 |
| 7.48170 | 4.07914 | 8.437 | 3.66671 | 8.418 | 3.28022 | 8.377 | 2.91722 | 8.321 | 2.61870 | 8.252 | 2.38271 | 8.169 | 2.17126 | 8.097 |
| 7.56833 | 4.16769 | 8.430 | 3.74744 | 8.412 | 3.35337 | 8.372 | 2.98326 | 8.317 | 2.67684 | 8.249 | 2.43569 | 8.166 | 2.22102 | 8.095 |
| 7.65597 | 4.25820 | 8.422 | 3.83007 | 8.406 | 3.42779 | 8.367 | 3.05085 | 8.313 | 2.73892 | 8.245 | 2.49339 | 8.164 | 2.27205 | 8.093 |
| 7.74462 | 4.34964 | 8.414 | 3.91298 | 8.400 | 3.50308 | 8.362 | 3.11914 | 8.309 | 2.79985 | 8.242 | 2.55089 | 8.161 | 2.32382 | 8.090 |
| 7.83430 | 4.44318 | 8.407 | 3.99821 | 8.393 | 3.58000 | 8.357 | 3.18859 | 8.305 | 2.86326 | 8.238 | 2.60891 | 8.157 | 2.37687 | 8.088 |
| 7.92501 | 4.53905 | 8.398 | 4.08653 | 8.387 | 3.65960 | 8.351 | 3.25896 | 8.301 | 2.92660 | 8.235 | 2.66811 | 8.154 | 2.43145 | 8.085 |
| 8.01678 | 4.63567 | 8.390 | 4.17626 | 8.380 | 3.74131 | 8.346 | 3.33149 | 8.296 | 2.99180 | 8.231 | 2.72769 | 8.151 | 2.48706 | 8.082 |
| 8.10961 | 4.73525 | 8.382 | 4.26652 | 8.373 | 3.82388 | 8.340 | 3.40606 | 8.291 | 3.06037 | 8.227 | 2.79021 | 8.147 | 2.54343 | 8.080 |
| 8.20352 | 4.83639 | 8.374 | 4.35915 | 8.366 | 3.90771 | 8.335 | 3.48298 | 8.287 | 3.12906 | 8.223 | 2.85089 | 8.144 | 2.60079 | 8.077 |
| 8.29851 | 4.93951 | 8.365 | 4.45327 | 8.359 | 3.99388 | 8.329 | 3.55949 | 8.282 | 3.20074 | 8.219 | 2.91422 | 8.141 | 2.65947 | 8.074 |
| 8.39460 | 5.04520 | 8.356 | 4.54904 | 8.352 | 4.08137 | 8.323 | 3.63916 | 8.277 | 3.27145 | 8.215 | 2.97930 | 8.138 | 2.71946 | 8.071 |
| 8.49180 | 5.15255 | 8.347 | 4.64788 | 8.344 | 4.17087 | 8.316 | 3.71953 | 8.272 | 3.34392 | 8.211 | 3.04585 | 8.135 | 2.78080 | 8.068 |
| 8.59014 | 5.26219 | 8.338 | 4.74908 | 8.336 | 4.26277 | 8.310 | 3.80121 | 8.267 | 3.41897 | 8.207 | 3.11523 | 8.131 | 2.84403 | 8.065 |
| 8.68960 | 5.37336 | 8.329 | 4.85129 | 8.329 | 4.35545 | 8.304 | 3.88500 | 8.262 | 3.49595 | 8.202 | 3.18571 | 8.128 | 2.90889 | 8.063 |
| 8.79023 | 5.48650 | 8.319 | 4.95537 | 8.321 | 4.45035 | 8.297 | 3.97090 | 8.256 | 3.57405 | 8.198 | 3.25767 | 8.124 | 2.97471 | 8.059 |
| 8.89201 | 5.60266 | 8.310 | 5.06145 | 8.313 | 4.54823 | 8.291 | 4.05896 | 8.251 | 3.65416 | 8.193 | 3.33147 | 8.120 | 3.04190 | 8.056 |
| 8.99498 | 5.72089 | 8.300 | 5.17012 | 8.304 | 4.64698 | 8.284 | 4.14826 | 8.245 | 3.73448 | 8.188 | 3.40659 | 8.116 | 3.11094 | 8.053 |
| 9.09913 | 5.84067 | 8.290 | 5.28082 | 8.296 | 4.74737 | 8.277 | 4.23977 | 8.239 | 3.81982 | 8.184 | 3.48347 | 8.113 | 3.18117 | 8.050 |
| 9.20450 | 5.96387 | 8.279 | 5.39315 | 8.288 | 4.85079 | 8.270 | 4.33368 | 8.234 | 3.90305 | 8.179 | 3.56251 | 8.108 | 3.25246 | 8.047 |
| 9.31108 | 6.08871 | 8.269 | 5.50903 | 8.279 | 4.95676 | 8.263 | 4.42899 | 8.228 | 3.99079 | 8.174 | 3.64136 | 8.104 | 3.32507 | 8.043 |
| 9.41890 | 6.21560 | 8.259 | 5.62603 | 8.270 | 5.06335 | 8.256 | 4.52565 | 8.222 | 4.07843 | 8.169 | 3.72145 | 8.100 | 3.39924 | 8.040 |
| 9.52796 | 6.34586 | 8.248 | 5.74503 | 8.261 | 5.17287 | 8.248 | 4.62567 | 8.216 | 4.16947 | 8.163 | 3.80461 | 8.095 | 3.47605 | 8.036 |
| 9.63829 | 6.47748 | 8.237 | 5.86681 | 8.252 | 5.28576 | 8.241 | 4.72580 | 8.210 | 4.26267 | 8.158 | 3.88873 | 8.091 | 3.55479 | 8.032 |
| 9.74990 | 6.61158 | 8.226 | 5.99138 | 8.243 | 5.39921 | 8.233 | 4.83050 | 8.203 | 4.35799 | 8.153 | 3.97495 | 8.087 | 3.63423 | 8.029 |
| 9.86279 | 6.74865 | 8.214 | 6.11660 | 8.233 | 5.51509 | 8.225 | 4.93537 | 8.196 | 4.45187 | 8.147 | 4.06466 | 8.082 | 3.71525 | 8.025 |
| 9.97700 | 6.88724 | 8.203 | 6.24655 | 8.223 | 5.63415 | 8.217 | 5.04496 | 8.190 | 4.55100 | 8.142 | 4.15558 | 8.078 | 3.79857 | 8.021 |
| 10.0925 | 7.0294 | 8.191 | 6.37733 | 8.214 | 5.75535 | 8.209 | 5.15430 | 8.183 | 4.64999 | 8.136 | 4.24839 | 8.074 | 3.88378 | 8.017 |
| 10.2094 | 7.1738 | 8.179 | 6.51249 | 8.204 | 5.87817 | 8.201 | 5.26609 | 8.176 | 4.75312 | 8.130 | 4.34250 | 8.069 | 3.97099 | 8.013 |
| 10.3276 | 7.3203 | 8.167 | 6.64892 | 8.193 | 6.00304 | 8.192 | 5.37942 | 8.169 | 4.85838 | 8.124 | 4.43880 | 8.064 | 4.05953 | 8.009 |
| 10.4472 | 7.4691 | 8.155 | 6.78856 | 8.183 | 6.13022 | 8.184 | 5.49542 | 8.162 | 4.96482 | 8.118 | 4.53708 | 8.059 | 4.15030 | 8.004 |
| 10.5682 | 7.6213 | 8.142 | 6.92913 | 8.172 | 6.26162 | 8.175 | 5.61423 | 8.155 | 5.07396 | 8.112 | 4.63817 | 8.054 | 4.24408 | 8.000 |
| 10.6905 | 7.7759 | 8.129 | 7.07348 | 8.161 | 6.39517 | 8.166 | 5.73579 | 8.147 | 5.18550 | 8.106 | 4.74206 | 8.049 | 4.33913 | 7.996 |
| 10.8143 | 7.9345 | 8.116 | 7.21928 | 8.151 | 6.52974 | 8.157 | 5.85717 | 8.140 | 5.30028 | 8.099 | 4.84534 | 8.043 | 4.43474 | 7.992 |
| 10.9396 | 8.0946 | 8.103 | 7.36813 | 8.139 | 6.66795 | 8.147 | 5.98586 | 8.132 | 5.41438 | 8.093 | 4.95170 | 8.038 | 4.53170 | 7.987 |
| 11.0662 | 8.2567 | 8.090 | 7.51958 | 8.128 | 6.80946 | 8.138 | 6.11464 | 8.124 | 5.53238 | 8.086 | 5.06172 | 8.032 | 4.63224 | 7.982 |
| 11.1944 | 8.4219 | 8.076 | 7.67461 | 8.117 | 6.95328 | 8.128 | 6.24485 | 8.116 | 5.65349 | 8.079 | 5.17423 | 8.026 | 4.73607 | 7.978 |
| 11.3240 | 8.5890 | 8.063 | 7.83411 | 8.105 | 7.09889 | 8.118 | 6.37834 | 8.108 | 5.77614 | 8.072 | 5.28679 | 8.020 | 4.84078 | 7.973 |
| 11.4551 | 8.7623 | 8.049 | 7.99325 | 8.093 | 7.24704 | 8.108 | 6.51562 | 8.100 | 5.90370 | 8.065 | 5.40092 | 8.015 | 4.94691 | 7.968 |
| 11.5878 | 8.9365 | 8.034 | 8.15564 | 8.081 | 7.39807 | 8.098 | 6.65379 | 8.091 | 6.03060 | 8.058 | 5.51977 | 8.009 | 5.05602 | 7.962 |
| 11.7220 | 9.1130 | 8.020 | 8.32367 | 8.068 | 7.55306 | 8.088 | 6.79412 | 8.082 | 6.16224 | 8.051 | 5.63970 | 8.003 | 5.16886 | 7.957 |
| 11.8577 | 9.2942 | 8.005 | 8.49345 | 8.056 | 7.71061 | 8.077 | 6.94018 | 8.073 | 6.29536 | 8.043 | 5.76185 | 7.997 | 5.28351 | 7.952 |
| 11.9950 | 9.4775 | 7.990 | 8.66483 | 8.043 | 7.86964 | 8.066 | 7.08712 | 8.064 | 6.43082 | 8.036 | 5.88772 | 7.990 | 5.39946 | 7.947 |
| 12.1339 | 9.6623 | 7.975 | 8.84068 | 8.030 | 8.03391 | 8.055 | 7.23811 | 8.055 | 6.56835 | 8.028 | 6.01603 | 7.984 | 5.51762 | 7.941 |
| 12.2744 | 9.8518 | 7.960 | 9.01957 | 8.017 | 8.20135 | 8.044 | 7.38879 | 8.046 | 6.70914 | 8.020 | 6.14734 | 7.977 | 5.63836 | 7.936 |
| 12.4165 | 10.0439 | 7.944 | 9.20173 | 8.004 | 8.36989 | 8.033 | 7.54605 | 8.036 | 6.85423 | 8.012 | 6.28190 | 7.970 | 5.76220 | 7.930 |
| 12.5603 | 10.2395 | 7.928 | 9.38390 | 7.990 | 8.54146 | 8.021 | 7.70485 | 8.027 | 7.00396 | 8.003 | 6.41839 | 7.963 | 5.88881 | 7.924 |
| 12.7057 | 10.4373 | 7.912 | 9.56978 | 7.977 | 8.71526 | 8.010 | 7.86426 | 8.017 | 7.15297 | 7.995 | 6.55568 | 7.956 | 6.01828 | 7.917 |
| 12.8529 | 10.6380 | 7.896 | 9.76033 | 7.963 | 8.89498 | 7.998 | 8.03193 | 8.007 | 7.30591 | 7.986 | 6.69569 | 7.949 | 6.15006 | 7.911 |
| 13.0017 | 10.8429 | 7.879 | 9.95539 | 7.949 | 9.07769 | 7.986 | 8.20127 | 7.997 | 7.46306 | 7.978 | 6.84087 | 7.942 | 6.28356 | 7.905 |
| 13.1522 | 11.0503 | 7.862 | 10.1533 | 7.934 | 9.26345 | 7.973 | 8.37246 | 7.986 | 7.62227 | 7.969 | 6.99043 | 7.935 | 6.42048 | 7.899 |
| 13.3045 | 11.2609 | 7.845 | 10.3524 | 7.919 | 9.45289 | 7.961 | 8.54631 | 7.976 | 7.78105 | 7.959 | 7.13984 | 7.927 | 6.56226 | 7.892 |
| 13.4586 | 11.4759 | 7.828 | 10.5558 | 7.904 | 9.64430 | 7.948 | 8.72588 | 7.965 | 7.94892 | 7.951 | 7.29198 | 7.919 | 6.70578 | 7.886 |
| 13.6144 | 11.6949 | 7.811 | 10.7623 | 7.889 | 9.83873 | 7.935 | 8.90228 | 7.954 | 8.11760 | 7.941 | 7.44879 | 7.911 | 6.84952 | 7.879 |
| 13.7721 | 11.9135 | 7.793 | 10.9729 | 7.873 | ####### | 7.922 | 9.08626 | 7.943 | 8.29000 | 7.931 | 7.60724 | 7.904 | 6.99741 | 7.872 |
| 13.9316 | 12.1377 | 7.775 | 11.1883 | 7.858 | 10.2382 | 7.909 | 9.27746 | 7.932 | 8.46482 | 7.922 | 7.76981 | 7.896 | 7.15015 | 7.865 |
| 14.0929 | 12.3654 | 7.757 | 11.4032 | 7.842 | 10.4436 | 7.895 | 9.46553 | 7.920 | 8.64273 | 7.912 | 7.93967 | 7.887 | 7.30402 | 7.858 |
| 14.2561 | 12.5951 | 7.739 | 11.6213 | 7.826 | 10.6528 | 7.881 | 9.66061 | 7.908 | 8.82256 | 7.902 | 8.11023 | 7.878 | 7.46080 | 7.851 |
| 14.4212 | 12.8299 | 7.720 | 11.8492 | 7.810 | 10.8644 | 7.867 | 9.86038 | 7.896 | 9.00687 | 7.891 | 8.28246 | 7.869 | 7.62233 | 7.843 |
| 14.5881 | 13.0667 | 7.701 | 12.0737 | 7.794 | 11.0803 | 7.853 | 10.0606 | 7.884 | 9.19568 | 7.881 | 8.45717 | 7.860 | 7.78644 | 7.836 |
| 14.7571 | 13.3063 | 7.682 | 12.3022 | 7.777 | 11.2999 | 7.839 | 10.2617 | 7.871 | 9.38649 | 7.870 | 8.63256 | 7.852 | 7.95414 | 7.828 |
| 14.9279 | 13.5511 | 7.663 | 12.5434 | 7.760 | 11.5230 | 7.824 | 10.4729 | 7.859 | 9.58041 | 7.859 | 8.81215 | 7.843 | 8.12667 | 7.820 |
| 15.1008 | 13.7979 | 7.643 | 12.7818 | 7.743 | 11.7481 | 7.809 | 10.6869 | 7.846 | 9.78486 | 7.849 | 8.99896 | 7.834 | 8.30008 | 7.812 |
| 15.2757 | 14.0504 | 7.623 | 13.0207 | 7.725 | 11.9774 | 7.794 | 10.9031 | 7.833 | 9.9858 | 7.837 | 9.18956 | 7.824 | 8.47536 | 7.804 |
| 15.4525 | 14.3032 | 7.603 | 13.2644 | 7.708 | 12.2114 | 7.779 | 11.1210 | 7.820 | 10.1918 | 7.826 | 9.38694 | 7.814 | 8.65847 | 7.796 |
| 15.6315 | 14.5602 | 7.583 | 13.5131 | 7.690 | 12.4485 | 7.763 | 11.3452 | 7.807 | 10.3998 | 7.814 | 9.58407 | 7.805 | 8.84451 | 7.788 |
| 15.8125 | 14.8213 | 7.562 | 13.7631 | 7.672 | 12.6891 | 7.748 | 11.5703 | 7.793 | 10.6159 | 7.802 | 9.78366 | 7.794 | 9.03074 | 7.779 |
| 15.9956 | 15.0848 | 7.542 | 14.0208 | 7.654 | 12.9355 | 7.731 | 11.8040 | 7.779 | 10.8312 | 7.790 | 9.9889 | 7.784 | 9.22188 | 7.770 |
| 16.1808 | 15.3535 | 7.520 | 14.2822 | 7.635 | 13.1866 | 7.715 | 12.0352 | 7.765 | 11.0514 | 7.779 | 10.1933 | 7.774 | 9.41770 | 7.761 |
| 16.3682 | 15.6245 | 7.499 | 14.5418 | 7.616 | 13.4376 | 7.698 | 12.2760 | 7.751 | 11.2766 | 7.766 | 10.4050 | 7.763 | 9.61463 | 7.752 |
| 16.5577 | 15.8997 | 7.477 | 14.8119 | 7.597 | 13.6937 | 7.682 | 12.5188 | 7.737 | 11.5036 | 7.753 | 10.6195 | 7.753 | 9.81693 | 7.743 |
| 16.7494 | 16.1784 | 7.456 | 15.0809 | 7.578 | 13.9545 | 7.665 | 12.7656 | 7.722 | 11.7415 | 7.741 | 10.8382 | 7.742 | 10.0245 | 7.734 |
| 16.9434 | 16.4596 | 7.434 | 15.3529 | 7.558 | 14.2193 | 7.647 | 13.0168 | 7.707 | 11.9805 | 7.727 | 11.0604 | 7.730 | 10.2344 | 7.724 |
| 17.1396 | 16.7423 | 7.412 | 15.6336 | 7.538 | 14.4893 | 7.630 | 13.2732 | 7.692 | 12.2139 | 7.714 | 11.2905 | 7.719 | 10.4503 | 7.714 |
| 17.3380 | 17.0335 | 7.390 | 15.9156 | 7.518 | 14.7622 | 7.612 | 13.5314 | 7.677 | 12.4675 | 7.701 | 11.5212 | 7.708 | 10.6681 | 7.704 |
| 17.5388 | 17.3236 | 7.367 | 16.2002 | 7.497 | 15.0388 | 7.594 | 13.7965 | 7.661 | 12.7144 | 7.687 | 11.7569 | 7.696 | 10.8886 | 7.694 |
| 17.7419 | 17.6198 | 7.344 | 16.4916 | 7.477 | 15.3204 | 7.577 | 14.0603 | 7.646 | 12.9653 | 7.673 | 11.9992 | 7.684 | 11.1147 | 7.684 |
| 17.9473 | 17.9199 | 7.321 | 16.7797 | 7.457 | 15.6045 | 7.558 | 14.3326 | 7.629 | 13.2311 | 7.659 | 12.2425 | 7.672 | 11.3433 | 7.674 |
| 18.1552 | 18.2190 | 7.298 | 17.0833 | 7.436 | 15.8940 | 7.539 | 14.6055 | 7.613 | 13.4884 | 7.645 | 12.4895 | 7.660 | 11.5774 | 7.663 |
| 18.3654 | 18.5246 | 7.275 | 17.3824 | 7.415 | 16.1878 | 7.520 | 14.8834 | 7.596 | 13.7593 | 7.631 | 12.7443 | 7.648 | 11.8168 | 7.652 |
| 18.5780 | 18.8344 | 7.251 | 17.6821 | 7.393 | 16.4827 | 7.501 | 15.1684 | 7.579 | 14.0290 | 7.615 | 13.0022 | 7.635 | 12.0582 | 7.641 |

| | | | | | | | | | | | | | |
|---|---|---|---|---|---|---|---|---|---|---|---|---|---|
| 18.7932 | 19.1455 | 7.227 | 17.9940 | 7.371 | 16.7827 | 7.482 | 15.4617 | 7.562 | 14.3036 | 7.601 | 13.2639 | 7.622 | 12.3061 | 7.630 |
| 19.0108 | 19.4668 | 7.203 | 18.2986 | 7.349 | 17.0895 | 7.463 | 15.7478 | 7.546 | 14.5882 | 7.585 | 13.5253 | 7.610 | 12.5579 | 7.618 |
| 19.2309 | 19.7835 | 7.179 | 18.6244 | 7.327 | 17.4013 | 7.443 | 16.0501 | 7.528 | 14.8602 | 7.570 | 13.7932 | 7.597 | 12.8161 | 7.606 |
| 19.4536 | 20.1081 | 7.154 | 18.9377 | 7.305 | 17.7117 | 7.423 | 16.3488 | 7.511 | 15.1600 | 7.555 | 14.0629 | 7.583 | 13.0797 | 7.595 |
| 19.6789 | 20.4326 | 7.130 | 19.2617 | 7.282 | 18.0278 | 7.402 | 16.6582 | 7.492 | 15.4558 | 7.539 | 14.3405 | 7.570 | 13.3451 | 7.583 |
| 19.9067 | 20.7626 | 7.105 | 19.5945 | 7.260 | 18.3501 | 7.382 | 16.9635 | 7.475 | 15.7478 | 7.523 | 14.6268 | 7.555 | 13.6164 | 7.571 |
| 20.1372 | 21.0958 | 7.080 | 19.9198 | 7.237 | 18.6796 | 7.361 | 17.2814 | 7.456 | 16.0577 | 7.507 | 14.9170 | 7.541 | 13.8895 | 7.558 |
| 20.3704 | 21.4314 | 7.055 | 20.2614 | 7.214 | 19.0117 | 7.340 | 17.5980 | 7.438 | 16.3534 | 7.490 | 15.2122 | 7.527 | 14.1685 | 7.545 |
| 20.6063 | 21.7735 | 7.029 | 20.5900 | 7.191 | 19.3430 | 7.319 | 17.9304 | 7.419 | 16.6814 | 7.475 | 15.5099 | 7.512 | 14.4526 | 7.533 |
| 20.8449 | 22.1132 | 7.004 | 20.9372 | 7.167 | 19.6839 | 7.298 | 18.2532 | 7.400 | 16.9844 | 7.457 | 15.8126 | 7.498 | 14.7411 | 7.520 |
| 21.0863 | 22.4650 | 6.978 | 21.2795 | 7.144 | 20.0219 | 7.278 | 18.5774 | 7.382 | 17.3226 | 7.440 | 16.1277 | 7.482 | 15.0341 | 7.507 |
| 21.3304 | 22.8104 | 6.952 | 21.6362 | 7.120 | 20.3612 | 7.258 | 18.9144 | 7.361 | 17.6360 | 7.422 | 16.4400 | 7.467 | 15.3293 | 7.494 |
| 21.5774 | 23.1640 | 6.926 | 21.9906 | 7.096 | 20.7130 | 7.236 | 19.2555 | 7.342 | 17.9775 | 7.405 | 16.7550 | 7.451 | 15.6342 | 7.480 |
| 21.8273 | 23.5197 | 6.900 | 22.3524 | 7.071 | 21.0670 | 7.214 | 19.6046 | 7.322 | 18.3161 | 7.386 | 17.0777 | 7.435 | 15.9472 | 7.466 |
| 22.0800 | 23.8785 | 6.873 | 22.7180 | 7.047 | 21.4237 | 7.192 | 19.9593 | 7.302 | 18.6588 | 7.369 | 17.4089 | 7.420 | 16.2669 | 7.452 |
| 22.3357 | 24.2441 | 6.847 | 23.0773 | 7.022 | 21.7890 | 7.170 | 20.3081 | 7.281 | 19.0013 | 7.350 | 17.7423 | 7.404 | 16.5851 | 7.438 |
| 22.5944 | 24.6062 | 6.820 | 23.4484 | 6.998 | 22.1610 | 7.147 | 20.6742 | 7.261 | 19.3549 | 7.332 | 18.0812 | 7.388 | 16.9087 | 7.423 |
| 22.8560 | 24.9732 | 6.793 | 23.8122 | 6.972 | 22.5355 | 7.124 | 21.0376 | 7.240 | 19.7204 | 7.314 | 18.4208 | 7.371 | 17.2398 | 7.408 |
| 23.1206 | 25.3439 | 6.766 | 24.1929 | 6.948 | 22.9118 | 7.101 | 21.4055 | 7.219 | 20.0786 | 7.294 | 18.7688 | 7.354 | 17.5751 | 7.393 |
| 23.3884 | 25.7232 | 6.739 | 24.5736 | 6.922 | 23.2863 | 7.078 | 21.7820 | 7.198 | 20.4479 | 7.275 | 19.1222 | 7.337 | 17.9145 | 7.378 |
| 23.6592 | 26.0995 | 6.712 | 24.9660 | 6.897 | 23.6676 | 7.055 | 22.1611 | 7.176 | 20.8166 | 7.256 | 19.4806 | 7.320 | 18.2595 | 7.363 |
| 23.9332 | 26.4782 | 6.684 | 25.3511 | 6.871 | 24.0539 | 7.031 | 22.5388 | 7.155 | 21.2037 | 7.236 | 19.8460 | 7.303 | 18.6090 | 7.348 |
| 24.2103 | 26.8584 | 6.657 | 25.7527 | 6.845 | 24.4458 | 7.007 | 22.9306 | 7.133 | 21.5721 | 7.217 | 20.2170 | 7.285 | 18.9604 | 7.333 |
| 24.4906 | 27.2454 | 6.629 | 26.1399 | 6.819 | 24.8411 | 6.984 | 23.3216 | 7.110 | 21.9672 | 7.196 | 20.5920 | 7.267 | 19.3248 | 7.316 |
| 24.7742 | 27.6404 | 6.601 | 26.5378 | 6.793 | 25.2481 | 6.960 | 23.7264 | 7.088 | 22.3445 | 7.177 | 20.9671 | 7.250 | 19.6959 | 7.300 |
| 25.0611 | 28.0336 | 6.574 | 26.9318 | 6.766 | 25.6523 | 6.935 | 24.1116 | 7.065 | 22.7570 | 7.155 | 21.3516 | 7.231 | 20.0701 | 7.284 |
| 25.3513 | 28.4271 | 6.546 | 27.3436 | 6.740 | 26.0631 | 6.911 | 24.5270 | 7.043 | 23.1503 | 7.135 | 21.7456 | 7.212 | 20.4509 | 7.267 |
| 25.6448 | 28.8229 | 6.518 | 27.7522 | 6.713 | 26.4823 | 6.887 | 24.9249 | 7.021 | 23.5638 | 7.113 | 22.1473 | 7.194 | 20.8350 | 7.251 |
| 25.9418 | 29.2302 | 6.489 | 28.1654 | 6.688 | 26.9001 | 6.862 | 25.3601 | 6.996 | 23.9804 | 7.094 | 22.5437 | 7.175 | 21.2229 | 7.234 |
| 26.2422 | 29.6255 | 6.462 | 28.5907 | 6.660 | 27.3222 | 6.836 | 25.7848 | 6.973 | 24.3834 | 7.071 | 22.9595 | 7.155 | 21.6232 | 7.216 |
| 26.5461 | 30.0367 | 6.433 | 28.9872 | 6.634 | 27.7472 | 6.811 | 26.2124 | 6.949 | 24.8247 | 7.052 | 23.3655 | 7.136 | 22.0245 | 7.199 |
| 26.8534 | 30.4387 | 6.405 | 29.4268 | 6.606 | 28.1765 | 6.785 | 26.6327 | 6.926 | 25.2478 | 7.029 | 23.7824 | 7.116 | 22.4330 | 7.181 |
| 27.1644 | 30.8594 | 6.376 | 29.8345 | 6.579 | 28.6073 | 6.759 | 27.0567 | 6.902 | 25.6786 | 7.006 | 24.2223 | 7.096 | 22.8439 | 7.163 |
| 27.4789 | 31.2698 | 6.348 | 30.2677 | 6.551 | 29.0532 | 6.733 | 27.5082 | 6.877 | 26.1205 | 6.984 | 24.6584 | 7.076 | 23.2610 | 7.146 |
| 27.7971 | 31.6828 | 6.320 | 30.6967 | 6.523 | 29.4930 | 6.707 | 27.9550 | 6.852 | 26.5509 | 6.961 | 25.0845 | 7.056 | 23.6872 | 7.127 |
| 28.1190 | 32.1129 | 6.290 | 31.1088 | 6.496 | 29.9346 | 6.681 | 28.3925 | 6.828 | 27.0296 | 6.940 | 25.5264 | 7.034 | 24.1095 | 7.108 |
| 28.4446 | 32.5292 | 6.262 | 31.5713 | 6.468 | 30.3881 | 6.655 | 28.8539 | 6.804 | 27.4493 | 6.916 | 25.9855 | 7.013 | 24.5504 | 7.090 |
| 28.7740 | 32.9551 | 6.233 | 32.0007 | 6.440 | 30.8377 | 6.629 | 29.2896 | 6.778 | 27.9234 | 6.893 | 26.4193 | 6.994 | 24.9862 | 7.070 |
| 29.1072 | 33.3837 | 6.204 | 32.4429 | 6.413 | 31.2939 | 6.602 | 29.7626 | 6.754 | 28.3839 | 6.871 | 26.8732 | 6.973 | 25.4255 | 7.050 |
| 29.4442 | 33.8042 | 6.176 | 32.9079 | 6.384 | 31.7592 | 6.574 | 30.2277 | 6.730 | 28.8386 | 6.846 | 27.3346 | 6.951 | 25.8841 | 7.031 |
| 29.7852 | 34.2440 | 6.146 | 33.3403 | 6.356 | 32.2200 | 6.547 | 30.6896 | 6.703 | 29.3506 | 6.824 | 27.7936 | 6.929 | 26.3372 | 7.011 |
| 30.1301 | 34.6676 | 6.118 | 33.7989 | 6.328 | 32.6850 | 6.521 | 31.1814 | 6.678 | 29.8055 | 6.801 | 28.2636 | 6.908 | 26.8007 | 6.991 |
| 30.4789 | 35.1087 | 6.089 | 34.2589 | 6.300 | 33.1471 | 6.493 | 31.6657 | 6.653 | 30.3044 | 6.775 | 28.7424 | 6.885 | 27.2760 | 6.971 |
| 30.8319 | 35.5577 | 6.060 | 34.7027 | 6.273 | 33.6139 | 6.467 | 32.1479 | 6.627 | 30.8039 | 6.752 | 29.2273 | 6.863 | 27.7520 | 6.951 |
| 31.1889 | 35.9886 | 6.031 | 35.1777 | 6.244 | 34.0929 | 6.441 | 32.6343 | 6.601 | 31.2729 | 6.729 | 29.7163 | 6.840 | 28.2430 | 6.931 |
| 31.5500 | 36.4294 | 6.003 | 35.6384 | 6.215 | 34.5743 | 6.412 | 33.1296 | 6.575 | 31.7847 | 6.702 | 30.2185 | 6.819 | 28.7324 | 6.909 |
| 31.9154 | 36.8799 | 5.973 | 36.0792 | 6.189 | 35.0663 | 6.384 | 33.5934 | 6.550 | 32.3025 | 6.679 | 30.7041 | 6.796 | 29.2230 | 6.889 |

| | | | | | | | | | | | | | |
|---|---|---|---|---|---|---|---|---|---|---|---|---|---|
| 32.2849 | 37.3186 | 5.945 | 36.5823 | 6.162 | 35.5587 | 6.358 | 34.1129 | 6.524 | 32.7794 | 6.655 | 31.2064 | 6.772 | 29.7319 | 6.867 |
| 32.6588 | 37.7618 | 5.916 | 37.0606 | 6.132 | 36.0402 | 6.330 | 34.6156 | 6.497 | 33.3166 | 6.629 | 31.7316 | 6.748 | 30.2286 | 6.844 |
| 33.0370 | 38.2228 | 5.886 | 37.4884 | 6.104 | 36.5305 | 6.303 | 35.0996 | 6.468 | 33.8597 | 6.602 | 32.2441 | 6.726 | 30.7253 | 6.822 |
| 33.4195 | 38.6678 | 5.857 | 37.9747 | 6.077 | 37.0426 | 6.275 | 35.6662 | 6.443 | 34.3720 | 6.579 | 32.7682 | 6.701 | 31.2644 | 6.801 |
| 33.8065 | 39.1153 | 5.830 | 38.5058 | 6.047 | 37.5475 | 6.247 | 36.1622 | 6.416 | 34.8622 | 6.553 | 33.3229 | 6.677 | 31.8045 | 6.778 |
| 34.1979 | 39.5810 | 5.801 | 38.9682 | 6.021 | 38.0477 | 6.219 | 36.6886 | 6.388 | 35.4242 | 6.526 | 33.8516 | 6.654 | 32.3078 | 6.755 |
| 34.5939 | 40.0455 | 5.771 | 39.4401 | 5.991 | 38.5598 | 6.192 | 37.2125 | 6.361 | 35.9968 | 6.498 | 34.3766 | 6.630 | 32.8290 | 6.734 |
| 34.9945 | 40.5017 | 5.743 | 39.9579 | 5.964 | 39.0754 | 6.166 | 37.7610 | 6.333 | 36.5063 | 6.475 | 34.9367 | 6.605 | 33.3863 | 6.712 |
| 35.3997 | 40.9592 | 5.716 | 40.4718 | 5.936 | 39.5877 | 6.137 | 38.2883 | 6.307 | 37.0176 | 6.448 | 35.4927 | 6.579 | 33.9336 | 6.688 |
| 35.8096 | 41.4322 | 5.686 | 40.9615 | 5.905 | 40.0978 | 6.109 | 38.7918 | 6.281 | 37.6351 | 6.422 | 36.0267 | 6.555 | 34.4901 | 6.663 |
| 36.2243 | 41.9019 | 5.656 | 41.4431 | 5.876 | 40.6062 | 6.080 | 39.3762 | 6.251 | 38.2090 | 6.394 | 36.5883 | 6.530 | 35.0924 | 6.642 |
| 36.6438 | 42.3618 | 5.628 | 41.9806 | 5.848 | 41.1367 | 6.050 | 39.9040 | 6.226 | 38.7856 | 6.368 | 37.1681 | 6.504 | 35.6690 | 6.619 |
| 37.0681 | 42.8207 | 5.601 | 42.5023 | 5.821 | 41.6888 | 6.021 | 40.4532 | 6.199 | 39.4314 | 6.340 | 37.7113 | 6.478 | 36.2244 | 6.593 |
| 37.4973 | 43.3026 | 5.573 | 43.0063 | 5.792 | 42.2290 | 5.994 | 40.9739 | 6.169 | 40.0021 | 6.312 | 38.3135 | 6.454 | 36.8083 | 6.568 |
| 37.9315 | 43.7942 | 5.544 | 43.4770 | 5.765 | 42.7558 | 5.964 | 41.5523 | 6.135 | 40.6222 | 6.284 | 38.8932 | 6.427 | 37.4159 | 6.541 |
| 38.3707 | 44.2694 | 5.515 | 44.0070 | 5.738 | 43.3145 | 5.934 | 42.0817 | 6.116 | 41.2217 | 6.262 | 39.4911 | 6.400 | 38.0165 | 6.516 |
| 38.8150 | 44.7375 | 5.486 | 44.5139 | 5.708 | 43.8658 | 5.907 | 42.6157 | 6.085 | 41.7551 | 6.236 | 40.1361 | 6.375 | 38.6337 | 6.492 |
| 39.2645 | 45.2152 | 5.458 | 45.0734 | 5.678 | 44.3757 | 5.879 | 43.1910 | 6.058 | 42.3629 | 6.208 | 40.7285 | 6.350 | 39.2343 | 6.469 |
| 39.7192 | 45.6953 | 5.433 | 45.5368 | 5.650 | 44.9114 | 5.851 | 43.7377 | 6.031 | 42.9408 | 6.185 | 41.3357 | 6.326 | 39.8887 | 6.445 |
| 40.1791 | 46.1870 | 5.404 | 46.0497 | 5.622 | 45.4460 | 5.822 | 44.2990 | 6.003 | 43.5317 | 6.161 | 41.9675 | 6.300 | 40.5293 | 6.421 |
| 40.6443 | 46.6791 | 5.374 | 46.5913 | 5.592 | 45.9791 | 5.795 | 44.8388 | 5.977 | 44.0702 | 6.134 | 42.5740 | 6.273 | 41.1726 | 6.395 |
| 41.1150 | 47.1853 | 5.346 | 47.0823 | 5.563 | 46.5119 | 5.766 | 45.4245 | 5.944 | 44.6617 | 6.108 | 43.1935 | 6.248 | 41.8070 | 6.371 |
| 41.5911 | 47.6722 | 5.318 | 47.5922 | 5.537 | 47.0513 | 5.737 | 46.0807 | 5.915 | 45.2702 | 6.083 | 43.8151 | 6.222 | 42.4367 | 6.347 |
| 42.0727 | 48.1672 | 5.290 | 48.1345 | 5.509 | 47.6290 | 5.707 | 46.6589 | 5.885 | 45.8717 | 6.054 | 44.4448 | 6.197 | 43.0748 | 6.323 |
| 42.5598 | 48.6611 | 5.263 | 48.6653 | 5.481 | 48.1820 | 5.678 | 47.2376 | 5.861 | 46.4555 | 6.023 | 45.0699 | 6.171 | 43.7182 | 6.299 |
| 43.0527 | 49.1439 | 5.236 | 49.1902 | 5.452 | 48.7404 | 5.650 | 47.8401 | 5.832 | 47.0638 | 5.998 | 45.6796 | 6.144 | 44.3822 | 6.273 |
| 43.5512 | 49.6362 | 5.210 | 49.7060 | 5.425 | 49.3160 | 5.622 | 48.3779 | 5.804 | 47.6995 | 5.969 | 46.3192 | 6.117 | 45.0687 | 6.246 |
| 44.0555 | 50.1372 | 5.182 | 50.2649 | 5.395 | 49.8847 | 5.594 | 49.0126 | 5.775 | 48.3299 | 5.941 | 46.9640 | 6.091 | 45.7391 | 6.221 |
| 44.5656 | 50.6323 | 5.154 | 50.7671 | 5.370 | 50.4614 | 5.566 | 49.6013 | 5.748 | 48.9677 | 5.914 | 47.6089 | 6.064 | 46.4167 | 6.195 |
| 45.0817 | 51.1369 | 5.125 | 51.2920 | 5.343 | 51.0496 | 5.539 | 50.2058 | 5.719 | 49.5902 | 5.886 | 48.2594 | 6.038 | 47.0952 | 6.169 |
| 45.6037 | 51.6443 | 5.097 | 51.8290 | 5.310 | 51.6160 | 5.511 | 50.8168 | 5.691 | 50.2346 | 5.858 | 48.9104 | 6.011 | 47.7756 | 6.143 |
| 46.1318 | 52.1447 | 5.068 | 52.3421 | 5.283 | 52.1952 | 5.483 | 51.4386 | 5.663 | 50.8604 | 5.830 | 49.5746 | 5.984 | 48.4612 | 6.117 |
| 46.6659 | 52.6457 | 5.041 | 52.8958 | 5.255 | 52.7750 | 5.456 | 52.0429 | 5.635 | 51.5096 | 5.803 | 50.2411 | 5.956 | 49.1500 | 6.091 |
| 47.2063 | 53.1780 | 5.015 | 53.3721 | 5.231 | 53.3762 | 5.428 | 52.6431 | 5.607 | 52.1597 | 5.775 | 50.9072 | 5.929 | 49.8409 | 6.065 |
| 47.7529 | 53.6443 | 4.990 | 53.9454 | 5.202 | 53.9967 | 5.398 | 53.2722 | 5.579 | 52.8057 | 5.748 | 51.5868 | 5.902 | 50.5347 | 6.039 |
| 48.3059 | 54.1788 | 4.965 | 54.4959 | 5.175 | 54.5829 | 5.371 | 53.8745 | 5.551 | 53.4631 | 5.720 | 52.2625 | 5.874 | 51.2236 | 6.013 |
| 48.8652 | 54.6961 | 4.936 | 55.0298 | 5.148 | 55.1565 | 5.342 | 54.5032 | 5.523 | 54.1105 | 5.693 | 52.9439 | 5.848 | 51.9114 | 5.989 |
| 49.4311 | 55.2075 | 4.912 | 55.6215 | 5.120 | 55.7279 | 5.316 | 55.1324 | 5.492 | 54.7356 | 5.669 | 53.6271 | 5.823 | 52.6205 | 5.963 |
| 50.0035 | 55.7391 | 4.885 | 56.2361 | 5.090 | 56.3274 | 5.290 | 55.7210 | 5.469 | 55.4520 | 5.642 | 54.3004 | 5.799 | 53.3108 | 5.939 |
| 51.1682 | 56.7207 | 4.833 | 57.4650 | 5.026 | 57.5147 | 5.237 | 57.2155 | 5.410 | 56.9024 | 5.593 | 55.9847 | 5.753 | 54.6367 | 5.896 |
| 52.3600 | 57.7571 | 4.783 | 58.5662 | 4.972 | 58.6949 | 5.182 | 58.5322 | 5.351 | 58.3002 | 5.533 | 57.4473 | 5.695 | 56.2050 | 5.838 |
| 53.5797 | 58.8219 | 4.732 | 59.6065 | 4.921 | 59.9260 | 5.129 | 59.8431 | 5.299 | 59.7062 | 5.479 | 58.9512 | 5.639 | 57.6178 | 5.790 |
| 54.8277 | 59.8054 | 4.684 | 60.6962 | 4.869 | 61.1128 | 5.078 | 61.0675 | 5.246 | 61.1067 | 5.423 | 60.4090 | 5.588 | 59.0964 | 5.737 |
| 56.1048 | 60.8527 | 4.638 | 61.8294 | 4.819 | 62.3729 | 5.019 | 62.3177 | 5.193 | 62.5168 | 5.367 | 61.8440 | 5.538 | 60.6634 | 5.685 |
| 57.4116 | 61.9034 | 4.589 | 62.9978 | 4.769 | 63.6460 | 4.969 | 63.7168 | 5.136 | 63.7750 | 5.315 | 63.3081 | 5.485 | 62.1491 | 5.629 |
| 58.7489 | 63.0157 | 4.542 | 64.1710 | 4.718 | 64.9556 | 4.917 | 64.9589 | 5.082 | 65.1831 | 5.254 | 64.6823 | 5.425 | 63.6363 | 5.577 |
| 60.1174 | 64.1745 | 4.500 | 65.2202 | 4.669 | 66.1204 | 4.865 | 66.4374 | 5.038 | 66.5240 | 5.200 | 66.0126 | 5.368 | 65.1671 | 5.520 |

| | | | | | | | | | | | | | |
|---|---|---|---|---|---|---|---|---|---|---|---|---|---|
| 61.5177 | 65.2181 | 4.450 | 66.3829 | 4.618 | 67.5332 | 4.811 | 67.8419 | 4.988 | 68.0060 | 5.149 | 67.5518 | 5.338 | 66.7354 | 5.469 |
| 62.2300 | 65.7726 | 4.430 | 66.8738 | 4.596 | 68.1365 | 4.780 | 68.4705 | 4.962 | 68.7208 | 5.116 | 68.1938 | 5.277 | 67.4013 | 5.442 |
| 62.9506 | 66.2231 | 4.408 | 67.4139 | 4.574 | 68.7583 | 4.759 | 69.1598 | 4.931 | 69.3255 | 5.099 | 68.9688 | 5.258 | 68.1936 | 5.416 |
| 63.6796 | 66.8427 | 4.385 | 68.1513 | 4.547 | 69.4941 | 4.735 | 69.9336 | 4.909 | 70.1639 | 5.072 | 69.8957 | 5.233 | 69.0328 | 5.391 |
| 64.4169 | 67.3254 | 4.357 | 68.7266 | 4.524 | 70.1105 | 4.708 | 70.5751 | 4.879 | 70.8746 | 5.049 | 70.6041 | 5.213 | 69.8657 | 5.361 |
| 65.1628 | 67.8153 | 4.338 | 69.2189 | 4.501 | 70.6902 | 4.685 | 71.2060 | 4.854 | 71.5118 | 5.023 | 71.2935 | 5.183 | 70.7295 | 5.337 |
| 65.9174 | 68.3911 | 4.316 | 69.8413 | 4.477 | 71.3506 | 4.659 | 71.9179 | 4.828 | 72.3015 | 4.996 | 72.1148 | 5.158 | 71.5662 | 5.311 |
| 66.6807 | 68.9446 | 4.292 | 70.4590 | 4.454 | 72.0083 | 4.635 | 72.6239 | 4.805 | 73.0401 | 4.972 | 72.9232 | 5.136 | 72.4156 | 5.287 |
| 67.4528 | 69.3922 | 4.269 | 70.9937 | 4.433 | 72.5721 | 4.611 | 73.2335 | 4.778 | 73.7144 | 4.945 | 73.6306 | 5.109 | 73.1819 | 5.259 |
| 68.2339 | 69.9478 | 4.250 | 71.5778 | 4.413 | 73.1928 | 4.591 | 73.9214 | 4.757 | 74.4430 | 4.923 | 74.4094 | 5.086 | 73.9842 | 5.236 |
| 69.0240 | 70.5148 | 4.228 | 72.2274 | 4.389 | 73.8917 | 4.567 | 74.6479 | 4.731 | 75.2312 | 4.897 | 75.2421 | 5.059 | 74.8601 | 5.210 |
| 69.8232 | 71.0630 | 4.205 | 72.7845 | 4.365 | 74.5159 | 4.540 | 75.3112 | 4.704 | 75.9404 | 4.869 | 76.0033 | 5.028 | 75.6733 | 5.178 |
| 70.6318 | 71.4642 | 4.186 | 73.2520 | 4.344 | 75.0486 | 4.520 | 75.8470 | 4.682 | 76.5546 | 4.844 | 76.6640 | 5.004 | 76.3831 | 5.152 |
| 71.4496 | 72.0844 | 4.164 | 73.9161 | 4.321 | 75.7078 | 4.495 | 76.6441 | 4.656 | 77.3942 | 4.819 | 77.5979 | 4.980 | 77.3119 | 5.132 |
| 72.2770 | 72.6623 | 4.140 | 74.5540 | 4.298 | 76.4248 | 4.471 | 77.3806 | 4.631 | 78.1660 | 4.794 | 78.3578 | 4.955 | 78.1416 | 5.104 |
| 73.1139 | 72.9446 | 4.120 | 74.9091 | 4.276 | 76.8666 | 4.446 | 77.9141 | 4.603 | 78.7013 | 4.765 | 78.9370 | 4.922 | 78.7868 | 5.072 |
| 73.9605 | 73.6592 | 4.101 | 75.6605 | 4.256 | 77.6086 | 4.426 | 78.6717 | 4.584 | 79.5722 | 4.745 | 79.8978 | 4.904 | 79.7465 | 5.052 |
| 74.8170 | 74.1944 | 4.080 | 76.2485 | 4.232 | 78.2685 | 4.401 | 79.3807 | 4.558 | 80.3237 | 4.717 | 80.6747 | 4.876 | 80.6195 | 5.023 |
| 75.6833 | 74.7551 | 4.061 | 76.8176 | 4.214 | 78.9007 | 4.381 | 80.0572 | 4.537 | 81.0097 | 4.697 | 81.4093 | 4.853 | 81.3692 | 4.998 |
| 76.5597 | 75.3068 | 4.042 | 77.3847 | 4.191 | 79.5074 | 4.358 | 80.7309 | 4.513 | 81.7441 | 4.671 | 82.1788 | 4.827 | 82.2276 | 4.973 |
| 77.4462 | 75.8140 | 4.019 | 77.9515 | 4.169 | 80.1156 | 4.333 | 81.4219 | 4.487 | 82.5078 | 4.645 | 82.9240 | 4.802 | 83.0460 | 4.948 |
| 78.3430 | 76.3182 | 4.002 | 78.5127 | 4.151 | 80.7312 | 4.314 | 82.0464 | 4.467 | 83.1751 | 4.623 | 83.7074 | 4.778 | 83.8322 | 4.922 |
| 79.2501 | 76.7308 | 3.983 | 79.0943 | 4.127 | 81.3468 | 4.292 | 82.6609 | 4.443 | 83.8842 | 4.596 | 84.4429 | 4.746 | 84.6237 | 4.890 |
| 80.1678 | 77.3610 | 3.964 | 79.6367 | 4.107 | 81.9747 | 4.269 | 83.3878 | 4.420 | 84.6155 | 4.572 | 85.2930 | 4.724 | 85.5316 | 4.870 |
| 81.0961 | 77.8419 | 3.944 | 80.1994 | 4.086 | 82.5786 | 4.247 | 84.0482 | 4.396 | 85.3191 | 4.548 | 86.0313 | 4.701 | 86.3098 | 4.844 |
| 82.0352 | 78.4014 | 3.927 | 80.7510 | 4.068 | 83.1980 | 4.226 | 84.7389 | 4.375 | 86.0348 | 4.527 | 86.8396 | 4.678 | 87.1536 | 4.823 |
| 82.9851 | 78.9587 | 3.907 | 81.4104 | 4.048 | 83.8897 | 4.206 | 85.4688 | 4.354 | 86.8587 | 4.505 | 87.6877 | 4.658 | 88.0569 | 4.802 |
| 83.9460 | 79.5081 | 3.890 | 82.0181 | 4.028 | 84.5803 | 4.185 | 86.2262 | 4.333 | 87.6725 | 4.484 | 88.5540 | 4.636 | 88.9799 | 4.780 |
| 84.9180 | 80.1251 | 3.869 | 82.6532 | 4.007 | 85.2366 | 4.161 | 86.9392 | 4.309 | 88.4334 | 4.457 | 89.3368 | 4.609 | 89.8223 | 4.751 |
| 85.9014 | 80.6890 | 3.853 | 83.2533 | 3.988 | 85.8733 | 4.142 | 87.6082 | 4.288 | 89.1933 | 4.439 | 90.1086 | 4.587 | 90.6940 | 4.729 |
| 86.8960 | 81.2549 | 3.830 | 83.8783 | 3.963 | 86.5518 | 4.117 | 88.3869 | 4.261 | 89.9321 | 4.410 | 91.0551 | 4.568 | 91.6562 | 4.706 |
| 87.9023 | 81.7021 | 3.815 | 84.3572 | 3.947 | 87.1199 | 4.100 | 88.9660 | 4.243 | 90.5997 | 4.388 | 91.6925 | 4.535 | 92.3352 | 4.675 |
| 88.9201 | 82.2896 | 3.799 | 84.9335 | 3.930 | 87.7326 | 4.081 | 89.6213 | 4.223 | 91.3135 | 4.366 | 92.4784 | 4.511 | 93.1436 | 4.651 |
| 89.9498 | 82.8788 | 3.779 | 85.5955 | 3.909 | 88.4550 | 4.058 | 90.4242 | 4.199 | 92.1683 | 4.343 | 93.3762 | 4.488 | 94.1167 | 4.628 |
| 90.9913 | 83.2958 | 3.762 | 86.0660 | 3.891 | 88.9973 | 4.038 | 91.0025 | 4.177 | 92.7978 | 4.321 | 94.0537 | 4.464 | 94.8811 | 4.602 |
| 92.0450 | 83.9239 | 3.745 | 86.7550 | 3.872 | 89.7075 | 4.020 | 91.8008 | 4.160 | 93.6265 | 4.302 | 94.9686 | 4.446 | 95.8276 | 4.583 |
| 93.1108 | 84.4156 | 3.729 | 87.3049 | 3.855 | 90.3356 | 4.001 | 92.4729 | 4.139 | 94.3755 | 4.279 | 95.7442 | 4.421 | 96.6528 | 4.556 |
| 94.1890 | 84.9316 | 3.710 | 87.8490 | 3.835 | 90.9172 | 3.980 | 93.1000 | 4.116 | 95.0412 | 4.256 | 96.4805 | 4.396 | 97.4529 | 4.532 |
| 95.2796 | 85.6223 | 3.694 | 88.5428 | 3.818 | 91.6702 | 3.960 | 93.9156 | 4.096 | 95.9148 | 4.234 | 97.3781 | 4.374 | 98.4291 | 4.510 |
| 96.3829 | 86.1555 | 3.678 | 89.1596 | 3.801 | 92.3569 | 3.942 | 94.6610 | 4.076 | 96.7583 | 4.214 | 98.2948 | 4.351 | 99.3795 | 4.486 |
| 97.4990 | 86.6786 | 3.661 | 89.7209 | 3.782 | 92.9197 | 3.922 | 95.2678 | 4.056 | 97.3945 | 4.193 | 98.9579 | 4.330 | 100.090 | 4.463 |
| 98.6279 | 87.2366 | 3.644 | 90.2974 | 3.763 | 93.5856 | 3.902 | 95.9735 | 4.033 | 98.1615 | 4.168 | 99.7965 | 4.302 | 101.024 | 4.435 |
| 99.7700 | 87.8901 | 3.627 | 90.9866 | 3.746 | 94.3431 | 3.883 | 96.8069 | 4.013 | 99.0297 | 4.147 | 100.719 | 4.283 | 102.014 | 4.415 |
| 100.925 | 88.3054 | 3.610 | 91.5092 | 3.728 | 94.9003 | 3.865 | 97.3885 | 3.994 | 99.7113 | 4.128 | 101.437 | 4.261 | 102.778 | 4.392 |
| 102.094 | 89.0203 | 3.592 | 92.3071 | 3.708 | 95.7409 | 3.843 | 98.3580 | 3.972 | 100.722 | 4.108 | 102.603 | 4.237 | 103.818 | 4.372 |
| 103.276 | 89.5222 | 3.578 | 92.8031 | 3.692 | 96.3045 | 3.827 | 98.9722 | 3.954 | 101.330 | 4.087 | 103.162 | 4.219 | 104.604 | 4.347 |
| 104.472 | 90.0758 | 3.562 | 93.3600 | 3.675 | 96.9309 | 3.808 | 99.6110 | 3.935 | 102.009 | 4.068 | 103.853 | 4.196 | 105.374 | 4.324 |

| | | | | | | | | | | | | | |
|---|---|---|---|---|---|---|---|---|---|---|---|---|---|
| 105.682 | 90.5711 | 3.546 | 93.9211 | 3.657 | 97.5359 | 3.789 | 100.257 | 3.913 | 102.835 | 4.040 | 104.834 | 4.170 | 106.389 | 4.299 |
| 106.905 | 91.2532 | 3.529 | 94.6888 | 3.640 | 98.3851 | 3.769 | 101.145 | 3.893 | 103.647 | 4.015 | 105.876 | 4.144 | 107.303 | 4.284 |
| 108.143 | 91.5279 | 3.517 | 94.9042 | 3.627 | 98.6567 | 3.756 | 101.555 | 3.877 | 104.060 | 4.001 | 106.193 | 4.127 | 107.830 | 4.253 |
| 109.396 | 92.2139 | 3.498 | 95.6495 | 3.608 | 99.4422 | 3.735 | 102.370 | 3.856 | 104.986 | 3.979 | 107.124 | 4.105 | 108.900 | 4.230 |
| 110.662 | 92.7626 | 3.480 | 96.5724 | 3.595 | 100.286 | 3.727 | 103.334 | 3.847 | 105.936 | 3.968 | 108.131 | 4.088 | 109.872 | 4.207 |
| 111.944 | 93.4207 | 3.466 | 97.2958 | 3.580 | 101.035 | 3.711 | 104.146 | 3.829 | 106.817 | 3.950 | 109.054 | 4.069 | 110.845 | 4.186 |
| 113.240 | 93.8203 | 3.452 | 97.7582 | 3.564 | 101.562 | 3.694 | 104.736 | 3.811 | 107.459 | 3.931 | 109.745 | 4.049 | 111.606 | 4.165 |
| 114.551 | 94.3423 | 3.436 | 98.2859 | 3.548 | 102.177 | 3.676 | 105.407 | 3.793 | 108.187 | 3.911 | 110.538 | 4.029 | 112.438 | 4.144 |
| 115.878 | 94.9919 | 3.421 | 99.0141 | 3.531 | 102.918 | 3.658 | 106.207 | 3.773 | 109.038 | 3.891 | 111.445 | 4.007 | 113.401 | 4.121 |
| 117.220 | 95.5447 | 3.407 | 99.6049 | 3.516 | 103.559 | 3.642 | 106.932 | 3.757 | 109.802 | 3.873 | 112.275 | 3.989 | 114.273 | 4.101 |
| 118.577 | 96.0427 | 3.394 | 100.146 | 3.501 | 104.188 | 3.625 | 107.593 | 3.739 | 110.538 | 3.855 | 113.087 | 3.969 | 115.160 | 4.081 |
| 119.950 | 96.5766 | 3.380 | 100.736 | 3.487 | 104.790 | 3.610 | 108.223 | 3.723 | 111.239 | 3.837 | 113.811 | 3.950 | 115.968 | 4.062 |
| 121.339 | 97.1573 | 3.364 | 101.376 | 3.470 | 105.515 | 3.593 | 109.012 | 3.704 | 112.064 | 3.818 | 114.698 | 3.930 | 116.885 | 4.040 |
| 122.744 | 97.7231 | 3.351 | 101.987 | 3.455 | 106.197 | 3.576 | 109.768 | 3.686 | 112.887 | 3.798 | 115.592 | 3.909 | 117.848 | 4.019 |
| 124.165 | 98.4715 | 3.340 | 102.795 | 3.442 | 107.012 | 3.562 | 110.669 | 3.669 | 113.876 | 3.780 | 116.590 | 3.891 | 119.000 | 4.000 |
| 125.603 | 98.9221 | 3.324 | 103.285 | 3.426 | 107.560 | 3.545 | 111.209 | 3.652 | 114.457 | 3.763 | 117.276 | 3.872 | 119.625 | 3.980 |
| 127.057 | 99.3871 | 3.311 | 103.777 | 3.413 | 108.210 | 3.530 | 111.932 | 3.636 | 115.213 | 3.746 | 118.120 | 3.853 | 120.519 | 3.960 |
| 128.529 | 99.9973 | 3.298 | 104.491 | 3.398 | 108.934 | 3.515 | 112.708 | 3.620 | 116.048 | 3.728 | 118.977 | 3.836 | 121.491 | 3.941 |
| 130.017 | 100.543 | 3.284 | 105.072 | 3.383 | 109.547 | 3.498 | 113.370 | 3.603 | 116.815 | 3.711 | 119.765 | 3.817 | 122.308 | 3.922 |
| 131.522 | 101.031 | 3.270 | 105.600 | 3.369 | 110.176 | 3.483 | 113.991 | 3.587 | 117.435 | 3.693 | 120.490 | 3.798 | 123.083 | 3.902 |
| 133.045 | 101.749 | 3.257 | 106.397 | 3.353 | 110.998 | 3.467 | 114.953 | 3.569 | 118.459 | 3.674 | 121.581 | 3.778 | 124.218 | 3.882 |
| 134.586 | 102.311 | 3.243 | 107.008 | 3.339 | 111.617 | 3.452 | 115.628 | 3.552 | 119.211 | 3.657 | 122.385 | 3.761 | 125.065 | 3.863 |
| 136.144 | 102.868 | 3.232 | 107.628 | 3.327 | 112.331 | 3.438 | 116.422 | 3.538 | 120.031 | 3.641 | 123.272 | 3.744 | 125.972 | 3.846 |
| 137.721 | 103.495 | 3.219 | 108.249 | 3.313 | 113.073 | 3.422 | 117.142 | 3.521 | 120.806 | 3.624 | 124.087 | 3.726 | 126.902 | 3.827 |
| 139.316 | 104.013 | 3.206 | 108.855 | 3.299 | 113.685 | 3.408 | 117.900 | 3.505 | 121.656 | 3.607 | 124.969 | 3.708 | 127.861 | 3.808 |
| 140.929 | 104.671 | 3.194 | 109.526 | 3.286 | 114.432 | 3.394 | 118.671 | 3.491 | 122.464 | 3.591 | 125.913 | 3.691 | 128.839 | 3.789 |
| 142.561 | 105.203 | 3.182 | 110.084 | 3.273 | 115.102 | 3.379 | 119.334 | 3.475 | 123.263 | 3.574 | 126.682 | 3.673 | 129.761 | 3.771 |
| 144.212 | 105.823 | 3.169 | 110.803 | 3.260 | 115.779 | 3.365 | 120.094 | 3.460 | 124.035 | 3.558 | 127.559 | 3.656 | 130.595 | 3.752 |
| 145.881 | 106.405 | 3.156 | 111.412 | 3.245 | 116.465 | 3.350 | 120.820 | 3.444 | 124.801 | 3.541 | 128.410 | 3.638 | 131.541 | 3.734 |
| 147.571 | 106.900 | 3.144 | 111.935 | 3.232 | 117.128 | 3.336 | 121.518 | 3.429 | 125.537 | 3.525 | 129.203 | 3.621 | 132.373 | 3.717 |
| 149.279 | 107.512 | 3.133 | 112.611 | 3.220 | 117.815 | 3.322 | 122.327 | 3.414 | 126.405 | 3.510 | 130.078 | 3.605 | 133.339 | 3.699 |
| 151.008 | 108.097 | 3.121 | 113.295 | 3.206 | 118.516 | 3.308 | 123.051 | 3.399 | 127.220 | 3.493 | 130.965 | 3.587 | 134.290 | 3.680 |
| 152.757 | 108.709 | 3.109 | 113.967 | 3.194 | 119.197 | 3.295 | 123.834 | 3.384 | 128.040 | 3.478 | 131.797 | 3.572 | 135.223 | 3.663 |
| 154.525 | 109.292 | 3.098 | 114.603 | 3.182 | 119.898 | 3.281 | 124.601 | 3.370 | 128.849 | 3.463 | 132.708 | 3.555 | 136.160 | 3.646 |
| 156.315 | 109.860 | 3.086 | 115.232 | 3.169 | 120.640 | 3.267 | 125.414 | 3.355 | 129.689 | 3.446 | 133.661 | 3.537 | 137.128 | 3.628 |
| 158.125 | 110.326 | 3.076 | 115.650 | 3.159 | 121.103 | 3.256 | 125.899 | 3.343 | 130.311 | 3.433 | 134.242 | 3.524 | 137.862 | 3.612 |
| 159.956 | 110.951 | 3.064 | 116.429 | 3.145 | 122.071 | 3.241 | 126.894 | 3.326 | 131.348 | 3.415 | 135.378 | 3.505 | 139.007 | 3.595 |
| 161.808 | 111.756 | 3.055 | 117.377 | 3.138 | 122.943 | 3.226 | 127.594 | 3.313 | 131.776 | 3.398 | 135.864 | 3.487 | 139.472 | 3.576 |
| 163.682 | 112.322 | 3.044 | 118.000 | 3.126 | 123.677 | 3.213 | 128.424 | 3.300 | 132.967 | 3.383 | 137.050 | 3.470 | 140.743 | 3.560 |
| 165.577 | 112.916 | 3.032 | 118.638 | 3.112 | 124.294 | 3.199 | 129.105 | 3.286 | 133.623 | 3.370 | 138.037 | 3.457 | 141.879 | 3.547 |
| 167.494 | 113.560 | 3.021 | 119.347 | 3.100 | 125.095 | 3.186 | 129.789 | 3.272 | 134.509 | 3.352 | 138.610 | 3.439 | 142.563 | 3.524 |
| 169.434 | 114.147 | 3.011 | 119.982 | 3.089 | 125.776 | 3.174 | 130.654 | 3.257 | 135.249 | 3.336 | 139.670 | 3.419 | 143.386 | 3.508 |
| 171.396 | 114.684 | 2.999 | 120.565 | 3.077 | 126.393 | 3.160 | 131.317 | 3.244 | 135.852 | 3.322 | 140.049 | 3.407 | 143.752 | 3.486 |
| 173.380 | 115.298 | 2.989 | 121.233 | 3.066 | 127.131 | 3.149 | 132.078 | 3.232 | 136.729 | 3.309 | 141.169 | 3.390 | 145.066 | 3.477 |
| 175.388 | 115.875 | 2.979 | 121.819 | 3.055 | 127.762 | 3.136 | 132.768 | 3.218 | 137.621 | 3.294 | 142.138 | 3.379 | 146.226 | 3.464 |
| 177.419 | 116.494 | 2.968 | 122.505 | 3.044 | 128.490 | 3.125 | 133.661 | 3.204 | 138.384 | 3.280 | 142.782 | 3.361 | 147.165 | 3.443 |
| 179.473 | 117.043 | 2.957 | 123.058 | 3.032 | 129.149 | 3.111 | 134.189 | 3.191 | 139.087 | 3.265 | 143.824 | 3.348 | 147.775 | 3.431 |

| | | | | | | | | | | | | | |
|---|---|---|---|---|---|---|---|---|---|---|---|---|---|
| 181.552 | 117.609 | 2.948 | 123.782 | 3.021 | 129.893 | 3.101 | 135.118 | 3.179 | 140.115 | 3.255 | 144.820 | 3.337 | 149.144 | 3.412 |
| 183.654 | 118.276 | 2.939 | 124.448 | 3.010 | 130.603 | 3.089 | 135.857 | 3.167 | 140.737 | 3.242 | 145.407 | 3.319 | 149.565 | 3.400 |
| 185.780 | 118.960 | 2.928 | 125.195 | 2.998 | 131.345 | 3.075 | 136.638 | 3.152 | 141.611 | 3.225 | 146.425 | 3.306 | 150.506 | 3.386 |
| 187.932 | 119.455 | 2.918 | 125.745 | 2.989 | 132.058 | 3.065 | 137.356 | 3.141 | 142.273 | 3.215 | 147.069 | 3.294 | 151.487 | 3.373 |
| 190.108 | 120.142 | 2.909 | 126.411 | 2.979 | 132.672 | 3.055 | 137.992 | 3.130 | 143.129 | 3.203 | 148.009 | 3.279 | 152.373 | 3.358 |
| 192.309 | 120.741 | 2.899 | 127.145 | 2.967 | 133.454 | 3.043 | 138.698 | 3.118 | 144.147 | 3.189 | 149.374 | 3.266 | 153.934 | 3.342 |
| 194.536 | 121.215 | 2.890 | 127.636 | 2.957 | 134.098 | 3.032 | 139.597 | 3.105 | 144.906 | 3.176 | 150.096 | 3.253 | 154.705 | 3.328 |
| 196.789 | 121.914 | 2.880 | 128.259 | 2.947 | 134.710 | 3.022 | 140.397 | 3.094 | 145.723 | 3.163 | 150.781 | 3.240 | 155.640 | 3.314 |
| 199.067 | 122.541 | 2.871 | 129.058 | 2.938 | 135.495 | 3.011 | 141.036 | 3.083 | 146.278 | 3.152 | 151.779 | 3.226 | 156.303 | 3.303 |
| 201.372 | 123.210 | 2.861 | 129.743 | 2.927 | 136.301 | 2.999 | 141.967 | 3.069 | 147.578 | 3.138 | 152.997 | 3.210 | 157.551 | 3.282 |
| 203.704 | 123.847 | 2.852 | 130.494 | 2.916 | 137.164 | 2.988 | 142.769 | 3.057 | 148.273 | 3.126 | 153.597 | 3.199 | 158.747 | 3.266 |
| 206.063 | 124.398 | 2.843 | 131.078 | 2.907 | 137.798 | 2.977 | 143.685 | 3.046 | 149.216 | 3.113 | 154.703 | 3.183 | 159.693 | 3.252 |
| 208.449 | 125.076 | 2.834 | 131.763 | 2.896 | 138.452 | 2.966 | 144.439 | 3.034 | 150.097 | 3.099 | 155.529 | 3.168 | 160.179 | 3.239 |
| 210.863 | 125.634 | 2.825 | 132.293 | 2.887 | 138.927 | 2.957 | 145.105 | 3.024 | 150.826 | 3.089 | 156.353 | 3.159 | 161.217 | 3.226 |
| 213.304 | 126.267 | 2.816 | 133.109 | 2.877 | 139.855 | 2.945 | 145.801 | 3.011 | 151.492 | 3.076 | 157.145 | 3.145 | 162.293 | 3.213 |
| 215.774 | 127.047 | 2.810 | 133.887 | 2.869 | 140.652 | 2.933 | 146.829 | 2.997 | 152.842 | 3.064 | 157.975 | 3.129 | 162.414 | 3.203 |
| 218.273 | 127.674 | 2.801 | 134.577 | 2.859 | 141.409 | 2.923 | 147.640 | 2.986 | 153.601 | 3.053 | 158.888 | 3.116 | 163.347 | 3.190 |
| 220.800 | 128.343 | 2.791 | 135.259 | 2.849 | 142.073 | 2.912 | 148.415 | 2.975 | 154.467 | 3.040 | 159.761 | 3.103 | 164.065 | 3.174 |
| 223.357 | 128.972 | 2.783 | 135.921 | 2.840 | 142.793 | 2.902 | 149.216 | 2.964 | 155.371 | 3.029 | 160.670 | 3.090 | 165.355 | 3.158 |
| 225.944 | 129.646 | 2.774 | 136.681 | 2.831 | 143.558 | 2.893 | 150.008 | 2.954 | 156.198 | 3.017 | 161.445 | 3.077 | 166.117 | 3.147 |
| 228.560 | 130.165 | 2.766 | 137.312 | 2.822 | 144.347 | 2.883 | 150.863 | 2.943 | 157.048 | 3.006 | 162.325 | 3.066 | 167.179 | 3.133 |
| 231.206 | 130.838 | 2.758 | 137.937 | 2.812 | 144.972 | 2.873 | 151.642 | 2.932 | 157.926 | 2.994 | 163.478 | 3.054 | 168.038 | 3.123 |
| 233.884 | 131.399 | 2.750 | 138.538 | 2.804 | 145.669 | 2.863 | 152.305 | 2.922 | 158.745 | 2.984 | 164.312 | 3.043 | 169.333 | 3.113 |
| 236.592 | 132.114 | 2.742 | 139.282 | 2.795 | 146.473 | 2.853 | 153.120 | 2.912 | 159.548 | 2.972 | 165.161 | 3.030 | 170.140 | 3.100 |
| 239.332 | 132.707 | 2.733 | 139.949 | 2.786 | 147.158 | 2.844 | 153.816 | 2.901 | 160.199 | 2.961 | 165.828 | 3.019 | 171.110 | 3.085 |
| 242.103 | 133.598 | 2.724 | 141.052 | 2.776 | 148.296 | 2.833 | 155.077 | 2.889 | 161.518 | 2.948 | 167.066 | 3.005 | 172.078 | 3.072 |
| 244.906 | 133.938 | 2.719 | 141.391 | 2.769 | 148.765 | 2.825 | 155.503 | 2.881 | 162.157 | 2.939 | 167.960 | 2.994 | 173.109 | 3.063 |
| 247.742 | 134.742 | 2.709 | 142.209 | 2.759 | 149.634 | 2.814 | 156.375 | 2.871 | 163.047 | 2.928 | 168.809 | 2.982 | 174.700 | 3.046 |
| 250.611 | 135.411 | 2.701 | 143.033 | 2.751 | 150.463 | 2.805 | 157.520 | 2.860 | 164.186 | 2.917 | 170.123 | 2.972 | 175.171 | 3.037 |
| 253.513 | 136.181 | 2.692 | 143.687 | 2.742 | 151.059 | 2.797 | 157.880 | 2.852 | 164.704 | 2.908 | 170.391 | 2.963 | 176.287 | 3.029 |
| 256.448 | 136.521 | 2.685 | 144.022 | 2.733 | 151.485 | 2.788 | 158.687 | 2.840 | 165.691 | 2.895 | 171.787 | 2.949 | 177.585 | 3.010 |
| 259.418 | 137.514 | 2.677 | 145.113 | 2.725 | 152.676 | 2.777 | 159.712 | 2.831 | 166.554 | 2.886 | 172.622 | 2.939 | 177.897 | 2.999 |
| 262.422 | 137.966 | 2.671 | 145.676 | 2.717 | 153.399 | 2.769 | 160.530 | 2.821 | 167.433 | 2.875 | 173.693 | 2.926 | 179.205 | 2.988 |
| 265.461 | 138.682 | 2.661 | 146.390 | 2.708 | 154.021 | 2.760 | 161.251 | 2.811 | 168.219 | 2.865 | 174.298 | 2.917 | 179.777 | 2.977 |
| 268.534 | 139.267 | 2.655 | 147.131 | 2.700 | 154.764 | 2.751 | 161.974 | 2.803 | 168.930 | 2.855 | 174.997 | 2.906 | 180.811 | 2.965 |
| 271.644 | 139.825 | 2.648 | 147.648 | 2.693 | 155.494 | 2.743 | 162.786 | 2.794 | 169.854 | 2.846 | 176.154 | 2.898 | 181.968 | 2.955 |
| 274.789 | 140.650 | 2.641 | 148.565 | 2.685 | 156.477 | 2.734 | 163.995 | 2.783 | 171.293 | 2.834 | 177.889 | 2.884 | 183.442 | 2.943 |
| 277.971 | 141.307 | 2.635 | 149.243 | 2.677 | 157.199 | 2.726 | 164.634 | 2.775 | 171.604 | 2.829 | 178.318 | 2.876 | 184.690 | 2.926 |
| 281.190 | 141.707 | 2.627 | 149.626 | 2.670 | 157.466 | 2.718 | 165.068 | 2.767 | 172.347 | 2.818 | 178.723 | 2.865 | 185.099 | 2.916 |
| 284.446 | 142.403 | 2.620 | 150.432 | 2.662 | 158.342 | 2.710 | 165.862 | 2.759 | 173.016 | 2.810 | 179.629 | 2.855 | 185.364 | 2.907 |
| 287.740 | 143.063 | 2.615 | 151.114 | 2.657 | 159.190 | 2.703 | 166.954 | 2.750 | 174.122 | 2.801 | 180.846 | 2.847 | 186.578 | 2.901 |
| 291.072 | 143.750 | 2.607 | 151.869 | 2.649 | 159.831 | 2.695 | 167.615 | 2.741 | 174.900 | 2.791 | 181.769 | 2.836 | 187.611 | 2.893 |
| 294.442 | 144.348 | 2.599 | 152.727 | 2.640 | 161.038 | 2.684 | 168.966 | 2.730 | 176.550 | 2.779 | 183.239 | 2.823 | 189.807 | 2.872 |
| 297.852 | 145.105 | 2.594 | 153.230 | 2.634 | 161.523 | 2.679 | 169.143 | 2.725 | 176.362 | 2.773 | 183.186 | 2.815 | 189.383 | 2.869 |
| 301.301 | 145.721 | 2.587 | 153.984 | 2.627 | 162.510 | 2.671 | 170.389 | 2.716 | 177.681 | 2.764 | 184.692 | 2.808 | 191.043 | 2.857 |
| 304.789 | 146.536 | 2.581 | 154.821 | 2.620 | 163.068 | 2.664 | 170.959 | 2.708 | 178.525 | 2.757 | 185.478 | 2.799 | 191.158 | 2.848 |
| 308.319 | 146.960 | 2.576 | 155.240 | 2.614 | 163.682 | 2.657 | 171.695 | 2.699 | 179.480 | 2.747 | 186.561 | 2.786 | 192.036 | 2.835 |

| | | | | | | | | | | | | | |
|---|---|---|---|---|---|---|---|---|---|---|---|---|---|
| 311.889 | 147.833 | 2.568 | 156.076 | 2.606 | 164.497 | 2.649 | 172.462 | 2.692 | 180.176 | 2.739 | 186.895 | 2.781 | 193.012 | 2.831 |
| 315.500 | 148.821 | 2.562 | 157.158 | 2.599 | 165.652 | 2.641 | 173.706 | 2.684 | 181.390 | 2.729 | 188.474 | 2.772 | 193.803 | 2.818 |
| 319.154 | 149.342 | 2.555 | 157.926 | 2.592 | 166.477 | 2.633 | 174.600 | 2.676 | 182.407 | 2.721 | 189.321 | 2.759 | 194.620 | 2.807 |
| 322.849 | 149.940 | 2.549 | 158.572 | 2.587 | 167.142 | 2.627 | 175.409 | 2.668 | 182.202 | 2.705 | 189.846 | 2.750 | 196.598 | 2.795 |
| 326.588 | 150.701 | 2.544 | 159.437 | 2.579 | 167.847 | 2.618 | 176.448 | 2.659 | 183.217 | 2.699 | 190.885 | 2.743 | 197.681 | 2.790 |
| 330.370 | 151.212 | 2.538 | 160.118 | 2.572 | 168.571 | 2.611 | 176.693 | 2.650 | 184.312 | 2.690 | 192.210 | 2.734 | 198.900 | 2.780 |
| 334.195 | 151.897 | 2.531 | 160.742 | 2.565 | 169.483 | 2.604 | 177.941 | 2.640 | 184.797 | 2.679 | 193.358 | 2.726 | 199.977 | 2.771 |
| 338.065 | 152.763 | 2.527 | 161.502 | 2.560 | 170.274 | 2.598 | 178.896 | 2.636 | 185.918 | 2.676 | 194.170 | 2.719 | 201.059 | 2.763 |
| 341.979 | 153.011 | 2.520 | 162.014 | 2.554 | 170.665 | 2.591 | 179.602 | 2.627 | 186.884 | 2.668 | 195.129 | 2.713 | 202.420 | 2.756 |
| 345.939 | 153.985 | 2.514 | 162.591 | 2.549 | 171.778 | 2.585 | 180.284 | 2.619 | 187.630 | 2.659 | 196.101 | 2.702 | 202.822 | 2.745 |
| 349.945 | 155.017 | 2.508 | 163.805 | 2.543 | 172.790 | 2.577 | 181.634 | 2.613 | 188.936 | 2.650 | 197.364 | 2.693 | 204.359 | 2.736 |
| 353.997 | 155.609 | 2.503 | 164.934 | 2.535 | 173.571 | 2.569 | 182.336 | 2.604 | 190.035 | 2.643 | 198.247 | 2.683 | 205.325 | 2.725 |
| 358.096 | 156.546 | 2.498 | 165.355 | 2.529 | 174.329 | 2.563 | 183.297 | 2.596 | 190.872 | 2.637 | 199.041 | 2.676 | 206.134 | 2.718 |
| 362.243 | 157.445 | 2.493 | 166.406 | 2.523 | 175.088 | 2.555 | 184.170 | 2.588 | 191.541 | 2.627 | 200.156 | 2.667 | 207.457 | 2.709 |
| 366.438 | 157.945 | 2.486 | 166.817 | 2.516 | 175.790 | 2.549 | 184.750 | 2.582 | 192.655 | 2.621 | 200.840 | 2.657 | 208.001 | 2.697 |
| 370.681 | 158.538 | 2.482 | 167.249 | 2.513 | 175.961 | 2.547 | 185.001 | 2.579 | 193.516 | 2.614 | 201.847 | 2.652 | 209.346 | 2.694 |
| 374.973 | 159.392 | 2.477 | 168.584 | 2.509 | 177.544 | 2.540 | 186.151 | 2.572 | 194.491 | 2.606 | 202.998 | 2.641 | 210.479 | 2.680 |
| 379.315 | 160.578 | 2.470 | 169.745 | 2.501 | 178.428 | 2.533 | 187.474 | 2.565 | 195.567 | 2.599 | 203.679 | 2.635 | 211.172 | 2.673 |
| 383.707 | 161.151 | 2.465 | 170.228 | 2.494 | 179.147 | 2.525 | 188.116 | 2.559 | 196.339 | 2.591 | 204.862 | 2.626 | 212.527 | 2.665 |
| 388.150 | 162.105 | 2.461 | 171.423 | 2.490 | 180.353 | 2.520 | 189.256 | 2.550 | 197.686 | 2.583 | 206.211 | 2.618 | 214.020 | 2.655 |
| 392.645 | 162.760 | 2.455 | 172.055 | 2.483 | 180.997 | 2.514 | 189.676 | 2.544 | 198.606 | 2.576 | 207.099 | 2.611 | 214.651 | 2.649 |
| 397.192 | 163.162 | 2.450 | 172.321 | 2.479 | 181.568 | 2.510 | 190.811 | 2.540 | 199.387 | 2.569 | 207.932 | 2.604 | 215.943 | 2.640 |
| 401.791 | 164.319 | 2.443 | 173.609 | 2.472 | 182.873 | 2.501 | 191.821 | 2.531 | 200.545 | 2.562 | 209.222 | 2.596 | 216.987 | 2.632 |
| 406.443 | 164.998 | 2.440 | 174.400 | 2.467 | 183.533 | 2.499 | 192.924 | 2.527 | 201.401 | 2.555 | 210.095 | 2.589 | 218.337 | 2.625 |
| 411.150 | 165.957 | 2.434 | 175.511 | 2.461 | 184.434 | 2.490 | 193.460 | 2.519 | 202.556 | 2.548 | 211.317 | 2.580 | 219.277 | 2.617 |
| 415.911 | 166.884 | 2.429 | 176.351 | 2.458 | 185.865 | 2.486 | 194.733 | 2.513 | 203.458 | 2.539 | 212.623 | 2.574 | 220.513 | 2.608 |
| 420.727 | 167.597 | 2.425 | 176.907 | 2.450 | 186.217 | 2.480 | 195.336 | 2.510 | 204.480 | 2.533 | 213.415 | 2.567 | 221.545 | 2.601 |
| 425.598 | 168.501 | 2.419 | 177.995 | 2.445 | 187.189 | 2.473 | 196.111 | 2.501 | 205.358 | 2.527 | 214.352 | 2.558 | 222.514 | 2.592 |
| 430.527 | 169.226 | 2.415 | 178.805 | 2.442 | 188.392 | 2.468 | 197.659 | 2.496 | 206.110 | 2.519 | 215.627 | 2.553 | 223.781 | 2.586 |
| 435.512 | 170.155 | 2.409 | 179.813 | 2.435 | 189.490 | 2.463 | 198.387 | 2.491 | 206.964 | 2.512 | 216.384 | 2.546 | 224.679 | 2.578 |
| 440.555 | 170.917 | 2.405 | 180.531 | 2.431 | 190.374 | 2.459 | 199.731 | 2.485 | 208.424 | 2.508 | 217.358 | 2.539 | 225.670 | 2.571 |
| 445.656 | 171.669 | 2.400 | 181.546 | 2.425 | 191.336 | 2.452 | 200.670 | 2.478 | 209.136 | 2.501 | 218.653 | 2.531 | 226.906 | 2.563 |
| 450.817 | 172.782 | 2.395 | 182.553 | 2.419 | 192.132 | 2.444 | 201.630 | 2.471 | 210.253 | 2.495 | 219.489 | 2.524 | 228.069 | 2.556 |
| 456.037 | 173.446 | 2.390 | 183.348 | 2.415 | 192.951 | 2.441 | 202.714 | 2.467 | 211.461 | 2.488 | 220.830 | 2.518 | 229.380 | 2.548 |
| 461.318 | 174.491 | 2.386 | 184.280 | 2.410 | 194.063 | 2.436 | 203.775 | 2.461 | 212.359 | 2.481 | 222.100 | 2.511 | 230.678 | 2.542 |
| 466.659 | 175.122 | 2.381 | 185.331 | 2.405 | 195.104 | 2.430 | 204.746 | 2.454 | 213.523 | 2.476 | 223.168 | 2.505 | 231.756 | 2.535 |
| 472.063 | 176.222 | 2.376 | 186.305 | 2.399 | 195.971 | 2.424 | 205.679 | 2.448 | 214.012 | 2.470 | 224.328 | 2.498 | 232.980 | 2.528 |
| 477.529 | 177.134 | 2.371 | 187.284 | 2.395 | 196.985 | 2.418 | 206.679 | 2.443 | 215.507 | 2.464 | 225.192 | 2.492 | 233.804 | 2.521 |
| 483.059 | 177.847 | 2.365 | 187.072 | 2.388 | 197.507 | 2.410 | 206.849 | 2.432 | 216.297 | 2.458 | 225.607 | 2.479 | 235.181 | 2.512 |
| 488.652 | 178.647 | 2.361 | 188.762 | 2.384 | 198.887 | 2.406 | 208.188 | 2.428 | 217.156 | 2.452 | 226.666 | 2.476 | 236.555 | 2.505 |
| 494.311 | 179.710 | 2.355 | 189.139 | 2.377 | 199.512 | 2.398 | 208.424 | 2.422 | 218.675 | 2.445 | 228.005 | 2.469 | 237.551 | 2.498 |
| 500.035 | 180.580 | 2.352 | 190.481 | 2.374 | 200.588 | 2.395 | 209.938 | 2.416 | 219.551 | 2.440 | 228.844 | 2.465 | 238.473 | 2.492 |
| 505.825 | 181.431 | 2.348 | 191.018 | 2.369 | 201.300 | 2.390 | 211.029 | 2.411 | 220.863 | 2.434 | 230.055 | 2.458 | 239.130 | 2.483 |
| 511.682 | 182.609 | 2.342 | 192.662 | 2.364 | 202.570 | 2.383 | 211.919 | 2.405 | 221.917 | 2.428 | 231.608 | 2.451 | 239.859 | 2.477 |
| 517.607 | 183.445 | 2.339 | 193.412 | 2.360 | 203.581 | 2.379 | 213.167 | 2.400 | 222.740 | 2.422 | 232.213 | 2.447 | 241.256 | 2.472 |
| 523.600 | 184.549 | 2.334 | 194.298 | 2.355 | 205.071 | 2.374 | 214.638 | 2.395 | 224.327 | 2.416 | 233.651 | 2.438 | 243.017 | 2.463 |
| 529.663 | 185.716 | 2.330 | 195.648 | 2.349 | 205.212 | 2.369 | 215.793 | 2.390 | 225.535 | 2.411 | 235.027 | 2.434 | 243.956 | 2.459 |

| | | | | | | | | | | | | | |
|---|---|---|---|---|---|---|---|---|---|---|---|---|---|
| 535.797 | 186.575 | 2.325 | 196.451 | 2.345 | 206.574 | 2.364 | 216.177 | 2.383 | 226.197 | 2.404 | 235.710 | 2.426 | 244.956 | 2.452 |
| 542.001 | 187.572 | 2.321 | 197.072 | 2.341 | 207.747 | 2.359 | 217.353 | 2.379 | 227.108 | 2.399 | 237.086 | 2.421 | 246.304 | 2.446 |
| 548.277 | 188.499 | 2.318 | 198.670 | 2.337 | 208.645 | 2.356 | 218.840 | 2.374 | 228.413 | 2.395 | 238.643 | 2.417 | 247.786 | 2.441 |
| 554.626 | 189.139 | 2.313 | 199.274 | 2.332 | 210.130 | 2.350 | 219.561 | 2.368 | 229.894 | 2.389 | 239.443 | 2.410 | 249.275 | 2.433 |
| 561.048 | 190.466 | 2.309 | 200.213 | 2.327 | 210.868 | 2.344 | 220.929 | 2.364 | 230.962 | 2.384 | 241.003 | 2.405 | 250.017 | 2.428 |
| 567.545 | 191.201 | 2.306 | 201.671 | 2.323 | 211.799 | 2.341 | 221.672 | 2.359 | 231.596 | 2.378 | 241.790 | 2.399 | 251.193 | 2.422 |
| 574.116 | 192.036 | 2.300 | 202.363 | 2.318 | 213.337 | 2.336 | 223.179 | 2.354 | 232.792 | 2.373 | 243.159 | 2.393 | 252.516 | 2.417 |
| 580.764 | 193.432 | 2.297 | 203.260 | 2.314 | 213.885 | 2.331 | 223.998 | 2.349 | 234.074 | 2.369 | 244.056 | 2.388 | 254.136 | 2.411 |
| 587.489 | 194.599 | 2.293 | 204.337 | 2.310 | 215.063 | 2.327 | 225.170 | 2.345 | 235.658 | 2.363 | 245.158 | 2.382 | 255.181 | 2.405 |
| 594.292 | 195.563 | 2.290 | 205.994 | 2.306 | 216.757 | 2.323 | 226.977 | 2.340 | 236.991 | 2.359 | 246.513 | 2.373 | 256.796 | 2.395 |
| 601.174 | 196.404 | 2.285 | 206.619 | 2.302 | 217.254 | 2.317 | 227.437 | 2.334 | 238.026 | 2.352 | 247.782 | 2.371 | 257.913 | 2.394 |
| 608.135 | 197.438 | 2.281 | 207.408 | 2.297 | 218.797 | 2.313 | 229.068 | 2.329 | 239.130 | 2.347 | 249.236 | 2.366 | 258.682 | 2.387 |
| 615.177 | 198.702 | 2.277 | 208.584 | 2.292 | 219.229 | 2.307 | 229.924 | 2.323 | 240.598 | 2.341 | 250.410 | 2.361 | 259.905 | 2.381 |
| 622.300 | 199.496 | 2.273 | 209.689 | 2.289 | 220.481 | 2.303 | 230.894 | 2.319 | 241.300 | 2.336 | 251.940 | 2.354 | 261.312 | 2.377 |
| 629.506 | 200.734 | 2.269 | 210.721 | 2.285 | 221.283 | 2.299 | 232.869 | 2.314 | 242.650 | 2.331 | 252.944 | 2.349 | 263.295 | 2.371 |
| 636.796 | 201.518 | 2.264 | 212.283 | 2.279 | 223.053 | 2.294 | 233.100 | 2.310 | 243.997 | 2.326 | 254.797 | 2.345 | 263.363 | 2.364 |
| 644.169 | 202.661 | 2.261 | 213.118 | 2.276 | 224.112 | 2.291 | 234.609 | 2.306 | 245.416 | 2.322 | 255.127 | 2.339 | 265.255 | 2.359 |
| 651.628 | 203.610 | 2.257 | 214.141 | 2.272 | 224.880 | 2.285 | 235.858 | 2.301 | 246.933 | 2.316 | 256.544 | 2.335 | 267.234 | 2.353 |
| 659.174 | 205.060 | 2.254 | 215.640 | 2.268 | 226.146 | 2.282 | 236.878 | 2.297 | 247.900 | 2.312 | 258.446 | 2.329 | 268.076 | 2.349 |
| 666.807 | 205.847 | 2.250 | 216.650 | 2.262 | 227.210 | 2.276 | 237.416 | 2.292 | 248.297 | 2.307 | 258.615 | 2.322 | 269.140 | 2.344 |
| 674.528 | 207.460 | 2.247 | 217.431 | 2.259 | 228.304 | 2.272 | 239.866 | 2.286 | 250.187 | 2.300 | 260.994 | 2.317 | 270.914 | 2.338 |
| 682.339 | 207.974 | 2.244 | 218.154 | 2.257 | 229.767 | 2.270 | 240.421 | 2.284 | 251.580 | 2.299 | 262.311 | 2.315 | 271.896 | 2.334 |
| 690.240 | 209.281 | 2.240 | 219.965 | 2.253 | 231.153 | 2.267 | 242.066 | 2.280 | 252.362 | 2.295 | 263.039 | 2.310 | 273.930 | 2.329 |
| 698.232 | 210.376 | 2.237 | 221.241 | 2.249 | 231.757 | 2.263 | 243.349 | 2.276 | 254.497 | 2.289 | 263.674 | 2.304 | 274.929 | 2.323 |
| 706.318 | 211.464 | 2.234 | 222.133 | 2.246 | 233.639 | 2.258 | 244.290 | 2.272 | 255.371 | 2.285 | 265.379 | 2.301 | 276.513 | 2.318 |
| 714.496 | 212.815 | 2.230 | 223.445 | 2.242 | 234.655 | 2.255 | 245.730 | 2.267 | 256.042 | 2.281 | 267.156 | 2.294 | 277.370 | 2.314 |
| 722.770 | 213.381 | 2.226 | 224.450 | 2.237 | 236.003 | 2.250 | 246.901 | 2.263 | 256.699 | 2.276 | 268.092 | 2.288 | 278.831 | 2.310 |
| 731.139 | 214.826 | 2.223 | 225.295 | 2.234 | 237.160 | 2.247 | 248.167 | 2.259 | 258.786 | 2.271 | 269.445 | 2.284 | 279.378 | 2.302 |
| 739.605 | 216.222 | 2.220 | 227.224 | 2.230 | 237.860 | 2.244 | 249.583 | 2.257 | 259.355 | 2.268 | 270.831 | 2.279 | 281.872 | 2.299 |
| 748.170 | 217.256 | 2.216 | 228.041 | 2.228 | 238.420 | 2.239 | 250.285 | 2.253 | 261.008 | 2.262 | 273.334 | 2.274 | 284.121 | 2.294 |
| 756.833 | 219.316 | 2.212 | 229.419 | 2.224 | 241.211 | 2.237 | 251.292 | 2.248 | 263.042 | 2.260 | 273.721 | 2.270 | 284.545 | 2.290 |
| 765.597 | 218.100 | 2.208 | 232.260 | 2.221 | 242.016 | 2.232 | 254.275 | 2.246 | 263.126 | 2.257 | 274.831 | 2.265 | 285.459 | 2.283 |
| 774.462 | 220.260 | 2.202 | 232.134 | 2.213 | 242.743 | 2.228 | 253.813 | 2.241 | 264.731 | 2.254 | 274.546 | 2.265 | 284.995 | 2.280 |
| 783.430 | 221.883 | 2.199 | 234.136 | 2.213 | 247.107 | 2.227 | 256.438 | 2.241 | 267.983 | 2.253 | 279.666 | 2.260 | 287.984 | 2.276 |
| 792.501 | 223.523 | 2.198 | 235.395 | 2.210 | 245.907 | 2.221 | 257.003 | 2.236 | 268.056 | 2.243 | 278.229 | 2.257 | 289.500 | 2.271 |
| 801.678 | 224.653 | 2.189 | 236.682 | 2.204 | 246.451 | 2.219 | 258.996 | 2.229 | 267.866 | 2.239 | 280.226 | 2.254 | 289.488 | 2.266 |
| 810.961 | 226.820 | 2.189 | 237.339 | 2.201 | 247.811 | 2.213 | 260.654 | 2.225 | 269.570 | 2.234 | 281.613 | 2.247 | 292.523 | 2.258 |
| 820.352 | 227.559 | 2.184 | 239.789 | 2.198 | 250.447 | 2.209 | 260.984 | 2.223 | 272.198 | 2.232 | 282.306 | 2.244 | 292.178 | 2.255 |
| 829.851 | 228.669 | 2.183 | 239.735 | 2.195 | 250.486 | 2.205 | 263.699 | 2.221 | 273.247 | 2.228 | 284.308 | 2.236 | 294.608 | 2.248 |
| 839.460 | 229.914 | 2.176 | 242.083 | 2.192 | 252.380 | 2.202 | 264.011 | 2.216 | 274.812 | 2.224 | 286.315 | 2.231 | 295.694 | 2.245 |
| 849.180 | 232.279 | 2.176 | 243.153 | 2.186 | 254.114 | 2.195 | 265.705 | 2.210 | 276.381 | 2.217 | 287.585 | 2.229 | 297.055 | 2.245 |
| 859.014 | 232.798 | 2.170 | 244.334 | 2.185 | 255.276 | 2.193 | 266.562 | 2.207 | 277.942 | 2.212 | 288.299 | 2.226 | 298.310 | 2.240 |
| 868.960 | 234.082 | 2.168 | 244.857 | 2.181 | 257.002 | 2.191 | 268.653 | 2.204 | 278.923 | 2.212 | 289.626 | 2.222 | 299.555 | 2.235 |
| 879.023 | 235.346 | 2.165 | 246.941 | 2.178 | 257.502 | 2.187 | 269.405 | 2.202 | 279.533 | 2.208 | 291.642 | 2.218 | 301.577 | 2.231 |
| 889.201 | 236.886 | 2.159 | 247.289 | 2.175 | 259.308 | 2.184 | 271.172 | 2.195 | 281.924 | 2.203 | 293.302 | 2.214 | 303.835 | 2.229 |
| 899.498 | 238.398 | 2.159 | 250.414 | 2.170 | 259.717 | 2.180 | 272.748 | 2.193 | 282.702 | 2.200 | 294.859 | 2.209 | 305.943 | 2.222 |
| 909.913 | 239.379 | 2.155 | 251.728 | 2.167 | 261.515 | 2.178 | 274.270 | 2.190 | 284.620 | 2.194 | 295.454 | 2.207 | 307.155 | 2.216 |

| | | | | | | | | | | | | |
|---|---|---|---|---|---|---|---|---|---|---|---|---|
| 920.450 | 241.805 | 2.151 | 252.918 | 2.165 | 263.336 | 2.173 | 275.998 | 2.185 | 287.124 | 2.191 | 297.538 | 2.202 | 308.792 | 2.212 |
| 931.108 | 242.984 | 2.150 | 254.190 | 2.161 | 265.372 | 2.170 | 277.545 | 2.181 | 288.039 | 2.186 | 299.846 | 2.199 | 309.604 | 2.208 |
| 941.890 | 243.696 | 2.147 | 255.462 | 2.158 | 266.415 | 2.168 | 279.335 | 2.179 | 290.070 | 2.185 | 300.435 | 2.194 | 312.380 | 2.205 |
| 952.796 | 244.468 | 2.144 | 257.619 | 2.156 | 268.516 | 2.164 | 280.405 | 2.175 | 291.202 | 2.180 | 302.388 | 2.188 | 313.413 | 2.203 |
| 963.829 | 246.638 | 2.142 | 258.962 | 2.151 | 270.367 | 2.160 | 282.327 | 2.169 | 292.155 | 2.176 | 303.606 | 2.186 | 314.961 | 2.198 |
| 974.990 | 248.239 | 2.136 | 260.147 | 2.148 | 271.572 | 2.156 | 283.103 | 2.168 | 293.215 | 2.172 | 304.494 | 2.186 | 317.871 | 2.194 |
| 986.279 | 249.579 | 2.133 | 262.255 | 2.145 | 272.941 | 2.155 | 283.602 | 2.164 | 295.956 | 2.169 | 306.787 | 2.182 | 316.441 | 2.192 |
| 997.700 | 250.695 | 2.130 | 262.938 | 2.145 | 271.423 | 2.151 | 285.117 | 2.162 | 298.175 | 2.163 | 307.462 | 2.175 | 321.027 | 2.191 |
| 1009.25 | 253.067 | 2.127 | 263.455 | 2.141 | 273.749 | 2.146 | 288.098 | 2.159 | 298.113 | 2.164 | 310.880 | 2.173 | 320.640 | 2.182 |
| 1020.94 | 254.566 | 2.125 | 265.765 | 2.138 | 277.204 | 2.141 | 290.042 | 2.153 | 299.656 | 2.161 | 311.139 | 2.169 | 320.835 | 2.181 |
| 1032.76 | 255.218 | 2.126 | 267.283 | 2.133 | 278.889 | 2.139 | 291.920 | 2.148 | 302.822 | 2.156 | 312.146 | 2.163 | 324.263 | 2.175 |
| 1044.72 | 256.339 | 2.115 | 267.569 | 2.132 | 280.016 | 2.137 | 291.982 | 2.149 | 304.038 | 2.155 | 317.019 | 2.161 | 326.608 | 2.171 |
| 1056.82 | 257.291 | 2.112 | 270.254 | 2.130 | 279.266 | 2.136 | 293.878 | 2.142 | 306.030 | 2.150 | 316.630 | 2.162 | 328.904 | 2.167 |
| 1069.05 | 259.449 | 2.112 | 270.845 | 2.122 | 283.448 | 2.131 | 296.412 | 2.137 | 307.500 | 2.145 | 317.371 | 2.159 | 328.349 | 2.162 |
| 1081.43 | 261.299 | 2.114 | 272.629 | 2.119 | 284.531 | 2.131 | 297.642 | 2.135 | 309.208 | 2.142 | 321.427 | 2.150 | 332.971 | 2.157 |
| 1093.96 | 263.065 | 2.111 | 273.814 | 2.117 | 286.214 | 2.129 | 301.343 | 2.134 | 309.977 | 2.141 | 322.316 | 2.151 | 333.115 | 2.154 |
| 1106.62 | 264.701 | 2.106 | 275.326 | 2.117 | 288.691 | 2.123 | 303.195 | 2.131 | 311.372 | 2.139 | 321.427 | 2.145 | 337.444 | 2.152 |
| 1119.44 | 266.327 | 2.101 | 276.479 | 2.113 | 290.543 | 2.125 | 303.507 | 2.129 | 312.912 | 2.137 | 325.563 | 2.142 | 337.444 | 2.150 |
| 1124.60 | 266.651 | 2.098 | 276.571 | 2.110 | 291.313 | 2.122 | 306.129 | 2.127 | 314.307 | 2.135 | 326.188 | 2.142 | 338.527 | 2.148 |

| Freq (GHz) | Water at 65 ºC a (cm$^{-1}$) | n | Water at 70 ºC a (cm$^{-1}$) | n | Water at 75 ºC a (cm$^{-1}$) | n | Water at 80 ºC a (cm$^{-1}$) | n | Water at 85 ºC a (cm$^{-1}$) | n | Water at 90 ºC a (cm$^{-1}$) | n | Water at 95 ºC a (cm$^{-1}$) | n |
|---|---|---|---|---|---|---|---|---|---|---|---|---|---|---|
| 0.05000 | 0.00016 | 8.123 | 0.00014 | 8.017 | 0.00015 | 7.924 | 0.00008 | 7.815 | 0.00030 | 7.718 | 0.00015 | 7.630 | 0.00027 | 7.540 |
| 0.05058 | 0.00018 | 8.124 | 0.00014 | 8.016 | 0.00020 | 7.924 | 0.00006 | 7.815 | 0.00034 | 7.720 | 0.00014 | 7.632 | 0.00023 | 7.540 |
| 0.05117 | 0.00020 | 8.125 | 0.00013 | 8.015 | 0.00023 | 7.924 | 0.00005 | 7.814 | 0.00038 | 7.722 | 0.00015 | 7.633 | 0.00018 | 7.540 |
| 0.05176 | 0.00021 | 8.126 | 0.00013 | 8.015 | 0.00019 | 7.924 | 0.00005 | 7.814 | 0.00042 | 7.724 | 0.00016 | 7.634 | 0.00014 | 7.540 |
| 0.05236 | 0.00022 | 8.126 | 0.00012 | 8.014 | 0.00015 | 7.924 | 0.00006 | 7.813 | 0.00044 | 7.724 | 0.00016 | 7.635 | 0.00014 | 7.540 |
| 0.05297 | 0.00022 | 8.126 | 0.00012 | 8.014 | 0.00011 | 7.923 | 0.00009 | 7.812 | 0.00044 | 7.724 | 0.00016 | 7.636 | 0.00015 | 7.540 |
| 0.05358 | 0.00022 | 8.127 | 0.00011 | 8.014 | 0.00007 | 7.922 | 0.00012 | 7.811 | 0.00044 | 7.723 | 0.00015 | 7.637 | 0.00016 | 7.540 |
| 0.05420 | 0.00021 | 8.126 | 0.00011 | 8.013 | 0.00010 | 7.921 | 0.00014 | 7.811 | 0.00045 | 7.722 | 0.00014 | 7.638 | 0.00016 | 7.541 |
| 0.05483 | 0.00020 | 8.125 | 0.00010 | 8.014 | 0.00014 | 7.920 | 0.00015 | 7.811 | 0.00048 | 7.722 | 0.00013 | 7.638 | 0.00016 | 7.541 |
| 0.05546 | 0.00018 | 8.124 | 0.00010 | 8.014 | 0.00018 | 7.921 | 0.00016 | 7.810 | 0.00055 | 7.722 | 0.00014 | 7.638 | 0.00015 | 7.542 |
| 0.05610 | 0.00016 | 8.123 | 0.00010 | 8.015 | 0.00021 | 7.921 | 0.00017 | 7.810 | 0.00063 | 7.723 | 0.00018 | 7.638 | 0.00014 | 7.543 |
| 0.05675 | 0.00016 | 8.122 | 0.00010 | 8.016 | 0.00019 | 7.921 | 0.00018 | 7.811 | 0.00071 | 7.723 | 0.00023 | 7.637 | 0.00013 | 7.544 |
| 0.05741 | 0.00017 | 8.122 | 0.00011 | 8.015 | 0.00016 | 7.921 | 0.00021 | 7.812 | 0.00075 | 7.724 | 0.00029 | 7.637 | 0.00013 | 7.545 |
| 0.05808 | 0.00017 | 8.121 | 0.00015 | 8.014 | 0.00014 | 7.920 | 0.00025 | 7.812 | 0.00073 | 7.724 | 0.00031 | 7.636 | 0.00018 | 7.545 |
| 0.05875 | 0.00018 | 8.121 | 0.00021 | 8.014 | 0.00012 | 7.919 | 0.00029 | 7.813 | 0.00069 | 7.723 | 0.00028 | 7.636 | 0.00023 | 7.545 |
| 0.05943 | 0.00018 | 8.121 | 0.00027 | 8.013 | 0.00020 | 7.918 | 0.00030 | 7.813 | 0.00066 | 7.723 | 0.00022 | 7.635 | 0.00029 | 7.546 |
| 0.06012 | 0.00017 | 8.120 | 0.00031 | 8.015 | 0.00029 | 7.917 | 0.00026 | 7.813 | 0.00062 | 7.723 | 0.00016 | 7.635 | 0.00034 | 7.546 |
| 0.06081 | 0.00016 | 8.120 | 0.00029 | 8.017 | 0.00038 | 7.917 | 0.00018 | 7.812 | 0.00059 | 7.722 | 0.00013 | 7.634 | 0.00028 | 7.547 |
| 0.06152 | 0.00015 | 8.120 | 0.00025 | 8.019 | 0.00044 | 7.916 | 0.00009 | 7.812 | 0.00056 | 7.721 | 0.00018 | 7.634 | 0.00022 | 7.547 |
| 0.06223 | 0.00013 | 8.121 | 0.00021 | 8.019 | 0.00042 | 7.915 | 0.00004 | 7.813 | 0.00052 | 7.721 | 0.00024 | 7.634 | 0.00016 | 7.548 |
| 0.06295 | 0.00010 | 8.121 | 0.00021 | 8.018 | 0.00040 | 7.916 | 0.00002 | 7.813 | 0.00051 | 7.722 | 0.00031 | 7.634 | 0.00011 | 7.549 |
| 0.06368 | 0.00008 | 8.122 | 0.00023 | 8.016 | 0.00038 | 7.918 | 0.00002 | 7.814 | 0.00050 | 7.723 | 0.00032 | 7.634 | 0.00017 | 7.551 |
| 0.06442 | 0.00008 | 8.122 | 0.00026 | 8.015 | 0.00036 | 7.920 | 0.00002 | 7.815 | 0.00050 | 7.725 | 0.00027 | 7.634 | 0.00024 | 7.554 |
| 0.06516 | 0.00008 | 8.121 | 0.00027 | 8.015 | 0.00034 | 7.921 | 0.00002 | 7.817 | 0.00049 | 7.726 | 0.00021 | 7.634 | 0.00030 | 7.556 |
| 0.06592 | 0.00008 | 8.120 | 0.00022 | 8.016 | 0.00032 | 7.920 | 0.00002 | 7.818 | 0.00048 | 7.725 | 0.00016 | 7.634 | 0.00032 | 7.558 |
| 0.06668 | 0.00008 | 8.120 | 0.00017 | 8.017 | 0.00030 | 7.918 | 0.00002 | 7.819 | 0.00047 | 7.724 | 0.00014 | 7.636 | 0.00027 | 7.557 |
| 0.06745 | 0.00007 | 8.120 | 0.00011 | 8.017 | 0.00025 | 7.917 | 0.00004 | 7.819 | 0.00045 | 7.723 | 0.00013 | 7.637 | 0.00021 | 7.556 |
| 0.06823 | 0.00007 | 8.120 | 0.00012 | 8.016 | 0.00021 | 7.918 | 0.00008 | 7.818 | 0.00045 | 7.723 | 0.00012 | 7.637 | 0.00015 | 7.556 |
| 0.06902 | 0.00008 | 8.120 | 0.00014 | 8.015 | 0.00018 | 7.919 | 0.00012 | 7.817 | 0.00045 | 7.723 | 0.00012 | 7.638 | 0.00014 | 7.555 |
| 0.06982 | 0.00012 | 8.121 | 0.00016 | 8.014 | 0.00023 | 7.920 | 0.00015 | 7.816 | 0.00045 | 7.723 | 0.00012 | 7.638 | 0.00013 | 7.554 |
| 0.07063 | 0.00016 | 8.121 | 0.00018 | 8.014 | 0.00028 | 7.919 | 0.00014 | 7.816 | 0.00046 | 7.723 | 0.00011 | 7.638 | 0.00012 | 7.553 |
| 0.07145 | 0.00017 | 8.122 | 0.00018 | 8.014 | 0.00033 | 7.918 | 0.00012 | 7.816 | 0.00048 | 7.723 | 0.00012 | 7.638 | 0.00012 | 7.551 |
| 0.07228 | 0.00016 | 8.122 | 0.00019 | 8.013 | 0.00033 | 7.917 | 0.00010 | 7.817 | 0.00051 | 7.722 | 0.00013 | 7.637 | 0.00012 | 7.550 |
| 0.07311 | 0.00014 | 8.121 | 0.00017 | 8.012 | 0.00031 | 7.916 | 0.00007 | 7.817 | 0.00052 | 7.722 | 0.00014 | 7.637 | 0.00011 | 7.548 |
| 0.07396 | 0.00015 | 8.122 | 0.00013 | 8.011 | 0.00029 | 7.916 | 0.00004 | 7.818 | 0.00049 | 7.723 | 0.00015 | 7.636 | 0.00011 | 7.546 |
| 0.07482 | 0.00019 | 8.122 | 0.00009 | 8.011 | 0.00029 | 7.915 | 0.00002 | 7.818 | 0.00045 | 7.723 | 0.00015 | 7.635 | 0.00013 | 7.548 |
| 0.07568 | 0.00024 | 8.123 | 0.00010 | 8.011 | 0.00029 | 7.915 | 0.00005 | 7.817 | 0.00043 | 7.723 | 0.00014 | 7.635 | 0.00014 | 7.550 |
| 0.07656 | 0.00025 | 8.124 | 0.00013 | 8.012 | 0.00029 | 7.915 | 0.00012 | 7.816 | 0.00043 | 7.722 | 0.00014 | 7.635 | 0.00016 | 7.552 |
| 0.07745 | 0.00025 | 8.124 | 0.00017 | 8.012 | 0.00034 | 7.915 | 0.00017 | 7.816 | 0.00044 | 7.721 | 0.00013 | 7.636 | 0.00015 | 7.552 |
| 0.07834 | 0.00024 | 8.124 | 0.00016 | 8.013 | 0.00039 | 7.915 | 0.00016 | 7.816 | 0.00046 | 7.721 | 0.00013 | 7.636 | 0.00014 | 7.552 |
| 0.07925 | 0.00023 | 8.124 | 0.00014 | 8.013 | 0.00045 | 7.916 | 0.00010 | 7.816 | 0.00051 | 7.722 | 0.00015 | 7.636 | 0.00013 | 7.553 |
| 0.08017 | 0.00022 | 8.125 | 0.00012 | 8.014 | 0.00044 | 7.916 | 0.00005 | 7.816 | 0.00057 | 7.723 | 0.00024 | 7.635 | 0.00013 | 7.553 |
| 0.08110 | 0.00021 | 8.126 | 0.00011 | 8.014 | 0.00043 | 7.917 | 0.00005 | 7.815 | 0.00058 | 7.723 | 0.00036 | 7.634 | 0.00013 | 7.553 |
| 0.08204 | 0.00022 | 8.126 | 0.00010 | 8.014 | 0.00041 | 7.917 | 0.00007 | 7.815 | 0.00054 | 7.722 | 0.00044 | 7.633 | 0.00013 | 7.553 |
| 0.08299 | 0.00023 | 8.126 | 0.00011 | 8.014 | 0.00040 | 7.917 | 0.00007 | 7.815 | 0.00047 | 7.720 | 0.00045 | 7.632 | 0.00024 | 7.553 |

| | | | | | | | | | | | | |
|---|---|---|---|---|---|---|---|---|---|---|---|---|
| 0.08395 | 0.00025 | 8.126 | 0.00015 | 8.014 | 0.00039 | 7.916 | 0.00006 | 7.816 | 0.00045 | 7.720 | 0.00042 | 7.632 | 0.00036 | 7.552 |
| 0.08492 | 0.00028 | 8.126 | 0.00020 | 8.014 | 0.00038 | 7.915 | 0.00003 | 7.817 | 0.00046 | 7.721 | 0.00040 | 7.632 | 0.00048 | 7.552 |
| 0.08590 | 0.00030 | 8.125 | 0.00020 | 8.014 | 0.00038 | 7.916 | 0.00004 | 7.816 | 0.00047 | 7.722 | 0.00039 | 7.633 | 0.00045 | 7.554 |
| 0.08690 | 0.00028 | 8.125 | 0.00018 | 8.014 | 0.00037 | 7.917 | 0.00010 | 7.816 | 0.00050 | 7.723 | 0.00039 | 7.633 | 0.00042 | 7.556 |
| 0.08790 | 0.00025 | 8.125 | 0.00016 | 8.015 | 0.00035 | 7.917 | 0.00016 | 7.816 | 0.00053 | 7.723 | 0.00046 | 7.633 | 0.00040 | 7.557 |
| 0.08892 | 0.00025 | 8.125 | 0.00016 | 8.016 | 0.00032 | 7.917 | 0.00017 | 7.816 | 0.00055 | 7.723 | 0.00059 | 7.632 | 0.00039 | 7.555 |
| 0.08995 | 0.00027 | 8.124 | 0.00015 | 8.016 | 0.00032 | 7.917 | 0.00016 | 7.816 | 0.00054 | 7.723 | 0.00067 | 7.632 | 0.00038 | 7.554 |
| 0.09099 | 0.00029 | 8.124 | 0.00018 | 8.016 | 0.00038 | 7.916 | 0.00014 | 7.816 | 0.00053 | 7.724 | 0.00066 | 7.633 | 0.00045 | 7.553 |
| 0.09204 | 0.00030 | 8.124 | 0.00023 | 8.016 | 0.00044 | 7.916 | 0.00014 | 7.816 | 0.00046 | 7.724 | 0.00060 | 7.633 | 0.00059 | 7.553 |
| 0.09311 | 0.00031 | 8.124 | 0.00026 | 8.016 | 0.00042 | 7.915 | 0.00015 | 7.816 | 0.00037 | 7.723 | 0.00051 | 7.632 | 0.00072 | 7.553 |
| 0.09419 | 0.00033 | 8.125 | 0.00026 | 8.016 | 0.00041 | 7.916 | 0.00020 | 7.816 | 0.00035 | 7.723 | 0.00039 | 7.632 | 0.00066 | 7.553 |
| 0.09528 | 0.00034 | 8.125 | 0.00026 | 8.016 | 0.00041 | 7.917 | 0.00029 | 7.816 | 0.00043 | 7.723 | 0.00033 | 7.632 | 0.00060 | 7.553 |
| 0.09638 | 0.00035 | 8.125 | 0.00026 | 8.016 | 0.00042 | 7.917 | 0.00032 | 7.816 | 0.00054 | 7.723 | 0.00037 | 7.632 | 0.00051 | 7.552 |
| 0.09750 | 0.00034 | 8.125 | 0.00027 | 8.016 | 0.00043 | 7.918 | 0.00030 | 7.815 | 0.00055 | 7.723 | 0.00046 | 7.632 | 0.00039 | 7.551 |
| 0.09863 | 0.00033 | 8.125 | 0.00027 | 8.016 | 0.00045 | 7.918 | 0.00027 | 7.814 | 0.00052 | 7.723 | 0.00048 | 7.632 | 0.00027 | 7.551 |
| 0.09977 | 0.00032 | 8.125 | 0.00026 | 8.016 | 0.00047 | 7.918 | 0.00024 | 7.814 | 0.00049 | 7.723 | 0.00044 | 7.632 | 0.00037 | 7.551 |
| 0.10093 | 0.00030 | 8.125 | 0.00026 | 8.016 | 0.00048 | 7.917 | 0.00021 | 7.814 | 0.00046 | 7.723 | 0.00040 | 7.631 | 0.00047 | 7.551 |
| 0.10209 | 0.00030 | 8.125 | 0.00026 | 8.016 | 0.00049 | 7.917 | 0.00021 | 7.815 | 0.00044 | 7.723 | 0.00036 | 7.631 | 0.00048 | 7.551 |
| 0.10328 | 0.00031 | 8.125 | 0.00026 | 8.016 | 0.00050 | 7.916 | 0.00024 | 7.815 | 0.00043 | 7.722 | 0.00033 | 7.631 | 0.00044 | 7.551 |
| 0.10447 | 0.00033 | 8.125 | 0.00026 | 8.016 | 0.00051 | 7.916 | 0.00027 | 7.816 | 0.00043 | 7.722 | 0.00031 | 7.631 | 0.00040 | 7.551 |
| 0.10568 | 0.00034 | 8.124 | 0.00025 | 8.016 | 0.00053 | 7.917 | 0.00030 | 7.817 | 0.00044 | 7.722 | 0.00031 | 7.631 | 0.00036 | 7.551 |
| 0.10691 | 0.00036 | 8.124 | 0.00024 | 8.016 | 0.00050 | 7.917 | 0.00033 | 7.817 | 0.00044 | 7.722 | 0.00032 | 7.631 | 0.00032 | 7.550 |
| 0.10814 | 0.00035 | 8.124 | 0.00023 | 8.016 | 0.00046 | 7.917 | 0.00036 | 7.817 | 0.00045 | 7.722 | 0.00032 | 7.631 | 0.00031 | 7.550 |
| 0.10940 | 0.00035 | 8.124 | 0.00023 | 8.016 | 0.00043 | 7.917 | 0.00038 | 7.817 | 0.00047 | 7.722 | 0.00033 | 7.631 | 0.00031 | 7.551 |
| 0.11066 | 0.00034 | 8.124 | 0.00029 | 8.016 | 0.00039 | 7.916 | 0.00041 | 7.817 | 0.00050 | 7.722 | 0.00037 | 7.630 | 0.00032 | 7.551 |
| 0.11194 | 0.00033 | 8.124 | 0.00035 | 8.016 | 0.00037 | 7.916 | 0.00043 | 7.817 | 0.00053 | 7.722 | 0.00043 | 7.630 | 0.00032 | 7.551 |
| 0.11324 | 0.00033 | 8.124 | 0.00041 | 8.016 | 0.00042 | 7.915 | 0.00044 | 7.817 | 0.00056 | 7.722 | 0.00048 | 7.629 | 0.00032 | 7.551 |
| 0.11455 | 0.00033 | 8.123 | 0.00047 | 8.016 | 0.00046 | 7.915 | 0.00042 | 7.817 | 0.00060 | 7.722 | 0.00053 | 7.629 | 0.00037 | 7.552 |
| 0.11588 | 0.00033 | 8.123 | 0.00053 | 8.016 | 0.00050 | 7.914 | 0.00040 | 7.816 | 0.00065 | 7.722 | 0.00055 | 7.629 | 0.00042 | 7.552 |
| 0.11722 | 0.00034 | 8.123 | 0.00058 | 8.016 | 0.00054 | 7.914 | 0.00038 | 7.816 | 0.00069 | 7.721 | 0.00051 | 7.628 | 0.00048 | 7.553 |
| 0.11858 | 0.00035 | 8.123 | 0.00064 | 8.016 | 0.00055 | 7.914 | 0.00036 | 7.816 | 0.00074 | 7.721 | 0.00048 | 7.628 | 0.00053 | 7.553 |
| 0.11995 | 0.00037 | 8.123 | 0.00069 | 8.016 | 0.00057 | 7.914 | 0.00034 | 7.816 | 0.00075 | 7.721 | 0.00044 | 7.627 | 0.00055 | 7.554 |
| 0.12134 | 0.00039 | 8.123 | 0.00067 | 8.016 | 0.00058 | 7.914 | 0.00034 | 7.816 | 0.00075 | 7.721 | 0.00043 | 7.627 | 0.00051 | 7.554 |
| 0.12274 | 0.00041 | 8.123 | 0.00062 | 8.015 | 0.00059 | 7.914 | 0.00033 | 7.817 | 0.00073 | 7.721 | 0.00047 | 7.627 | 0.00047 | 7.555 |
| 0.12417 | 0.00047 | 8.123 | 0.00057 | 8.015 | 0.00062 | 7.914 | 0.00033 | 7.816 | 0.00071 | 7.721 | 0.00053 | 7.627 | 0.00044 | 7.555 |
| 0.12560 | 0.00054 | 8.123 | 0.00051 | 8.015 | 0.00065 | 7.914 | 0.00034 | 7.816 | 0.00071 | 7.721 | 0.00059 | 7.627 | 0.00041 | 7.555 |
| 0.12706 | 0.00062 | 8.123 | 0.00050 | 8.015 | 0.00068 | 7.914 | 0.00038 | 7.816 | 0.00073 | 7.721 | 0.00065 | 7.627 | 0.00047 | 7.554 |
| 0.12853 | 0.00069 | 8.123 | 0.00049 | 8.015 | 0.00070 | 7.914 | 0.00041 | 7.816 | 0.00075 | 7.720 | 0.00069 | 7.627 | 0.00053 | 7.552 |
| 0.13002 | 0.00069 | 8.123 | 0.00049 | 8.015 | 0.00072 | 7.914 | 0.00043 | 7.816 | 0.00078 | 7.720 | 0.00074 | 7.627 | 0.00059 | 7.551 |
| 0.13152 | 0.00068 | 8.123 | 0.00049 | 8.016 | 0.00073 | 7.914 | 0.00042 | 7.816 | 0.00079 | 7.720 | 0.00078 | 7.628 | 0.00065 | 7.550 |
| 0.13305 | 0.00067 | 8.123 | 0.00051 | 8.016 | 0.00074 | 7.914 | 0.00040 | 7.816 | 0.00079 | 7.721 | 0.00078 | 7.627 | 0.00069 | 7.550 |
| 0.13459 | 0.00066 | 8.123 | 0.00053 | 8.017 | 0.00076 | 7.914 | 0.00038 | 7.816 | 0.00078 | 7.721 | 0.00072 | 7.627 | 0.00073 | 7.549 |
| 0.13614 | 0.00066 | 8.122 | 0.00055 | 8.017 | 0.00079 | 7.914 | 0.00036 | 7.816 | 0.00077 | 7.721 | 0.00063 | 7.627 | 0.00078 | 7.549 |
| 0.13772 | 0.00065 | 8.122 | 0.00059 | 8.017 | 0.00082 | 7.914 | 0.00034 | 7.816 | 0.00079 | 7.721 | 0.00055 | 7.628 | 0.00080 | 7.549 |
| 0.13932 | 0.00065 | 8.122 | 0.00064 | 8.018 | 0.00085 | 7.914 | 0.00033 | 7.816 | 0.00081 | 7.721 | 0.00051 | 7.628 | 0.00072 | 7.548 |
| 0.14093 | 0.00067 | 8.122 | 0.00070 | 8.018 | 0.00083 | 7.914 | 0.00031 | 7.816 | 0.00084 | 7.722 | 0.00051 | 7.629 | 0.00063 | 7.548 |
| 0.14256 | 0.00071 | 8.122 | 0.00074 | 8.018 | 0.00079 | 7.914 | 0.00031 | 7.816 | 0.00087 | 7.722 | 0.00052 | 7.629 | 0.00055 | 7.548 |

| | | | | | | | | | | | | | |
|---|---|---|---|---|---|---|---|---|---|---|---|---|---|
| 0.14421 | 0.00074 | 8.122 | 0.00076 | 8.018 | 0.00076 | 7.914 | 0.00030 | 7.816 | 0.00090 | 7.722 | 0.00053 | 7.629 | 0.00050 | 7.548 |
| 0.14588 | 0.00079 | 8.122 | 0.00076 | 8.018 | 0.00075 | 7.914 | 0.00030 | 7.816 | 0.00094 | 7.723 | 0.00054 | 7.630 | 0.00051 | 7.548 |
| 0.14757 | 0.00084 | 8.122 | 0.00077 | 8.018 | 0.00079 | 7.915 | 0.00029 | 7.816 | 0.00098 | 7.724 | 0.00056 | 7.630 | 0.00052 | 7.547 |
| 0.14928 | 0.00089 | 8.122 | 0.00079 | 8.018 | 0.00082 | 7.914 | 0.00028 | 7.816 | 0.00098 | 7.724 | 0.00057 | 7.631 | 0.00053 | 7.547 |
| 0.15101 | 0.00093 | 8.122 | 0.00081 | 8.018 | 0.00086 | 7.914 | 0.00027 | 7.816 | 0.00095 | 7.724 | 0.00060 | 7.631 | 0.00054 | 7.547 |
| 0.15276 | 0.00096 | 8.122 | 0.00083 | 8.018 | 0.00089 | 7.914 | 0.00027 | 7.816 | 0.00092 | 7.724 | 0.00063 | 7.631 | 0.00055 | 7.547 |
| 0.15453 | 0.00100 | 8.122 | 0.00085 | 8.018 | 0.00092 | 7.914 | 0.00030 | 7.816 | 0.00093 | 7.723 | 0.00067 | 7.631 | 0.00057 | 7.548 |
| 0.15631 | 0.00101 | 8.122 | 0.00086 | 8.019 | 0.00095 | 7.914 | 0.00034 | 7.815 | 0.00097 | 7.724 | 0.00070 | 7.631 | 0.00059 | 7.548 |
| 0.15812 | 0.00100 | 8.123 | 0.00088 | 8.019 | 0.00101 | 7.914 | 0.00038 | 7.815 | 0.00102 | 7.724 | 0.00071 | 7.631 | 0.00063 | 7.549 |
| 0.15996 | 0.00099 | 8.123 | 0.00088 | 8.019 | 0.00108 | 7.914 | 0.00044 | 7.815 | 0.00106 | 7.724 | 0.00073 | 7.631 | 0.00067 | 7.550 |
| 0.16181 | 0.00098 | 8.123 | 0.00089 | 8.019 | 0.00116 | 7.914 | 0.00051 | 7.815 | 0.00110 | 7.724 | 0.00074 | 7.631 | 0.00070 | 7.551 |
| 0.16368 | 0.00098 | 8.123 | 0.00089 | 8.019 | 0.00117 | 7.915 | 0.00057 | 7.815 | 0.00113 | 7.724 | 0.00075 | 7.631 | 0.00071 | 7.551 |
| 0.16558 | 0.00097 | 8.123 | 0.00089 | 8.019 | 0.00117 | 7.915 | 0.00062 | 7.815 | 0.00117 | 7.724 | 0.00076 | 7.631 | 0.00072 | 7.551 |
| 0.16749 | 0.00100 | 8.123 | 0.00089 | 8.018 | 0.00117 | 7.914 | 0.00066 | 7.815 | 0.00120 | 7.724 | 0.00077 | 7.631 | 0.00073 | 7.551 |
| 0.16943 | 0.00105 | 8.123 | 0.00090 | 8.018 | 0.00112 | 7.914 | 0.00068 | 7.814 | 0.00124 | 7.724 | 0.00081 | 7.631 | 0.00074 | 7.552 |
| 0.17140 | 0.00111 | 8.123 | 0.00083 | 8.019 | 0.00106 | 7.914 | 0.00063 | 7.814 | 0.00122 | 7.724 | 0.00086 | 7.631 | 0.00075 | 7.552 |
| 0.17338 | 0.00112 | 8.123 | 0.00075 | 8.019 | 0.00101 | 7.915 | 0.00053 | 7.813 | 0.00112 | 7.723 | 0.00090 | 7.631 | 0.00076 | 7.553 |
| 0.17539 | 0.00112 | 8.123 | 0.00069 | 8.019 | 0.00113 | 7.915 | 0.00048 | 7.814 | 0.00099 | 7.723 | 0.00092 | 7.631 | 0.00081 | 7.553 |
| 0.17742 | 0.00115 | 8.124 | 0.00085 | 8.019 | 0.00126 | 7.915 | 0.00055 | 7.814 | 0.00096 | 7.723 | 0.00094 | 7.631 | 0.00085 | 7.554 |
| 0.17947 | 0.00124 | 8.123 | 0.00106 | 8.019 | 0.00135 | 7.915 | 0.00070 | 7.814 | 0.00102 | 7.723 | 0.00096 | 7.631 | 0.00090 | 7.554 |
| 0.18155 | 0.00134 | 8.123 | 0.00121 | 8.019 | 0.00130 | 7.915 | 0.00077 | 7.814 | 0.00111 | 7.723 | 0.00098 | 7.631 | 0.00092 | 7.554 |
| 0.18365 | 0.00138 | 8.122 | 0.00127 | 8.020 | 0.00126 | 7.915 | 0.00070 | 7.814 | 0.00115 | 7.723 | 0.00101 | 7.631 | 0.00093 | 7.554 |
| 0.18578 | 0.00138 | 8.122 | 0.00133 | 8.020 | 0.00124 | 7.915 | 0.00059 | 7.814 | 0.00115 | 7.723 | 0.00106 | 7.631 | 0.00095 | 7.555 |
| 0.18793 | 0.00137 | 8.122 | 0.00136 | 8.020 | 0.00125 | 7.915 | 0.00058 | 7.814 | 0.00115 | 7.723 | 0.00113 | 7.632 | 0.00098 | 7.555 |
| 0.19011 | 0.00139 | 8.122 | 0.00139 | 8.020 | 0.00127 | 7.915 | 0.00071 | 7.815 | 0.00125 | 7.723 | 0.00121 | 7.632 | 0.00101 | 7.555 |
| 0.19231 | 0.00142 | 8.122 | 0.00140 | 8.020 | 0.00130 | 7.915 | 0.00081 | 7.815 | 0.00139 | 7.723 | 0.00123 | 7.632 | 0.00105 | 7.555 |
| 0.19454 | 0.00144 | 8.123 | 0.00136 | 8.021 | 0.00134 | 7.915 | 0.00085 | 7.814 | 0.00144 | 7.723 | 0.00121 | 7.632 | 0.00113 | 7.555 |
| 0.19679 | 0.00144 | 8.123 | 0.00132 | 8.021 | 0.00141 | 7.915 | 0.00084 | 7.814 | 0.00140 | 7.723 | 0.00121 | 7.632 | 0.00121 | 7.555 |
| 0.19907 | 0.00145 | 8.123 | 0.00126 | 8.020 | 0.00151 | 7.915 | 0.00088 | 7.815 | 0.00133 | 7.723 | 0.00125 | 7.632 | 0.00122 | 7.554 |
| 0.20137 | 0.00147 | 8.123 | 0.00119 | 8.020 | 0.00161 | 7.915 | 0.00101 | 7.815 | 0.00137 | 7.723 | 0.00131 | 7.632 | 0.00120 | 7.554 |
| 0.20370 | 0.00149 | 8.123 | 0.00118 | 8.020 | 0.00161 | 7.915 | 0.00112 | 7.815 | 0.00147 | 7.722 | 0.00135 | 7.632 | 0.00118 | 7.554 |
| 0.20606 | 0.00151 | 8.123 | 0.00122 | 8.020 | 0.00159 | 7.915 | 0.00113 | 7.814 | 0.00155 | 7.722 | 0.00137 | 7.631 | 0.00124 | 7.554 |
| 0.20845 | 0.00153 | 8.123 | 0.00128 | 8.020 | 0.00161 | 7.915 | 0.00108 | 7.814 | 0.00161 | 7.722 | 0.00138 | 7.631 | 0.00130 | 7.555 |
| 0.21086 | 0.00156 | 8.123 | 0.00134 | 8.020 | 0.00166 | 7.915 | 0.00106 | 7.814 | 0.00167 | 7.722 | 0.00137 | 7.631 | 0.00134 | 7.555 |
| 0.21330 | 0.00162 | 8.123 | 0.00140 | 8.020 | 0.00171 | 7.915 | 0.00107 | 7.814 | 0.00169 | 7.722 | 0.00136 | 7.631 | 0.00136 | 7.556 |
| 0.21577 | 0.00168 | 8.123 | 0.00144 | 8.020 | 0.00176 | 7.915 | 0.00108 | 7.814 | 0.00171 | 7.722 | 0.00136 | 7.631 | 0.00138 | 7.556 |
| 0.21827 | 0.00174 | 8.123 | 0.00147 | 8.020 | 0.00181 | 7.915 | 0.00109 | 7.814 | 0.00172 | 7.722 | 0.00136 | 7.631 | 0.00137 | 7.556 |
| 0.22080 | 0.00180 | 8.123 | 0.00150 | 8.020 | 0.00182 | 7.915 | 0.00110 | 7.814 | 0.00174 | 7.722 | 0.00137 | 7.631 | 0.00135 | 7.555 |
| 0.22336 | 0.00186 | 8.123 | 0.00152 | 8.020 | 0.00181 | 7.915 | 0.00113 | 7.814 | 0.00176 | 7.722 | 0.00138 | 7.631 | 0.00135 | 7.555 |
| 0.22594 | 0.00189 | 8.122 | 0.00155 | 8.020 | 0.00180 | 7.915 | 0.00119 | 7.814 | 0.00181 | 7.722 | 0.00139 | 7.631 | 0.00136 | 7.555 |
| 0.22856 | 0.00193 | 8.122 | 0.00158 | 8.020 | 0.00179 | 7.915 | 0.00125 | 7.814 | 0.00188 | 7.722 | 0.00142 | 7.631 | 0.00137 | 7.555 |
| 0.23121 | 0.00197 | 8.122 | 0.00163 | 8.020 | 0.00178 | 7.915 | 0.00131 | 7.814 | 0.00195 | 7.722 | 0.00147 | 7.631 | 0.00137 | 7.555 |
| 0.23388 | 0.00201 | 8.122 | 0.00167 | 8.020 | 0.00178 | 7.915 | 0.00136 | 7.814 | 0.00202 | 7.722 | 0.00152 | 7.631 | 0.00138 | 7.555 |
| 0.23659 | 0.00203 | 8.122 | 0.00172 | 8.019 | 0.00177 | 7.915 | 0.00140 | 7.814 | 0.00206 | 7.722 | 0.00157 | 7.631 | 0.00142 | 7.555 |
| 0.23933 | 0.00206 | 8.122 | 0.00177 | 8.019 | 0.00176 | 7.915 | 0.00142 | 7.814 | 0.00205 | 7.721 | 0.00162 | 7.631 | 0.00146 | 7.556 |
| 0.24210 | 0.00208 | 8.122 | 0.00187 | 8.019 | 0.00175 | 7.915 | 0.00145 | 7.814 | 0.00203 | 7.721 | 0.00165 | 7.631 | 0.00151 | 7.556 |
| 0.24491 | 0.00211 | 8.122 | 0.00197 | 8.019 | 0.00178 | 7.915 | 0.00147 | 7.814 | 0.00201 | 7.721 | 0.00168 | 7.631 | 0.00156 | 7.556 |

| | | | | | | | | | | | | |
|---|---|---|---|---|---|---|---|---|---|---|---|---|
| 0.24774 | 0.00219 | 8.122 | 0.00207 | 8.019 | 0.00188 | 7.915 | 0.00150 | 7.814 | 0.00198 | 7.721 | 0.00171 | 7.631 | 0.00161 | 7.556 |
| 0.25061 | 0.00230 | 8.122 | 0.00217 | 8.019 | 0.00198 | 7.915 | 0.00153 | 7.814 | 0.00205 | 7.721 | 0.00174 | 7.631 | 0.00164 | 7.556 |
| 0.25351 | 0.00241 | 8.122 | 0.00223 | 8.019 | 0.00208 | 7.915 | 0.00156 | 7.815 | 0.00215 | 7.721 | 0.00175 | 7.631 | 0.00167 | 7.556 |
| 0.25645 | 0.00253 | 8.122 | 0.00226 | 8.020 | 0.00218 | 7.916 | 0.00159 | 7.815 | 0.00226 | 7.721 | 0.00175 | 7.631 | 0.00170 | 7.556 |
| 0.25942 | 0.00260 | 8.122 | 0.00230 | 8.020 | 0.00226 | 7.916 | 0.00164 | 7.815 | 0.00237 | 7.721 | 0.00174 | 7.631 | 0.00173 | 7.556 |
| 0.26242 | 0.00265 | 8.122 | 0.00234 | 8.020 | 0.00234 | 7.916 | 0.00172 | 7.815 | 0.00244 | 7.721 | 0.00173 | 7.631 | 0.00175 | 7.556 |
| 0.26546 | 0.00270 | 8.122 | 0.00241 | 8.020 | 0.00242 | 7.916 | 0.00182 | 7.815 | 0.00247 | 7.721 | 0.00174 | 7.631 | 0.00174 | 7.556 |
| 0.26853 | 0.00275 | 8.122 | 0.00250 | 8.020 | 0.00250 | 7.916 | 0.00192 | 7.815 | 0.00249 | 7.721 | 0.00177 | 7.631 | 0.00173 | 7.556 |
| 0.27164 | 0.00282 | 8.122 | 0.00258 | 8.020 | 0.00254 | 7.916 | 0.00200 | 7.815 | 0.00252 | 7.721 | 0.00181 | 7.630 | 0.00172 | 7.556 |
| 0.27479 | 0.00290 | 8.122 | 0.00267 | 8.020 | 0.00258 | 7.915 | 0.00203 | 7.815 | 0.00254 | 7.720 | 0.00185 | 7.630 | 0.00172 | 7.556 |
| 0.27797 | 0.00298 | 8.122 | 0.00275 | 8.020 | 0.00262 | 7.915 | 0.00204 | 7.814 | 0.00254 | 7.720 | 0.00189 | 7.630 | 0.00176 | 7.556 |
| 0.28119 | 0.00306 | 8.122 | 0.00282 | 8.019 | 0.00266 | 7.915 | 0.00205 | 7.814 | 0.00255 | 7.720 | 0.00194 | 7.630 | 0.00180 | 7.556 |
| 0.28445 | 0.00314 | 8.122 | 0.00290 | 8.019 | 0.00268 | 7.915 | 0.00207 | 7.814 | 0.00256 | 7.720 | 0.00200 | 7.630 | 0.00184 | 7.555 |
| 0.28774 | 0.00322 | 8.122 | 0.00297 | 8.019 | 0.00270 | 7.915 | 0.00213 | 7.814 | 0.00262 | 7.720 | 0.00205 | 7.631 | 0.00188 | 7.555 |
| 0.29107 | 0.00330 | 8.123 | 0.00302 | 8.019 | 0.00273 | 7.915 | 0.00221 | 7.814 | 0.00271 | 7.720 | 0.00208 | 7.631 | 0.00193 | 7.555 |
| 0.29444 | 0.00339 | 8.123 | 0.00305 | 8.020 | 0.00278 | 7.915 | 0.00228 | 7.814 | 0.00282 | 7.721 | 0.00206 | 7.631 | 0.00199 | 7.555 |
| 0.29785 | 0.00347 | 8.123 | 0.00309 | 8.020 | 0.00287 | 7.915 | 0.00236 | 7.814 | 0.00293 | 7.721 | 0.00203 | 7.631 | 0.00204 | 7.555 |
| 0.30130 | 0.00355 | 8.123 | 0.00312 | 8.020 | 0.00296 | 7.915 | 0.00242 | 7.814 | 0.00298 | 7.721 | 0.00200 | 7.631 | 0.00208 | 7.555 |
| 0.30479 | 0.00363 | 8.123 | 0.00314 | 8.020 | 0.00305 | 7.915 | 0.00249 | 7.815 | 0.00302 | 7.721 | 0.00201 | 7.631 | 0.00205 | 7.556 |
| 0.30832 | 0.00369 | 8.123 | 0.00316 | 8.020 | 0.00313 | 7.915 | 0.00256 | 7.815 | 0.00305 | 7.721 | 0.00205 | 7.631 | 0.00203 | 7.556 |
| 0.31189 | 0.00374 | 8.123 | 0.00320 | 8.020 | 0.00320 | 7.915 | 0.00262 | 7.814 | 0.00307 | 7.721 | 0.00209 | 7.631 | 0.00200 | 7.556 |
| 0.31550 | 0.00379 | 8.123 | 0.00330 | 8.019 | 0.00328 | 7.915 | 0.00269 | 7.814 | 0.00306 | 7.721 | 0.00214 | 7.631 | 0.00200 | 7.556 |
| 0.31915 | 0.00390 | 8.123 | 0.00340 | 8.019 | 0.00334 | 7.914 | 0.00276 | 7.814 | 0.00305 | 7.720 | 0.00223 | 7.631 | 0.00204 | 7.555 |
| 0.32285 | 0.00404 | 8.123 | 0.00351 | 8.019 | 0.00339 | 7.914 | 0.00284 | 7.814 | 0.00304 | 7.720 | 0.00234 | 7.631 | 0.00208 | 7.555 |
| 0.32659 | 0.00418 | 8.123 | 0.00361 | 8.020 | 0.00345 | 7.915 | 0.00293 | 7.814 | 0.00312 | 7.720 | 0.00246 | 7.630 | 0.00212 | 7.555 |
| 0.33037 | 0.00425 | 8.123 | 0.00371 | 8.020 | 0.00351 | 7.915 | 0.00302 | 7.814 | 0.00325 | 7.720 | 0.00256 | 7.630 | 0.00222 | 7.555 |
| 0.33420 | 0.00428 | 8.123 | 0.00381 | 8.020 | 0.00363 | 7.915 | 0.00311 | 7.814 | 0.00338 | 7.720 | 0.00265 | 7.630 | 0.00233 | 7.555 |
| 0.33806 | 0.00429 | 8.122 | 0.00394 | 8.020 | 0.00376 | 7.915 | 0.00319 | 7.814 | 0.00345 | 7.720 | 0.00275 | 7.630 | 0.00244 | 7.555 |
| 0.34198 | 0.00437 | 8.122 | 0.00409 | 8.020 | 0.00388 | 7.915 | 0.00326 | 7.814 | 0.00348 | 7.720 | 0.00284 | 7.630 | 0.00255 | 7.555 |
| 0.34594 | 0.00449 | 8.122 | 0.00423 | 8.020 | 0.00390 | 7.915 | 0.00334 | 7.815 | 0.00351 | 7.720 | 0.00294 | 7.630 | 0.00264 | 7.555 |
| 0.34995 | 0.00463 | 8.122 | 0.00427 | 8.020 | 0.00391 | 7.915 | 0.00340 | 7.815 | 0.00357 | 7.720 | 0.00303 | 7.630 | 0.00274 | 7.555 |
| 0.35400 | 0.00475 | 8.122 | 0.00426 | 8.020 | 0.00392 | 7.915 | 0.00346 | 7.815 | 0.00369 | 7.720 | 0.00312 | 7.630 | 0.00283 | 7.555 |
| 0.35810 | 0.00487 | 8.123 | 0.00424 | 8.020 | 0.00403 | 7.915 | 0.00352 | 7.815 | 0.00382 | 7.720 | 0.00320 | 7.630 | 0.00292 | 7.555 |
| 0.36224 | 0.00500 | 8.123 | 0.00430 | 8.019 | 0.00417 | 7.915 | 0.00358 | 7.815 | 0.00392 | 7.720 | 0.00328 | 7.630 | 0.00301 | 7.555 |
| 0.36644 | 0.00513 | 8.123 | 0.00439 | 8.019 | 0.00430 | 7.915 | 0.00362 | 7.815 | 0.00396 | 7.720 | 0.00337 | 7.630 | 0.00310 | 7.555 |
| 0.37068 | 0.00528 | 8.122 | 0.00449 | 8.019 | 0.00440 | 7.915 | 0.00368 | 7.815 | 0.00398 | 7.720 | 0.00346 | 7.630 | 0.00319 | 7.555 |
| 0.37497 | 0.00542 | 8.122 | 0.00461 | 8.019 | 0.00451 | 7.915 | 0.00375 | 7.815 | 0.00406 | 7.719 | 0.00355 | 7.630 | 0.00327 | 7.555 |
| 0.37931 | 0.00553 | 8.122 | 0.00474 | 8.020 | 0.00461 | 7.915 | 0.00385 | 7.815 | 0.00421 | 7.719 | 0.00363 | 7.630 | 0.00335 | 7.555 |
| 0.38371 | 0.00563 | 8.122 | 0.00487 | 8.020 | 0.00465 | 7.915 | 0.00395 | 7.815 | 0.00439 | 7.719 | 0.00367 | 7.630 | 0.00345 | 7.556 |
| 0.38815 | 0.00573 | 8.122 | 0.00501 | 8.019 | 0.00469 | 7.915 | 0.00404 | 7.815 | 0.00449 | 7.719 | 0.00369 | 7.630 | 0.00354 | 7.556 |
| 0.39264 | 0.00586 | 8.122 | 0.00514 | 8.019 | 0.00475 | 7.915 | 0.00413 | 7.814 | 0.00450 | 7.719 | 0.00372 | 7.630 | 0.00363 | 7.557 |
| 0.39719 | 0.00599 | 8.122 | 0.00529 | 8.020 | 0.00484 | 7.915 | 0.00422 | 7.814 | 0.00449 | 7.719 | 0.00377 | 7.630 | 0.00365 | 7.556 |
| 0.40179 | 0.00616 | 8.122 | 0.00544 | 8.020 | 0.00493 | 7.915 | 0.00430 | 7.814 | 0.00454 | 7.719 | 0.00382 | 7.630 | 0.00367 | 7.556 |
| 0.40644 | 0.00637 | 8.122 | 0.00560 | 8.019 | 0.00509 | 7.915 | 0.00438 | 7.814 | 0.00462 | 7.719 | 0.00389 | 7.630 | 0.00370 | 7.556 |
| 0.41115 | 0.00657 | 8.122 | 0.00570 | 8.019 | 0.00529 | 7.915 | 0.00447 | 7.814 | 0.00471 | 7.719 | 0.00398 | 7.630 | 0.00375 | 7.556 |
| 0.41591 | 0.00669 | 8.122 | 0.00579 | 8.019 | 0.00549 | 7.915 | 0.00458 | 7.815 | 0.00484 | 7.719 | 0.00407 | 7.630 | 0.00381 | 7.556 |
| 0.42073 | 0.00680 | 8.122 | 0.00591 | 8.019 | 0.00563 | 7.915 | 0.00470 | 7.815 | 0.00499 | 7.719 | 0.00419 | 7.630 | 0.00387 | 7.556 |

| | | | | | | | | | | | | |
|---|---|---|---|---|---|---|---|---|---|---|---|---|
| 0.42560 | 0.00695 | 8.122 | 0.00610 | 8.019 | 0.00576 | 7.915 | 0.00483 | 7.815 | 0.00509 | 7.719 | 0.00433 | 7.629 | 0.00396 | 7.556 |
| 0.43053 | 0.00713 | 8.122 | 0.00631 | 8.020 | 0.00589 | 7.915 | 0.00497 | 7.815 | 0.00514 | 7.719 | 0.00442 | 7.629 | 0.00405 | 7.556 |
| 0.43551 | 0.00731 | 8.122 | 0.00645 | 8.020 | 0.00600 | 7.915 | 0.00512 | 7.815 | 0.00518 | 7.718 | 0.00440 | 7.630 | 0.00417 | 7.556 |
| 0.44055 | 0.00748 | 8.122 | 0.00656 | 8.020 | 0.00611 | 7.915 | 0.00528 | 7.815 | 0.00531 | 7.718 | 0.00435 | 7.630 | 0.00431 | 7.556 |
| 0.44566 | 0.00766 | 8.122 | 0.00669 | 8.020 | 0.00627 | 7.915 | 0.00542 | 7.815 | 0.00548 | 7.718 | 0.00443 | 7.630 | 0.00444 | 7.555 |
| 0.45082 | 0.00783 | 8.122 | 0.00683 | 8.020 | 0.00642 | 7.915 | 0.00551 | 7.815 | 0.00565 | 7.718 | 0.00462 | 7.630 | 0.00439 | 7.555 |
| 0.45604 | 0.00801 | 8.122 | 0.00697 | 8.020 | 0.00658 | 7.915 | 0.00556 | 7.815 | 0.00583 | 7.718 | 0.00479 | 7.630 | 0.00433 | 7.555 |
| 0.46132 | 0.00819 | 8.122 | 0.00711 | 8.020 | 0.00674 | 7.915 | 0.00567 | 7.815 | 0.00600 | 7.718 | 0.00495 | 7.629 | 0.00441 | 7.555 |
| 0.46666 | 0.00837 | 8.122 | 0.00726 | 8.020 | 0.00690 | 7.915 | 0.00584 | 7.815 | 0.00616 | 7.718 | 0.00509 | 7.629 | 0.00460 | 7.555 |
| 0.47206 | 0.00854 | 8.122 | 0.00743 | 8.019 | 0.00706 | 7.915 | 0.00602 | 7.815 | 0.00632 | 7.718 | 0.00518 | 7.629 | 0.00478 | 7.555 |
| 0.47753 | 0.00872 | 8.122 | 0.00763 | 8.019 | 0.00723 | 7.915 | 0.00620 | 7.815 | 0.00647 | 7.718 | 0.00525 | 7.629 | 0.00493 | 7.555 |
| 0.48306 | 0.00890 | 8.122 | 0.00783 | 8.019 | 0.00736 | 7.915 | 0.00638 | 7.815 | 0.00663 | 7.718 | 0.00531 | 7.629 | 0.00507 | 7.555 |
| 0.48865 | 0.00911 | 8.122 | 0.00804 | 8.019 | 0.00747 | 7.915 | 0.00657 | 7.815 | 0.00678 | 7.718 | 0.00538 | 7.629 | 0.00516 | 7.555 |
| 0.49431 | 0.00933 | 8.122 | 0.00825 | 8.019 | 0.00758 | 7.915 | 0.00678 | 7.815 | 0.00692 | 7.718 | 0.00547 | 7.629 | 0.00523 | 7.555 |
| 0.50003 | 0.00956 | 8.122 | 0.00846 | 8.019 | 0.00770 | 7.915 | 0.00699 | 7.815 | 0.00705 | 7.718 | 0.00570 | 7.629 | 0.00529 | 7.555 |
| 0.50582 | 0.00979 | 8.122 | 0.00867 | 8.019 | 0.00781 | 7.915 | 0.00720 | 7.815 | 0.00719 | 7.718 | 0.00604 | 7.629 | 0.00536 | 7.555 |
| 0.51168 | 0.01002 | 8.122 | 0.00888 | 8.019 | 0.00806 | 7.915 | 0.00740 | 7.815 | 0.00733 | 7.718 | 0.00638 | 7.629 | 0.00543 | 7.555 |
| 0.51761 | 0.01025 | 8.122 | 0.00910 | 8.019 | 0.00832 | 7.915 | 0.00756 | 7.815 | 0.00747 | 7.718 | 0.00672 | 7.629 | 0.00568 | 7.555 |
| 0.52360 | 0.01047 | 8.122 | 0.00930 | 8.019 | 0.00858 | 7.915 | 0.00769 | 7.815 | 0.00761 | 7.718 | 0.00698 | 7.629 | 0.00601 | 7.555 |
| 0.52966 | 0.01070 | 8.122 | 0.00948 | 8.019 | 0.00885 | 7.915 | 0.00781 | 7.814 | 0.00775 | 7.718 | 0.00705 | 7.629 | 0.00635 | 7.555 |
| 0.53580 | 0.01092 | 8.122 | 0.00966 | 8.019 | 0.00908 | 7.915 | 0.00794 | 7.814 | 0.00790 | 7.718 | 0.00702 | 7.629 | 0.00669 | 7.555 |
| 0.54200 | 0.01116 | 8.122 | 0.00984 | 8.019 | 0.00924 | 7.915 | 0.00809 | 7.814 | 0.00805 | 7.718 | 0.00700 | 7.629 | 0.00704 | 7.555 |
| 0.54828 | 0.01140 | 8.122 | 0.01002 | 8.019 | 0.00941 | 7.915 | 0.00827 | 7.814 | 0.00825 | 7.718 | 0.00697 | 7.629 | 0.00702 | 7.555 |
| 0.55463 | 0.01164 | 8.122 | 0.01029 | 8.019 | 0.00958 | 7.915 | 0.00846 | 7.814 | 0.00848 | 7.718 | 0.00704 | 7.629 | 0.00700 | 7.555 |
| 0.56105 | 0.01189 | 8.121 | 0.01058 | 8.019 | 0.00976 | 7.915 | 0.00865 | 7.814 | 0.00871 | 7.718 | 0.00722 | 7.629 | 0.00697 | 7.555 |
| 0.56754 | 0.01215 | 8.121 | 0.01088 | 8.019 | 0.01002 | 7.915 | 0.00884 | 7.814 | 0.00894 | 7.718 | 0.00740 | 7.629 | 0.00694 | 7.555 |
| 0.57412 | 0.01242 | 8.121 | 0.01118 | 8.019 | 0.01028 | 7.915 | 0.00904 | 7.814 | 0.00917 | 7.718 | 0.00758 | 7.629 | 0.00701 | 7.555 |
| 0.58076 | 0.01269 | 8.121 | 0.01137 | 8.019 | 0.01055 | 7.915 | 0.00924 | 7.814 | 0.00941 | 7.718 | 0.00777 | 7.628 | 0.00719 | 7.555 |
| 0.58749 | 0.01297 | 8.121 | 0.01153 | 8.019 | 0.01082 | 7.915 | 0.00944 | 7.814 | 0.00965 | 7.718 | 0.00797 | 7.628 | 0.00737 | 7.555 |
| 0.59429 | 0.01327 | 8.121 | 0.01169 | 8.019 | 0.01106 | 7.915 | 0.00966 | 7.814 | 0.00990 | 7.718 | 0.00816 | 7.628 | 0.00755 | 7.555 |
| 0.60117 | 0.01359 | 8.121 | 0.01187 | 8.019 | 0.01131 | 7.915 | 0.00991 | 7.814 | 0.01009 | 7.718 | 0.00836 | 7.628 | 0.00774 | 7.555 |
| 0.60814 | 0.01392 | 8.121 | 0.01224 | 8.019 | 0.01155 | 7.915 | 0.01017 | 7.814 | 0.01023 | 7.718 | 0.00856 | 7.628 | 0.00793 | 7.555 |
| 0.61518 | 0.01425 | 8.121 | 0.01265 | 8.019 | 0.01181 | 7.915 | 0.01043 | 7.814 | 0.01037 | 7.718 | 0.00876 | 7.628 | 0.00813 | 7.554 |
| 0.62230 | 0.01460 | 8.121 | 0.01307 | 8.019 | 0.01215 | 7.915 | 0.01067 | 7.814 | 0.01050 | 7.717 | 0.00896 | 7.628 | 0.00833 | 7.554 |
| 0.62951 | 0.01495 | 8.121 | 0.01346 | 8.019 | 0.01250 | 7.915 | 0.01088 | 7.814 | 0.01071 | 7.717 | 0.00916 | 7.628 | 0.00853 | 7.554 |
| 0.63680 | 0.01531 | 8.121 | 0.01378 | 8.019 | 0.01285 | 7.915 | 0.01106 | 7.814 | 0.01097 | 7.717 | 0.00937 | 7.628 | 0.00872 | 7.554 |
| 0.64417 | 0.01567 | 8.121 | 0.01409 | 8.019 | 0.01316 | 7.914 | 0.01125 | 7.814 | 0.01124 | 7.717 | 0.00958 | 7.628 | 0.00892 | 7.554 |
| 0.65163 | 0.01603 | 8.121 | 0.01440 | 8.018 | 0.01344 | 7.914 | 0.01150 | 7.814 | 0.01151 | 7.717 | 0.00980 | 7.628 | 0.00912 | 7.554 |
| 0.65917 | 0.01639 | 8.121 | 0.01471 | 8.018 | 0.01372 | 7.914 | 0.01184 | 7.814 | 0.01179 | 7.717 | 0.01002 | 7.628 | 0.00933 | 7.554 |
| 0.66681 | 0.01677 | 8.121 | 0.01503 | 8.018 | 0.01401 | 7.914 | 0.01218 | 7.814 | 0.01206 | 7.717 | 0.01026 | 7.628 | 0.00954 | 7.554 |
| 0.67453 | 0.01717 | 8.121 | 0.01535 | 8.018 | 0.01429 | 7.914 | 0.01253 | 7.813 | 0.01235 | 7.717 | 0.01052 | 7.628 | 0.00976 | 7.554 |
| 0.68234 | 0.01758 | 8.121 | 0.01567 | 8.018 | 0.01457 | 7.914 | 0.01285 | 7.814 | 0.01264 | 7.717 | 0.01079 | 7.628 | 0.00998 | 7.554 |
| 0.69024 | 0.01800 | 8.121 | 0.01599 | 8.018 | 0.01485 | 7.914 | 0.01316 | 7.814 | 0.01296 | 7.717 | 0.01107 | 7.628 | 0.01022 | 7.554 |
| 0.69823 | 0.01844 | 8.121 | 0.01631 | 8.018 | 0.01518 | 7.914 | 0.01347 | 7.814 | 0.01328 | 7.717 | 0.01140 | 7.628 | 0.01048 | 7.554 |
| 0.70632 | 0.01891 | 8.121 | 0.01664 | 8.018 | 0.01554 | 7.914 | 0.01378 | 7.814 | 0.01360 | 7.717 | 0.01175 | 7.628 | 0.01074 | 7.554 |
| 0.71450 | 0.01938 | 8.121 | 0.01713 | 8.018 | 0.01591 | 7.914 | 0.01408 | 7.814 | 0.01384 | 7.717 | 0.01211 | 7.628 | 0.01101 | 7.554 |
| 0.72277 | 0.01980 | 8.121 | 0.01767 | 8.018 | 0.01628 | 7.914 | 0.01437 | 7.813 | 0.01403 | 7.717 | 0.01242 | 7.628 | 0.01135 | 7.554 |

| | | | | | | | | | | | | | |
|---|---|---|---|---|---|---|---|---|---|---|---|---|---|
|0.73114|0.02020|8.121|0.01821|8.018|0.01657|7.914|0.01469|7.813|0.01423|7.717|0.01271|7.628|0.01170|7.554|
|0.73961|0.02060|8.121|0.01859|8.018|0.01686|7.914|0.01504|7.813|0.01453|7.717|0.01299|7.628|0.01206|7.554|
|0.74817|0.02109|8.121|0.01892|8.018|0.01717|7.914|0.01541|7.813|0.01490|7.717|0.01332|7.628|0.01237|7.554|
|0.75683|0.02165|8.121|0.01925|8.018|0.01763|7.914|0.01580|7.813|0.01530|7.717|0.01370|7.628|0.01265|7.554|
|0.76560|0.02223|8.121|0.01966|8.018|0.01810|7.914|0.01624|7.813|0.01566|7.717|0.01410|7.628|0.01294|7.554|
|0.77446|0.02273|8.121|0.02010|8.018|0.01858|7.914|0.01671|7.813|0.01600|7.716|0.01451|7.628|0.01325|7.554|
|0.78343|0.02319|8.121|0.02055|8.018|0.01902|7.913|0.01715|7.813|0.01633|7.716|0.01493|7.628|0.01364|7.554|
|0.79250|0.02365|8.121|0.02108|8.018|0.01946|7.913|0.01750|7.813|0.01668|7.716|0.01536|7.627|0.01404|7.554|
|0.80168|0.02409|8.121|0.02165|8.018|0.01990|7.913|0.01777|7.813|0.01708|7.716|0.01573|7.627|0.01444|7.554|
|0.81096|0.02454|8.121|0.02222|8.017|0.02029|7.913|0.01809|7.813|0.01749|7.716|0.01597|7.627|0.01487|7.554|
|0.82035|0.02501|8.121|0.02274|8.017|0.02067|7.913|0.01853|7.813|0.01787|7.716|0.01615|7.627|0.01529|7.554|
|0.82985|0.02558|8.121|0.02325|8.017|0.02107|7.913|0.01906|7.813|0.01818|7.716|0.01639|7.627|0.01573|7.554|
|0.83946|0.02618|8.121|0.02378|8.017|0.02158|7.913|0.01958|7.813|0.01849|7.716|0.01673|7.627|0.01591|7.554|
|0.84918|0.02679|8.121|0.02435|8.017|0.02208|7.913|0.02009|7.812|0.01889|7.715|0.01711|7.627|0.01609|7.554|
|0.85901|0.02741|8.121|0.02494|8.017|0.02258|7.913|0.02058|7.812|0.01937|7.715|0.01751|7.627|0.01628|7.553|
|0.86896|0.02804|8.121|0.02548|8.017|0.02303|7.913|0.02102|7.812|0.01988|7.715|0.01792|7.627|0.01666|7.553|
|0.87902|0.02858|8.121|0.02597|8.017|0.02349|7.913|0.02139|7.812|0.02026|7.715|0.01835|7.627|0.01704|7.553|
|0.88920|0.02908|8.121|0.02646|8.016|0.02406|7.913|0.02175|7.812|0.02059|7.715|0.01864|7.627|0.01743|7.553|
|0.89950|0.02959|8.121|0.02696|8.016|0.02470|7.912|0.02224|7.812|0.02092|7.715|0.01881|7.627|0.01785|7.553|
|0.90991|0.03013|8.120|0.02745|8.016|0.02534|7.912|0.02288|7.812|0.02130|7.714|0.01903|7.627|0.01827|7.553|
|0.92045|0.03068|8.121|0.02796|8.015|0.02592|7.912|0.02349|7.812|0.02170|7.714|0.01957|7.627|0.01858|7.553|
|0.93111|0.03150|8.120|0.02848|8.014|0.02651|7.912|0.02400|7.811|0.02206|7.714|0.02034|7.627|0.01873|7.553|
|0.94189|0.03245|8.120|0.02901|8.014|0.02702|7.911|0.02445|7.811|0.02239|7.713|0.02099|7.626|0.01889|7.553|
|0.95280|0.03341|8.120|0.02994|8.014|0.02747|7.911|0.02490|7.811|0.02272|7.713|0.02140|7.626|0.01948|7.552|
|0.96383|0.03432|8.120|0.03098|8.014|0.02792|7.910|0.02532|7.810|0.02318|7.713|0.02174|7.626|0.02026|7.552|
|0.97499|0.03524|8.120|0.03192|8.014|0.02853|7.910|0.02582|7.810|0.02369|7.713|0.02210|7.626|0.02099|7.552|
|0.98628|0.03616|8.120|0.03273|8.014|0.02915|7.909|0.02652|7.809|0.02422|7.714|0.02250|7.626|0.02131|7.552|
|0.99770|0.03710|8.120|0.03354|8.015|0.02981|7.909|0.02732|7.808|0.02478|7.714|0.02288|7.626|0.02165|7.551|
|1.00925|0.03809|8.120|0.03435|8.015|0.03049|7.908|0.02809|7.807|0.02536|7.715|0.02323|7.626|0.02201|7.551|
|1.02094|0.03916|8.120|0.03518|8.015|0.03118|7.907|0.02883|7.806|0.02601|7.715|0.02357|7.626|0.02241|7.551|
|1.03276|0.04026|8.120|0.03604|8.015|0.03187|7.908|0.02958|7.805|0.02668|7.715|0.02394|7.626|0.02280|7.551|
|1.04472|0.04138|8.120|0.03694|8.015|0.03257|7.908|0.03033|7.804|0.02736|7.715|0.02432|7.626|0.02313|7.550|
|1.05682|0.04250|8.120|0.03786|8.015|0.03337|7.909|0.03108|7.805|0.02805|7.715|0.02471|7.626|0.02347|7.550|
|1.06905|0.04359|8.120|0.03878|8.014|0.03423|7.910|0.03178|7.806|0.02876|7.715|0.02511|7.626|0.02384|7.549|
|1.08143|0.04465|8.120|0.03972|8.014|0.03510|7.910|0.03243|7.807|0.02951|7.715|0.02553|7.626|0.02422|7.548|
|1.09396|0.04571|8.120|0.04064|8.014|0.03598|7.911|0.03308|7.808|0.03027|7.714|0.02603|7.626|0.02461|7.547|
|1.10662|0.04679|8.120|0.04156|8.014|0.03687|7.911|0.03375|7.809|0.03105|7.714|0.02660|7.625|0.02501|7.547|
|1.11944|0.04788|8.120|0.04250|8.014|0.03790|7.911|0.03445|7.809|0.03183|7.714|0.02717|7.625|0.02541|7.546|
|1.13240|0.04900|8.120|0.04344|8.014|0.03895|7.912|0.03524|7.808|0.03258|7.713|0.02775|7.625|0.02592|7.546|
|1.14551|0.05014|8.120|0.04444|8.014|0.04001|7.912|0.03611|7.807|0.03328|7.713|0.02840|7.625|0.02648|7.545|
|1.15878|0.05128|8.120|0.04555|8.013|0.04109|7.913|0.03699|7.806|0.03398|7.713|0.02920|7.625|0.02705|7.545|
|1.17220|0.05244|8.120|0.04670|8.013|0.04218|7.913|0.03787|7.806|0.03468|7.712|0.03007|7.625|0.02763|7.544|
|1.18577|0.05365|8.119|0.04786|8.012|0.04331|7.914|0.03888|7.807|0.03539|7.712|0.03095|7.625|0.02821|7.544|
|1.19950|0.05488|8.119|0.04903|8.013|0.04445|7.914|0.04005|7.808|0.03641|7.712|0.03184|7.625|0.02907|7.544|
|1.21339|0.05612|8.119|0.05027|8.013|0.04560|7.914|0.04125|7.809|0.03757|7.712|0.03261|7.625|0.02994|7.545|
|1.22744|0.05738|8.119|0.05154|8.014|0.04675|7.913|0.04247|7.809|0.03873|7.712|0.03326|7.625|0.03082|7.545|
|1.24165|0.05866|8.119|0.05282|8.014|0.04782|7.913|0.04368|7.809|0.03991|7.712|0.03392|7.625|0.03171|7.546|

| | | | | | | | | | | | | |
|---|---|---|---|---|---|---|---|---|---|---|---|---|
| 1.25603 | 0.05995 | 8.119 | 0.05412 | 8.014 | 0.04889 | 7.912 | 0.04486 | 7.809 | 0.04104 | 7.712 | 0.03458 | 7.625 | 0.03247 | 7.546 |
| 1.27057 | 0.06126 | 8.119 | 0.05534 | 8.014 | 0.04999 | 7.910 | 0.04602 | 7.809 | 0.04212 | 7.712 | 0.03527 | 7.625 | 0.03312 | 7.546 |
| 1.28529 | 0.06259 | 8.119 | 0.05655 | 8.015 | 0.05113 | 7.909 | 0.04720 | 7.809 | 0.04320 | 7.713 | 0.03599 | 7.624 | 0.03378 | 7.545 |
| 1.30017 | 0.06396 | 8.119 | 0.05777 | 8.015 | 0.05256 | 7.907 | 0.04835 | 7.809 | 0.04430 | 7.713 | 0.03672 | 7.624 | 0.03444 | 7.545 |
| 1.31522 | 0.06535 | 8.119 | 0.05902 | 8.015 | 0.05400 | 7.906 | 0.04943 | 7.808 | 0.04543 | 7.713 | 0.03746 | 7.623 | 0.03512 | 7.544 |
| 1.33045 | 0.06675 | 8.119 | 0.06038 | 8.015 | 0.05545 | 7.907 | 0.05049 | 7.807 | 0.04659 | 7.713 | 0.03843 | 7.623 | 0.03584 | 7.544 |
| 1.34586 | 0.06822 | 8.119 | 0.06179 | 8.015 | 0.05684 | 7.907 | 0.05155 | 7.806 | 0.04776 | 7.712 | 0.03973 | 7.623 | 0.03656 | 7.543 |
| 1.36144 | 0.06985 | 8.119 | 0.06321 | 8.015 | 0.05796 | 7.908 | 0.05263 | 7.806 | 0.04895 | 7.712 | 0.04115 | 7.622 | 0.03730 | 7.542 |
| 1.37721 | 0.07153 | 8.119 | 0.06467 | 8.015 | 0.05909 | 7.908 | 0.05374 | 7.807 | 0.05019 | 7.712 | 0.04258 | 7.622 | 0.03816 | 7.542 |
| 1.39316 | 0.07323 | 8.119 | 0.06618 | 8.015 | 0.06023 | 7.908 | 0.05487 | 7.808 | 0.05148 | 7.711 | 0.04384 | 7.622 | 0.03955 | 7.543 |
| 1.40929 | 0.07498 | 8.118 | 0.06772 | 8.015 | 0.06147 | 7.909 | 0.05601 | 7.808 | 0.05280 | 7.711 | 0.04488 | 7.623 | 0.04097 | 7.544 |
| 1.42561 | 0.07676 | 8.118 | 0.06927 | 8.015 | 0.06280 | 7.909 | 0.05720 | 7.808 | 0.05412 | 7.711 | 0.04587 | 7.623 | 0.04240 | 7.545 |
| 1.44212 | 0.07856 | 8.118 | 0.07089 | 8.015 | 0.06415 | 7.909 | 0.05844 | 7.809 | 0.05539 | 7.711 | 0.04691 | 7.623 | 0.04371 | 7.545 |
| 1.45881 | 0.08039 | 8.118 | 0.07255 | 8.015 | 0.06551 | 7.909 | 0.05970 | 7.809 | 0.05665 | 7.711 | 0.04844 | 7.623 | 0.04469 | 7.545 |
| 1.47571 | 0.08223 | 8.118 | 0.07423 | 8.015 | 0.06700 | 7.909 | 0.06100 | 7.809 | 0.05793 | 7.711 | 0.05042 | 7.622 | 0.04568 | 7.544 |
| 1.49279 | 0.08408 | 8.118 | 0.07597 | 8.015 | 0.06851 | 7.910 | 0.06237 | 7.809 | 0.05921 | 7.711 | 0.05242 | 7.622 | 0.04669 | 7.544 |
| 1.51008 | 0.08597 | 8.118 | 0.07781 | 8.015 | 0.07005 | 7.910 | 0.06380 | 7.809 | 0.06049 | 7.711 | 0.05433 | 7.622 | 0.04823 | 7.544 |
| 1.52757 | 0.08796 | 8.118 | 0.07968 | 8.014 | 0.07168 | 7.910 | 0.06526 | 7.809 | 0.06177 | 7.711 | 0.05605 | 7.621 | 0.05020 | 7.544 |
| 1.54525 | 0.09000 | 8.118 | 0.08157 | 8.014 | 0.07342 | 7.910 | 0.06675 | 7.809 | 0.06308 | 7.711 | 0.05767 | 7.621 | 0.05219 | 7.545 |
| 1.56315 | 0.09207 | 8.118 | 0.08341 | 8.014 | 0.07519 | 7.909 | 0.06830 | 7.809 | 0.06451 | 7.711 | 0.05930 | 7.621 | 0.05421 | 7.546 |
| 1.58125 | 0.09418 | 8.118 | 0.08526 | 8.014 | 0.07695 | 7.909 | 0.06988 | 7.809 | 0.06600 | 7.711 | 0.06080 | 7.621 | 0.05581 | 7.543 |
| 1.59956 | 0.09631 | 8.117 | 0.08713 | 8.014 | 0.07869 | 7.909 | 0.07149 | 7.809 | 0.06751 | 7.712 | 0.06219 | 7.621 | 0.05743 | 7.541 |
| 1.61808 | 0.09847 | 8.117 | 0.08909 | 8.014 | 0.08045 | 7.909 | 0.07317 | 7.809 | 0.06902 | 7.711 | 0.06359 | 7.621 | 0.05906 | 7.539 |
| 1.63682 | 0.10070 | 8.117 | 0.09109 | 8.013 | 0.08223 | 7.909 | 0.07489 | 7.809 | 0.07055 | 7.711 | 0.06499 | 7.621 | 0.06055 | 7.538 |
| 1.65577 | 0.10300 | 8.117 | 0.09311 | 8.013 | 0.08416 | 7.909 | 0.07662 | 7.809 | 0.07209 | 7.711 | 0.06637 | 7.621 | 0.06193 | 7.539 |
| 1.67494 | 0.10532 | 8.117 | 0.09523 | 8.013 | 0.08612 | 7.909 | 0.07837 | 7.808 | 0.07369 | 7.711 | 0.06777 | 7.621 | 0.06333 | 7.540 |
| 1.69434 | 0.10763 | 8.117 | 0.09740 | 8.013 | 0.08811 | 7.909 | 0.08014 | 7.808 | 0.07536 | 7.711 | 0.06913 | 7.620 | 0.06473 | 7.541 |
| 1.71396 | 0.10995 | 8.117 | 0.09960 | 8.012 | 0.09007 | 7.908 | 0.08193 | 7.808 | 0.07706 | 7.711 | 0.07041 | 7.620 | 0.06610 | 7.542 |
| 1.73380 | 0.11229 | 8.117 | 0.10180 | 8.012 | 0.09206 | 7.908 | 0.08376 | 7.808 | 0.07873 | 7.711 | 0.07167 | 7.620 | 0.06749 | 7.542 |
| 1.75388 | 0.11448 | 8.117 | 0.10403 | 8.012 | 0.09407 | 7.908 | 0.08562 | 7.808 | 0.08036 | 7.710 | 0.07320 | 7.620 | 0.06888 | 7.543 |
| 1.77419 | 0.11663 | 8.116 | 0.10627 | 8.012 | 0.09618 | 7.908 | 0.08751 | 7.808 | 0.08199 | 7.710 | 0.07521 | 7.620 | 0.07012 | 7.540 |
| 1.79473 | 0.11883 | 8.116 | 0.10840 | 8.012 | 0.09832 | 7.907 | 0.08943 | 7.807 | 0.08371 | 7.710 | 0.07747 | 7.620 | 0.07137 | 7.538 |
| 1.81552 | 0.12135 | 8.116 | 0.11054 | 8.012 | 0.10050 | 7.907 | 0.09137 | 7.807 | 0.08551 | 7.710 | 0.07945 | 7.620 | 0.07268 | 7.536 |
| 1.83654 | 0.12399 | 8.116 | 0.11269 | 8.011 | 0.10279 | 7.907 | 0.09336 | 7.806 | 0.08735 | 7.710 | 0.08096 | 7.620 | 0.07490 | 7.536 |
| 1.85780 | 0.12664 | 8.116 | 0.11485 | 8.011 | 0.10511 | 7.906 | 0.09539 | 7.806 | 0.08900 | 7.710 | 0.08227 | 7.620 | 0.07714 | 7.537 |
| 1.87932 | 0.12931 | 8.115 | 0.11703 | 8.011 | 0.10743 | 7.905 | 0.09745 | 7.805 | 0.09050 | 7.710 | 0.08372 | 7.620 | 0.07934 | 7.537 |
| 1.90108 | 0.13200 | 8.115 | 0.11953 | 8.010 | 0.10973 | 7.904 | 0.09946 | 7.804 | 0.09198 | 7.711 | 0.08536 | 7.620 | 0.08063 | 7.538 |
| 1.92309 | 0.13530 | 8.114 | 0.12234 | 8.010 | 0.11205 | 7.903 | 0.10138 | 7.803 | 0.09388 | 7.711 | 0.08707 | 7.620 | 0.08194 | 7.538 |
| 1.94536 | 0.13892 | 8.114 | 0.12523 | 8.010 | 0.11400 | 7.903 | 0.10334 | 7.802 | 0.09600 | 7.711 | 0.08868 | 7.620 | 0.08333 | 7.539 |
| 1.96789 | 0.14282 | 8.113 | 0.12908 | 8.010 | 0.11571 | 7.906 | 0.10584 | 7.801 | 0.09808 | 7.711 | 0.09019 | 7.620 | 0.08501 | 7.539 |
| 1.99067 | 0.14725 | 8.113 | 0.13319 | 8.010 | 0.11755 | 7.910 | 0.10888 | 7.800 | 0.09993 | 7.711 | 0.09178 | 7.619 | 0.08671 | 7.539 |
| 2.01372 | 0.15185 | 8.113 | 0.13692 | 8.010 | 0.12048 | 7.912 | 0.11186 | 7.801 | 0.10171 | 7.712 | 0.09365 | 7.619 | 0.08832 | 7.539 |
| 2.03704 | 0.15595 | 8.113 | 0.13998 | 8.011 | 0.12348 | 7.912 | 0.11469 | 7.803 | 0.10379 | 7.713 | 0.09573 | 7.619 | 0.08982 | 7.539 |
| 2.06063 | 0.15982 | 8.113 | 0.14301 | 8.010 | 0.12670 | 7.913 | 0.11749 | 7.804 | 0.10615 | 7.713 | 0.09758 | 7.618 | 0.09134 | 7.538 |
| 2.08449 | 0.16379 | 8.113 | 0.14659 | 8.009 | 0.13008 | 7.914 | 0.12051 | 7.803 | 0.10869 | 7.713 | 0.09907 | 7.618 | 0.09326 | 7.538 |
| 2.10863 | 0.16793 | 8.113 | 0.15033 | 8.009 | 0.13356 | 7.916 | 0.12376 | 7.803 | 0.11193 | 7.713 | 0.10057 | 7.618 | 0.09533 | 7.537 |
| 2.13304 | 0.17214 | 8.113 | 0.15435 | 8.009 | 0.13764 | 7.915 | 0.12701 | 7.804 | 0.11547 | 7.712 | 0.10317 | 7.619 | 0.09730 | 7.537 |

| | | | | | | | | | | | | | |
|---|---|---|---|---|---|---|---|---|---|---|---|---|---|
| 2.15774 | 0.17639 | 8.112 | 0.15863 | 8.009 | 0.14179 | 7.913 | 0.13023 | 7.804 | 0.11915 | 7.712 | 0.10671 | 7.619 | 0.09866 | 7.535 |
| 2.18273 | 0.18070 | 8.112 | 0.16300 | 8.009 | 0.14576 | 7.912 | 0.13347 | 7.804 | 0.12296 | 7.712 | 0.10994 | 7.618 | 0.10004 | 7.534 |
| 2.20800 | 0.18508 | 8.112 | 0.16741 | 8.009 | 0.14967 | 7.910 | 0.13678 | 7.804 | 0.12668 | 7.711 | 0.11266 | 7.617 | 0.10274 | 7.534 |
| 2.23357 | 0.18953 | 8.112 | 0.17188 | 8.009 | 0.15362 | 7.908 | 0.14016 | 7.803 | 0.12993 | 7.711 | 0.11524 | 7.617 | 0.10626 | 7.535 |
| 2.25944 | 0.19404 | 8.112 | 0.17621 | 8.009 | 0.15759 | 7.908 | 0.14357 | 7.803 | 0.13305 | 7.712 | 0.11790 | 7.617 | 0.10964 | 7.535 |
| 2.28560 | 0.19860 | 8.112 | 0.18041 | 8.010 | 0.16157 | 7.908 | 0.14703 | 7.802 | 0.13620 | 7.712 | 0.12063 | 7.617 | 0.11218 | 7.535 |
| 2.31206 | 0.20321 | 8.112 | 0.18464 | 8.010 | 0.16540 | 7.909 | 0.15051 | 7.803 | 0.13939 | 7.712 | 0.12338 | 7.617 | 0.11475 | 7.535 |
| 2.33884 | 0.20778 | 8.112 | 0.18892 | 8.010 | 0.16917 | 7.910 | 0.15414 | 7.804 | 0.14276 | 7.712 | 0.12617 | 7.616 | 0.11741 | 7.535 |
| 2.36592 | 0.21235 | 8.112 | 0.19326 | 8.010 | 0.17298 | 7.911 | 0.15794 | 7.806 | 0.14660 | 7.712 | 0.12910 | 7.616 | 0.12013 | 7.535 |
| 2.39332 | 0.21695 | 8.112 | 0.19754 | 8.010 | 0.17687 | 7.910 | 0.16178 | 7.808 | 0.15063 | 7.712 | 0.13247 | 7.616 | 0.12287 | 7.535 |
| 2.42103 | 0.22161 | 8.111 | 0.20183 | 8.010 | 0.18081 | 7.909 | 0.16568 | 7.809 | 0.15470 | 7.712 | 0.13616 | 7.617 | 0.12565 | 7.534 |
| 2.44906 | 0.22638 | 8.111 | 0.20616 | 8.009 | 0.18534 | 7.909 | 0.16964 | 7.809 | 0.15883 | 7.712 | 0.13989 | 7.617 | 0.12846 | 7.534 |
| 2.47742 | 0.23139 | 8.111 | 0.21055 | 8.009 | 0.18995 | 7.908 | 0.17366 | 7.807 | 0.16299 | 7.712 | 0.14367 | 7.617 | 0.13191 | 7.535 |
| 2.50611 | 0.23652 | 8.110 | 0.21504 | 8.009 | 0.19464 | 7.907 | 0.17771 | 7.806 | 0.16720 | 7.712 | 0.14750 | 7.617 | 0.13559 | 7.535 |
| 2.53513 | 0.24170 | 8.110 | 0.21970 | 8.009 | 0.19939 | 7.905 | 0.18180 | 7.804 | 0.17145 | 7.712 | 0.15138 | 7.617 | 0.13931 | 7.536 |
| 2.56448 | 0.24693 | 8.109 | 0.22443 | 8.008 | 0.20412 | 7.903 | 0.18594 | 7.803 | 0.17575 | 7.712 | 0.15532 | 7.617 | 0.14307 | 7.536 |
| 2.59418 | 0.25183 | 8.109 | 0.22921 | 8.008 | 0.20881 | 7.901 | 0.19016 | 7.802 | 0.18009 | 7.713 | 0.15930 | 7.617 | 0.14688 | 7.537 |
| 2.62422 | 0.25664 | 8.109 | 0.23403 | 8.008 | 0.21354 | 7.900 | 0.19451 | 7.801 | 0.18440 | 7.713 | 0.16329 | 7.616 | 0.15075 | 7.537 |
| 2.65461 | 0.26151 | 8.109 | 0.23873 | 8.009 | 0.21828 | 7.900 | 0.19891 | 7.800 | 0.18873 | 7.713 | 0.16636 | 7.616 | 0.15467 | 7.538 |
| 2.68534 | 0.26645 | 8.108 | 0.24344 | 8.009 | 0.22297 | 7.900 | 0.20336 | 7.800 | 0.19311 | 7.714 | 0.16860 | 7.615 | 0.15863 | 7.539 |
| 2.71644 | 0.27283 | 8.108 | 0.24822 | 8.009 | 0.22734 | 7.900 | 0.20771 | 7.800 | 0.19754 | 7.714 | 0.17086 | 7.615 | 0.16264 | 7.540 |
| 2.74789 | 0.27989 | 8.108 | 0.25308 | 8.009 | 0.23175 | 7.900 | 0.21186 | 7.800 | 0.20171 | 7.714 | 0.17314 | 7.614 | 0.16567 | 7.540 |
| 2.77971 | 0.28708 | 8.108 | 0.25849 | 8.008 | 0.23623 | 7.900 | 0.21596 | 7.800 | 0.20574 | 7.714 | 0.17639 | 7.614 | 0.16790 | 7.540 |
| 2.81190 | 0.29438 | 8.108 | 0.26406 | 8.008 | 0.24076 | 7.900 | 0.22011 | 7.800 | 0.20977 | 7.714 | 0.18083 | 7.615 | 0.17016 | 7.539 |
| 2.84446 | 0.30242 | 8.108 | 0.26970 | 8.008 | 0.24532 | 7.900 | 0.22422 | 7.799 | 0.21385 | 7.714 | 0.18558 | 7.615 | 0.17243 | 7.539 |
| 2.87740 | 0.31079 | 8.108 | 0.27572 | 8.008 | 0.24992 | 7.900 | 0.22820 | 7.799 | 0.21855 | 7.714 | 0.19039 | 7.615 | 0.17540 | 7.539 |
| 2.91072 | 0.31926 | 8.108 | 0.28447 | 8.008 | 0.25452 | 7.902 | 0.23214 | 7.799 | 0.22377 | 7.713 | 0.19491 | 7.615 | 0.18008 | 7.540 |
| 2.94442 | 0.32752 | 8.108 | 0.29374 | 8.009 | 0.25940 | 7.904 | 0.23611 | 7.799 | 0.22914 | 7.713 | 0.19901 | 7.615 | 0.18481 | 7.541 |
| 2.97852 | 0.33506 | 8.107 | 0.30314 | 8.009 | 0.26507 | 7.906 | 0.24045 | 7.799 | 0.23458 | 7.713 | 0.20302 | 7.615 | 0.18960 | 7.542 |
| 3.01301 | 0.34246 | 8.107 | 0.31170 | 8.008 | 0.27084 | 7.906 | 0.24541 | 7.801 | 0.24073 | 7.713 | 0.20708 | 7.615 | 0.19424 | 7.542 |
| 3.04789 | 0.34994 | 8.106 | 0.31868 | 8.008 | 0.27673 | 7.906 | 0.25064 | 7.802 | 0.24735 | 7.712 | 0.21211 | 7.614 | 0.19819 | 7.541 |
| 3.08319 | 0.35771 | 8.106 | 0.32550 | 8.007 | 0.28385 | 7.905 | 0.25592 | 7.803 | 0.25411 | 7.712 | 0.21829 | 7.614 | 0.20218 | 7.539 |
| 3.11889 | 0.36572 | 8.106 | 0.33237 | 8.007 | 0.29211 | 7.904 | 0.26273 | 7.804 | 0.26074 | 7.712 | 0.22472 | 7.614 | 0.20622 | 7.538 |
| 3.15500 | 0.37385 | 8.105 | 0.33960 | 8.007 | 0.30048 | 7.904 | 0.27143 | 7.805 | 0.26547 | 7.712 | 0.23101 | 7.614 | 0.21102 | 7.538 |
| 3.19154 | 0.38253 | 8.105 | 0.34702 | 8.006 | 0.30890 | 7.903 | 0.28034 | 7.806 | 0.26950 | 7.711 | 0.23530 | 7.615 | 0.21736 | 7.539 |
| 3.22849 | 0.39245 | 8.104 | 0.35455 | 8.006 | 0.31611 | 7.902 | 0.28885 | 7.806 | 0.27358 | 7.711 | 0.23790 | 7.615 | 0.22377 | 7.540 |
| 3.26588 | 0.40280 | 8.104 | 0.36308 | 8.006 | 0.32329 | 7.902 | 0.29601 | 7.806 | 0.27838 | 7.711 | 0.24054 | 7.616 | 0.23025 | 7.541 |
| 3.30370 | 0.41327 | 8.103 | 0.37325 | 8.005 | 0.33053 | 7.903 | 0.30223 | 7.805 | 0.28408 | 7.711 | 0.24489 | 7.616 | 0.23438 | 7.542 |
| 3.34195 | 0.42369 | 8.103 | 0.38375 | 8.005 | 0.33747 | 7.903 | 0.30850 | 7.805 | 0.29007 | 7.711 | 0.25232 | 7.615 | 0.23697 | 7.544 |
| 3.38065 | 0.43419 | 8.103 | 0.39430 | 8.004 | 0.34414 | 7.903 | 0.31485 | 7.806 | 0.29639 | 7.711 | 0.26121 | 7.615 | 0.23959 | 7.546 |
| 3.41979 | 0.44479 | 8.102 | 0.40437 | 8.004 | 0.35087 | 7.904 | 0.32133 | 7.807 | 0.30496 | 7.710 | 0.27010 | 7.615 | 0.24248 | 7.547 |
| 3.45939 | 0.45461 | 8.102 | 0.41447 | 8.004 | 0.35846 | 7.905 | 0.32790 | 7.807 | 0.31445 | 7.710 | 0.27822 | 7.615 | 0.25123 | 7.547 |
| 3.49945 | 0.46405 | 8.101 | 0.42470 | 8.003 | 0.36872 | 7.906 | 0.33570 | 7.807 | 0.32406 | 7.709 | 0.28570 | 7.616 | 0.26008 | 7.547 |
| 3.53997 | 0.47356 | 8.100 | 0.43353 | 8.002 | 0.37916 | 7.905 | 0.34554 | 7.808 | 0.33287 | 7.709 | 0.29328 | 7.616 | 0.26904 | 7.547 |
| 3.58096 | 0.48316 | 8.100 | 0.44210 | 8.001 | 0.38975 | 7.904 | 0.35619 | 7.808 | 0.34121 | 7.709 | 0.30001 | 7.615 | 0.27706 | 7.547 |
| 3.62243 | 0.49288 | 8.099 | 0.45073 | 8.001 | 0.40064 | 7.904 | 0.36675 | 7.807 | 0.34957 | 7.709 | 0.30571 | 7.615 | 0.28451 | 7.547 |
| 3.66438 | 0.50271 | 8.099 | 0.45947 | 8.001 | 0.41169 | 7.902 | 0.37691 | 7.807 | 0.35578 | 7.708 | 0.31127 | 7.614 | 0.29205 | 7.548 |

| | | | | | | | | | | | | | |
|---|---|---|---|---|---|---|---|---|---|---|---|---|---|
| 3.70681 | 0.51341 | 8.098 | 0.46833 | 8.000 | 0.42291 | 7.901 | 0.38689 | 7.807 | 0.35986 | 7.707 | 0.31756 | 7.614 | 0.29897 | 7.547 |
| 3.74973 | 0.52465 | 8.098 | 0.47730 | 8.000 | 0.43502 | 7.900 | 0.39732 | 7.806 | 0.36349 | 7.706 | 0.32490 | 7.613 | 0.30444 | 7.545 |
| 3.79315 | 0.53610 | 8.097 | 0.48751 | 8.000 | 0.44733 | 7.898 | 0.40877 | 7.806 | 0.37137 | 7.706 | 0.33266 | 7.612 | 0.30998 | 7.542 |
| 3.83707 | 0.54870 | 8.096 | 0.49814 | 7.999 | 0.45946 | 7.897 | 0.42095 | 7.805 | 0.38418 | 7.706 | 0.34148 | 7.612 | 0.31585 | 7.541 |
| 3.88150 | 0.56182 | 8.096 | 0.50897 | 7.999 | 0.46753 | 7.895 | 0.43181 | 7.803 | 0.39841 | 7.707 | 0.35211 | 7.613 | 0.32351 | 7.542 |
| 3.92645 | 0.57524 | 8.095 | 0.52066 | 7.998 | 0.47547 | 7.895 | 0.43920 | 7.800 | 0.40924 | 7.707 | 0.36359 | 7.613 | 0.33125 | 7.543 |
| 3.97192 | 0.58929 | 8.095 | 0.53263 | 7.998 | 0.48411 | 7.894 | 0.44448 | 7.797 | 0.41651 | 7.706 | 0.37325 | 7.613 | 0.33934 | 7.544 |
| 4.01791 | 0.60366 | 8.094 | 0.54492 | 7.998 | 0.49579 | 7.893 | 0.45143 | 7.796 | 0.42297 | 7.705 | 0.37991 | 7.612 | 0.35063 | 7.543 |
| 4.06443 | 0.61842 | 8.093 | 0.55787 | 7.998 | 0.50765 | 7.892 | 0.46156 | 7.795 | 0.43144 | 7.705 | 0.38546 | 7.611 | 0.36205 | 7.542 |
| 4.11150 | 0.63369 | 8.093 | 0.57107 | 7.997 | 0.51929 | 7.890 | 0.47319 | 7.795 | 0.44141 | 7.705 | 0.39210 | 7.610 | 0.37287 | 7.540 |
| 4.15911 | 0.64921 | 8.092 | 0.58537 | 7.997 | 0.53050 | 7.889 | 0.48421 | 7.793 | 0.45176 | 7.705 | 0.40010 | 7.610 | 0.37834 | 7.539 |
| 4.20727 | 0.66439 | 8.091 | 0.60067 | 7.996 | 0.54181 | 7.889 | 0.49427 | 7.791 | 0.46082 | 7.705 | 0.40847 | 7.609 | 0.38386 | 7.537 |
| 4.25598 | 0.67948 | 8.090 | 0.61626 | 7.995 | 0.55149 | 7.888 | 0.50404 | 7.791 | 0.46941 | 7.704 | 0.41580 | 7.610 | 0.39012 | 7.536 |
| 4.30527 | 0.69508 | 8.090 | 0.63113 | 7.994 | 0.56051 | 7.889 | 0.51284 | 7.791 | 0.48022 | 7.704 | 0.42216 | 7.610 | 0.39840 | 7.535 |
| 4.35512 | 0.71157 | 8.089 | 0.64597 | 7.993 | 0.57001 | 7.890 | 0.52075 | 7.792 | 0.49715 | 7.703 | 0.42899 | 7.610 | 0.40677 | 7.534 |
| 4.40555 | 0.72845 | 8.088 | 0.66118 | 7.993 | 0.58255 | 7.891 | 0.52978 | 7.793 | 0.51631 | 7.703 | 0.43729 | 7.609 | 0.41415 | 7.535 |
| 4.45656 | 0.74604 | 8.087 | 0.67690 | 7.993 | 0.59528 | 7.891 | 0.54103 | 7.794 | 0.52916 | 7.702 | 0.44660 | 7.609 | 0.42045 | 7.537 |
| 4.50817 | 0.76406 | 8.087 | 0.69293 | 7.992 | 0.60879 | 7.891 | 0.55348 | 7.794 | 0.53662 | 7.702 | 0.45631 | 7.609 | 0.42682 | 7.540 |
| 4.56037 | 0.78202 | 8.086 | 0.70958 | 7.991 | 0.62291 | 7.891 | 0.56594 | 7.795 | 0.54340 | 7.701 | 0.46656 | 7.609 | 0.43546 | 7.538 |
| 4.61318 | 0.79993 | 8.085 | 0.72649 | 7.990 | 0.63727 | 7.892 | 0.57837 | 7.796 | 0.55251 | 7.700 | 0.47758 | 7.609 | 0.44475 | 7.535 |
| 4.66659 | 0.81797 | 8.084 | 0.74347 | 7.989 | 0.65229 | 7.892 | 0.59169 | 7.797 | 0.56254 | 7.700 | 0.49166 | 7.607 | 0.45430 | 7.533 |
| 4.72063 | 0.83619 | 8.083 | 0.76053 | 7.988 | 0.66750 | 7.892 | 0.60672 | 7.799 | 0.57496 | 7.700 | 0.50816 | 7.607 | 0.46463 | 7.535 |
| 4.77529 | 0.85463 | 8.082 | 0.77777 | 7.987 | 0.68684 | 7.893 | 0.62324 | 7.799 | 0.58938 | 7.701 | 0.52247 | 7.607 | 0.47508 | 7.538 |
| 4.83059 | 0.87345 | 8.082 | 0.79521 | 7.986 | 0.70781 | 7.893 | 0.64314 | 7.798 | 0.60438 | 7.701 | 0.53382 | 7.607 | 0.48947 | 7.538 |
| 4.88652 | 0.89272 | 8.081 | 0.81286 | 7.985 | 0.72902 | 7.893 | 0.66622 | 7.797 | 0.61965 | 7.700 | 0.54438 | 7.607 | 0.50595 | 7.537 |
| 4.94311 | 0.91225 | 8.080 | 0.83041 | 7.985 | 0.75074 | 7.891 | 0.68957 | 7.796 | 0.63512 | 7.700 | 0.55490 | 7.607 | 0.52126 | 7.536 |
| 5.00035 | 0.93204 | 8.079 | 0.84790 | 7.984 | 0.77281 | 7.888 | 0.71319 | 7.795 | 0.65077 | 7.699 | 0.56544 | 7.606 | 0.53165 | 7.536 |
| 5.05825 | 0.95202 | 8.078 | 0.86557 | 7.983 | 0.79107 | 7.884 | 0.73614 | 7.794 | 0.66660 | 7.698 | 0.57610 | 7.605 | 0.54216 | 7.536 |
| 5.11682 | 0.97449 | 8.076 | 0.88344 | 7.982 | 0.80802 | 7.881 | 0.75292 | 7.792 | 0.68255 | 7.697 | 0.58687 | 7.605 | 0.55258 | 7.536 |
| 5.17607 | 0.99856 | 8.075 | 0.90147 | 7.981 | 0.82516 | 7.877 | 0.76368 | 7.790 | 0.69853 | 7.697 | 0.59870 | 7.605 | 0.56308 | 7.535 |
| 5.23600 | 1.02311 | 8.074 | 0.92343 | 7.981 | 0.84214 | 7.877 | 0.77457 | 7.789 | 0.71463 | 7.696 | 0.61365 | 7.605 | 0.57369 | 7.535 |
| 5.29663 | 1.04794 | 8.073 | 0.94683 | 7.981 | 0.85924 | 7.877 | 0.78555 | 7.787 | 0.73093 | 7.695 | 0.63077 | 7.605 | 0.58442 | 7.535 |
| 5.35797 | 1.07320 | 8.072 | 0.97059 | 7.980 | 0.87556 | 7.877 | 0.79741 | 7.786 | 0.74741 | 7.694 | 0.64810 | 7.605 | 0.59529 | 7.535 |
| 5.42001 | 1.09856 | 8.076 | 0.99468 | 7.979 | 0.89202 | 7.877 | 0.81141 | 7.786 | 0.76515 | 7.693 | 0.66564 | 7.604 | 0.61106 | 7.534 |
| 5.48277 | 1.12483 | 8.076 | 1.01929 | 7.978 | 0.90865 | 7.877 | 0.82686 | 7.785 | 0.78433 | 7.693 | 0.68320 | 7.604 | 0.62811 | 7.533 |
| 5.54626 | 1.15164 | 8.074 | 1.04446 | 7.976 | 0.92545 | 7.878 | 0.84249 | 7.784 | 0.80404 | 7.692 | 0.70062 | 7.603 | 0.64537 | 7.532 |
| 5.61048 | 1.17871 | 8.073 | 1.06996 | 7.975 | 0.94525 | 7.878 | 0.85826 | 7.784 | 0.82398 | 7.692 | 0.71809 | 7.602 | 0.66282 | 7.531 |
| 5.67545 | 1.20514 | 8.071 | 1.09576 | 7.973 | 0.96802 | 7.878 | 0.87752 | 7.785 | 0.84407 | 7.691 | 0.73576 | 7.601 | 0.68042 | 7.530 |
| 5.74116 | 1.23155 | 8.070 | 1.12170 | 7.972 | 0.99109 | 7.878 | 0.90120 | 7.785 | 0.86340 | 7.691 | 0.75373 | 7.600 | 0.69762 | 7.530 |
| 5.80764 | 1.25826 | 8.068 | 1.14670 | 7.970 | 1.01456 | 7.878 | 0.92544 | 7.785 | 0.88257 | 7.690 | 0.77321 | 7.600 | 0.71501 | 7.530 |
| 5.87489 | 1.28532 | 8.066 | 1.17179 | 7.968 | 1.03913 | 7.877 | 0.95004 | 7.785 | 0.90196 | 7.690 | 0.79408 | 7.600 | 0.73261 | 7.531 |
| 5.94292 | 1.31423 | 8.065 | 1.19716 | 7.967 | 1.06662 | 7.876 | 0.97541 | 7.784 | 0.92158 | 7.689 | 0.81519 | 7.599 | 0.75041 | 7.531 |
| 6.01174 | 1.34408 | 8.064 | 1.22280 | 7.966 | 1.09445 | 7.875 | 1.00174 | 7.783 | 0.94072 | 7.689 | 0.83652 | 7.599 | 0.76992 | 7.531 |
| 6.08135 | 1.37432 | 8.063 | 1.24998 | 7.965 | 1.12276 | 7.873 | 1.02859 | 7.781 | 0.95966 | 7.688 | 0.85685 | 7.599 | 0.79070 | 7.530 |
| 6.15177 | 1.40508 | 8.062 | 1.27772 | 7.964 | 1.15114 | 7.871 | 1.05578 | 7.779 | 0.97875 | 7.688 | 0.87595 | 7.598 | 0.81172 | 7.529 |
| 6.22300 | 1.43833 | 8.061 | 1.30580 | 7.963 | 1.17862 | 7.868 | 1.08211 | 7.777 | 0.99806 | 7.687 | 0.89501 | 7.598 | 0.83298 | 7.528 |
| 6.29506 | 1.47264 | 8.060 | 1.33467 | 7.962 | 1.20636 | 7.867 | 1.10654 | 7.775 | 1.02118 | 7.687 | 0.91431 | 7.598 | 0.85353 | 7.528 |

| | | | | | | | | | | | | | |
|---|---|---|---|---|---|---|---|---|---|---|---|---|---|
| 6.36796 | 1.50736 | 8.059 | 1.36656 | 7.962 | 1.23431 | 7.865 | 1.13027 | 7.773 | 1.04720 | 7.687 | 0.93993 | 7.597 | 0.87230 | 7.528 |
| 6.44169 | 1.54264 | 8.057 | 1.39923 | 7.961 | 1.26195 | 7.864 | 1.15419 | 7.771 | 1.07400 | 7.687 | 0.97363 | 7.597 | 0.89128 | 7.528 |
| 6.51628 | 1.57870 | 8.055 | 1.43234 | 7.960 | 1.28833 | 7.863 | 1.17813 | 7.770 | 1.10112 | 7.686 | 1.00966 | 7.597 | 0.91048 | 7.528 |
| 6.59174 | 1.61527 | 8.054 | 1.46632 | 7.958 | 1.31497 | 7.862 | 1.20213 | 7.770 | 1.12951 | 7.685 | 1.04609 | 7.597 | 0.93390 | 7.528 |
| 6.66807 | 1.65227 | 8.052 | 1.50132 | 7.956 | 1.34185 | 7.861 | 1.22637 | 7.769 | 1.15877 | 7.684 | 1.07370 | 7.596 | 0.96937 | 7.527 |
| 6.74528 | 1.69001 | 8.050 | 1.53680 | 7.954 | 1.37141 | 7.861 | 1.25089 | 7.769 | 1.18844 | 7.683 | 1.09144 | 7.596 | 1.00525 | 7.527 |
| 6.82339 | 1.72839 | 8.048 | 1.57271 | 7.952 | 1.40314 | 7.860 | 1.27839 | 7.769 | 1.21810 | 7.682 | 1.10800 | 7.595 | 1.04154 | 7.526 |
| 6.90240 | 1.76724 | 8.046 | 1.60883 | 7.951 | 1.43534 | 7.859 | 1.30943 | 7.769 | 1.24585 | 7.681 | 1.12476 | 7.594 | 1.07064 | 7.525 |
| 6.98232 | 1.80701 | 8.044 | 1.64532 | 7.949 | 1.46805 | 7.858 | 1.34090 | 7.768 | 1.27309 | 7.680 | 1.14163 | 7.593 | 1.08693 | 7.523 |
| 7.06318 | 1.84822 | 8.042 | 1.68219 | 7.947 | 1.50346 | 7.857 | 1.37305 | 7.768 | 1.30065 | 7.679 | 1.15861 | 7.592 | 1.10341 | 7.521 |
| 7.14496 | 1.89018 | 8.041 | 1.71987 | 7.946 | 1.53946 | 7.855 | 1.40625 | 7.767 | 1.32831 | 7.677 | 1.17579 | 7.591 | 1.12008 | 7.519 |
| 7.22770 | 1.93273 | 8.039 | 1.75859 | 7.944 | 1.57591 | 7.854 | 1.44025 | 7.766 | 1.35605 | 7.676 | 1.19557 | 7.590 | 1.13684 | 7.519 |
| 7.31139 | 1.97717 | 8.037 | 1.79785 | 7.942 | 1.61318 | 7.852 | 1.47473 | 7.764 | 1.38405 | 7.675 | 1.21962 | 7.589 | 1.15375 | 7.519 |
| 7.39605 | 2.02258 | 8.035 | 1.83785 | 7.941 | 1.65119 | 7.850 | 1.51002 | 7.762 | 1.41267 | 7.674 | 1.24562 | 7.588 | 1.17085 | 7.519 |
| 7.48170 | 2.06853 | 8.033 | 1.88029 | 7.940 | 1.68965 | 7.849 | 1.54610 | 7.760 | 1.44345 | 7.673 | 1.27257 | 7.587 | 1.18881 | 7.519 |
| 7.56833 | 2.11636 | 8.031 | 1.92352 | 7.938 | 1.72844 | 7.847 | 1.58257 | 7.759 | 1.47527 | 7.671 | 1.30424 | 7.586 | 1.21440 | 7.518 |
| 7.65597 | 2.16537 | 8.029 | 1.96741 | 7.936 | 1.76737 | 7.845 | 1.61892 | 7.757 | 1.50745 | 7.669 | 1.33984 | 7.585 | 1.24029 | 7.517 |
| 7.74462 | 2.21501 | 8.027 | 2.01344 | 7.934 | 1.80669 | 7.844 | 1.65488 | 7.755 | 1.54548 | 7.668 | 1.37587 | 7.584 | 1.26648 | 7.516 |
| 7.83430 | 2.26565 | 8.024 | 2.06037 | 7.932 | 1.84665 | 7.842 | 1.69096 | 7.754 | 1.58706 | 7.667 | 1.41386 | 7.583 | 1.29859 | 7.515 |
| 7.92501 | 2.31711 | 8.022 | 2.10787 | 7.930 | 1.88868 | 7.841 | 1.72804 | 7.753 | 1.62951 | 7.666 | 1.45399 | 7.582 | 1.33403 | 7.515 |
| 8.01678 | 2.36920 | 8.020 | 2.15607 | 7.928 | 1.93121 | 7.839 | 1.76688 | 7.752 | 1.67072 | 7.665 | 1.49479 | 7.581 | 1.36989 | 7.514 |
| 8.10961 | 2.42303 | 8.017 | 2.20486 | 7.926 | 1.97443 | 7.838 | 1.80689 | 7.751 | 1.71090 | 7.663 | 1.53419 | 7.580 | 1.40742 | 7.514 |
| 8.20352 | 2.47805 | 8.015 | 2.25422 | 7.923 | 2.02066 | 7.836 | 1.84808 | 7.750 | 1.75121 | 7.661 | 1.57143 | 7.580 | 1.44762 | 7.514 |
| 8.29851 | 2.53382 | 8.012 | 2.30544 | 7.921 | 2.06768 | 7.834 | 1.89134 | 7.748 | 1.78964 | 7.660 | 1.60837 | 7.579 | 1.48829 | 7.514 |
| 8.39460 | 2.59164 | 8.010 | 2.35757 | 7.919 | 2.11537 | 7.832 | 1.93605 | 7.747 | 1.82609 | 7.658 | 1.64335 | 7.577 | 1.52846 | 7.514 |
| 8.49180 | 2.65063 | 8.007 | 2.41050 | 7.917 | 2.16486 | 7.830 | 1.98157 | 7.745 | 1.86236 | 7.657 | 1.67483 | 7.575 | 1.56474 | 7.512 |
| 8.59014 | 2.71050 | 8.005 | 2.46584 | 7.915 | 2.21491 | 7.828 | 2.02812 | 7.743 | 1.90191 | 7.656 | 1.70513 | 7.573 | 1.60143 | 7.509 |
| 8.68960 | 2.77173 | 8.002 | 2.52217 | 7.912 | 2.26566 | 7.826 | 2.07543 | 7.741 | 1.94459 | 7.655 | 1.73861 | 7.571 | 1.63764 | 7.507 |
| 8.79023 | 2.83387 | 7.999 | 2.57986 | 7.910 | 2.31745 | 7.824 | 2.12321 | 7.740 | 1.98836 | 7.655 | 1.77674 | 7.570 | 1.66751 | 7.504 |
| 8.89201 | 2.89772 | 7.996 | 2.63975 | 7.907 | 2.36986 | 7.822 | 2.17145 | 7.738 | 2.03555 | 7.654 | 1.81678 | 7.569 | 1.69771 | 7.502 |
| 8.99498 | 2.96346 | 7.993 | 2.70054 | 7.905 | 2.42372 | 7.820 | 2.22022 | 7.736 | 2.08520 | 7.653 | 1.85931 | 7.567 | 1.72962 | 7.500 |
| 9.09913 | 3.03026 | 7.990 | 2.76223 | 7.902 | 2.47926 | 7.818 | 2.27040 | 7.734 | 2.13574 | 7.652 | 1.90474 | 7.566 | 1.76902 | 7.500 |
| 9.20450 | 3.09913 | 7.987 | 2.82477 | 7.899 | 2.53544 | 7.816 | 2.32226 | 7.732 | 2.18738 | 7.650 | 1.95118 | 7.564 | 1.80887 | 7.499 |
| 9.31108 | 3.16932 | 7.984 | 2.88822 | 7.897 | 2.59379 | 7.814 | 2.37503 | 7.730 | 2.23982 | 7.647 | 1.99839 | 7.563 | 1.85065 | 7.499 |
| 9.41890 | 3.24092 | 7.981 | 2.95423 | 7.894 | 2.65337 | 7.811 | 2.42931 | 7.729 | 2.29187 | 7.645 | 2.04639 | 7.561 | 1.89638 | 7.498 |
| 9.52796 | 3.31430 | 7.978 | 3.02136 | 7.891 | 2.71384 | 7.809 | 2.48508 | 7.727 | 2.34217 | 7.643 | 2.09442 | 7.560 | 1.94265 | 7.498 |
| 9.63829 | 3.38881 | 7.974 | 3.08956 | 7.888 | 2.77608 | 7.806 | 2.54199 | 7.725 | 2.39228 | 7.641 | 2.14150 | 7.559 | 1.98968 | 7.496 |
| 9.74990 | 3.46473 | 7.971 | 3.15887 | 7.885 | 2.83904 | 7.804 | 2.60045 | 7.723 | 2.44449 | 7.639 | 2.18816 | 7.557 | 2.03747 | 7.494 |
| 9.86279 | 3.54174 | 7.968 | 3.22931 | 7.882 | 2.90378 | 7.801 | 2.66000 | 7.720 | 2.49849 | 7.637 | 2.23506 | 7.556 | 2.08581 | 7.491 |
| 9.97700 | 3.62135 | 7.964 | 3.30283 | 7.879 | 2.96992 | 7.798 | 2.72053 | 7.718 | 2.55399 | 7.636 | 2.28212 | 7.555 | 2.13222 | 7.489 |
| 10.0925 | 3.70372 | 7.961 | 3.37766 | 7.876 | 3.03705 | 7.796 | 2.78211 | 7.716 | 2.61298 | 7.635 | 2.33051 | 7.553 | 2.17869 | 7.488 |
| 10.2094 | 3.78748 | 7.957 | 3.45459 | 7.873 | 3.10611 | 7.793 | 2.84504 | 7.714 | 2.67369 | 7.633 | 2.38339 | 7.551 | 2.22551 | 7.486 |
| 10.3276 | 3.87220 | 7.953 | 3.53329 | 7.869 | 3.17599 | 7.790 | 2.90987 | 7.712 | 2.73657 | 7.632 | 2.43985 | 7.549 | 2.27225 | 7.485 |
| 10.4472 | 3.95791 | 7.949 | 3.61301 | 7.866 | 3.24914 | 7.787 | 2.97622 | 7.709 | 2.80127 | 7.629 | 2.49902 | 7.547 | 2.31954 | 7.484 |
| 10.5682 | 4.04647 | 7.945 | 3.69365 | 7.863 | 3.32384 | 7.785 | 3.04465 | 7.707 | 2.86678 | 7.627 | 2.56144 | 7.545 | 2.37285 | 7.483 |
| 10.6905 | 4.13815 | 7.941 | 3.77528 | 7.859 | 3.39940 | 7.782 | 3.11503 | 7.704 | 2.93273 | 7.625 | 2.62424 | 7.543 | 2.42904 | 7.483 |
| 10.8143 | 4.23137 | 7.937 | 3.86026 | 7.855 | 3.47592 | 7.779 | 3.18623 | 7.702 | 2.99934 | 7.622 | 2.68508 | 7.542 | 2.48721 | 7.482 |

| | | | | | | | | | | | | | |
|---|---|---|---|---|---|---|---|---|---|---|---|---|---|
| 10.9396 | 4.32567 | 7.933 | 3.94794 | 7.852 | 3.55335 | 7.776 | 3.25826 | 7.699 | 3.06673 | 7.620 | 2.74499 | 7.540 | 2.55001 | 7.481 |
| 11.0662 | 4.42106 | 7.929 | 4.03686 | 7.848 | 3.63440 | 7.773 | 3.33147 | 7.697 | 3.13490 | 7.617 | 2.80560 | 7.538 | 2.61353 | 7.479 |
| 11.1944 | 4.52011 | 7.924 | 4.12681 | 7.844 | 3.71719 | 7.769 | 3.40720 | 7.694 | 3.20510 | 7.615 | 2.86692 | 7.536 | 2.67322 | 7.477 |
| 11.3240 | 4.62170 | 7.920 | 4.21794 | 7.841 | 3.80095 | 7.766 | 3.48529 | 7.691 | 3.27894 | 7.613 | 2.93048 | 7.534 | 2.73287 | 7.474 |
| 11.4551 | 4.72463 | 7.915 | 4.31339 | 7.836 | 3.88577 | 7.763 | 3.56428 | 7.689 | 3.35456 | 7.610 | 2.99905 | 7.531 | 2.79320 | 7.472 |
| 11.5878 | 4.82875 | 7.911 | 4.41082 | 7.832 | 3.97189 | 7.759 | 3.64422 | 7.686 | 3.43105 | 7.608 | 3.07106 | 7.529 | 2.85424 | 7.470 |
| 11.7220 | 4.93520 | 7.906 | 4.50944 | 7.828 | 4.06289 | 7.756 | 3.72602 | 7.683 | 3.50843 | 7.605 | 3.14390 | 7.527 | 2.91598 | 7.467 |
| 11.8577 | 5.04542 | 7.901 | 4.60925 | 7.824 | 4.15508 | 7.753 | 3.81087 | 7.680 | 3.58807 | 7.603 | 3.21762 | 7.525 | 2.98564 | 7.466 |
| 11.9950 | 5.15764 | 7.896 | 4.71173 | 7.820 | 4.24845 | 7.749 | 3.89790 | 7.677 | 3.67005 | 7.600 | 3.29203 | 7.523 | 3.05734 | 7.464 |
| 12.1339 | 5.27116 | 7.891 | 4.81768 | 7.815 | 4.34299 | 7.745 | 3.98594 | 7.674 | 3.75332 | 7.597 | 3.36697 | 7.520 | 3.12986 | 7.462 |
| 12.2744 | 5.38623 | 7.886 | 4.92518 | 7.811 | 4.44100 | 7.741 | 4.07504 | 7.670 | 3.83754 | 7.595 | 3.44263 | 7.518 | 3.20322 | 7.461 |
| 12.4165 | 5.50587 | 7.881 | 5.03395 | 7.806 | 4.54210 | 7.738 | 4.16654 | 7.667 | 3.92310 | 7.592 | 3.51918 | 7.515 | 3.27736 | 7.459 |
| 12.5603 | 5.62793 | 7.875 | 5.14461 | 7.802 | 4.64442 | 7.734 | 4.26070 | 7.664 | 4.01218 | 7.589 | 3.59676 | 7.513 | 3.35182 | 7.456 |
| 12.7057 | 5.75140 | 7.870 | 5.25940 | 7.797 | 4.74803 | 7.729 | 4.35600 | 7.660 | 4.10325 | 7.586 | 3.67648 | 7.510 | 3.42714 | 7.453 |
| 12.8529 | 5.87645 | 7.865 | 5.37597 | 7.792 | 4.85371 | 7.725 | 4.45245 | 7.657 | 4.19538 | 7.583 | 3.75815 | 7.508 | 3.50334 | 7.450 |
| 13.0017 | 6.00657 | 7.859 | 5.49391 | 7.787 | 4.96287 | 7.721 | 4.55173 | 7.653 | 4.28858 | 7.580 | 3.84077 | 7.505 | 3.58042 | 7.447 |
| 13.1522 | 6.13952 | 7.853 | 5.61376 | 7.782 | 5.07332 | 7.717 | 4.65474 | 7.650 | 4.38522 | 7.577 | 3.92442 | 7.502 | 3.65992 | 7.445 |
| 13.3045 | 6.27405 | 7.847 | 5.73866 | 7.776 | 5.18520 | 7.712 | 4.75974 | 7.646 | 4.48431 | 7.574 | 4.01253 | 7.499 | 3.74122 | 7.443 |
| 13.4586 | 6.41050 | 7.841 | 5.86566 | 7.771 | 5.29949 | 7.708 | 4.86597 | 7.642 | 4.58471 | 7.570 | 4.10543 | 7.496 | 3.82346 | 7.441 |
| 13.6144 | 6.55139 | 7.835 | 5.99418 | 7.766 | 5.41921 | 7.703 | 4.97483 | 7.638 | 4.68627 | 7.567 | 4.19994 | 7.494 | 3.90665 | 7.439 |
| 13.7721 | 6.69481 | 7.829 | 6.12521 | 7.760 | 5.54037 | 7.699 | 5.08734 | 7.634 | 4.79189 | 7.564 | 4.29559 | 7.491 | 3.99366 | 7.437 |
| 13.9316 | 6.83989 | 7.823 | 6.26154 | 7.754 | 5.66304 | 7.694 | 5.20206 | 7.630 | 4.90066 | 7.560 | 4.39355 | 7.487 | 4.08666 | 7.435 |
| 14.0929 | 6.98796 | 7.816 | 6.40002 | 7.749 | 5.78852 | 7.689 | 5.31819 | 7.626 | 5.01099 | 7.557 | 4.49403 | 7.484 | 4.18074 | 7.432 |
| 14.2561 | 7.14017 | 7.810 | 6.54016 | 7.743 | 5.91832 | 7.684 | 5.43710 | 7.622 | 5.12261 | 7.553 | 4.59597 | 7.481 | 4.27591 | 7.429 |
| 14.4212 | 7.29481 | 7.803 | 6.68333 | 7.737 | 6.04982 | 7.679 | 5.55941 | 7.618 | 5.23787 | 7.549 | 4.69916 | 7.478 | 4.37304 | 7.426 |
| 14.5881 | 7.45123 | 7.796 | 6.82991 | 7.731 | 6.18300 | 7.674 | 5.68366 | 7.614 | 5.35567 | 7.545 | 4.80617 | 7.474 | 4.47334 | 7.424 |
| 14.7571 | 7.61278 | 7.789 | 6.97844 | 7.725 | 6.32040 | 7.668 | 5.80950 | 7.609 | 5.47492 | 7.541 | 4.91711 | 7.471 | 4.57480 | 7.421 |
| 14.9279 | 7.77798 | 7.781 | 7.12901 | 7.719 | 6.46115 | 7.663 | 5.93882 | 7.604 | 5.59624 | 7.538 | 5.02960 | 7.467 | 4.67745 | 7.418 |
| 15.1008 | 7.94532 | 7.774 | 7.28543 | 7.712 | 6.60364 | 7.657 | 6.07183 | 7.600 | 5.72226 | 7.533 | 5.14363 | 7.464 | 4.78357 | 7.415 |
| 15.2757 | 8.11577 | 7.767 | 7.44474 | 7.705 | 6.74817 | 7.652 | 6.20641 | 7.595 | 5.85095 | 7.529 | 5.26005 | 7.460 | 4.89424 | 7.412 |
| 15.4525 | 8.29038 | 7.759 | 7.60609 | 7.699 | 6.89923 | 7.646 | 6.34368 | 7.590 | 5.98113 | 7.525 | 5.37861 | 7.457 | 5.00620 | 7.409 |
| 15.6315 | 8.46758 | 7.751 | 7.77066 | 7.692 | 7.05232 | 7.640 | 6.48499 | 7.585 | 6.11423 | 7.520 | 5.49858 | 7.453 | 5.11945 | 7.406 |
| 15.8125 | 8.64732 | 7.743 | 7.93869 | 7.685 | 7.20728 | 7.634 | 6.62934 | 7.580 | 6.25033 | 7.516 | 5.62110 | 7.449 | 5.23547 | 7.402 |
| 15.9956 | 8.83337 | 7.735 | 8.10892 | 7.678 | 7.36562 | 7.628 | 6.77549 | 7.575 | 6.38836 | 7.511 | 5.74680 | 7.445 | 5.35347 | 7.399 |
| 16.1808 | 9.02293 | 7.727 | 8.28209 | 7.670 | 7.52690 | 7.622 | 6.92443 | 7.570 | 6.52867 | 7.507 | 5.87461 | 7.441 | 5.47285 | 7.395 |
| 16.3682 | 9.21472 | 7.718 | 8.46093 | 7.663 | 7.69030 | 7.616 | 7.07619 | 7.564 | 6.67383 | 7.502 | 6.00436 | 7.437 | 5.59403 | 7.391 |
| 16.5577 | 9.41071 | 7.710 | 8.64243 | 7.655 | 7.85704 | 7.609 | 7.22981 | 7.559 | 6.82184 | 7.497 | 6.13760 | 7.433 | 5.71978 | 7.388 |
| 16.7494 | 9.60983 | 7.702 | 8.82645 | 7.648 | 8.02883 | 7.603 | 7.38757 | 7.553 | 6.97157 | 7.492 | 6.27389 | 7.429 | 5.84697 | 7.384 |
| 16.9434 | 9.81133 | 7.693 | 9.01649 | 7.640 | 8.20272 | 7.596 | 7.55060 | 7.547 | 7.12529 | 7.487 | 6.41185 | 7.424 | 5.97564 | 7.381 |
| 17.1396 | 10.0193 | 7.684 | 9.20955 | 7.632 | 8.37935 | 7.589 | 7.71646 | 7.542 | 7.28196 | 7.482 | 6.55348 | 7.420 | 6.10853 | 7.377 |
| 17.3380 | 10.2317 | 7.675 | 9.40508 | 7.624 | 8.56257 | 7.583 | 7.88527 | 7.536 | 7.44057 | 7.477 | 6.69887 | 7.416 | 6.24415 | 7.373 |
| 17.5388 | 10.4469 | 7.665 | 9.60580 | 7.616 | 8.74840 | 7.575 | 8.05801 | 7.530 | 7.60361 | 7.472 | 6.84633 | 7.411 | 6.38135 | 7.369 |
| 17.7419 | 10.6672 | 7.655 | 9.80954 | 7.607 | 8.93671 | 7.568 | 8.23377 | 7.524 | 7.77062 | 7.467 | 6.99843 | 7.406 | 6.52193 | 7.365 |
| 17.9473 | 10.8914 | 7.646 | 10.0163 | 7.599 | 9.13001 | 7.561 | 8.41313 | 7.517 | 7.93998 | 7.461 | 7.15599 | 7.401 | 6.66683 | 7.361 |
| 18.1552 | 11.1184 | 7.636 | 10.2302 | 7.590 | 9.32586 | 7.553 | 8.59754 | 7.511 | 8.11377 | 7.456 | 7.31641 | 7.396 | 6.81341 | 7.357 |
| 18.3654 | 11.3507 | 7.626 | 10.4475 | 7.581 | 9.52490 | 7.545 | 8.78570 | 7.504 | 8.29185 | 7.450 | 7.47918 | 7.391 | 6.96342 | 7.352 |
| 18.5780 | 11.5866 | 7.616 | 10.6676 | 7.572 | 9.73305 | 7.538 | 8.97727 | 7.498 | 8.47253 | 7.444 | 7.64452 | 7.386 | 7.12120 | 7.348 |

| | | | | | | | | | | | | | |
|---|---|---|---|---|---|---|---|---|---|---|---|---|---|
| 18.7932 | 11.8260 | 7.606 | 10.8923 | 7.563 | 9.94374 | 7.530 | 9.17334 | 7.491 | 8.65833 | 7.438 | 7.81209 | 7.380 | 7.28080 | 7.344 |
| 19.0108 | 12.0705 | 7.595 | 11.1202 | 7.554 | 10.1572 | 7.522 | 9.37275 | 7.484 | 8.84883 | 7.432 | 7.98325 | 7.375 | 7.44249 | 7.339 |
| 19.2309 | 12.3184 | 7.584 | 11.3526 | 7.544 | 10.3742 | 7.513 | 9.57527 | 7.477 | 9.04208 | 7.426 | 8.15856 | 7.370 | 7.60723 | 7.334 |
| 19.4536 | 12.5706 | 7.573 | 11.5903 | 7.534 | 10.5940 | 7.505 | 9.78148 | 7.470 | 9.23831 | 7.419 | 8.33672 | 7.364 | 7.77389 | 7.330 |
| 19.6789 | 12.8272 | 7.562 | 11.8312 | 7.525 | 10.8198 | 7.496 | 9.9908 | 7.463 | 9.43726 | 7.412 | 8.52066 | 7.358 | 7.94338 | 7.325 |
| 19.9067 | 13.0872 | 7.551 | 12.0771 | 7.515 | 11.0519 | 7.487 | 10.2054 | 7.455 | 9.63889 | 7.406 | 8.71082 | 7.352 | 8.11844 | 7.320 |
| 20.1372 | 13.3556 | 7.539 | 12.3273 | 7.505 | 11.2869 | 7.478 | 10.4256 | 7.447 | 9.84775 | 7.399 | 8.90397 | 7.346 | 8.29553 | 7.315 |
| 20.3704 | 13.6290 | 7.527 | 12.5815 | 7.494 | 11.5279 | 7.469 | 10.6491 | 7.440 | 10.0609 | 7.392 | 9.10112 | 7.341 | 8.47756 | 7.310 |
| 20.6063 | 13.9065 | 7.515 | 12.8409 | 7.483 | 11.7724 | 7.460 | 10.8768 | 7.432 | 10.2774 | 7.385 | 9.30227 | 7.334 | 8.66740 | 7.304 |
| 20.8449 | 14.1885 | 7.504 | 13.1034 | 7.472 | 12.0204 | 7.451 | 11.1088 | 7.424 | 10.4981 | 7.378 | 9.50602 | 7.328 | 8.85944 | 7.299 |
| 21.0863 | 14.4742 | 7.492 | 13.3719 | 7.461 | 12.2741 | 7.441 | 11.3453 | 7.415 | 10.7220 | 7.371 | 9.71209 | 7.321 | 9.05565 | 7.293 |
| 21.3304 | 14.7677 | 7.479 | 13.6469 | 7.451 | 12.5311 | 7.432 | 11.5872 | 7.406 | 10.9502 | 7.363 | 9.92019 | 7.315 | 9.25562 | 7.288 |
| 21.5774 | 15.0664 | 7.466 | 13.9267 | 7.439 | 12.7926 | 7.422 | 11.8328 | 7.398 | 11.1822 | 7.356 | 10.1346 | 7.308 | 9.45792 | 7.282 |
| 21.8273 | 15.3692 | 7.454 | 14.2137 | 7.428 | 13.0584 | 7.412 | 12.0830 | 7.389 | 11.4186 | 7.348 | 10.3566 | 7.302 | 9.66272 | 7.276 |
| 22.0800 | 15.6763 | 7.440 | 14.5050 | 7.416 | 13.3287 | 7.402 | 12.3385 | 7.380 | 11.6632 | 7.341 | 10.5827 | 7.295 | 9.86991 | 7.270 |
| 22.3357 | 15.9870 | 7.427 | 14.8011 | 7.404 | 13.6088 | 7.391 | 12.5991 | 7.371 | 11.9124 | 7.333 | 10.8113 | 7.288 | 10.0823 | 7.264 |
| 22.5944 | 16.3014 | 7.414 | 15.1017 | 7.392 | 13.8924 | 7.381 | 12.8659 | 7.362 | 12.1650 | 7.325 | 11.0423 | 7.281 | 10.3041 | 7.258 |
| 22.8560 | 16.6194 | 7.400 | 15.4058 | 7.380 | 14.1840 | 7.370 | 13.1376 | 7.353 | 12.4207 | 7.316 | 11.2797 | 7.273 | 10.5286 | 7.251 |
| 23.1206 | 16.9460 | 7.386 | 15.7135 | 7.368 | 14.4800 | 7.359 | 13.4152 | 7.343 | 12.6818 | 7.307 | 11.5243 | 7.266 | 10.7556 | 7.245 |
| 23.3884 | 17.2813 | 7.372 | 16.0256 | 7.356 | 14.7795 | 7.348 | 13.6984 | 7.334 | 12.9505 | 7.299 | 11.7734 | 7.258 | 10.9852 | 7.239 |
| 23.6592 | 17.6216 | 7.357 | 16.3459 | 7.343 | 15.0830 | 7.336 | 13.9848 | 7.324 | 13.2236 | 7.290 | 12.0272 | 7.250 | 11.2201 | 7.232 |
| 23.9332 | 17.9659 | 7.343 | 16.6727 | 7.329 | 15.3901 | 7.325 | 14.2747 | 7.314 | 13.4999 | 7.281 | 12.2848 | 7.242 | 11.4642 | 7.226 |
| 24.2103 | 18.3141 | 7.328 | 17.0035 | 7.316 | 15.7062 | 7.313 | 14.5693 | 7.304 | 13.7794 | 7.272 | 12.5454 | 7.234 | 11.7111 | 7.219 |
| 24.4906 | 18.6704 | 7.313 | 17.3382 | 7.303 | 16.0271 | 7.302 | 14.8714 | 7.294 | 14.0648 | 7.263 | 12.8090 | 7.226 | 11.9640 | 7.212 |
| 24.7742 | 19.0328 | 7.298 | 17.6771 | 7.289 | 16.3519 | 7.290 | 15.1800 | 7.283 | 14.3587 | 7.254 | 13.0788 | 7.218 | 12.2201 | 7.204 |
| 25.0611 | 19.3996 | 7.283 | 18.0282 | 7.276 | 16.6813 | 7.277 | 15.4922 | 7.272 | 14.6577 | 7.245 | 13.3588 | 7.209 | 12.4792 | 7.197 |
| 25.3513 | 19.7706 | 7.267 | 18.3855 | 7.262 | 17.0156 | 7.265 | 15.8082 | 7.261 | 14.9602 | 7.235 | 13.6461 | 7.201 | 12.7413 | 7.190 |
| 25.6448 | 20.1473 | 7.251 | 18.7469 | 7.248 | 17.3633 | 7.252 | 16.1305 | 7.250 | 15.2661 | 7.226 | 13.9367 | 7.192 | 13.0065 | 7.182 |
| 25.9418 | 20.5310 | 7.235 | 19.1134 | 7.234 | 17.7152 | 7.240 | 16.4616 | 7.239 | 15.5808 | 7.216 | 14.2312 | 7.183 | 13.2874 | 7.175 |
| 26.2422 | 20.9199 | 7.219 | 19.4868 | 7.219 | 18.0719 | 7.227 | 16.7991 | 7.228 | 15.9038 | 7.205 | 14.5324 | 7.174 | 13.5731 | 7.167 |
| 26.5461 | 21.3133 | 7.202 | 19.8668 | 7.204 | 18.4332 | 7.213 | 17.1406 | 7.216 | 16.2317 | 7.194 | 14.8411 | 7.165 | 13.8621 | 7.159 |
| 26.8534 | 21.7118 | 7.185 | 20.2516 | 7.188 | 18.8025 | 7.200 | 17.4865 | 7.204 | 16.5633 | 7.184 | 15.1548 | 7.155 | 14.1544 | 7.151 |
| 27.1644 | 22.1186 | 7.168 | 20.6407 | 7.173 | 19.1787 | 7.186 | 17.8411 | 7.192 | 16.8998 | 7.173 | 15.4722 | 7.145 | 14.4516 | 7.143 |
| 27.4789 | 22.5314 | 7.151 | 21.0349 | 7.157 | 19.5594 | 7.172 | 18.2043 | 7.180 | 17.2455 | 7.162 | 15.7942 | 7.135 | 14.7598 | 7.134 |
| 27.7971 | 22.9489 | 7.134 | 21.4380 | 7.142 | 19.9446 | 7.158 | 18.5717 | 7.168 | 17.5972 | 7.151 | 16.1243 | 7.125 | 15.0716 | 7.125 |
| 28.1190 | 23.3715 | 7.116 | 21.8465 | 7.126 | 20.3361 | 7.144 | 18.9435 | 7.155 | 17.9529 | 7.140 | 16.4615 | 7.115 | 15.3871 | 7.117 |
| 28.4446 | 23.8037 | 7.098 | 22.2600 | 7.110 | 20.7362 | 7.130 | 19.3223 | 7.143 | 18.3127 | 7.129 | 16.8026 | 7.105 | 15.7062 | 7.108 |
| 28.7740 | 24.2424 | 7.080 | 22.6804 | 7.094 | 21.1412 | 7.116 | 19.7091 | 7.130 | 18.6851 | 7.118 | 17.1481 | 7.095 | 16.0347 | 7.098 |
| 29.1072 | 24.6862 | 7.062 | 23.1099 | 7.077 | 21.5519 | 7.101 | 20.1013 | 7.117 | 19.0660 | 7.106 | 17.5030 | 7.084 | 16.3699 | 7.089 |
| 29.4442 | 25.1349 | 7.043 | 23.5452 | 7.060 | 21.9684 | 7.086 | 20.4986 | 7.104 | 19.4517 | 7.094 | 17.8674 | 7.073 | 16.7089 | 7.079 |
| 29.7852 | 25.5881 | 7.024 | 23.9853 | 7.043 | 22.3929 | 7.071 | 20.9027 | 7.090 | 19.8418 | 7.082 | 18.2366 | 7.062 | 17.0519 | 7.069 |
| 30.1301 | 26.0463 | 7.005 | 24.4312 | 7.026 | 22.8223 | 7.056 | 21.3145 | 7.076 | 20.2391 | 7.069 | 18.6104 | 7.051 | 17.4036 | 7.059 |
| 30.4789 | 26.5098 | 6.986 | 24.8859 | 7.009 | 23.2566 | 7.040 | 21.7321 | 7.062 | 20.6426 | 7.057 | 18.9938 | 7.040 | 17.7663 | 7.050 |
| 30.8319 | 26.9790 | 6.967 | 25.3464 | 6.991 | 23.6979 | 7.025 | 22.1547 | 7.048 | 21.0510 | 7.044 | 19.3871 | 7.028 | 18.1332 | 7.040 |
| 31.1889 | 27.4542 | 6.947 | 25.8135 | 6.973 | 24.1477 | 7.009 | 22.5845 | 7.034 | 21.4640 | 7.031 | 19.7859 | 7.017 | 18.5043 | 7.030 |
| 31.5500 | 27.9350 | 6.927 | 26.2897 | 6.955 | 24.6040 | 6.993 | 23.0224 | 7.020 | 21.8889 | 7.018 | 20.1904 | 7.004 | 18.8838 | 7.020 |
| 31.9154 | 28.4213 | 6.906 | 26.7741 | 6.936 | 25.0663 | 6.976 | 23.4661 | 7.005 | 22.3218 | 7.005 | 20.6008 | 6.992 | 19.2756 | 7.009 |

| | | | | | | | | | | | | | |
|---|---|---|---|---|---|---|---|---|---|---|---|---|---|
| 32.2849 | 28.9171 | 6.886 | 27.2644 | 6.917 | 25.5420 | 6.960 | 23.9169 | 6.990 | 22.7599 | 6.992 | 21.0158 | 6.979 | 19.6719 | 6.998 |
| 32.6588 | 29.4205 | 6.867 | 27.7592 | 6.899 | 26.0276 | 6.943 | 24.3786 | 6.974 | 23.2037 | 6.978 | 21.4355 | 6.966 | 20.0728 | 6.987 |
| 33.0370 | 29.9298 | 6.847 | 28.2577 | 6.880 | 26.5192 | 6.926 | 24.8499 | 6.958 | 23.6553 | 6.965 | 21.8606 | 6.954 | 20.4792 | 6.976 |
| 33.4195 | 30.4485 | 6.826 | 28.7614 | 6.860 | 27.0165 | 6.909 | 25.3252 | 6.942 | 24.1130 | 6.951 | 22.2967 | 6.941 | 20.8916 | 6.965 |
| 33.8065 | 30.9786 | 6.804 | 29.2716 | 6.841 | 27.5185 | 6.891 | 25.8040 | 6.927 | 24.5760 | 6.937 | 22.7429 | 6.929 | 21.3087 | 6.953 |
| 34.1979 | 31.5163 | 6.782 | 29.7965 | 6.821 | 28.0256 | 6.874 | 26.2891 | 6.912 | 25.0492 | 6.922 | 23.1958 | 6.916 | 21.7306 | 6.942 |
| 34.5939 | 32.0566 | 6.760 | 30.3366 | 6.801 | 28.5381 | 6.856 | 26.7813 | 6.897 | 25.5325 | 6.907 | 23.6585 | 6.902 | 22.1658 | 6.930 |
| 34.9945 | 32.5806 | 6.739 | 30.8837 | 6.781 | 29.0542 | 6.839 | 27.2826 | 6.881 | 26.0223 | 6.892 | 24.1309 | 6.887 | 22.6092 | 6.917 |
| 35.3997 | 33.1037 | 6.717 | 31.4342 | 6.761 | 29.5756 | 6.821 | 27.7944 | 6.864 | 26.5175 | 6.877 | 24.6094 | 6.873 | 23.0577 | 6.905 |
| 35.8096 | 33.6335 | 6.696 | 31.9858 | 6.741 | 30.1039 | 6.802 | 28.3136 | 6.846 | 27.0174 | 6.862 | 25.0934 | 6.859 | 23.5132 | 6.892 |
| 36.2243 | 34.1932 | 6.673 | 32.5430 | 6.721 | 30.6392 | 6.783 | 28.8376 | 6.829 | 27.5227 | 6.847 | 25.5890 | 6.846 | 23.9834 | 6.880 |
| 36.6438 | 34.7687 | 6.651 | 33.1069 | 6.700 | 31.1819 | 6.764 | 29.3628 | 6.812 | 28.0338 | 6.832 | 26.0958 | 6.832 | 24.4590 | 6.868 |
| 37.0681 | 35.3510 | 6.628 | 33.6831 | 6.679 | 31.7302 | 6.745 | 29.8895 | 6.795 | 28.5646 | 6.816 | 26.6113 | 6.818 | 24.9402 | 6.856 |
| 37.4973 | 35.9301 | 6.605 | 34.2670 | 6.659 | 32.2844 | 6.726 | 30.4223 | 6.778 | 29.1080 | 6.800 | 27.1287 | 6.802 | 25.4350 | 6.843 |
| 37.9315 | 36.4818 | 6.587 | 34.8549 | 6.638 | 32.8426 | 6.707 | 30.9689 | 6.759 | 29.6580 | 6.784 | 27.6453 | 6.786 | 25.9384 | 6.830 |
| 38.3707 | 37.0544 | 6.566 | 35.4437 | 6.617 | 33.4097 | 6.687 | 31.5331 | 6.741 | 30.2022 | 6.768 | 28.1669 | 6.771 | 26.4476 | 6.816 |
| 38.8150 | 37.6455 | 6.544 | 36.0382 | 6.596 | 33.9841 | 6.666 | 32.1081 | 6.723 | 30.7439 | 6.751 | 28.7031 | 6.755 | 26.9589 | 6.803 |
| 39.2645 | 38.2785 | 6.520 | 36.6403 | 6.575 | 34.5694 | 6.645 | 32.6842 | 6.704 | 31.2902 | 6.734 | 29.2572 | 6.739 | 27.4714 | 6.789 |
| 39.7192 | 38.8870 | 6.498 | 37.2550 | 6.553 | 35.1619 | 6.625 | 33.2563 | 6.685 | 31.8562 | 6.717 | 29.8209 | 6.723 | 27.9898 | 6.774 |
| 40.1791 | 39.5290 | 6.479 | 37.8775 | 6.531 | 35.7617 | 6.603 | 33.8303 | 6.666 | 32.4397 | 6.700 | 30.3896 | 6.707 | 28.5201 | 6.760 |
| 40.6443 | 40.1272 | 6.459 | 38.5034 | 6.509 | 36.3710 | 6.582 | 34.4202 | 6.646 | 33.0323 | 6.683 | 30.9624 | 6.691 | 29.0731 | 6.746 |
| 41.1150 | 40.7623 | 6.439 | 39.1261 | 6.487 | 36.9865 | 6.561 | 35.0324 | 6.626 | 33.6232 | 6.665 | 31.5420 | 6.674 | 29.6324 | 6.731 |
| 41.5911 | 41.4494 | 6.413 | 39.7547 | 6.465 | 37.6104 | 6.539 | 35.6576 | 6.607 | 34.2144 | 6.646 | 32.1340 | 6.657 | 30.1973 | 6.717 |
| 42.0727 | 42.1014 | 6.387 | 40.3931 | 6.442 | 38.2451 | 6.517 | 36.2831 | 6.587 | 34.8112 | 6.628 | 32.7389 | 6.639 | 30.7651 | 6.701 |
| 42.5598 | 42.7460 | 6.367 | 41.0434 | 6.420 | 38.8876 | 6.495 | 36.9066 | 6.566 | 35.4312 | 6.609 | 33.3516 | 6.622 | 31.3395 | 6.686 |
| 43.0527 | 43.3944 | 6.343 | 41.7011 | 6.397 | 39.5371 | 6.473 | 37.5348 | 6.546 | 36.0671 | 6.590 | 33.9723 | 6.604 | 31.9239 | 6.670 |
| 43.5512 | 44.0734 | 6.319 | 42.3594 | 6.373 | 40.1912 | 6.451 | 38.1781 | 6.525 | 36.7104 | 6.571 | 34.6021 | 6.587 | 32.5259 | 6.655 |
| 44.0555 | 44.7470 | 6.292 | 43.0214 | 6.350 | 40.8524 | 6.428 | 38.8382 | 6.505 | 37.3513 | 6.552 | 35.2401 | 6.569 | 33.1348 | 6.639 |
| 44.5656 | 45.4456 | 6.270 | 43.6890 | 6.326 | 41.5563 | 6.406 | 39.5052 | 6.484 | 37.9959 | 6.533 | 35.8857 | 6.551 | 33.7515 | 6.623 |
| 45.0817 | 46.0920 | 6.249 | 44.3674 | 6.303 | 42.2402 | 6.384 | 40.1750 | 6.463 | 38.6526 | 6.513 | 36.5384 | 6.533 | 34.3767 | 6.607 |
| 45.6037 | 46.7889 | 6.225 | 45.0539 | 6.279 | 42.9604 | 6.361 | 40.8813 | 6.442 | 39.3255 | 6.493 | 37.1994 | 6.514 | 35.0092 | 6.590 |
| 46.1318 | 47.5244 | 6.199 | 45.7448 | 6.255 | 43.6555 | 6.338 | 41.5988 | 6.421 | 40.0087 | 6.473 | 37.8710 | 6.495 | 35.6494 | 6.573 |
| 46.6659 | 48.2569 | 6.175 | 46.4402 | 6.230 | 44.3894 | 6.316 | 42.3298 | 6.400 | 40.6939 | 6.453 | 38.5532 | 6.477 | 36.2972 | 6.556 |
| 47.2063 | 48.9551 | 6.151 | 47.1404 | 6.206 | 45.0999 | 6.293 | 43.0377 | 6.378 | 41.3834 | 6.433 | 39.2478 | 6.457 | 36.9526 | 6.539 |
| 47.7529 | 49.6725 | 6.124 | 47.8519 | 6.181 | 45.8483 | 6.270 | 43.7583 | 6.356 | 42.0823 | 6.412 | 39.9172 | 6.444 | 37.6203 | 6.521 |
| 48.3059 | 50.3778 | 6.101 | 48.5748 | 6.156 | 46.5768 | 6.246 | 44.4163 | 6.345 | 42.7948 | 6.391 | 40.5905 | 6.431 | 38.2963 | 6.503 |
| 48.8652 | 51.1607 | 6.078 | 49.2624 | 6.135 | 47.3023 | 6.228 | 45.1635 | 6.328 | 43.4681 | 6.378 | 41.2744 | 6.416 | 38.9833 | 6.485 |
| 49.4311 | 51.8492 | 6.051 | 50.0093 | 6.115 | 48.0578 | 6.206 | 45.9059 | 6.308 | 44.1571 | 6.363 | 42.0088 | 6.395 | 39.6852 | 6.467 |
| 50.0035 | 52.4646 | 6.030 | 50.7306 | 6.094 | 48.7932 | 6.186 | 46.6379 | 6.291 | 44.9541 | 6.338 | 42.6847 | 6.374 | 40.3953 | 6.449 |
| 51.1682 | 53.5091 | 5.978 | 52.2592 | 6.048 | 50.2875 | 6.142 | 48.1159 | 6.256 | 46.3531 | 6.308 | 44.2068 | 6.343 | 41.8938 | 6.420 |
| 52.3600 | 55.1278 | 5.924 | 53.7805 | 6.000 | 51.8220 | 6.095 | 49.6031 | 6.215 | 47.7592 | 6.279 | 45.6602 | 6.311 | 43.3004 | 6.396 |
| 53.5797 | 56.7567 | 5.873 | 55.3100 | 5.959 | 53.3756 | 6.051 | 51.1275 | 6.174 | 49.3240 | 6.235 | 47.1724 | 6.279 | 44.8123 | 6.366 |
| 54.8277 | 58.3840 | 5.823 | 56.9143 | 5.910 | 54.9453 | 6.008 | 52.6858 | 6.135 | 50.7947 | 6.202 | 48.7580 | 6.245 | 46.4500 | 6.326 |
| 56.1048 | 59.9729 | 5.771 | 58.5964 | 5.860 | 56.5958 | 5.960 | 54.2724 | 6.093 | 52.2794 | 6.170 | 50.4218 | 6.209 | 48.1497 | 6.288 |
| 57.4116 | 61.7497 | 5.713 | 60.1639 | 5.819 | 58.1810 | 5.917 | 55.9439 | 6.047 | 53.8912 | 6.127 | 52.1544 | 6.171 | 49.9322 | 6.249 |
| 58.7489 | 63.3531 | 5.667 | 61.8391 | 5.766 | 59.7483 | 5.872 | 57.6379 | 5.998 | 55.5541 | 6.083 | 53.7997 | 6.136 | 51.6847 | 6.212 |
| 60.1174 | 64.9297 | 5.615 | 63.5284 | 5.718 | 61.4446 | 5.834 | 59.3309 | 5.948 | 57.2023 | 6.039 | 55.4102 | 6.101 | 53.4289 | 6.174 |

| | | | | | | | | | | | | | |
|---|---|---|---|---|---|---|---|---|---|---|---|---|---|
| 61.5177 | 66.4099 | 5.565 | 65.1493 | 5.670 | 63.1419 | 5.791 | 61.0519 | 5.907 | 58.8533 | 6.005 | 57.0427 | 6.070 | 55.1727 | 6.155 |
| 62.2300 | 67.2774 | 5.534 | 65.9549 | 5.649 | 63.9114 | 5.774 | 61.8796 | 5.878 | 59.7200 | 5.979 | 58.0149 | 6.045 | 56.3416 | 6.113 |
| 62.9506 | 68.0769 | 5.507 | 66.7675 | 5.621 | 64.7188 | 5.751 | 62.7394 | 5.852 | 60.6128 | 5.956 | 58.9757 | 6.019 | 57.2676 | 6.096 |
| 63.6796 | 68.8719 | 5.484 | 67.4987 | 5.602 | 65.5896 | 5.729 | 63.5804 | 5.829 | 61.5446 | 5.934 | 59.8989 | 5.996 | 58.2223 | 6.076 |
| 64.4169 | 69.7178 | 5.461 | 68.2903 | 5.585 | 66.4379 | 5.707 | 64.4592 | 5.809 | 62.5139 | 5.910 | 60.8811 | 5.978 | 59.1673 | 6.054 |
| 65.1628 | 70.5346 | 5.432 | 69.0989 | 5.562 | 67.2962 | 5.683 | 65.3761 | 5.783 | 63.4368 | 5.886 | 61.8142 | 5.958 | 60.1113 | 6.035 |
| 65.9174 | 71.4092 | 5.411 | 69.8019 | 5.535 | 68.2341 | 5.658 | 66.3310 | 5.760 | 64.3943 | 5.864 | 62.7673 | 5.934 | 61.1081 | 6.012 |
| 66.6807 | 72.2839 | 5.386 | 70.6839 | 5.500 | 69.1450 | 5.633 | 67.3073 | 5.737 | 65.3506 | 5.841 | 63.7505 | 5.910 | 62.0803 | 5.991 |
| 67.4528 | 73.1503 | 5.361 | 71.6444 | 5.475 | 70.1557 | 5.607 | 68.2945 | 5.714 | 66.4218 | 5.818 | 64.7427 | 5.890 | 63.0521 | 5.972 |
| 68.2339 | 74.0477 | 5.334 | 72.5014 | 5.457 | 71.0368 | 5.584 | 69.2428 | 5.688 | 67.3826 | 5.794 | 65.6785 | 5.865 | 64.0581 | 5.948 |
| 69.0240 | 74.9021 | 5.308 | 73.4387 | 5.438 | 72.0077 | 5.558 | 70.2187 | 5.663 | 68.3500 | 5.769 | 66.6695 | 5.843 | 65.0360 | 5.927 |
| 69.8232 | 75.7989 | 5.279 | 74.3133 | 5.403 | 72.9004 | 5.531 | 71.1433 | 5.637 | 69.3355 | 5.744 | 67.6686 | 5.818 | 66.0529 | 5.903 |
| 70.6318 | 76.6553 | 5.255 | 75.2626 | 5.381 | 73.8794 | 5.506 | 72.1382 | 5.611 | 70.3125 | 5.719 | 68.6383 | 5.795 | 67.0453 | 5.881 |
| 71.4496 | 77.5819 | 5.228 | 76.2076 | 5.351 | 74.8445 | 5.479 | 73.1401 | 5.584 | 71.3613 | 5.694 | 69.6624 | 5.771 | 68.0190 | 5.858 |
| 72.2770 | 78.3959 | 5.199 | 77.0326 | 5.323 | 75.7077 | 5.453 | 74.0320 | 5.558 | 72.2847 | 5.668 | 70.6280 | 5.745 | 69.0338 | 5.833 |
| 73.1139 | 79.2910 | 5.174 | 77.9619 | 5.299 | 76.6741 | 5.427 | 75.0290 | 5.533 | 73.2929 | 5.643 | 71.6491 | 5.721 | 70.0417 | 5.810 |
| 73.9605 | 80.1818 | 5.150 | 78.9176 | 5.279 | 77.6580 | 5.401 | 76.0405 | 5.509 | 74.3093 | 5.618 | 72.6513 | 5.698 | 71.0380 | 5.787 |
| 74.8170 | 81.0147 | 5.122 | 79.8129 | 5.246 | 78.5855 | 5.375 | 76.9751 | 5.481 | 75.2590 | 5.592 | 73.6271 | 5.673 | 72.0375 | 5.763 |
| 75.6833 | 81.8769 | 5.097 | 80.7249 | 5.225 | 79.5371 | 5.349 | 77.9347 | 5.457 | 76.2641 | 5.568 | 74.6253 | 5.649 | 73.0393 | 5.740 |
| 76.5597 | 82.7803 | 5.072 | 81.6503 | 5.201 | 80.4978 | 5.323 | 78.9631 | 5.431 | 77.2966 | 5.542 | 75.6703 | 5.625 | 74.0559 | 5.716 |
| 77.4462 | 83.6493 | 5.047 | 82.5723 | 5.174 | 81.4421 | 5.297 | 79.9517 | 5.405 | 78.2967 | 5.517 | 76.7114 | 5.600 | 75.0958 | 5.693 |
| 78.3430 | 84.3183 | 5.020 | 83.3307 | 5.150 | 82.2561 | 5.271 | 80.8474 | 5.372 | 79.3206 | 5.481 | 78.0719 | 5.578 | 76.1631 | 5.667 |
| 79.2501 | 85.3754 | 4.993 | 84.3725 | 5.120 | 83.3083 | 5.245 | 81.9869 | 5.348 | 80.4211 | 5.456 | 79.1619 | 5.555 | 77.1513 | 5.644 |
| 80.1678 | 86.2405 | 4.971 | 85.2815 | 5.096 | 84.2854 | 5.221 | 82.9410 | 5.325 | 81.4252 | 5.433 | 80.1776 | 5.532 | 78.1650 | 5.621 |
| 81.0961 | 87.1584 | 4.945 | 86.1854 | 5.065 | 85.2099 | 5.193 | 83.9534 | 5.296 | 82.4624 | 5.407 | 81.2682 | 5.508 | 79.2288 | 5.598 |
| 82.0352 | 88.1058 | 4.927 | 87.2063 | 5.044 | 86.2760 | 5.168 | 84.9998 | 5.276 | 83.4849 | 5.384 | 82.3122 | 5.486 | 80.2993 | 5.575 |
| 82.9851 | 88.7858 | 4.892 | 87.9747 | 5.017 | 87.1125 | 5.147 | 85.8575 | 5.250 | 84.4494 | 5.358 | 83.3179 | 5.458 | 81.3852 | 5.549 |
| 83.9460 | 89.7245 | 4.869 | 88.9407 | 4.992 | 88.0972 | 5.122 | 86.9351 | 5.228 | 85.5340 | 5.335 | 84.4126 | 5.435 | 82.5023 | 5.526 |
| 84.9180 | 90.9611 | 4.850 | 90.4594 | 4.960 | 89.1201 | 5.085 | 88.3283 | 5.193 | 86.9396 | 5.309 | 85.8038 | 5.407 | 83.6230 | 5.501 |
| 85.9014 | 91.7194 | 4.829 | 90.9252 | 4.947 | 90.2310 | 5.067 | 89.1012 | 5.175 | 87.6806 | 5.284 | 86.5935 | 5.386 | 84.6902 | 5.479 |
| 86.8960 | 92.4505 | 4.798 | 91.8548 | 4.921 | 91.1402 | 5.046 | 90.0327 | 5.153 | 88.6842 | 5.260 | 87.6306 | 5.363 | 85.8346 | 5.457 |
| 87.9023 | 93.6590 | 4.775 | 92.8489 | 4.895 | 92.1723 | 5.016 | 91.2247 | 5.127 | 89.8755 | 5.236 | 88.8503 | 5.340 | 86.9906 | 5.434 |
| 88.9201 | 93.9273 | 4.746 | 93.6443 | 4.856 | 92.9288 | 4.993 | 92.1279 | 5.092 | 90.8465 | 5.203 | 89.8901 | 5.308 | 88.2622 | 5.405 |
| 89.9498 | 95.1586 | 4.725 | 94.6362 | 4.845 | 94.0758 | 4.970 | 93.1181 | 5.077 | 91.8894 | 5.186 | 90.9422 | 5.289 | 89.3760 | 5.387 |
| 90.9913 | 95.9676 | 4.696 | 95.5433 | 4.822 | 95.0308 | 4.948 | 94.0658 | 5.054 | 92.9324 | 5.160 | 92.0257 | 5.265 | 90.5666 | 5.362 |
| 92.0450 | 97.1394 | 4.680 | 96.6432 | 4.796 | 96.1284 | 4.917 | 95.4146 | 5.029 | 94.2133 | 5.138 | 93.3326 | 5.246 | 91.7541 | 5.341 |
| 93.1108 | 97.7386 | 4.647 | 97.3996 | 4.773 | 97.0283 | 4.899 | 96.1444 | 5.004 | 95.1095 | 5.111 | 94.2410 | 5.216 | 93.0742 | 5.308 |
| 94.1890 | 98.7715 | 4.626 | 98.5233 | 4.750 | 98.1560 | 4.873 | 97.3147 | 4.980 | 96.2119 | 5.088 | 95.4133 | 5.193 | 94.2369 | 5.285 |
| 95.2796 | 99.7963 | 4.599 | 99.4926 | 4.722 | 99.1283 | 4.846 | 98.4159 | 4.954 | 97.3492 | 5.062 | 96.5811 | 5.168 | 95.4215 | 5.260 |
| 96.3829 | 100.635 | 4.581 | 100.391 | 4.701 | 100.151 | 4.825 | 99.4088 | 4.932 | 98.4363 | 5.040 | 97.7031 | 5.146 | 96.5297 | 5.239 |
| 97.4990 | 101.531 | 4.556 | 101.354 | 4.677 | 101.151 | 4.800 | 100.470 | 4.907 | 99.5278 | 5.015 | 98.8548 | 5.120 | 97.7734 | 5.213 |
| 98.6279 | 102.474 | 4.530 | 102.313 | 4.652 | 102.108 | 4.776 | 101.567 | 4.881 | 100.638 | 4.991 | 100.002 | 5.095 | 98.9329 | 5.190 |
| 99.7700 | 103.332 | 4.511 | 103.439 | 4.625 | 103.460 | 4.747 | 102.822 | 4.860 | 101.978 | 4.964 | 101.417 | 5.066 | 100.182 | 5.160 |
| 100.925 | 104.029 | 4.485 | 104.491 | 4.601 | 104.612 | 4.724 | 103.943 | 4.836 | 103.363 | 4.937 | 102.785 | 5.038 | 101.489 | 5.132 |
| 102.094 | 105.017 | 4.461 | 105.409 | 4.577 | 105.545 | 4.701 | 104.933 | 4.815 | 104.222 | 4.913 | 103.960 | 5.014 | 102.694 | 5.108 |
| 103.276 | 105.845 | 4.440 | 106.474 | 4.556 | 106.751 | 4.678 | 106.153 | 4.790 | 105.637 | 4.892 | 105.095 | 4.992 | 103.870 | 5.086 |
| 104.472 | 106.899 | 4.421 | 107.538 | 4.534 | 107.785 | 4.657 | 107.272 | 4.770 | 106.796 | 4.869 | 106.352 | 4.969 | 105.129 | 5.063 |

| | | | | | | | | | | | | | |
|---|---|---|---|---|---|---|---|---|---|---|---|---|---|
| 105.682 | 107.830 | 4.394 | 108.497 | 4.509 | 108.727 | 4.632 | 108.294 | 4.745 | 107.839 | 4.844 | 107.567 | 4.943 | 106.402 | 5.038 |
| 106.905 | 108.769 | 4.373 | 109.696 | 4.486 | 110.030 | 4.607 | 109.592 | 4.718 | 109.150 | 4.819 | 108.852 | 4.921 | 107.731 | 5.015 |
| 108.143 | 109.874 | 4.352 | 110.527 | 4.463 | 110.907 | 4.583 | 110.563 | 4.694 | 110.359 | 4.795 | 109.936 | 4.893 | 108.869 | 4.986 |
| 109.396 | 110.705 | 4.319 | 111.624 | 4.434 | 112.084 | 4.553 | 111.760 | 4.665 | 111.614 | 4.767 | 111.457 | 4.870 | 110.363 | 4.967 |
| 110.662 | 111.881 | 4.307 | 112.780 | 4.415 | 113.234 | 4.536 | 112.918 | 4.650 | 112.651 | 4.750 | 112.573 | 4.848 | 111.526 | 4.943 |
| 111.944 | 112.686 | 4.281 | 113.677 | 4.391 | 114.231 | 4.509 | 114.023 | 4.621 | 113.835 | 4.721 | 113.772 | 4.820 | 112.835 | 4.917 |
| 113.240 | 113.803 | 4.261 | 114.832 | 4.369 | 115.453 | 4.486 | 115.264 | 4.596 | 115.271 | 4.698 | 115.049 | 4.797 | 114.108 | 4.892 |
| 114.551 | 114.710 | 4.238 | 115.782 | 4.346 | 116.406 | 4.464 | 116.310 | 4.574 | 116.222 | 4.674 | 116.213 | 4.772 | 115.341 | 4.867 |
| 115.878 | 115.844 | 4.214 | 116.921 | 4.320 | 117.634 | 4.437 | 117.489 | 4.550 | 117.545 | 4.651 | 117.533 | 4.750 | 116.740 | 4.844 |
| 117.220 | 116.644 | 4.192 | 117.771 | 4.300 | 118.630 | 4.416 | 118.567 | 4.524 | 118.687 | 4.623 | 118.756 | 4.723 | 117.994 | 4.818 |
| 118.577 | 117.574 | 4.171 | 118.694 | 4.276 | 119.500 | 4.392 | 119.571 | 4.503 | 119.510 | 4.602 | 119.869 | 4.700 | 119.146 | 4.794 |
| 119.950 | 118.409 | 4.150 | 119.782 | 4.254 | 120.761 | 4.368 | 120.848 | 4.476 | 121.008 | 4.577 | 121.114 | 4.675 | 120.437 | 4.770 |
| 121.339 | 119.444 | 4.128 | 120.776 | 4.233 | 121.760 | 4.348 | 121.838 | 4.457 | 122.132 | 4.553 | 122.290 | 4.651 | 121.673 | 4.746 |
| 122.744 | 120.086 | 4.108 | 121.587 | 4.211 | 122.621 | 4.323 | 122.788 | 4.431 | 123.241 | 4.530 | 123.357 | 4.626 | 122.781 | 4.720 |
| 124.165 | 121.290 | 4.092 | 122.756 | 4.192 | 123.903 | 4.303 | 124.182 | 4.406 | 124.668 | 4.504 | 124.714 | 4.605 | 124.134 | 4.699 |
| 125.603 | 122.058 | 4.065 | 123.455 | 4.168 | 124.676 | 4.280 | 125.020 | 4.386 | 125.177 | 4.480 | 125.844 | 4.580 | 125.350 | 4.674 |
| 127.057 | 123.006 | 4.044 | 124.579 | 4.144 | 125.764 | 4.255 | 126.146 | 4.362 | 126.533 | 4.459 | 126.995 | 4.555 | 126.559 | 4.649 |
| 128.529 | 123.811 | 4.026 | 125.581 | 4.128 | 126.898 | 4.237 | 127.277 | 4.341 | 127.918 | 4.435 | 128.266 | 4.532 | 127.874 | 4.626 |
| 130.017 | 124.796 | 4.009 | 126.488 | 4.107 | 127.773 | 4.217 | 128.277 | 4.323 | 128.778 | 4.416 | 129.396 | 4.511 | 129.056 | 4.604 |
| 131.522 | 125.770 | 3.987 | 127.589 | 4.084 | 128.927 | 4.194 | 129.485 | 4.299 | 130.108 | 4.393 | 130.644 | 4.488 | 130.362 | 4.580 |
| 133.045 | 126.799 | 3.972 | 128.600 | 4.067 | 130.040 | 4.176 | 130.719 | 4.280 | 131.511 | 4.372 | 131.949 | 4.468 | 131.697 | 4.560 |
| 134.586 | 127.682 | 3.952 | 129.573 | 4.047 | 131.102 | 4.156 | 131.854 | 4.259 | 132.647 | 4.352 | 133.223 | 4.446 | 133.024 | 4.538 |
| 136.144 | 128.765 | 3.928 | 130.724 | 4.025 | 132.323 | 4.130 | 133.022 | 4.233 | 133.795 | 4.327 | 134.547 | 4.421 | 134.372 | 4.512 |
| 137.721 | 129.592 | 3.908 | 131.677 | 4.004 | 133.410 | 4.109 | 134.156 | 4.212 | 135.106 | 4.306 | 135.793 | 4.397 | 135.674 | 4.487 |
| 139.316 | 130.603 | 3.894 | 132.622 | 3.988 | 134.285 | 4.092 | 135.203 | 4.195 | 136.205 | 4.283 | 136.886 | 4.376 | 136.889 | 4.467 |
| 140.929 | 131.530 | 3.873 | 133.705 | 3.967 | 135.419 | 4.070 | 136.357 | 4.169 | 137.434 | 4.258 | 138.121 | 4.348 | 138.202 | 4.441 |
| 142.561 | 132.529 | 3.852 | 134.790 | 3.944 | 136.489 | 4.047 | 137.457 | 4.146 | 138.627 | 4.235 | 139.378 | 4.328 | 139.575 | 4.418 |
| 144.212 | 133.431 | 3.839 | 135.658 | 3.927 | 137.532 | 4.031 | 138.670 | 4.131 | 139.708 | 4.219 | 140.671 | 4.311 | 140.856 | 4.402 |
| 145.881 | 134.638 | 3.817 | 136.848 | 3.906 | 138.831 | 4.010 | 139.967 | 4.106 | 141.053 | 4.196 | 142.027 | 4.288 | 142.239 | 4.377 |
| 147.571 | 135.568 | 3.795 | 137.896 | 3.885 | 139.818 | 3.989 | 140.913 | 4.088 | 142.392 | 4.178 | 143.293 | 4.270 | 143.493 | 4.357 |
| 149.279 | 136.489 | 3.784 | 138.748 | 3.867 | 140.930 | 3.970 | 142.205 | 4.065 | 143.627 | 4.154 | 144.629 | 4.246 | 144.930 | 4.336 |
| 151.008 | 137.215 | 3.759 | 139.726 | 3.848 | 141.948 | 3.949 | 143.216 | 4.044 | 144.540 | 4.133 | 145.707 | 4.222 | 146.123 | 4.311 |
| 152.757 | 138.185 | 3.742 | 140.762 | 3.831 | 143.155 | 3.928 | 144.435 | 4.021 | 145.925 | 4.112 | 147.159 | 4.201 | 147.591 | 4.289 |
| 154.525 | 139.443 | 3.722 | 141.887 | 3.811 | 144.265 | 3.908 | 145.719 | 4.004 | 147.171 | 4.090 | 148.202 | 4.175 | 148.769 | 4.261 |
| 156.315 | 140.182 | 3.710 | 142.806 | 3.793 | 145.158 | 3.890 | 146.925 | 3.981 | 148.351 | 4.066 | 149.481 | 4.151 | 149.991 | 4.237 |
| 158.125 | 141.278 | 3.688 | 144.054 | 3.774 | 146.425 | 3.870 | 147.982 | 3.963 | 149.567 | 4.049 | 150.990 | 4.138 | 151.621 | 4.224 |
| 159.956 | 142.249 | 3.673 | 145.048 | 3.759 | 147.497 | 3.850 | 149.333 | 3.939 | 150.783 | 4.024 | 151.795 | 4.103 | 152.520 | 4.184 |
| 161.808 | 143.197 | 3.649 | 146.079 | 3.736 | 148.559 | 3.824 | 150.546 | 3.912 | 152.042 | 3.993 | 153.238 | 4.079 | 154.115 | 4.161 |
| 163.682 | 144.225 | 3.637 | 147.199 | 3.721 | 149.720 | 3.809 | 151.771 | 3.897 | 153.358 | 3.976 | 154.430 | 4.063 | 155.257 | 4.144 |
| 165.577 | 145.238 | 3.614 | 148.212 | 3.699 | 150.852 | 3.787 | 153.008 | 3.875 | 154.648 | 3.954 | 156.028 | 4.041 | 156.866 | 4.120 |
| 167.494 | 146.009 | 3.601 | 149.385 | 3.685 | 151.950 | 3.771 | 154.157 | 3.857 | 155.831 | 3.936 | 157.067 | 4.020 | 157.962 | 4.099 |
| 169.434 | 147.130 | 3.587 | 150.161 | 3.670 | 152.826 | 3.755 | 155.016 | 3.840 | 156.705 | 3.919 | 158.135 | 4.001 | 159.083 | 4.078 |
| 171.396 | 148.218 | 3.567 | 151.539 | 3.649 | 154.366 | 3.734 | 156.707 | 3.819 | 158.526 | 3.893 | 160.010 | 3.979 | 161.082 | 4.059 |
| 173.380 | 149.153 | 3.552 | 152.384 | 3.631 | 155.244 | 3.715 | 157.730 | 3.798 | 159.631 | 3.875 | 161.085 | 3.958 | 162.166 | 4.036 |
| 175.388 | 150.133 | 3.534 | 153.465 | 3.613 | 156.423 | 3.696 | 158.862 | 3.779 | 160.854 | 3.854 | 162.288 | 3.938 | 163.485 | 4.016 |
| 177.419 | 151.109 | 3.518 | 154.538 | 3.598 | 157.484 | 3.680 | 159.999 | 3.762 | 162.060 | 3.838 | 163.611 | 3.918 | 164.959 | 3.996 |
| 179.473 | 152.338 | 3.500 | 155.650 | 3.578 | 158.685 | 3.660 | 161.264 | 3.745 | 163.339 | 3.820 | 165.100 | 3.903 | 166.438 | 3.979 |

| | | | | | | | | | | | | | |
|---|---|---|---|---|---|---|---|---|---|---|---|---|---|
| 181.552 | 153.185 | 3.485 | 156.751 | 3.561 | 159.839 | 3.641 | 162.455 | 3.722 | 164.585 | 3.795 | 166.363 | 3.877 | 167.925 | 3.953 |
| 183.654 | 154.248 | 3.474 | 157.901 | 3.549 | 161.123 | 3.629 | 163.792 | 3.708 | 165.941 | 3.779 | 167.579 | 3.862 | 169.053 | 3.937 |
| 185.780 | 155.185 | 3.456 | 158.843 | 3.531 | 162.061 | 3.611 | 164.789 | 3.691 | 167.073 | 3.762 | 168.870 | 3.843 | 170.402 | 3.918 |
| 187.932 | 156.150 | 3.437 | 159.932 | 3.510 | 163.307 | 3.587 | 166.162 | 3.667 | 168.461 | 3.738 | 170.110 | 3.819 | 171.738 | 3.891 |
| 190.108 | 157.141 | 3.426 | 160.922 | 3.497 | 164.270 | 3.575 | 167.182 | 3.653 | 169.612 | 3.724 | 171.496 | 3.803 | 173.168 | 3.877 |
| 192.309 | 157.924 | 3.402 | 161.791 | 3.475 | 165.284 | 3.551 | 168.237 | 3.627 | 170.755 | 3.701 | 172.660 | 3.775 | 174.417 | 3.847 |
| 194.536 | 159.363 | 3.390 | 163.286 | 3.461 | 166.777 | 3.536 | 169.800 | 3.613 | 172.267 | 3.685 | 174.321 | 3.760 | 176.176 | 3.832 |
| 196.789 | 160.151 | 3.370 | 164.154 | 3.441 | 167.765 | 3.515 | 170.883 | 3.591 | 173.363 | 3.665 | 175.656 | 3.738 | 177.536 | 3.808 |
| 199.067 | 161.112 | 3.361 | 165.075 | 3.430 | 168.712 | 3.505 | 171.720 | 3.580 | 174.430 | 3.647 | 176.631 | 3.726 | 178.573 | 3.797 |
| 201.372 | 162.094 | 3.341 | 166.331 | 3.412 | 169.927 | 3.486 | 173.096 | 3.560 | 175.772 | 3.631 | 178.083 | 3.703 | 180.135 | 3.772 |
| 203.704 | 163.356 | 3.328 | 167.461 | 3.396 | 171.207 | 3.470 | 174.532 | 3.543 | 177.160 | 3.615 | 179.608 | 3.682 | 181.588 | 3.753 |
| 206.063 | 164.399 | 3.311 | 168.467 | 3.379 | 172.227 | 3.452 | 175.559 | 3.523 | 178.372 | 3.591 | 180.827 | 3.664 | 182.995 | 3.733 |
| 208.449 | 165.082 | 3.301 | 169.330 | 3.366 | 173.145 | 3.437 | 176.650 | 3.507 | 179.527 | 3.572 | 181.862 | 3.650 | 183.988 | 3.719 |
| 210.863 | 166.306 | 3.284 | 170.630 | 3.352 | 174.552 | 3.423 | 178.025 | 3.493 | 180.922 | 3.563 | 183.605 | 3.626 | 185.823 | 3.693 |
| 213.304 | 167.392 | 3.268 | 171.712 | 3.334 | 175.730 | 3.402 | 179.285 | 3.471 | 182.399 | 3.539 | 185.016 | 3.608 | 187.229 | 3.675 |
| 215.774 | 167.841 | 3.251 | 171.917 | 3.316 | 175.903 | 3.384 | 179.470 | 3.451 | 182.699 | 3.520 | 185.605 | 3.591 | 188.047 | 3.662 |
| 218.273 | 168.948 | 3.236 | 172.833 | 3.301 | 176.910 | 3.369 | 180.493 | 3.436 | 183.801 | 3.504 | 186.781 | 3.573 | 189.209 | 3.644 |
| 220.800 | 170.086 | 3.221 | 174.095 | 3.285 | 178.132 | 3.353 | 181.860 | 3.419 | 185.242 | 3.486 | 188.241 | 3.556 | 190.731 | 3.626 |
| 223.357 | 171.015 | 3.208 | 174.975 | 3.274 | 179.207 | 3.340 | 182.965 | 3.405 | 186.304 | 3.473 | 189.451 | 3.540 | 192.058 | 3.610 |
| 225.944 | 171.965 | 3.196 | 176.100 | 3.259 | 180.328 | 3.324 | 184.076 | 3.388 | 187.639 | 3.455 | 190.715 | 3.521 | 193.513 | 3.591 |
| 228.560 | 173.074 | 3.182 | 177.316 | 3.245 | 181.628 | 3.310 | 185.400 | 3.374 | 189.031 | 3.440 | 192.113 | 3.505 | 195.040 | 3.574 |
| 231.206 | 174.027 | 3.170 | 178.261 | 3.231 | 182.540 | 3.295 | 186.545 | 3.358 | 190.138 | 3.423 | 193.354 | 3.488 | 196.297 | 3.555 |
| 233.884 | 174.778 | 3.158 | 179.347 | 3.218 | 183.810 | 3.281 | 187.713 | 3.342 | 191.425 | 3.408 | 194.835 | 3.471 | 197.742 | 3.538 |
| 236.592 | 176.105 | 3.143 | 180.647 | 3.204 | 185.036 | 3.267 | 189.171 | 3.328 | 192.808 | 3.393 | 196.205 | 3.457 | 199.098 | 3.524 |
| 239.332 | 176.677 | 3.130 | 181.686 | 3.189 | 186.262 | 3.251 | 190.341 | 3.312 | 194.096 | 3.376 | 197.583 | 3.439 | 200.781 | 3.506 |
| 242.103 | 178.171 | 3.116 | 182.854 | 3.178 | 187.285 | 3.237 | 191.537 | 3.297 | 195.430 | 3.361 | 199.065 | 3.423 | 202.022 | 3.488 |
| 244.906 | 179.151 | 3.108 | 183.642 | 3.165 | 188.320 | 3.225 | 192.452 | 3.285 | 196.497 | 3.347 | 200.164 | 3.409 | 203.400 | 3.473 |
| 247.742 | 180.099 | 3.090 | 184.868 | 3.151 | 189.708 | 3.209 | 193.875 | 3.269 | 197.882 | 3.331 | 201.525 | 3.392 | 204.859 | 3.455 |
| 250.611 | 181.294 | 3.078 | 186.096 | 3.137 | 190.918 | 3.195 | 195.284 | 3.254 | 199.247 | 3.315 | 203.068 | 3.376 | 206.351 | 3.439 |
| 253.513 | 182.282 | 3.067 | 187.144 | 3.124 | 192.033 | 3.182 | 196.403 | 3.241 | 200.470 | 3.301 | 204.372 | 3.361 | 207.798 | 3.423 |
| 256.448 | 183.054 | 3.052 | 188.273 | 3.109 | 193.110 | 3.166 | 197.680 | 3.224 | 201.878 | 3.284 | 205.781 | 3.343 | 209.179 | 3.405 |
| 259.418 | 183.763 | 3.042 | 189.225 | 3.097 | 194.164 | 3.154 | 198.684 | 3.211 | 203.063 | 3.270 | 206.887 | 3.329 | 210.430 | 3.390 |
| 262.422 | 185.477 | 3.029 | 190.345 | 3.085 | 195.481 | 3.142 | 199.955 | 3.198 | 204.254 | 3.257 | 208.232 | 3.314 | 211.900 | 3.375 |
| 265.461 | 186.292 | 3.020 | 191.415 | 3.073 | 196.498 | 3.128 | 201.221 | 3.183 | 205.684 | 3.242 | 209.664 | 3.299 | 213.408 | 3.359 |
| 268.534 | 187.495 | 3.007 | 192.689 | 3.060 | 197.717 | 3.115 | 202.495 | 3.170 | 206.983 | 3.228 | 211.183 | 3.284 | 214.936 | 3.344 |
| 271.644 | 188.299 | 2.996 | 193.771 | 3.048 | 198.905 | 3.102 | 203.691 | 3.157 | 208.168 | 3.214 | 212.395 | 3.270 | 216.349 | 3.329 |
| 274.789 | 189.484 | 2.984 | 194.916 | 3.035 | 200.098 | 3.088 | 204.981 | 3.142 | 209.579 | 3.199 | 213.853 | 3.254 | 217.600 | 3.313 |
| 277.971 | 190.624 | 2.972 | 195.961 | 3.023 | 201.257 | 3.076 | 206.141 | 3.129 | 210.803 | 3.185 | 215.130 | 3.240 | 219.193 | 3.297 |
| 281.190 | 191.839 | 2.960 | 197.141 | 3.011 | 202.537 | 3.063 | 207.494 | 3.115 | 212.179 | 3.171 | 216.570 | 3.225 | 220.752 | 3.283 |
| 284.446 | 192.588 | 2.949 | 198.146 | 2.999 | 203.686 | 3.051 | 208.777 | 3.103 | 213.485 | 3.158 | 218.021 | 3.211 | 222.248 | 3.268 |
| 287.740 | 193.776 | 2.938 | 199.365 | 2.987 | 204.874 | 3.038 | 209.956 | 3.089 | 214.663 | 3.143 | 219.350 | 3.196 | 223.507 | 3.252 |
| 291.072 | 194.694 | 2.926 | 200.322 | 2.975 | 206.020 | 3.026 | 211.165 | 3.076 | 215.920 | 3.130 | 220.588 | 3.182 | 224.767 | 3.237 |
| 294.442 | 195.825 | 2.916 | 201.440 | 2.965 | 207.186 | 3.015 | 212.379 | 3.062 | 217.495 | 3.117 | 222.022 | 3.169 | 226.333 | 3.224 |
| 297.852 | 197.214 | 2.904 | 202.685 | 2.953 | 208.482 | 3.002 | 213.554 | 3.051 | 218.754 | 3.104 | 223.404 | 3.155 | 227.901 | 3.208 |
| 301.301 | 197.750 | 2.893 | 204.021 | 2.941 | 209.556 | 2.990 | 214.962 | 3.039 | 220.116 | 3.091 | 224.857 | 3.141 | 229.307 | 3.195 |
| 304.789 | 199.181 | 2.884 | 204.851 | 2.931 | 210.781 | 2.979 | 216.195 | 3.025 | 221.426 | 3.078 | 226.183 | 3.126 | 230.672 | 3.180 |
| 308.319 | 200.062 | 2.873 | 205.888 | 2.918 | 211.982 | 2.966 | 217.544 | 3.013 | 222.573 | 3.064 | 227.671 | 3.113 | 232.106 | 3.165 |

| 311.889 | 201.255 | 2.861 | 207.164 | 2.907 | 213.181 | 2.954 | 218.563 | 3.001 | 224.058 | 3.051 | 229.069 | 3.099 | 233.616 | 3.150 |
|---|---|---|---|---|---|---|---|---|---|---|---|---|---|---|
| 315.500 | 202.098 | 2.852 | 208.321 | 2.897 | 214.532 | 2.943 | 219.929 | 2.989 | 225.359 | 3.039 | 230.257 | 3.088 | 234.928 | 3.138 |
| 319.154 | 203.036 | 2.843 | 209.501 | 2.886 | 215.583 | 2.932 | 221.147 | 2.978 | 226.604 | 3.027 | 231.674 | 3.074 | 236.342 | 3.124 |
| 322.849 | 204.198 | 2.832 | 210.622 | 2.875 | 216.967 | 2.920 | 222.587 | 2.966 | 227.921 | 3.014 | 233.152 | 3.061 | 237.772 | 3.110 |
| 326.588 | 205.656 | 2.821 | 212.028 | 2.865 | 218.225 | 2.910 | 223.879 | 2.955 | 229.330 | 3.002 | 234.769 | 3.049 | 239.535 | 3.097 |
| 330.370 | 205.975 | 2.813 | 212.766 | 2.854 | 218.904 | 2.898 | 224.800 | 2.943 | 230.500 | 2.990 | 235.864 | 3.036 | 240.572 | 3.084 |
| 334.195 | 207.142 | 2.803 | 213.875 | 2.844 | 220.002 | 2.887 | 225.886 | 2.931 | 231.723 | 2.978 | 237.126 | 3.023 | 241.927 | 3.071 |
| 338.065 | 208.098 | 2.791 | 214.375 | 2.831 | 220.594 | 2.872 | 226.654 | 2.917 | 231.857 | 2.960 | 237.535 | 3.003 | 242.401 | 3.045 |
| 341.979 | 209.031 | 2.783 | 216.045 | 2.821 | 222.233 | 2.860 | 228.322 | 2.905 | 233.521 | 2.948 | 239.369 | 2.991 | 244.370 | 3.033 |
| 345.939 | 210.338 | 2.772 | 217.116 | 2.811 | 223.164 | 2.850 | 229.244 | 2.892 | 234.630 | 2.935 | 240.625 | 2.977 | 245.539 | 3.018 |
| 349.945 | 211.231 | 2.761 | 217.859 | 2.799 | 224.301 | 2.839 | 230.318 | 2.883 | 235.908 | 2.924 | 241.699 | 2.965 | 246.903 | 3.005 |
| 353.997 | 212.176 | 2.753 | 219.195 | 2.791 | 225.513 | 2.831 | 231.688 | 2.873 | 237.210 | 2.914 | 243.162 | 2.955 | 248.411 | 2.995 |
| 358.096 | 213.669 | 2.744 | 220.764 | 2.779 | 227.156 | 2.817 | 233.503 | 2.861 | 238.995 | 2.900 | 244.985 | 2.941 | 250.352 | 2.981 |
| 362.243 | 214.448 | 2.735 | 221.463 | 2.772 | 227.966 | 2.811 | 234.596 | 2.851 | 240.024 | 2.892 | 246.195 | 2.932 | 251.514 | 2.970 |
| 366.438 | 216.123 | 2.726 | 222.670 | 2.761 | 229.049 | 2.797 | 235.364 | 2.838 | 241.114 | 2.880 | 247.178 | 2.919 | 252.520 | 2.958 |
| 370.681 | 216.941 | 2.718 | 224.154 | 2.752 | 230.834 | 2.789 | 237.296 | 2.830 | 243.053 | 2.870 | 249.254 | 2.910 | 254.795 | 2.949 |
| 374.973 | 217.785 | 2.712 | 225.227 | 2.744 | 231.877 | 2.780 | 238.721 | 2.821 | 244.511 | 2.859 | 250.821 | 2.899 | 256.366 | 2.936 |
| 379.315 | 219.502 | 2.700 | 226.452 | 2.735 | 233.260 | 2.770 | 239.933 | 2.811 | 245.681 | 2.849 | 252.130 | 2.887 | 257.713 | 2.925 |
| 383.707 | 220.323 | 2.691 | 227.778 | 2.727 | 234.564 | 2.762 | 241.102 | 2.802 | 246.973 | 2.840 | 253.386 | 2.878 | 259.224 | 2.914 |
| 388.150 | 221.653 | 2.682 | 228.627 | 2.716 | 235.484 | 2.751 | 242.231 | 2.789 | 248.200 | 2.827 | 254.750 | 2.864 | 260.418 | 2.901 |
| 392.645 | 222.332 | 2.674 | 229.639 | 2.708 | 236.513 | 2.742 | 243.218 | 2.780 | 249.364 | 2.817 | 255.911 | 2.853 | 261.725 | 2.889 |
| 397.192 | 224.040 | 2.665 | 231.242 | 2.698 | 238.134 | 2.732 | 245.036 | 2.771 | 251.179 | 2.806 | 257.846 | 2.843 | 263.701 | 2.878 |
| 401.791 | 224.709 | 2.656 | 232.238 | 2.690 | 239.117 | 2.724 | 246.047 | 2.761 | 252.204 | 2.796 | 259.048 | 2.833 | 265.005 | 2.866 |
| 406.443 | 226.129 | 2.649 | 233.135 | 2.682 | 240.234 | 2.715 | 247.165 | 2.752 | 253.395 | 2.788 | 260.118 | 2.823 | 266.129 | 2.858 |
| 411.150 | 226.858 | 2.641 | 234.382 | 2.673 | 241.428 | 2.707 | 248.260 | 2.741 | 254.418 | 2.778 | 261.232 | 2.813 | 267.234 | 2.847 |
| 415.911 | 228.453 | 2.632 | 236.074 | 2.663 | 243.300 | 2.696 | 250.292 | 2.732 | 256.629 | 2.766 | 263.555 | 2.801 | 269.601 | 2.834 |
| 420.727 | 229.612 | 2.624 | 237.271 | 2.654 | 244.608 | 2.687 | 251.746 | 2.722 | 258.102 | 2.755 | 265.006 | 2.790 | 271.195 | 2.822 |
| 425.598 | 230.534 | 2.616 | 238.384 | 2.646 | 245.563 | 2.678 | 252.881 | 2.713 | 259.175 | 2.746 | 266.204 | 2.780 | 272.433 | 2.812 |
| 430.527 | 231.637 | 2.608 | 239.722 | 2.637 | 247.008 | 2.669 | 254.311 | 2.703 | 260.826 | 2.736 | 267.825 | 2.769 | 274.093 | 2.800 |
| 435.512 | 232.808 | 2.600 | 240.890 | 2.629 | 248.304 | 2.660 | 255.799 | 2.693 | 262.400 | 2.726 | 269.583 | 2.759 | 275.967 | 2.790 |
| 440.555 | 234.012 | 2.592 | 241.865 | 2.622 | 249.473 | 2.653 | 256.781 | 2.686 | 263.539 | 2.718 | 270.820 | 2.750 | 277.192 | 2.781 |
| 445.656 | 235.433 | 2.585 | 243.180 | 2.614 | 250.599 | 2.644 | 258.198 | 2.677 | 264.828 | 2.707 | 272.012 | 2.739 | 278.404 | 2.770 |
| 450.817 | 236.379 | 2.578 | 244.385 | 2.605 | 251.884 | 2.635 | 259.414 | 2.667 | 266.116 | 2.698 | 273.333 | 2.730 | 279.881 | 2.760 |
| 456.037 | 237.571 | 2.572 | 245.602 | 2.598 | 253.372 | 2.628 | 261.106 | 2.660 | 267.925 | 2.689 | 275.318 | 2.721 | 281.892 | 2.750 |
| 461.318 | 238.770 | 2.563 | 247.043 | 2.589 | 254.662 | 2.617 | 262.527 | 2.647 | 269.528 | 2.678 | 276.628 | 2.708 | 283.207 | 2.738 |
| 466.659 | 240.272 | 2.555 | 248.225 | 2.582 | 255.907 | 2.611 | 263.906 | 2.643 | 270.750 | 2.671 | 278.082 | 2.702 | 284.775 | 2.731 |
| 472.063 | 241.293 | 2.547 | 249.334 | 2.574 | 257.177 | 2.603 | 264.772 | 2.633 | 271.775 | 2.661 | 279.333 | 2.692 | 286.011 | 2.720 |
| 477.529 | 242.315 | 2.541 | 250.526 | 2.567 | 258.494 | 2.596 | 266.260 | 2.625 | 273.375 | 2.653 | 280.861 | 2.683 | 287.599 | 2.711 |
| 483.059 | 243.681 | 2.533 | 251.752 | 2.559 | 259.669 | 2.587 | 267.525 | 2.615 | 275.285 | 2.645 | 282.390 | 2.672 | 289.144 | 2.700 |
| 488.652 | 244.909 | 2.525 | 253.115 | 2.551 | 260.928 | 2.578 | 268.890 | 2.605 | 276.616 | 2.635 | 283.559 | 2.662 | 290.493 | 2.690 |
| 494.311 | 245.568 | 2.525 | 254.146 | 2.547 | 261.247 | 2.577 | 269.505 | 2.599 | 276.618 | 2.624 | 284.706 | 2.652 | 292.560 | 2.679 |
| 500.035 | 246.812 | 2.518 | 254.742 | 2.538 | 263.420 | 2.567 | 271.156 | 2.591 | 278.687 | 2.616 | 286.661 | 2.641 | 293.770 | 2.667 |
| 505.825 | 248.112 | 2.512 | 256.818 | 2.532 | 264.471 | 2.563 | 272.408 | 2.585 | 279.933 | 2.609 | 288.457 | 2.634 | 295.796 | 2.656 |
| 511.682 | 249.649 | 2.506 | 258.013 | 2.524 | 265.948 | 2.549 | 273.776 | 2.575 | 281.384 | 2.599 | 289.663 | 2.623 | 297.315 | 2.646 |
| 517.607 | 250.358 | 2.496 | 259.286 | 2.516 | 266.428 | 2.539 | 274.822 | 2.566 | 282.797 | 2.590 | 290.378 | 2.615 | 298.478 | 2.637 |
| 523.600 | 250.779 | 2.488 | 259.824 | 2.508 | 268.084 | 2.536 | 275.679 | 2.557 | 283.364 | 2.582 | 291.924 | 2.607 | 299.562 | 2.628 |
| 529.663 | 252.634 | 2.483 | 261.089 | 2.502 | 269.478 | 2.526 | 276.738 | 2.553 | 285.013 | 2.574 | 293.445 | 2.598 | 301.067 | 2.621 |

| | | | | | | | | | | | | | |
|---|---|---|---|---|---|---|---|---|---|---|---|---|---|
| 535.797 | 253.441 | 2.476 | 262.088 | 2.495 | 270.317 | 2.518 | 278.465 | 2.544 | 286.970 | 2.569 | 295.224 | 2.591 | 303.471 | 2.612 |
| 542.001 | 254.610 | 2.473 | 263.538 | 2.491 | 271.562 | 2.517 | 279.435 | 2.539 | 287.099 | 2.562 | 296.128 | 2.585 | 303.721 | 2.605 |
| 548.277 | 256.162 | 2.465 | 265.042 | 2.483 | 273.042 | 2.507 | 281.501 | 2.530 | 289.108 | 2.553 | 298.375 | 2.576 | 306.090 | 2.596 |
| 554.626 | 257.613 | 2.457 | 265.901 | 2.476 | 274.370 | 2.498 | 282.037 | 2.520 | 291.064 | 2.543 | 298.993 | 2.567 | 307.506 | 2.588 |
| 561.048 | 258.584 | 2.450 | 266.959 | 2.469 | 275.976 | 2.494 | 284.126 | 2.513 | 292.168 | 2.537 | 300.265 | 2.558 | 308.200 | 2.577 |
| 567.545 | 259.628 | 2.443 | 269.079 | 2.464 | 277.322 | 2.485 | 285.063 | 2.508 | 293.652 | 2.531 | 302.395 | 2.552 | 310.766 | 2.571 |
| 574.116 | 260.802 | 2.436 | 269.492 | 2.455 | 278.523 | 2.477 | 286.708 | 2.500 | 295.211 | 2.522 | 303.985 | 2.542 | 312.084 | 2.562 |
| 580.764 | 262.131 | 2.432 | 271.226 | 2.452 | 279.767 | 2.472 | 287.958 | 2.493 | 296.620 | 2.515 | 305.150 | 2.536 | 313.355 | 2.555 |
| 587.489 | 263.372 | 2.425 | 272.553 | 2.444 | 280.527 | 2.466 | 289.674 | 2.488 | 297.761 | 2.510 | 307.322 | 2.531 | 314.947 | 2.549 |
| 594.292 | 265.202 | 2.417 | 274.035 | 2.438 | 282.650 | 2.458 | 291.138 | 2.481 | 299.579 | 2.501 | 308.882 | 2.521 | 317.250 | 2.540 |
| 601.174 | 266.050 | 2.410 | 275.716 | 2.431 | 284.118 | 2.452 | 292.579 | 2.473 | 300.896 | 2.494 | 309.573 | 2.513 | 318.373 | 2.531 |
| 608.135 | 268.067 | 2.406 | 276.655 | 2.424 | 285.463 | 2.444 | 293.404 | 2.464 | 302.910 | 2.486 | 312.288 | 2.505 | 320.597 | 2.523 |
| 615.177 | 269.577 | 2.400 | 277.945 | 2.418 | 286.894 | 2.437 | 295.956 | 2.456 | 304.327 | 2.478 | 313.747 | 2.498 | 321.940 | 2.516 |
| 622.300 | 270.465 | 2.393 | 279.439 | 2.410 | 289.321 | 2.429 | 297.768 | 2.449 | 306.522 | 2.469 | 314.343 | 2.488 | 323.520 | 2.506 |
| 629.506 | 271.691 | 2.388 | 280.439 | 2.404 | 289.119 | 2.422 | 298.156 | 2.441 | 306.959 | 2.462 | 315.834 | 2.481 | 324.154 | 2.500 |
| 636.796 | 273.500 | 2.380 | 281.789 | 2.401 | 291.552 | 2.418 | 300.610 | 2.438 | 309.157 | 2.458 | 318.356 | 2.475 | 327.299 | 2.492 |
| 644.169 | 274.349 | 2.377 | 283.086 | 2.394 | 292.510 | 2.415 | 301.456 | 2.433 | 310.197 | 2.452 | 319.202 | 2.470 | 328.295 | 2.485 |
| 651.628 | 275.825 | 2.369 | 285.096 | 2.389 | 293.420 | 2.407 | 303.078 | 2.425 | 312.303 | 2.443 | 321.473 | 2.461 | 329.851 | 2.477 |
| 659.174 | 276.995 | 2.364 | 286.438 | 2.383 | 295.736 | 2.402 | 304.308 | 2.419 | 313.728 | 2.437 | 323.356 | 2.454 | 331.202 | 2.470 |
| 666.807 | 278.208 | 2.358 | 287.825 | 2.374 | 296.690 | 2.394 | 306.757 | 2.412 | 315.098 | 2.429 | 325.155 | 2.447 | 333.339 | 2.461 |
| 674.528 | 279.353 | 2.351 | 289.072 | 2.370 | 298.492 | 2.386 | 307.613 | 2.405 | 316.301 | 2.423 | 326.159 | 2.439 | 334.549 | 2.456 |
| 682.339 | 281.379 | 2.349 | 290.676 | 2.366 | 299.557 | 2.383 | 309.261 | 2.401 | 318.325 | 2.418 | 327.420 | 2.434 | 336.566 | 2.448 |
| 690.240 | 282.470 | 2.344 | 291.550 | 2.360 | 300.654 | 2.378 | 310.101 | 2.393 | 319.923 | 2.411 | 328.971 | 2.427 | 338.240 | 2.441 |
| 698.232 | 283.819 | 2.339 | 294.098 | 2.355 | 302.923 | 2.370 | 312.063 | 2.386 | 321.460 | 2.403 | 330.973 | 2.419 | 340.584 | 2.435 |
| 706.318 | 285.312 | 2.334 | 294.987 | 2.348 | 304.188 | 2.366 | 313.587 | 2.380 | 323.055 | 2.397 | 331.564 | 2.412 | 341.737 | 2.427 |
| 714.496 | 286.594 | 2.326 | 296.343 | 2.342 | 305.671 | 2.358 | 314.768 | 2.369 | 324.993 | 2.389 | 333.994 | 2.404 | 342.806 | 2.421 |
| 722.770 | 287.844 | 2.323 | 297.039 | 2.334 | 307.366 | 2.351 | 316.724 | 2.362 | 325.811 | 2.381 | 335.246 | 2.398 | 344.156 | 2.415 |
| 731.139 | 288.745 | 2.311 | 299.055 | 2.333 | 307.498 | 2.344 | 318.113 | 2.355 | 327.319 | 2.374 | 336.174 | 2.393 | 346.083 | 2.409 |
| 739.605 | 290.588 | 2.307 | 299.935 | 2.325 | 309.656 | 2.346 | 319.529 | 2.348 | 327.569 | 2.365 | 338.821 | 2.383 | 347.253 | 2.403 |
| 748.170 | 291.101 | 2.302 | 300.846 | 2.318 | 311.199 | 2.335 | 320.759 | 2.341 | 329.375 | 2.359 | 338.782 | 2.376 | 349.761 | 2.396 |
| 756.833 | 293.419 | 2.300 | 304.125 | 2.314 | 313.331 | 2.336 | 322.077 | 2.332 | 331.681 | 2.353 | 342.898 | 2.371 | 351.879 | 2.388 |
| 765.597 | 293.847 | 2.298 | 305.387 | 2.308 | 313.914 | 2.313 | 327.444 | 2.330 | 334.801 | 2.348 | 343.891 | 2.364 | 354.063 | 2.378 |
| 774.462 | 294.769 | 2.293 | 304.776 | 2.304 | 316.566 | 2.316 | 326.061 | 2.326 | 337.637 | 2.344 | 343.931 | 2.360 | 356.801 | 2.372 |
| 783.430 | 296.608 | 2.292 | 308.777 | 2.296 | 318.540 | 2.308 | 329.070 | 2.324 | 337.997 | 2.337 | 348.572 | 2.352 | 356.591 | 2.367 |
| 792.501 | 299.374 | 2.282 | 305.849 | 2.290 | 322.577 | 2.297 | 331.360 | 2.311 | 339.188 | 2.328 | 350.812 | 2.341 | 358.768 | 2.351 |
| 801.678 | 299.782 | 2.274 | 310.248 | 2.286 | 320.895 | 2.295 | 331.671 | 2.308 | 341.881 | 2.317 | 351.053 | 2.336 | 361.872 | 2.349 |
| 810.961 | 301.499 | 2.273 | 312.352 | 2.283 | 324.016 | 2.287 | 334.249 | 2.301 | 344.934 | 2.315 | 353.010 | 2.333 | 363.252 | 2.342 |
| 820.352 | 304.280 | 2.266 | 313.783 | 2.275 | 325.117 | 2.284 | 336.571 | 2.295 | 345.616 | 2.309 | 355.103 | 2.324 | 364.283 | 2.338 |
| 829.851 | 304.295 | 2.259 | 313.942 | 2.269 | 325.652 | 2.281 | 335.686 | 2.291 | 346.724 | 2.300 | 355.819 | 2.319 | 363.395 | 2.325 |
| 839.460 | 306.338 | 2.259 | 317.290 | 2.271 | 328.133 | 2.281 | 338.307 | 2.287 | 349.513 | 2.297 | 359.063 | 2.314 | 368.719 | 2.324 |
| 849.180 | 309.238 | 2.250 | 318.049 | 2.263 | 328.583 | 2.268 | 340.758 | 2.277 | 350.863 | 2.289 | 359.120 | 2.307 | 370.061 | 2.314 |
| 859.014 | 308.756 | 2.244 | 319.683 | 2.255 | 331.236 | 2.263 | 341.418 | 2.271 | 352.618 | 2.281 | 361.566 | 2.298 | 371.331 | 2.307 |
| 868.960 | 310.625 | 2.240 | 321.229 | 2.247 | 330.369 | 2.260 | 344.000 | 2.267 | 353.053 | 2.275 | 361.823 | 2.295 | 371.848 | 2.305 |
| 879.023 | 310.856 | 2.236 | 322.490 | 2.248 | 332.999 | 2.260 | 343.613 | 2.266 | 354.037 | 2.273 | 364.041 | 2.289 | 374.282 | 2.303 |
| 889.201 | 313.741 | 2.233 | 324.229 | 2.245 | 336.047 | 2.256 | 345.548 | 2.259 | 355.576 | 2.269 | 366.637 | 2.284 | 377.017 | 2.298 |
| 899.498 | 314.657 | 2.230 | 326.387 | 2.243 | 337.874 | 2.249 | 348.526 | 2.255 | 358.562 | 2.264 | 370.366 | 2.279 | 377.792 | 2.289 |
| 909.913 | 318.145 | 2.224 | 328.627 | 2.239 | 338.271 | 2.242 | 350.158 | 2.247 | 360.426 | 2.258 | 370.392 | 2.272 | 379.902 | 2.285 |

| | | | | | | | | | | | | | |
|---|---|---|---|---|---|---|---|---|---|---|---|---|---|
|920.450|318.317|2.222|328.982|2.235|339.816|2.238|351.081|2.243|362.878|2.257|372.471|2.266|382.290|2.276|
|931.108|319.362|2.217|331.571|2.229|342.073|2.236|351.855|2.241|364.273|2.251|373.698|2.263|384.056|2.276|
|941.890|321.758|2.215|332.726|2.228|343.555|2.230|355.356|2.236|366.151|2.248|374.802|2.257|385.822|2.269|
|952.796|323.341|2.211|334.471|2.224|345.638|2.226|356.223|2.232|367.759|2.241|377.284|2.258|386.882|2.264|
|963.829|324.200|2.204|336.284|2.218|347.027|2.225|357.859|2.229|369.408|2.236|378.259|2.248|390.195|2.260|
|974.990|326.753|2.200|336.443|2.210|347.989|2.222|359.389|2.222|371.589|2.227|380.970|2.244|395.048|2.255|
|986.279|326.100|2.196|338.975|2.208|348.760|2.215|361.219|2.218|370.622|2.224|381.765|2.238|392.479|2.248|
|997.700|327.203|2.191|337.261|2.198|352.639|2.209|362.905|2.215|373.775|2.217|382.513|2.225|393.521|2.239|
|1009.25|329.978|2.188|339.378|2.197|353.017|2.211|363.593|2.212|374.311|2.213|386.565|2.227|397.272|2.242|
|1020.94|331.880|2.183|342.118|2.193|354.664|2.203|365.909|2.206|376.685|2.212|389.419|2.225|398.696|2.235|
|1032.76|333.824|2.184|343.540|2.192|358.288|2.199|367.812|2.204|379.299|2.209|389.922|2.219|401.840|2.231|
|1044.72|336.408|2.176|347.120|2.190|360.551|2.196|369.742|2.199|382.030|2.204|390.997|2.220|402.483|2.226|
|1056.82|338.531|2.177|348.551|2.184|361.925|2.191|374.029|2.193|381.664|2.198|393.948|2.213|403.768|2.222|
|1069.05|338.398|2.171|349.518|2.176|362.894|2.185|373.397|2.187|384.631|2.189|395.551|2.203|405.375|2.209|
|1081.43|340.636|2.171|350.735|2.179|365.617|2.181|374.974|2.181|386.491|2.187|397.518|2.205|405.436|2.205|
|1093.96|342.287|2.171|352.962|2.178|367.988|2.178|377.001|2.175|388.798|2.184|398.681|2.187|406.478|2.205|
|1106.62|343.394|2.170|355.552|2.173|369.824|2.175|379.165|2.169|389.120|2.174|400.764|2.194|411.450|2.193|
|1119.44|346.415|2.168|358.159|2.168|372.390|2.170|382.113|2.164|391.552|2.175|403.249|2.184|412.009|2.187|
|1124.60|346.439|2.166|359.976|2.167|373.834|2.168|382.603|2.162|392.596|2.172|405.089|2.186|413.112|2.176|